\documentclass[preprint,3p,times]{elsarticle}
\RequirePackage{multirow,booktabs,subfigure,color,array,hhline,makecell}
\usepackage{amssymb}
\usepackage{amsmath}
\usepackage{graphicx}
\usepackage{amsthm}
\usepackage{mathrsfs}
\usepackage{indentfirst}
\usepackage[colorlinks,citecolor=blue,urlcolor=blue]{hyperref}
\allowdisplaybreaks
\usepackage[table]{xcolor}
\usepackage{tikz-network}
\usepackage{pgf}
\usepackage{tikz}
\usepackage{subfigure}
\usepackage{bbm}
\usetikzlibrary{arrows, decorations.pathmorphing, backgrounds, positioning, fit, petri, automata}
\usetikzlibrary{shadows,arrows,positioning}
\RequirePackage{algorithm,algpseudocode}
\usepackage{matlab-prettifier}
\allowdisplaybreaks
\usepackage{changes}
\usepackage{caption}
\usepackage{setspace}
\newtheorem{thm}{Theorem}
\newtheorem{defin}{Definition}
\newtheorem{lem}{Lemma}
\newtheorem{assum}{Assumption}
\newtheorem{rem}{Remark}
\newtheorem{cor}{Corollary}

\newtheorem{prop}{Proposition}

\allowdisplaybreaks[4]

\AtBeginDocument{%
	\providecommand\BibTeX{{%
			\normalfont B\kern-0.5em{\scshape i\kern-0.25em b}\kern-0.8em\TeX}}}
\journal{~}
\begin{document}
\captionsetup[figure]{labelfont={bf},labelformat={default},labelsep=period,name={Fig.}}
\begin{frontmatter}
\title{Community detection in multi-layer networks by regularized debiased spectral clustering}
\author[label1]{Huan Qing\corref{cor1}}
\ead{qinghuan@u.nus.edu;qinghuan@cqut.edu.cn;qinghuan07131995@163.com}
\cortext[cor1]{Corresponding author.}
\address[label1]{School of Economics and Finance, Chongqing University of Technology, Chongqing, 400054, China}
\begin{abstract}
Community detection is a crucial problem in the analysis of multi-layer networks. While regularized spectral clustering methods using the classical regularized Laplacian matrix have shown great potential in handling sparse single-layer networks, to our knowledge, their potential in multi-layer network community detection remains unexplored. To address this gap, in this work, we introduce a new method, called regularized debiased sum of squared adjacency matrices (RDSoS), to detect communities in multi-layer networks. RDSoS is developed based on a novel regularized Laplacian matrix that regularizes the debiased sum of squared adjacency matrices. In contrast, the classical regularized Laplacian matrix typically regularizes the adjacency matrix of a single-layer network. Therefore, at a high level, our regularized Laplacian matrix extends the classical one to multi-layer networks. We establish the consistency property of RDSoS under the multi-layer stochastic block model (MLSBM) and further extend RDSoS and its theoretical results to the degree-corrected version of the MLSBM model. Additionally, we introduce a sum of squared adjacency matrices modularity (SoS-modularity) to measure the quality of community partitions in multi-layer networks and estimate the number of communities by maximizing this metric. Our methods offer promising applications for predicting gene functions, improving recommender systems, detecting medical insurance fraud, and facilitating link prediction. Experimental results demonstrate that our methods exhibit insensitivity to the selection of the regularizer, generally outperform state-of-the-art techniques, uncover the assortative property of real networks, and that our SoS-modularity provides a more accurate assessment of community quality compared to the average of the Newman-Girvan modularity across layers.
\end{abstract}
\end{frontmatter}
\section{Introduction}\label{sec1}
Multi-layer (dynamic) networks are extensively collected in various fields like social, biological, and computer sciences, providing a rich and significant framework for describing complex systems characterized by diverse interactions and relationships. These networks are typically represented by edges among nodes in distinct layers \citep{mucha2010community,kivela2014multilayer,boccaletti2014structure}. For instance, in social sciences, individuals are connected through various social platforms such as Facebook, Twitter, Instagram, and emails, forming a multi-layer social network with each layer denoting a distinct type of social relationship, ranging from friendships to professional connections \citep{ansari2011modeling,oselio2014multi}. Similarly, in biological sciences, proteins engage in interactions through various biological processes or different stages of development, resulting in a multi-layer protein-protein interaction network where different layers signify different types of biological interactions or different times \citep{bakken2016comprehensive,zhang2017finding,lei2023bias}. Another tangible example is transportation networks, where different modes of transportation, such as buses, trains, and subways, can be depicted as separate layers interconnected by transfer stations or stops.

Community detection is a fundamental problem in analyzing multi-layer networks. It involves finding groups (blocks/communities/clusters) of nodes that are densely connected within the group, but sparsely connected to nodes in other groups \citep{newman2004finding,jin2021survey,huang2021survey}. It can reveal underlying structural patterns and functional modules within networks, providing valuable insights into the organization and dynamics of complex systems \citep{fortunato2010community,fortunato2016community}. For instance, communities may stand for groups of individuals who share common interests or behaviors in social networks, while in biological networks, they may correspond to functional modules. Community detection methods are versatile and can be applied to various domains: identifying a firm's network memberships in business \citep{kumar2022much}, analyzing stock network structures for stock return prediction \citep{wang2023knowledge,ding2024stock}, discovering tourism communities in interactive networks of tourism destinations to aid regional tourism \citep{xu2022tourism}, uncovering groups in co-citation and co-authorship networks \citep{ji2022co,schafermeier2023research}, predicting gene functions through gene co-clusters in gene networks \citep{sun2021bipartite}, enhancing traditional collaborative filtering in recommender systems \citep{cao2015improved,satuluri2020simclusters,guan2021community,ni2023community}, inferring and evolving community structures in social media \citep{zhang2014extracting,bedi2016community}, helping feature subset selection in data mining \citep{moradi2015graph,rostami2021novel}, identifying medical insurance gang fraud \citep{cheng2024research}, improving task allocation problem in sparse mobile crowdsensing systems \citep{wang2024method}, and facilitating link prediction \citep{de2016discriminative,wu2017balanced,singh2020clp,karimi2021community,tang2022cold}. For more real-world applications of community detection, refer to survey papers such as \citep{fortunato2010community,papadopoulos2012community,malliaros2013clustering,fortunato2016community,javed2018community,huang2021survey,gasparetti2021community,jin2021survey,amira2023survey,rostami2023community}. Given that multi-layer networks have more edges than single-layer (static) networks, thus providing more community information \citep{kim2015community,paul2020spectral,huang2021survey,lei2023bias}, this work studies the community detection problem in multi-layer networks where layers share common nodes but have no cross-layer connections.

The multi-layer stochastic block model (MLSBM) is a powerful statistical model for describing the hidden community structure of multi-layer networks. This model extends the classical stochastic block model (SBM) \citep{holland1983stochastic} from static networks to multi-layer networks. Under the MLSBM, nodes are assumed to belong to latent communities, and edges among nodes are generated by specific probabilistic rules that may vary across different layers. Several methods have been developed to discover nodes' communities under the MLSBM, and their theoretical consistencies have been thoroughly investigated. For instance, \cite{han2015consistent} studied the asymptotic properties of a spectral method and a maximum-likelihood method when $T\rightarrow\infty$ and $n$ is fixed under the MLSBM, where $T$ denotes the number of layers and $n$ represents the number of vertices. \cite{paul2020spectral} established the theoretical consistency of spectral and matrix factorization approaches when $T\rightarrow\infty$ and $n\rightarrow\infty$ under the MLSBM model. \cite{lei2020consistent} proposed a least squares estimation approach to fit the MLSBM and established its theoretical analysis when $T\rightarrow\infty$ and $n\rightarrow\infty$. Recently, \cite{lei2023bias} developed an efficient spectral method via the debiased sum of squared adjacency matrices and studied its consistency as $T\rightarrow\infty$ and $n\rightarrow\infty$ within the framework of the MLSBM. However, a notable limitation of the MLSBM is that it does not consider the variety of nodes' degrees, a prevalent characteristic of real-world networks. The multi-layer degree-corrected stochastic block model (MLDCSBM) \citep{qingMLDCSBM} has been introduced to overcome MLSBM's limitation. MLDCSBM permits nodes within the identical group to exhibit varying expected degrees and extends the popular degree-corrected stochastic block model (DCSBM) \citep{karrer2011stochastic} from static networks to multi-layer networks. A debiased spectral clustering method with theoretical guarantees is developed to fit MLDCSBM in \citep{qingMLDCSBM}.
\begin{table}[h!]
\footnotesize
	\centering
	\caption{Literature review summary. Here, $A_{l}$ denotes the adjacency matrix of the $l$-th layer for a multi-layer network with $T$ layers, and $S$ denotes the debiased sum of squared adjacency matrices (defined later). Problems focused on community detection in single-layer networks are marked by ($\heartsuit$), and those in multi-layer networks are marked by ($\clubsuit$). Finally, methods related to the Laplacian matrix are marked by ($\spadesuit$).}
	\label{LiteratureReview}
	\resizebox{\columnwidth}{!}{
	\begin{tabular}{cccccccccccc}
\hline
References&Problems&Methods proposed and/or considered\\
\hline
\cite{ng2001spectral}&Clustering&Run K-means on the normalized eigenvectors of $D^{-1/2}AD^{-1/2}$ ($\spadesuit$)\\
\cite{BinYuAoS2011}&Community detection in SBM ($\heartsuit$)&Run K-means on $D^{-1/2}AD^{-1/2}$'s eigenvectors ($\spadesuit$)\\
\cite{qin2013regularized}&Community detection in DCSBM ($\heartsuit$)&Run K-means on the normalized version of $D^{-1/2}_{\tau}AD^{-1/2}_{\tau}$'s eigenvectors ($\spadesuit$)\\
\cite{SCORE}&Community detection in DCSBM ($\heartsuit$)&Run K-means on the row-wise ratio matrix of $A$'s eigenvectors\\
\cite{lei2015consistency}&Consistency of spectral clustering in SBM and DCSBM ($\heartsuit$)&Run approximated K-means on (and the normalized version of) $A$'s eigenvectors\\
\cite{han2015consistent}&Community detection in MLSBM ($\clubsuit$)&Run K-means on the eigenvectors of $\sum_{l\in[T]}A_{l}$&\\
\cite{joseph2016impact}&Impact of regularization on spectral clustering in SBM ($\heartsuit$)&Run K-means on $D^{-1/2}_{\tau}(A+\tau J)D^{-1/2}_{\tau}$'s eigenvectors ($\spadesuit$)\\
\cite{rohe2016co}&Community detection in directed networks ($\heartsuit$)&Run K-means on the singular vectors of the directed version of $D^{-1/2}_{\tau}AD^{-1/2}_{\tau}$ ($\spadesuit$)\\
\cite{binkiewicz2017covariate}&Community detection with covariates ($\heartsuit$)&Run K-means on the eigenvectors of the combination of $D^{-1/2}_{\tau}AD^{-1/2}_{\tau}$ with a covariate matrix ($\spadesuit$)\\
\cite{JiamingAoS2018}&Community detection in DCSBM ($\heartsuit$)&Convexified modularity maximization approach\\
\cite{ali2018improved}&Existence of $\alpha$'s optimal value in $D^{-\alpha}AD^{-\alpha}$ in SBM ($\heartsuit$)&Run K-means on a variant of the eigenvectors of $D^{-\alpha}AD^{-\alpha}$ ($\spadesuit$)\\
\cite{zhang2018discovering}&Political engagement analysis on Facebook ($\heartsuit$)&Run K-means on the normalized eigenvectors of a variant of $D^{-1/2}_{\tau}AD^{-1/2}_{\tau}$ ($\spadesuit$)\\
\cite{su2019strong}&Strong consistency of spectral clustering in SBM ($\heartsuit$)&Run K-means on the eigenvectors of  $D^{-1/2}AD^{-1/2}$ ($\spadesuit$)\\
\cite{wang2020spectral}&Community detection in directed networks ($\heartsuit$)&Directed version of the algorithm introduced in \cite{SCORE} ($\spadesuit$)\\
\cite{zhang2020detecting}&Mixed memberships estimation ($\heartsuit$)&Run K-medians on the normalized version of $A$'s eigenvectors\\
\cite{paul2020spectral}&Consistency of some methods in MLSBM ($\clubsuit$)&Spectral and matrix factorization methods\\
\cite{lei2020consistent}&Community detection in MLSBM ($\clubsuit$)&A least squares estimation method\\
\cite{jin2021improvements}&Community detection in DCSBM ($\heartsuit$)&Improved version of the algorithm introduced in \cite{SCORE} using $D^{-1/2}_{\tau}AD^{-1/2}_{\tau}$ ($\spadesuit$)\\
\cite{jing2021community}&Community detection in mixture multilayer SBM ($\clubsuit$)&Run K-means on eigenvectors from the Tucker decomposition of an adjacency tensor\\
\cite{cucuringu2021regularized}&Community detection in signed networks ($\heartsuit$)&Run K-means on the eigenvectors of the symmetric signed Laplacian matrix ($\spadesuit$)\\
\cite{mao2021estimating}&Mixed memberships estimation ($\heartsuit$)&Run the successive projection algorithm (SPA) \cite{gillis2013fast} on $A$'s eigenvectors\\
\cite{arroyo2021inference}&Community detection in a multiple graph model ($\clubsuit$)&Run K-means on a matrix got from the adjacency spectral embedding of each layer\\
\cite{wu2023distributed}&Community detection for large-scale networks in SBM ($\heartsuit$)&Run K-means on the normalized eigenvectors of $D^{-1/2}_{\tau}AD^{-1/2}_{\tau}$ and its variant ($\spadesuit$)\\
\cite{qing2023community}&Community detection in weighted bipartite networks ($\heartsuit$)&Run K-means on $A$'s singular vectors\\
\cite{qing2023regularized}&Mixed memberships estimation ($\heartsuit$)&Run SPA on a variant of $D^{-1/2}_{\tau}AD^{-1/2}_{\tau}$'s eigenvectors ($\spadesuit$)\\
\cite{lei2023bias}&Community detection in MLSBM ($\clubsuit$)&Run K-means on $S$'s eigenvectors\\
\cite{xu2023covariate}&Community detection with covariates ($\clubsuit$)&Run K-means on eigenvectors from the Tucker decomposition of an augmented adjacency tensor\\
\cite{qing2024applications}&Community detection in DCSBM ($\heartsuit$)&Run K-means on the eigenvectors of a dual version of $D^{-1/2}_{\tau}AD^{-1/2}_{\tau}$ ($\spadesuit$)\\
\cite{serrano2024community}&Community detection in DCSBM ($\heartsuit$)&Mixed integer programming\\
\cite{qingMLDCSBM}&Community detection in MLDCSBM ($\clubsuit$)&Run K-means on the normalized eigenvectors of $S$\\
\cite{jin2024mixed}&Mixed memberships estimation ($\heartsuit$)&Run a vertex hunting algorithm on $A$'s eigenvectors\\
\cite{qing2024bipartite}&Mixed memberships estimation in bipartite weighted networks ($\heartsuit$)&Run SPA on the singular vectors of the adjacency matrix\\
\cite{qing2024finding}&Mixed memberships estimation in categorical data&Run SPA on the eigenvectors of a regularized response matrix ($\spadesuit$)\\
\cite{ke2024optimal}&Mixed memberships estimation ($\heartsuit$)&Run a vertex hunting algorithm on $D^{-1/2}_{\tau}AD^{-1/2}_{\tau}$'s eigenvectors ($\spadesuit$)\\
\cite{su2024spectral}&Community detection in directed version of MLSBM ($\clubsuit$)&Run K-means on the singular vectors of $S$'s directed version\\
This work&Community detection in MLSBM and MLDCSBM ($\clubsuit$)&Regularized debiased spectral clustering using $S$'s regularization ($\spadesuit$)\\
\hline
\end{tabular}}
\end{table}

In single-layer networks, using the classical regularized Laplacian matrix $D^{-\frac{1}{2}}_{\tau}AD^{-\frac{1}{2}}_{\tau}$ (or $D^{-\frac{1}{2}}_{\tau}(A+\tau J)D^{-\frac{1}{2}}_{\tau}$) to design spectral methods has been shown great potential in handling sparse networks \citep{qin2013regularized,joseph2016impact,cucuringu2021regularized} under the SBM and the DCSBM. Here, $A$ denotes the adjacency matrix of a static network, $D_{\tau}=\tau I+D$ for $\tau\geq0$, $D$ is a diagonal matrix containing nodes' degrees, $J$ denotes a matrix with all elements being 1, and $I$ represents the identity matrix. The classical regularized Laplacian matrix incorporates a regularization term $\tau$ that helps mitigate the impact of noise and sparsity, ultimately enhancing performance in community detection. Substantial works have been developed based on the classical regularized Laplacian matrix. To name a few, \cite{rohe2016co} proposed a spectral co-clustering algorithm designed based on the directed version of this matrix to detect communities in single-layer directed networks. \cite{binkiewicz2017covariate} developed a covariate-assisted spectral clustering algorithm using this matrix to uncover latent communities in single-layer networks. \cite{ali2018improved} proved the existence of an optimal value of $\alpha$ in spectral clustering using $D^{-\alpha}AD^{-\alpha}$ under SBM. \cite{zhang2018discovering} proposed a graph contextualization method using this matrix to study political engagement on Facebook. \cite{su2019strong} proved the strong consistency of regularized spectral clustering methods under both SBM and DCSBM. \cite{qing2023regularized} proposed spectral methods using the classical regularized Laplacian matrix to estimate nodes' mixed memberships in overlapping networks. \cite{wu2023distributed} developed a distributed spectral clustering algorithm using this matrix to discover community structure in large-scale single-layer networks. Table \ref{LiteratureReview} summarizes the literature on various community detection methods in networks or those associated with the regularized Laplacian matrix. Note that the table does not exhaustively list all significant research in this field, readers are encouraged to explore more on references in this table and those cited within these references.

However, despite these advances in single-layer networks, a critical scientific question remains: how can we extend the idea of regularized spectral clustering in static networks to dynamic networks for community detection? Despite the progress made in spectral methods for multi-layer networks under the MLSBM and the MLDCSBM \citep{han2015consistent,paul2020spectral,lei2023bias,qingMLDCSBM}, methods using the regularized Laplacian matrix have not been proposed. This gap in the literature motivates the present study, which aims to develop novel methods for discovering nodes' communities in multi-layer networks that harness the strengths of regularized spectral clustering. For convenience, we provide a list of main abbreviations and notations with their descriptions in Table \ref{Abbr}.

The key contributions of this work are enumerated below:

\begin{enumerate}
  \item By applying regularization to the debiased sum of squared adjacency matrices introduced in \citep{lei2023bias}, we propose a novel regularized Laplacian matrix, denoted as $L_{\tau}$ in Equation (\ref{DefineLaplacian}). This matrix extends the concept of the classical regularized Laplacian matrix to multi-layer networks, incorporating the advantages of the debiased spectral clustering presented in \citep{lei2023bias}.

  \item Leveraging the proposed regularized Laplacian matrix, we develop a new method, termed regularized debiased sum of squared adjacency matrices (RDSoS), for detecting communities in multi-layer networks under the MLSBM. RDSoS identifies communities by applying the K-means algorithm to an eigenvector matrix constructed from the leading eigenvectors of our regularized Laplacian matrix $L_{\tau}$. Additionally, we propose a degree-corrected RDSoS (DC-RDSoS) method for detecting communities within the MLDCSBM. DC-RDSoS is an enhanced version of RDSoS tailored for multi-layer networks where nodes exhibit varying degrees.

  \item We conduct a rigorous analysis of the consistency of RDSoS and DC-RDSoS as $T \rightarrow \infty$ and $n \rightarrow \infty$ under the MLSBM and MLDCSBM, respectively. Our theoretical findings suggest that a moderate value of the regularization parameter $\tau$ is preferable.
  \item We propose a novel metric, the sum of squared adjacency matrices modularity (SoS-modularity), to measure the quality of community partitions in multi-layer networks. We then estimate the number of communities in multi-layer networks by maximizing this metric.

  \item We conduct numerical and empirical studies to show that our proposed methods exhibit robustness to the selection of the regularizer $\tau$ and achieve at least competitive performance compared to the current state-of-the-art methods. The experimental results also demonstrate that our methods for estimating the number of communities via maximizing the SoS-modularity are more accurate than those that maximize the average of the Newman-Girvan modularity \citep{newman2004finding,paul2021null}, which highlights the effectiveness of our SoS-modularity for assessing community quality in multi-layer networks.
\end{enumerate}

\begin{table}[h!]
\footnotesize
	\centering
	\caption{List of main abbreviations and notations with their descriptions}
	\label{Abbr}
\resizebox{\columnwidth}{!}{
	\begin{tabular}{cc|cccccccccc}
\hline
Abbreviation and notation&Description&Abbreviation and notation&Description\\
\hline
SBM&Stochastic block model&MLSBM&Multi-layer stochastic block model\\
DCSBM&Degree-corrected stochastic block model&MLDCSBM&Multi-layer degree-corrected stochastic block model\\
SoS&Sum of squared adjacency matrices&SoS-modularity&Sum of squared adjacency matrices modularity\\
RDSoS&Regularized debiased sum of squared adjacency matrices&DC-RDSoS&Degree-corrected regularized debiased sum of squared adjacency matrices\\
RSoS&Regularized sum of squared adjacency matrices&DC-RSoS&Degree-corrected regularized sum of squared adjacency matrices\\
RSum&Regularized sum of adjacency matrices&DC-RSum&Degree-corrected regularized sum of adjacency matrices\\
NDSoSA&Normalized spectral clustering based on the debiased sum of squared adjacency matrices&MASE&Multiple adjacency spectral embedding\\
SoS-Debias&Bias-adjusted sum of squared adjacency matrices&MNavrg-modularity&Multi-normalized average modularity\\
\hline
$n$&Nodes count&$T$&Layers count\\
$A$&Adjacency matrix of a single-layer network&$A_{l}$&Adjacency matrix of the $l$-th layer in a multi-layer network\\
$\tau$&Regularization parameter&$J$&Matrix with all entries being 1\\
$D$&Diagonal matrix with entries being the ``degree"&$I$& Identity matrix with proper dimensions\\
$D_{\tau}$&$D+\tau I$&
$[m]$&Set $\{1,2,\ldots,m\}$ for any integer $m$\\
$M'$&Transpose of any matrix $M$&$M(i,:)$&$i$-th row of any matrix $M$\\
$\|M\|_{F}$&Frobenius norm of any matrix $M$&$\|M\|$&Spectral norm of any matrix $M$\\
$\|M\|_{1}$&$l_{1}$ norm of any matrix $M$&$\lambda_{m}(M)$& $m$-th largest eigenvalue in magnitude of any matrix $M$\\
$\mathrm{rank}(M)$&Rank of any matrix $M$&$\textbf{1}_{m}$& $m$-dimensional vector with all entries being 1 for any integer $m$\\
$\mathbb{P}(\cdot)$&Probability&$\mathbb{E}(\cdot)$&Expectation&\\
$K$&Communities count&$\mathcal{P}_{K}$&Set of all permutations of $[K]$\\
$\mathcal{C}_{k}$&$k$-th community&$Z$&Community membership matrix\\
$\ell$&Community label vector&$B_{l}$&$l$-th connectivity matrix\\
$\Omega_{l}$&$A_{l}$'s expectation&$\rho$&Sparsity parameter\\
$\tilde{B}_{l}$&$\frac{B_{l}}{\rho}$&$n_{k}$& Nodes count in the $k$-th community\\
$\hat{\ell}$&Estimated label vector&$\mathcal{E}_{k}$&$k$-th estimated community\\
$\mathcal{S}$&$\sum_{l\in[T]}\Omega^{2}_{l}$&$\mathcal{D}$&Diagonal matrix with $i$-th entry being $\sum_{j\in[n]}\mathcal{S}(i,j)$ for $i\in[n]$\\
$\mathcal{D}_{\tau}$&$\tau I+\mathcal{D}$&$\mathcal{L}_{\tau}$&$\mathcal{D}^{-\frac{1}{2}}_{\tau}\mathcal{S}\mathcal{D}^{-\frac{1}{2}}_{\tau}$\\
$U$&Matrix of $\mathcal{L}_{\tau}$'s leading $K$ eigenvectors&$\Sigma$&Diagonal matrix with $k$-th entry being $\lambda_{k}(\mathcal{L}_{\tau})$ for $k\in[K]$\\
$X$&Matrix satisfying $U=ZX$&$D_{l}$&Diagonal matrix with entries being nodes' degrees in the $l$-th layer for $l\in[T]$\\
$S$&Debiased sum of squared adjacency matrices $\sum_{l\in[T]}(A^{2}_{l}-D_{l})$&$L_{\tau}$&Regularized Laplacian matrix $D^{-\frac{1}{2}}_{\tau}SD^{-\frac{1}{2}}_{\tau}$\\
$\hat{f}_{\mathcal{M}}$&Clustering error for method $\mathcal{M}$&$\hat{U}$&Matrix of $L_{\tau}$'s leading $K$ eigenvectors\\
$\hat{\Sigma}$&Diagonal matrix with $k$-th entry being $\lambda_{k}(L_{\tau})$ for $k\in[K]$&$n_{\mathrm{max}}$&Maximum community size\\
$n_{\mathrm{min}}$&Minimum community size&$\delta_{\mathrm{max}}$&$\mathrm{max}_{i\in[n]}\mathcal{D}(i,i)$\\
$\delta_{\mathrm{min}}$&$\mathrm{min}_{i\in[n]}\mathcal{D}(i,i)$&$\theta$&Degree heterogeneity vector\\
$\Theta$&Diagonal matrix with $i$-th diagonal entry being $\theta(i)$ for $i\in[n]$&$U_{*}$&Row-normalized version of $U$\\
$Y$&Matrix satisfying $U_{*}=ZY$&$\hat{U}_{*}$&Row-normalized version of $\hat{U}$\\
$\theta_{\mathrm{min}}$&$\mathrm{min}_{i\in[n]}\theta(i)$&$\theta_{\mathrm{max}}$&$\mathrm{max}_{i\in[n]}\theta(i)$\\
$d_{l}$&Vector with $d_{l}(i)=D_{l}(i,i)$ for $i\in[n], l\in[T]$&$e_{l}$&$\sum_{i\in[n]}d_{l}(i)/2$\\
$\tilde{S}$&Sum of squared adjacency matrices $\sum_{l\in[T]}A^{2}_{l}$&$A_{\mathrm{sum}}$&Sum of adjacency matrices $\sum_{l\in[T]}A_{l}$\\
$Q_{SoS}$&SoS-modularity score&$K_{SoS,\mathcal{M}}$&Estimated communities count for method $\mathcal{M}$ via maximizing $Q_{SoS}$\\
$Q_{SoS,\mathcal{M}}$&SoS-modularity score computed using $K_{SoS,\mathcal{M}}$ for method $\mathcal{M}$&$d_{SoS}$&Vector with $d_{SoS}(i)=\sum_{j\in[n]}\tilde{S}$ for $i\in[n]$\\
$m_{SoS}$&$\sum_{i\in[n]}d_{SoS}(i)/2$&Unif(0,1)&Standard uniform distribution with minimum 0 and maximum 1\\
$\mathbbm{1}$&Indicator function&$Q_{MNavrg}$&MNavrg-modularity score\\
$K_{MNavrg,\mathcal{M}}$&Estimated communities count for method $\mathcal{M}$ via maximizing $Q_{MNavrg}$&$Q_{MNavrg,\mathcal{M}}$&MNavrg-modularity score computed using $K_{MNavrg,\mathcal{M}}$ for method $\mathcal{M}$ \\
$\mathbb{M}_{n,K}$ &Set of all $n \times K$ matrices where each row contains exactly one 1 and $(K-1)$ 0's&$\mathbb{R}_{K,K}$&Set of all $K\times K$ real matrices\\
\hline
\end{tabular}
}
\end{table}
\section{Multi-layer stochastic block model}\label{sec2}
This part introduces the multi-layer stochastic block model, our first regularized debiased spectral method, and its theoretical properties. We consider the following definition for the model.
\begin{defin}\label{DefinMLSBM}
(\textbf{MLSBM}) Consider a multi-layer network with $n$ common nodes and $T$ layers. Let the $n\times n$ symmetric matrix  $A_{l}$ be the adjacency matrix of the $l$-th layer network for $l\in[T]$, where $A_{l}(i,j)=1$ if there is an edge between nodes $i$ and $j$ and $0$ otherwise for all nodes, i.e., we consider undirected unweighted networks in this paper. Assume that all nodes are partitioned into $K$ communities $\{\mathcal{C}_{k}\}^{K}_{k=1}$, where we assume that $K$ is known in this work. Let $Z\in\{0,1\}^{n\times K}$, where $Z(i,k)=1$ if node $i$ belongs to the $k$-th community $\mathcal{C}_{k}$ and 0 otherwise for $i\in[n], k\in[K]$. Suppose that these communities are disjoint and there is no empty community, ensuring $\mathrm{rank}(Z)=K$. Let the $n$-by-$1$ vector $\ell$ record nodes' community information such that its $i$-th entry is $k$ if $Z(i,k)=1$ for $i\in[n], k\in[K]$. For $l\in[T]$, let the $K\times K$ symmetric matrix $B_{l}$ be the $l$-th connectivity matrix such that all entries of $B_{l}$ locate in $[0,1]$. Given $(Z,\{B_{l}\}^{T}_{l=1})$, the multi-layer stochastic block model (MLSBM) assumes that for $i\in[n], j\in[n], l\in[T]$, each entry of $A_{l}$ is generated independently as
\begin{align}\label{AGeneratedMLSBM}
A_{l}(i,j)=A_{l}(j,i)\sim\mathrm{Bernoulli}(B_{l}(\ell(i),\ell(j))).
\end{align}
\end{defin}
The MLSBM becomes the classical SBM when only one layer, i.e., $T=1$.
\begin{rem}\label{MLSBMADiS}
It is worth noting that MLSBM merely requires that the elements of the $K$-by-$K$ symmetric matrix $B_{l}$ should fall within the interval $[0,1]$, which aligns with the fact that the probability of edge formation between any pair of nodes also lies within this range for $l\in[L]$. Real-world social networks often exhibit assortativity, where nodes within the same community are more densely connected than those across different communities \citep{newman2002assortative,newman2003mixing,radicchi2004defining}. To generate assortative multi-layer networks, MLSBM can be configured such that the smallest diagonal elements of $B_{l}$ exceed its largest off-diagonal elements for $l\in[L]$. Conversely, dis-assortative networks feature more connections between nodes in different communities than within the same. MLSBM can model dis-assortative multi-layer networks by setting the largest diagonal element of $B_{l}$ to be less than its smallest off-diagonal elements for $l\in[L]$. Given that MLSBM imposes no additional constraints on $B_{l}$'s elements across all $l\in[L]$, it is a powerful statistical model to generate complex multi-layer networks encompassing ground-truth community structures beyond mere assortative or dis-assortative patterns \citep{jusup2022social}.
\end{rem}
The following proposition guarantees that the MLSBM model is identifiable (i.e., well-defined) up to a label permutation. Thus, we can design algorithms to detect nodes' communities when a multi-layer network is generated from the MLSBM model with true membership matrix $Z$.
\begin{prop}\label{idMLSBM}
(Identifiability of MLSBM). The MLSBM model is identifiable: For eligible $(Z, \{B_{l}\}^{T}_{l=1})$ and $(\breve{Z},\{\breve{B}_{l}\}^{T}_{l=1})$, if $ZB_{l}Z'=\breve{Z}\breve{B}_{l}\breve{Z}'$ for $l\in[T]$, then $(Z, \{B_{l}\}^{T}_{l=1})$ and $(\breve{Z},\{\breve{B}_{l}\}^{T}_{l=1})$ are identical up to a label permutation.
\end{prop}
Define $\Omega_{l}=ZB_{l}Z'$ for $l\in[T]$. We have
\begin{align*}
\mathbb{E}(A_{l})=\Omega_{l},\qquad l\in[T].
\end{align*}
Thus, $\Omega_{l}$ is the $l$-th population adjacency matrix for $l\in[T]$. Our algorithm for fitting MLSBM is designed based on further analysis of the structure of $\{\Omega_{l}\}^{T}_{l=1}$. Set $\rho=\mathrm{max}_{l\in[T]}\mathrm{max}_{k\in[K], \tilde{k}\in[K]}B_{l}(k,\tilde{k})$. We see that $\rho\in(0,1]$. Set $\tilde{B}_{l}=\frac{B_{l}}{\rho}$ for $l\in[T]$. We have $\tilde{B}_{l}=\tilde{B}'_{l}\in(0,1]^{K\times K}$ and $\Omega_{l}=\rho Z\tilde{B}_{l}Z'$ for $l\in[T]$. Meanwhile, for $i\in[n], j\in[n], l\in[T]$, Equation (\ref{AGeneratedMLSBM}) gives
\begin{align*}
\mathbb{P}(A_{l}(i,j)=0)=1-\rho\tilde{B}_{l}(\ell(i),\ell(j)).
\end{align*}

\begin{figure}
\centering
\includegraphics[width=0.3\columnwidth]{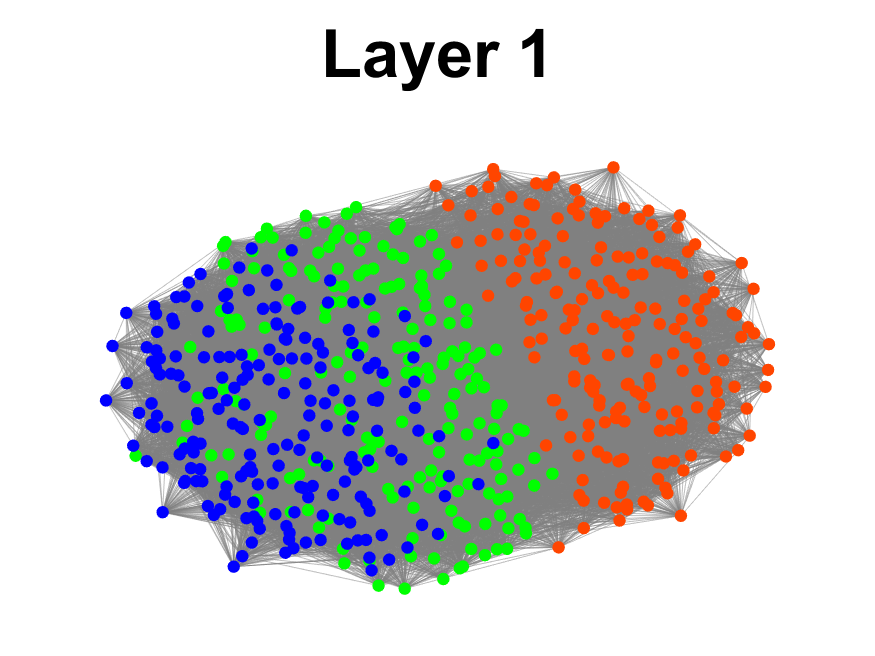}
\includegraphics[width=0.3\columnwidth]{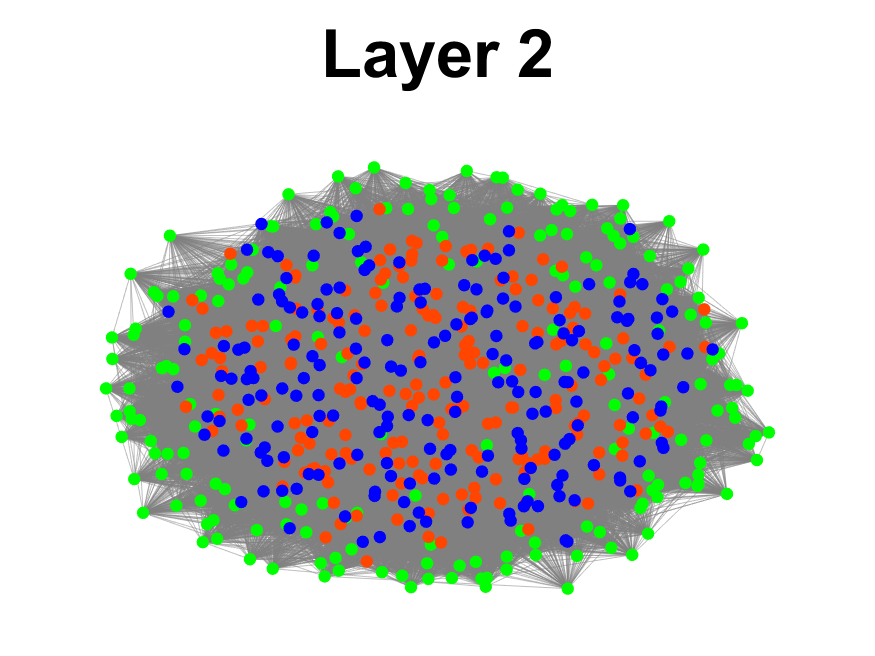}
\includegraphics[width=0.3\columnwidth]{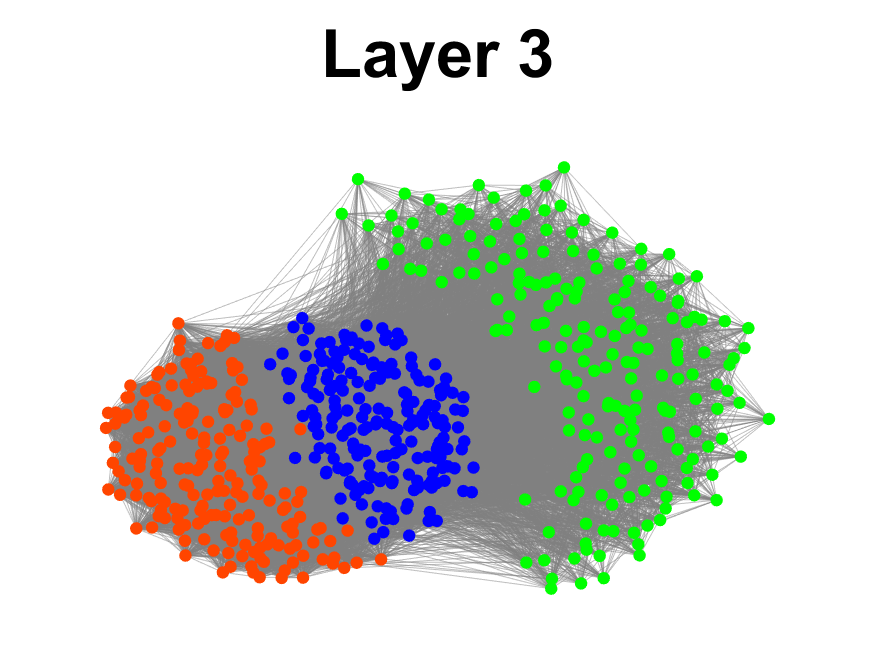}
\caption{A multi-layer network generated from the MLSBM model with 600 nodes, 3 layers, 3 balanced communities, sparsity parameter $\rho=0.3$, and $B_{l}$'s elements being random values generated from Unif(0,1). Colors indicate communities. The distinct structure of each layer suggests that nodes exhibit varying connection patterns $\{B_{l}\}^{T}_{l=1}$.}
\label{NSim} 
\end{figure}

Thus, the network becomes denser (i.e., more edges are generated) as $\rho$ grows, which implies that $\rho$ governs the network's sparsity. For convenience, call $\rho$ the sparsity parameter. Since real networks are usually sparse, analyzing the capability of an algorithm in multi-layer networks under different levels of sparsity \citep{lei2015consistency} is important. For convenience, we use ``MLSBM parameterized by $\{Z, \rho, \{\tilde{B}_{l}\}^{T}_{l=1}\}$'' to denote the MLSBM defined in Definition \ref{DefinMLSBM}. Given the model parameters $\{Z, \rho, \{\tilde{B}_{l}\}^{T}_{l=1}\}$, the $T$ adjacency matrices $\{A_{l}\}^{T}_{l=1}$ can be generated by the MLSBM model through Equation (\ref{AGeneratedMLSBM}). Fig.~\ref{NSim} displays a simulated multi-layer network derived from the MLSBM model. Community detection aims to estimate the node label vectors $\ell$ by leveraging the $T$ adjacency matrices $\{A_{l}\}^{T}_{l=1}$. Specifically, the objective in Fig.~\ref{NSim} is to identify nodes that belong to the same community across the three layers, even when initial community (color) assignments are not provided.

For $k\in[K]$, let $n_{k}$ be the size of the $k$-th community $\mathcal{C}_{k}$. Let the superscript $^{c}$ represent a complement set and $\mathcal{P}_{K}$ denote the set collecting all permutations of $\{1,2,\ldots, K\}$. Set the $n\times 1$ vector $\hat{\ell}$ as an estimated label vector returned by any method (say method $\mathcal{M}$) when there are $K$ communities. For $k\in[K]$, define $\mathcal{E}_{k}$ as the $k$-th estimated community such that node $i$ belongs to $\mathcal{E}_{k}$ if $\hat{\ell}(i)=k$. To quantity the difference between the true community partitions $\{\mathcal{C}_{k}\}^{K}_{k=1}$ and the estimated community partitions $\{\mathcal{E}_{k}\}^{K}_{k=1}$, we use the metric \emph{Clustering error} defined in Equation (6) of \citep{joseph2016impact}. The Clustering error can be computed in the following way:
\begin{align}\label{ErrorRate}
\hat{f}_{\mathcal{M}}=\mathrm{min}_{p\in \mathcal{P}_{K}}\mathrm{max}_{k\in[K]}\frac{|\mathcal{C}_{k}\cap \mathcal{E}^{c}_{p(k)}|+|\mathcal{C}^{c}_{k}\cap \mathcal{E}_{p(k)}|}{n_{k}},
\end{align}
where we use $\hat{f}_{\mathcal{M}}$ to represent the Clustering error of the method $\mathcal{M}$ when the estimated community label $\hat{\ell}$ is returned from $\mathcal{M}$. To establish the estimation consistency of the proposed methods, we will provide the theoretical upper bounds of their Clustering errors by considering the influence of the model parameters in this work.
\subsection{The RDSoS algorithm}\label{sec3}
Under the MLSBM model, the heuristic for designing an efficient method to detect nodes' communities is to consider the oracle scenario when the $T$ population adjacency matrices $\{\Omega_{l}\}^{T}_{l=1}$ are provided. Ideally, a proficient community detection method should accurately recover the community partitions in this oracle case.

Define an $n\times n$ aggregation matrix $\mathcal{S}$ as $\mathcal{S}=\sum_{l\in[T]}\Omega^{2}_{l}$, which is the sum of the $T$ squared population adjacency matrices. Define $\mathcal{D}$ as a diagonal matrix with $\mathcal{D}(i,i)=\sum_{j\in[n]}\mathcal{S}(i,j)$ for $i\in[n]$. Set $\mathcal{D}_{\tau}=\tau I+\mathcal{D}$, where $\tau$ is the nonnegative regularizer. This work introduces a population regularized Laplacian matrix $\mathcal{L}_{\tau}$ as
\begin{align}\label{DefinePopulationLaplacian}
\mathcal{L}_{\tau}=\mathcal{D}^{-\frac{1}{2}}_{\tau}\mathcal{S}\mathcal{D}^{-\frac{1}{2}}_{\tau}.
\end{align}
\begin{rem}\label{DifferencePopulationLaplacian}
At first glance, our population regularized Laplacian matrix $\mathcal{L}_{\tau}$ defined in Equation (\ref{DefinePopulationLaplacian}) shares a similar form as the population Laplacian matrix considered in \citep{qin2013regularized, joseph2016impact} because both matrices are computed by regularizing certain matrices. However, there is a huge difference between their definitions: our $\mathcal{L}_{\tau}$ is defined based on the aggregation matrix $\mathcal{S}$ computed from the $L$ population adjacency matrices $\{\Omega_{l}\}^{T}_{l=1}$ while the population Laplacian matrix studied in \citep{qin2013regularized, joseph2016impact} is designed based on a single-layer population adjacency matrix. This difference causes that our population regularized Laplacian matrix $\mathcal{L}_{\tau}$ can be used for discovering communities in multi-layer networks whereas the population Laplacian matrix used in \citep{qin2013regularized, joseph2016impact} only works for a single-layer network.
\end{rem}
The subsequent lemma demonstrates that the compact eigenvector matrix of the population regularized Laplacian matrix $\mathcal{L}_{\tau}$ can reveal the true communities.
\begin{lem}\label{PopulationLtauMLSBM}
Under the MLSBM parameterized by $(Z,\rho, \{\tilde{B}_{l}\}^{T}_{l=1})$. When the rank of the sum of squared connectivity matrices $\sum_{l\in[T]}\tilde{B}^{2}_{l}$ is $K$, let the leading $K$ eigen-decomposition of $\mathcal{L}_{\tau}$ be $U\Sigma U'$, where $U'U=I$ and $\Sigma$ is a $K\times K$ diagonal matrix. We have: (a) if $\ell(i)=\ell(\bar{i})$ for $i\in[n], \bar{i}\in[n]$, then $U(i,:)=U(\bar{i},:)$; (b) there exits a matrix $X$ such that $U=ZX$, where $X$ satisfies $\|X(\tilde{k},:)-X(k,:)\|_{F}=\sqrt{\frac{1}{n_{\tilde{k}}}+\frac{1}{n_{k}}}$ for $\tilde{k}\neq k, \tilde{k}\in[K], k\in[K]$.
\end{lem}
\begin{figure}
\centering
\subfigure{\includegraphics[width=1\textwidth]{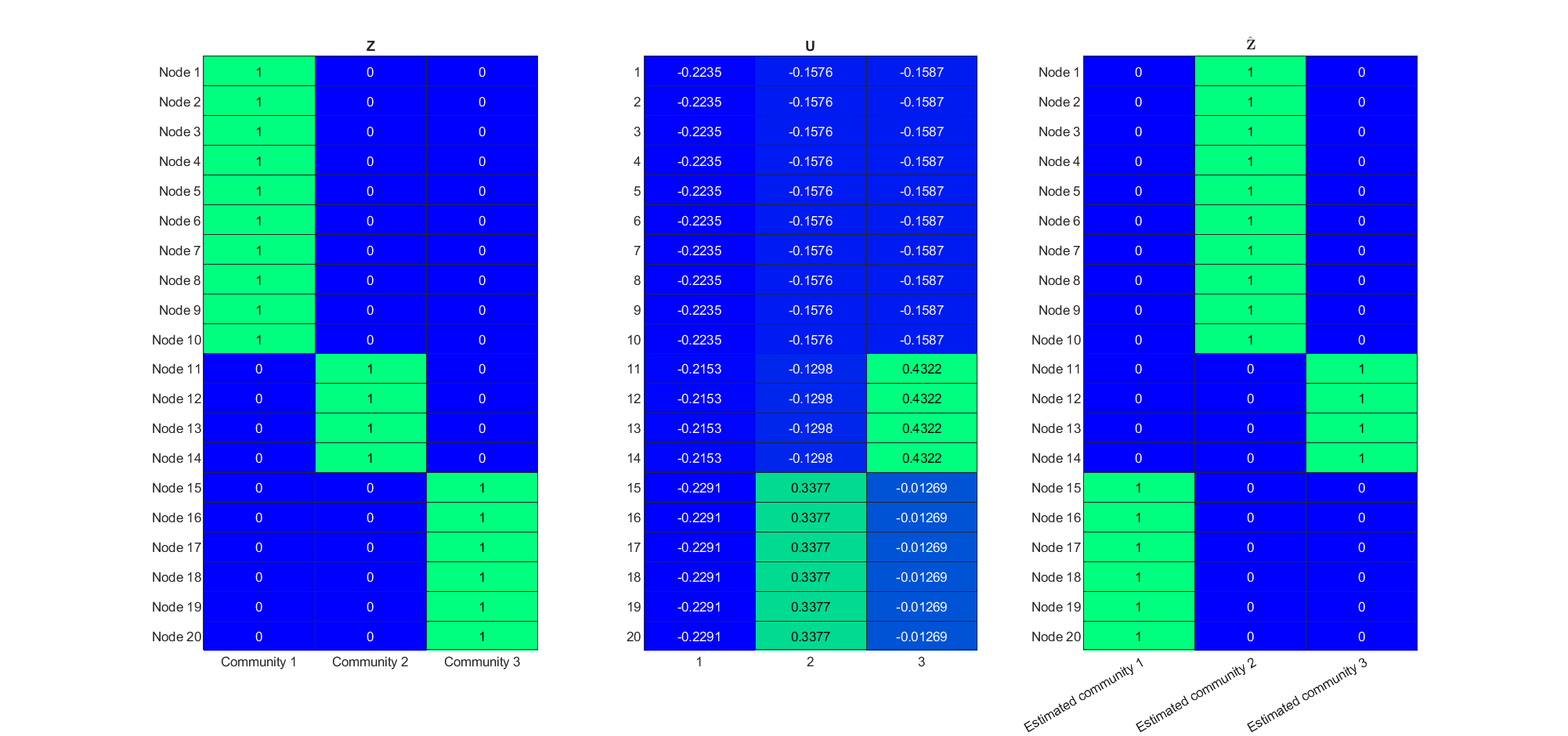}}
\caption{A toy example showing that running K-means clustering on $U$ recovers the true membership matrix $Z$ up to a label permutation. Left: the true membership matrix $Z$, where nodes 1-10 belong to Community 1, nodes 11-14 to Community 2, and nodes 15-20 to Community 3. Middle: the eigenvector matrix $U$ of $\mathcal{L}_{\tau}$ with $\tau=\frac{\sum_{i\in[n]}\mathcal{D}(i,i)}{10n}$. Right: $\hat{Z}$ obtained by applying K-means clustering defined in Equation (\ref{KmeansH}) to $U$ with 3 clusters. Here, we consider a multi-layer network with 20 nodes, 3 layers, and $Z$ the true membership matrix. For this example, we set the three connectivity matrices $\{B_{l}\}^{3}_{l=1}$ as:
$
B_{1}=\begin{bmatrix}
0.9 & 0.2 & 0.5 \\
0.2 & 0.6 & 0.4 \\
0.5 & 0.4 & 0.1 \\
\end{bmatrix},
$
$
B_{2}=\begin{bmatrix}
1 & 0.3 & 0.2 \\
0.3 & 0.8 & 0.1 \\
0.2 & 0.1 & 0.7 \\
\end{bmatrix},
$
and
$
B_{3}=\begin{bmatrix}
0.5 & 0.3 & 0.2 \\
0.3 & 0.7 & 0.2 \\
0.2 & 0.2 & 0.8 \\
\end{bmatrix}
$.}
\label{ZUZhatSimulated} 
\end{figure}

By Lemma \ref{PopulationLtauMLSBM}, we see that nodes within the same community have the same respective rows in the $n\times K$ eigenvector matrix $U$. Thus, employing the K-means algorithm to $U$ can perfectly recover nodes' communities up to a label permutation. For a better understanding of the exact recovery, similar to \citep{SCORE,lei2015consistency}, we define the K-means clustering considered in this study as
\begin{align}\label{KmeansH}
(\hat{Z},\hat{G})=\mathrm{arg~}_{\tilde{Z}\in\mathbb{M}_{n,K},\tilde{G}\in\mathbb{R}_{K,K}}\mathrm{min}\|\tilde{Z}\tilde{G}-\hat{H}\|^{2}_{F},
\end{align}
where $\hat{H}$ represents any $n\times K$ matrix serving as the input matrix of the K-means clustering, and the definitions of $\mathbb{M}_{n,K}$ and $\mathbb{R}_{K,K}$are provided in the last row of Table \ref{Abbr}. When we set the input matrix as the eigenvector matrix $U$, given that $U$  possesses exactly $K$ distinct rows by Lemma \ref{PopulationLtauMLSBM}, it is evident that $\|\tilde{Z}\tilde{G}-U\|^{2}_{F}$ reaches its minimum value of 0 only when $\tilde{Z}\tilde{G}\equiv U$. In fact, since $\|U(\tilde{k},:)-U(k,:)\|=\sqrt{\frac{1}{n_{k}}+\frac{1}{n_{\tilde{k}}}}$ for any two distinct rows $k$ and $\tilde{k}$ by Lemma \ref{PopulationLtauMLSBM}, the K-means clustering defined in Equation (\ref{KmeansH}) can always find the $K$ centers recorded in the $K$ rows of $X$ for $U$. Consequently, applying the K-means clustering to the rows of $U$ enables the exact recovery of $\ell$ up to a permutation of labels. Fig.~\ref{ZUZhatSimulated} illustrates a toy example demonstrating the satisfactory performance of the K-means estimations. Note that since $Z, \{B_{l}\}^{T}_{l=1}$, and $\mathcal{L}_{\tau}$ are known in this example, one can compute $\mathcal{L}_{\tau}$'s leading $K$ eigenvector matrix to verify the $U$ and $\hat{Z}$ in Fig.~\ref{ZUZhatSimulated}. In this example, we observe that $n_{1}=10, n_{2}=4$, and $n_{3}=6$. The validity of Lemma \ref{PopulationLtauMLSBM} can be easily verified through the following observations: (a) $U$ contains 3 distinct rows, and $U(i,:)=U(j,:)$ if $Z(i,:)=Z(j,:)$ for $i\in[20], j\in[20]$; (b) $\|U(1,:)-U(11,:)\|_{F}=0.5916=\sqrt{\frac{1}{n_{1}}+\frac{1}{n_{2}}}=\sqrt{\frac{1}{10}+\frac{1}{4}}, \|U(1,:)-U(15,:)\|_{F}=0.5164=\sqrt{\frac{1}{n_{1}}+\frac{1}{n_{3}}}=\sqrt{\frac{1}{10}+\frac{1}{6}}, $, and $\|U(11,:)-U(15,:)\|_{F}=0.6455=\sqrt{\frac{1}{n_{2}}+\frac{1}{n_{3}}}=\sqrt{\frac{1}{4}+\frac{1}{6}}$. Furthermore, examining the estimated membership matrix $\hat{Z}$ in Fig.~\ref{ZUZhatSimulated}, we observe that although $\hat{Z}\neq Z$, nodes sharing the same community in $Z$ are also grouped together in $\hat{Z}$, confirming that running K-means clustering to $U$ accurately recovers the true community assignments of nodes up to a label permutation. Based on this observation, it becomes evident that if we have a reliable estimation of the eigenvector matrix $U$, applying K-means clustering to this estimated eigenvector matrix should yield a good estimation of the nodes' communities. It is important to recall that the eigenvector matrix $U$ is derived from $\mathcal{L}_{\tau}$, which itself is constructed based on the aggregation matrix $\mathcal{S}$. Therefore, when $\{A_{l}\}^{T}_{l=1}$ are known while their population versions $\{\Omega_{l}\}^{T}_{l=1}$ are unavailable in the practical case, as long as we can find a good estimator of the aggregation matrix $\mathcal{S}$ from $\{A_{l}\}^{T}_{l=1}$, we can design an efficient method to detect nodes' communities. One may think that $\sum_{l\in[T]}A^{2}_{l}$ is a good estimator of the aggregation matrix $\mathcal{S}$ since $\mathbb{E}(A_{l})=\Omega_{l}$ for $l\in[T]$. However, as analyzed in \citep{lei2023bias}, $\sum_{l\in[T]}A^{2}_{l}$ is a biased estimation of $\mathcal{S}$. Instead, $S:\equiv\sum_{l\in[T]}(A^{2}_{l}-D_{l})$ is a debiased (bias-adjusted) estimator of $\mathcal{S}$, where $D_{l}(i,i)=\sum_{j\in[n]}A_{l}(i,j)$ and $D_{l}(i,\bar{i})=0$ for $i\in[n], \bar{i}\neq i, \bar{i}\in[n], l\in[T]$. $S$ is known as the debiased sum of squared adjacency matrices. Thus, when we design a community detection method, $S$ should be used instead of $\sum_{l\in[T]}A^{2}_{l}$. We are ready to define our regularized Laplacian matrix $L_{\tau}$:
\begin{align}\label{DefineLaplacian}
L_{\tau}=D^{-\frac{1}{2}}_{\tau}SD^{-\frac{1}{2}}_{\tau},
\end{align}
where $D_{\tau}=D+\tau I$ for $\tau\geq0$ and $D$ is a diagonal matrix with $D(i,i)=\sum_{j\in[n]}S(i,j)$ for $i\in[n]$. Note that, without causing any confusion, $D$ is considered as the degree matrix of the aggregation matrix $S$ here. It should be distinguished from the one derived from $A$ in the introduction section, and it differs significantly from $D_{l}$, which represents the degree matrix of $A_{l}$ for $l\in[T]$.
\begin{rem}
Similar to Remark \ref{DifferencePopulationLaplacian}, though our regularized Laplacian matrix $L_{\tau}$ defined in Equation (\ref{DefineLaplacian}) shares a similar form as the regularized Laplacian matrix studied in \citep{qin2013regularized,joseph2016impact}, they have a huge difference: our regularized Laplacian matrix $L_{\tau}$ is defined using the debiased sum of squared adjacency matrices $S$, causing that one can design community detection methods for multi-layer networks using our $L_{\tau}$. In contrast, the one used in \citep{qin2013regularized,joseph2016impact} is based on a single-layer adjacency matrix and can only be used for single-layer networks.
\end{rem}
Let $\hat{U}\hat{\Sigma}\hat{U}'$ be the leading $K$ eigen-decomposition of $L_{\tau}$ such that $\hat{U}'\hat{U}=I$ and $\hat{\Sigma}(k,k)=\lambda_{k}(L_{\tau})$ of the $K\times K$ diagonal matrix $\hat{\Sigma}$ for $k\in[K]$. Because $S$ is a good estimator of $\mathcal{S}$, we believe that $\hat{U}$ can be considered a good approximation of $U$. Therefore, running K-means to $\hat{U}$ with $K$ communities can provide a satisfactory estimation of nodes' communities. The above analysis is summarized by our first algorithm, called ``\textbf{r}egularized \textbf{d}ebiased \textbf{s}um \textbf{o}f \textbf{s}quared adjacency matrices" (RDSoS, Algorithm \ref{alg:RDSoS}). Our RDSoS is a regularized debiased spectral clustering method because it is designed by clustering the eigenvector matrix of our regularized Laplacian matrix $L_{\tau}$ calculated via Equation (\ref{DefineLaplacian}). In the final step of Algorithm \ref{alg:RDSoS}, we set $\hat{U}$ as the input matrix for the K-means clustering defined in Equation (\ref{KmeansH}), and $\hat{Z}$ as the resulting membership matrix. Subsequently, the estimated label vector $\hat{\ell}$ is derived by assigning $\hat{\ell}(i)=k$ if $\hat{Z}(i,k)=1$, for $i\in[n]$ and $k\in[K]$. The flowchart of RDSoS is shown in Fig.~\ref{RoadmapRDSoS}. It is worth noting that if the expectation adjacency matrices $\{\Omega_{l}\}^{T}_{l=1}$ are ideally known (meaning $\mathcal{S}$ is given ideally), substituting $\mathcal{S}$ for $S$ in RDSoS allows us to accurately recover the true node label vector $\ell$ up to a permutation of labels based on Lemma \ref{PopulationLtauMLSBM}. This further confirms the identifiability of the MLSBM model. In this paper, the built-in K-means package in MATLAB is used for our methods. Moreover, when the sum of squared adjacency matrices $\tilde{S} = \sum_{l \in [T]} A_{l}^2$ is substituted for the debiased sum of squared adjacency matrices $S$ in Algorithm \ref{alg:RDSoS}, we refer to the resulting method as "regularized sum of squared adjacency matrices" (abbreviated as RSoS).

The RDSoS method exhibits the following complexities for its steps: the complexities of Steps 1-5 are $O(Tn^{3})$, $O(n^{2})$, $O(n^{3})$, $O(Kn^{2})$, and $O(KnI_{ter})$, respectively, where $I_{ter}$ represents K-means' iteration number, which we have fixed at 100 in this paper. In this work, $K$ is set such that $K \ll n$. Thus, RDSoS's complexity is $O(Tn^{3})$.
\begin{algorithm}
\caption{\textbf{RDSoS}}
\label{alg:RDSoS}
\begin{algorithmic}[1]
\Require $\{A_{l}\}^{T}_{l=1}$, $K$ , and $\tau$.
\Ensure $\hat{\ell}$.
\State Compute $S=\sum_{l\in[T]}(A^{2}_{l}-D_{l})$.
\State Compute the diagonal matrix $D$ by $D(i,i)=\sum_{j\in[n]}S(i,j)$ for $i\in[n]$.
\State Compute $L_{\tau}=D^{-\frac{1}{2}}_{\tau}SD^{-\frac{1}{2}}_{\tau}$, where $D_{\tau}=\tau I+D$ (a default value of $\tau$ is $\frac{\sum_{i\in[n]}D(i,i)}{10n}$).
\State Obtain $\hat{U}\hat{\Sigma}\hat{U}'$, the leading $K$ eigen-decomposition of $L_{\tau}$.
\State Run the K-means algorithm to $\hat{U}$ with $K$ clusters to obtain the estimated node labels $\hat{\ell}$.
\end{algorithmic}
\end{algorithm}

\begin{figure}[htbp]
\centering
\begin{tikzpicture}[
    auto,
    node distance=1cm,
    decision/.style={diamond, draw=blue!80, fill=blue!20, text width=4.5em, align=flush center, font=\sffamily\bfseries, inner sep=4pt},
    block1/.style={rectangle, draw=black!80, fill=blue!10, text width=10em, align=center, rounded corners=8pt, font=\sffamily, minimum height=3em, shade, top color=blue!10, bottom color=blue!30},
    block2/.style={rectangle, draw=black!80, fill=green!10, text width=16em, align=center, rounded corners=8pt, font=\sffamily, minimum height=3em, shade, top color=green!10, bottom color=green!30},
    block3/.style={rectangle, draw=black!80, fill=orange!10, text width=21em, align=center, rounded corners=8pt, font=\sffamily, minimum height=3.5em, shade, top color=orange!10, bottom color=orange!30},
    block4/.style={rectangle, draw=black!80, fill=purple!10, text width=15em, align=center, rounded corners=8pt, font=\sffamily, minimum height=3em, shade, top color=purple!10, bottom color=purple!30},
    block5/.style={rectangle, draw=black!80, fill=red!10, text width=8.8em, align=center, rounded corners=8pt, font=\sffamily, minimum height=3em, shade, top color=red!10, bottom color=red!30},
    line/.style={draw, -latex', line width=1.5pt, shorten >=2pt},
    cloud/.style={draw=red!80, fill=red!20, ellipse, minimum height=3em, font=\sffamily\bfseries}
]
\matrix [row sep=1cm, column sep=1cm] {
    \node [block1] (s1) {\textbf{A multi-layer network}}; \\
    \node [block2] (s2) {\textbf{Calculate the aggregation matrix $S$}}; \\
    \node [block3] (s3) {\textbf{Calculate the regularized Laplacian matrix $L_{\tau}$}}; \\
    \node [block4] (s4) {\textbf{Obtain the eigenvector matrix $U$}}; \\
    \node [block5] (s5) {\textbf{Estimated clusters}};\\
};
\begin{scope}[every path/.style=line]
    \path (s1) -- node[left, font=\sffamily\itshape] {\textbf{Aggregation}} (s2);
    \path (s2) -- node[left, font=\sffamily\itshape] {\textbf{Regularization}} (s3);
    \path (s3) -- node[left, font=\sffamily\itshape] {\textbf{Eigen-decomposition}} (s4);
    \path (s4) -- node[left, font=\sffamily\itshape] {\textbf{K-means}} (s5);
\end{scope}
\end{tikzpicture}
\caption{The flowchart of the proposed RDSoS algorithm}
\label{RoadmapRDSoS}
\end{figure}
\subsection{Consistency of RDSoS in MLSBM}\label{sec4}
This subsection studies RDSoS's estimation consistency within the framework of MLSBM by allowing $n\rightarrow\infty, T\rightarrow\infty$, and $\rho\rightarrow0$. We want to show that the estimated label vector $\hat{\ell}$ returned by our RDSoS algorithm is close to the true label vector $\ell$. We achieve this by providing a theoretical upper bound of RDSoS's Clustering error $\hat{f}_{RDSoS}$ and proving that this theoretical bound will go to zero when we increase the number of nodes and/or layers under mild conditions. Three main technical components are needed to build RDSoS's theoretical error bound. First, we need to prove that the regularized Laplacian matrix $L_{\tau}$ concentrates around its population version $\mathcal{L}_{\tau}$. This is provided in Lemma \ref{BoundLMLDCSBM} given later, where the main technique we used for this lemma is the well-known matrix Bernstein inequality, a random matrix theory developed in \citep{tropp2012user}. Second, we should show that $L_{\tau}$'s eigenvector matrix $\hat{U}$ is also close to $\mathcal{L}_{\tau}$'s eigenvector matrix $U$. The main technique we used to bound the difference between $U$ and $\hat{U}$ is Lemma 5.1 of \citep{lei2015consistency}, a popular theoretical result used for eigenvector analysis. Lastly, we obtain an upper bound for $\hat{f}_{RDSoS}$ by combing the results of the first two technical components with the findings in Lemma \ref{PopulationLtauMLSBM} and Lemma 2.1 of \citep{joseph2016impact}. To establish RDSoS's theoretical guarantees, the next two technical assumptions are needed.
\begin{assum}\label{Assum1}
 $\rho^{2}n^{2}T\geq\mathrm{log}(n+T)$.
\end{assum}
\begin{assum}\label{Assum11}
 $|\lambda_{K}(\sum_{l\in[T]}\tilde{B}^{2}_{l})|\geq c_{1}T$ for some $c_{1}>0$.
\end{assum}
Both assumptions are mild. Assumption \ref{Assum1} only requires $\rho\geq\sqrt{\frac{\mathrm{log}(n+T)}{n^{2}T}}$. Thus, the sparsity parameter can be very small (i.e., the multi-layer network can be very sparse) for numerous nodes and/or layers. Assumption \ref{Assum11} only requires that the smallest nonzero eigenvalue in the magnitude of  $\sum_{l\in[T]}\tilde{B}^{2}_{l}$ has a linear growth with the number of layers $T$. Note that our Assumptions \ref{Assum1} and \ref{Assum11} are the same as the sparsity requirement in Theorem 1 of \citep{lei2023bias} and Assumption 1(b) in \citep{lei2023bias}, respectively. Set $n_{\mathrm{max}}=\mathrm{max}_{k\in[K]}n_{k}$ and $\mathrm{\delta}_{\mathrm{max}}=\mathrm{max}_{i\in[n]}\mathcal{D}(i,i)$. $n_{\mathrm{min}}$ and $\mathrm{\delta}_{\mathrm{min}}$ are defined similarly. Theorem \ref{MainMlSBM} below is our main technical result for RDSoS under the MLSBM model.
\begin{thm}\label{MainMlSBM}
Under the MLSBM parameterized by $(Z,\rho, \{\tilde{B}_{l}\}^{T}_{l=1})$, suppose that Assumptions \ref{Assum1} and \ref{Assum11} are satisfied, we have
\begin{align*}
\hat{f}_{RDSoS}=\frac{K^{2}n_{\mathrm{max}}(\tau+\delta_{\mathrm{max}})^{2}}{\rho^{4}n^{5}_{\mathrm{min}}T^{2}}(O(\frac{\rho^{2}n^{2}T\mathrm{log}(n+T)}{(\tau+\delta_{\mathrm{min}})^{2}})+O(\frac{\rho^{4}n^{2}T^{2}}{(\tau+\delta_{\mathrm{min}})^{2}})+O(\frac{\rho^{4}n^{4}T^{2}\mathrm{log}^{2}(n+T)}{(\tau+\delta_{\mathrm{min}})^{4}})+O(\frac{\rho^{8}n^{4}T^{4}}{(\tau+\delta_{\mathrm{min}})^{4}})),
\end{align*}
with probability at least $1-o(\frac{1}{n+T})$.
\end{thm}
The complex form of the bound in Theorem \ref{MainMlSBM} can be simplified by the following analysis.

\textbf{(Choice of $\tau$)} We let $\varrho(\tau)=\frac{K^{2}n_{\mathrm{max}}(\tau+\delta_{\mathrm{max}})^{2}}{\rho^{4}n^{5}_{\mathrm{min}}T^{2}}(O(\frac{\rho^{2}n^{2}T\mathrm{log}(n+T)}{(\tau+\delta_{\mathrm{min}})^{2}})+O(\frac{\rho^{4}n^{2}T^{2}}{(\tau+\delta_{\mathrm{min}})^{2}})+O(\frac{\rho^{4}n^{4}T^{2}\mathrm{log}^{2}(n+T)}{(\tau+\delta_{\mathrm{min}})^{4}})+O(\frac{\rho^{8}n^{4}T^{4}}{(\tau+\delta_{\mathrm{min}})^{4}}))$ be a function of the regularizer $\tau$. On the one hand, Theorem \ref{MainMlSBM} indicates that $\varrho(\tau)$ decreases as $\tau$ grows since $\delta_{\mathrm{min}}\leq\delta_{\mathrm{max}}$. This suggests that regularization (the case $\tau>0$) is theoretically better than non-regularization (the case $\tau=0$) for the RDSoS algorithm. Meanwhile, when the network is excessively sparse, several isolated nodes may exist without connections to other nodes across all layers. Suppose node $\tilde{i}$ is such an isolated node. We observe that $D^{-1/2}_{\tau}(\tilde{i},\tilde{i})$ does not exist for $\tau=0$. Consequently, non-regularization necessitates the presence of isolated nodes, whereas regularization does not. Therefore, we prefer $\tau>0$ rather than $\tau=0$. On the other hand, we know that there is an inverse proportional relationship between $|\lambda_{K}(L_{\tau})|$ and $\tau$ since $|\lambda_{K}(L_{\tau})|=|\lambda_{K}(D^{-0.5}_{\tau}SD^{-0.5}_{\tau})|=|\lambda_{K}(D^{-1}_{\tau}S)|\geq \lambda_{K}(D^{-1}_{\tau})|\lambda_{K}(S)|=\frac{|\lambda_{K}(S)|}{\tau+\mathrm{max}_{i\in[n]}D(i,i)}$ and $|\lambda_{1}(L_{\tau})|\leq\frac{|\lambda_{1}(S)|}{\tau+\mathrm{min}_{i\in[n]}D(i,i)}$. This implies that we can not set the regularization parameter $\tau$ too large otherwise the leading $K$ eigenvalues of $L_{\tau}$ may be close to zero, which causes $L_{\tau}$'s leading $K$ eigenvectors can not provide sufficient information about nodes' communities. Meanwhile, if $0\leq\tau_{1}<\tau_{2}$, though $\varrho(\tau_{1})>\varrho(\tau_{2})$, this does not mean that RDSoS-$\tau_{2}$ strictly outperforms RDSoS-$\tau_{1}$ since the leading $K$ eigenvectors of $L_{\tau_{2}}$ may contain lesser community information than that of $L_{\tau_{1}}$, where RDSoS-$\tau_{1}$ and RDSoS-$\tau_{2}$ represent the RDSoS algorithm using $\tau_{1}$ and $\tau_{2}$, respectively. Therefore, theoretically speaking, the above analysis suggests a moderate value of the regularizer $\tau$ should be preferred. When $\tau+\delta_{\mathrm{min}}\geq\mathrm{max}(\sqrt{\rho^{2}n^{2}T\mathrm{log}(n+T)},\rho^{2}nT)$, $\varrho(\tau)$ can be simplified as $\varrho(\tau)=\frac{K^{2}n_{\mathrm{max}}(\tau+\delta_{\mathrm{max}})^{2}}{\rho^{4}n^{5}_{\mathrm{min}}T^{2}}(O(\frac{\rho^{2}n^{2}T\mathrm{log}(n+T)}{(\tau+\delta_{\mathrm{min}})^{2}})+O(\frac{\rho^{4}n^{2}T^{2}}{(\tau+\delta_{\mathrm{min}})^{2}})$.
The form of $\varrho(\tau)$ can be further simplified as $\varrho(\tau)=\frac{K^{2}n_{\mathrm{max}}}{\rho^{2}n^{5}_{\mathrm{min}}T}(O(n^{2}\mathrm{log}(n+T))+O(\rho^{2}n^{2}T)$ if $\tau$ is set such that $\frac{\tau+\delta_{\mathrm{max}}}{\tau+\delta_{\mathrm{min}}}=O(1)$. Since $0<\delta_{\mathrm{min}}\leq \mathcal{D}(i,i)\leq\rho^{2}n^{2}T$, we have $D(i,i)=O(\rho^{2}n^{2}T)$ with high probability by Lemma \ref{BoundDii} for $i\in[n]$. Therefore, setting $\tau$ as a moderate value like $\frac{\sum_{i\in[n]}D(i,i)}{10n}$ can make $\varrho(t)$ be in its simple form. Sure, larger $\tau$ also makes $\varrho(\tau)$ in its simple form. In simulations presented in Section \ref{secSim}, we find that our methods demonstrate a certain level of robustness concerning the selection of the regularization parameter $\tau$ and achieve satisfactory performance when $\tau$ is a moderate value, such as the default setting adopted in this paper.

Setting $\tau$ as its default value, we know that
\begin{align}\label{Simple1}
\hat{f}_{RDSoS}=O(\frac{K^{2}n_{\mathrm{max}}n^{2}\mathrm{log}(n+T)}{\rho^{2}n^{5}_{\mathrm{min}}T})+O(\frac{K^{2}n_{\mathrm{max}}n^{2}}{n^{5}_{\mathrm{min}}}).
\end{align}

By Equation (\ref{Simple1}), we observe that (a) increasing the sparsity parameter $\rho$ or the number of layers $T$ improves RDSoS's performance; (b) when $\rho, T$, and $n$ are fixed, decreasing the minimize community size $n_{\mathrm{min}}$ and increasing the number of communities $K$ lead to a harder case for community detection. Furthermore, similar to Assumption 1 of \citep{lei2023bias}, if we add some conditions on $K, n_{\mathrm{min}}$, and $n_{\mathrm{max}}$, the bound in Equation (\ref{Simple1}) can be further simplified.
\begin{cor}\label{CorSimpleForm}
If $K$ is fixed and the size of each community is balanced, Theorem \ref{MainMlSBM} can be simplified as
\begin{align*}
\hat{f}_{RDSoS}=O(\frac{\mathrm{log}(n+T)}{\rho^{2}n^{2}T})+O(\frac{1}{n^{2}}).
\end{align*}
\end{cor}
The result of Corollary \ref{CorSimpleForm} is the same as that of Theorem 1 \citep{lei2023bias}, which validates the correctness of our theoretical analysis. This corollary also indicates that when $n\rightarrow\infty$ and/or $T\rightarrow\infty$, RDSoS's Clustering error decreases to 0, implying the estimation consistency of RDSoS.
\section{Multi-layer degree-corrected stochastic block model}
Under the MLSBM model, akin to the SBM model, nodes belonging to the same community exhibit stochastic equivalence. This equivalence arises because $\mathbb{E}(\sum_{j\in[n]}A_{l}(i,j))=\sum_{j\in[n]}\Omega_{l}(i,j)=Z(i,:)\sum_{j\in[n]}B_{l}Z'(j,:)\equiv\mathbb{E}(\sum_{j\in[n]}A_{l}(\bar{i},j))$ if $\ell(i)=\ell(\bar{i})$ for $i\in[n], \bar{i}\in[n]$, i.e., the expected degrees of two distinct nodes from the same community are identical. However, nodes usually have various degrees in real networks \citep{karrer2011stochastic}. For instance, hub nodes that have more links compared to other nodes within the same community are frequently observed in real-world networks \citep{zhang2020detecting}. Additionally, it has been observed that the degrees of nodes in real networks follow an approximate power-law distribution \citep{albert2002statistical,clauset2009power,goldenberg2010survey}.  Therefore, despite the mathematical simplicity of MLSBM due to its stochastic equivalence property, it typically fails to capture the degree heterogeneity of nodes in real-world networks. Similar to the DCSBM model introduced in \citep{karrer2011stochastic} that extends the SBM model by considering node-specific parameters, the MLDCSBM model introduced in \citep{qingMLDCSBM} extends the MLSBM model by considering the degree heterogeneity vector $\theta\in(0,1]^{n\times1}$ to model networks in which nodes have varying degrees even within the same community. Let $\Theta$ be a diagonal matrix with $\Theta(i,i)=\theta(i)$ for $i\in[n]$. We consider the following definition for the MLDCSBM model.
\begin{defin}\label{DefinMLDCSBM}
(\textbf{MLDCSBM}) Let $\{A_{l}\}^{T}_{l=1}, Z$, and $\{\tilde{B}_{l}\}^{T}_{l=1}$ be the same as that of MLSBM. Given $(Z,\Theta,\{\tilde{B}_{l}\}^{T}_{l=1})$, the MLDCSBM assumes that
\begin{align}\label{AGeneratedMLDCSBM}
A_{l}(i,j)=A_{l}(j,i)\sim\mathrm{Bernoulli}(\theta(i)\theta(j)\tilde{B}_{l}(\ell(i),\ell(j))),\qquad i\in[n], j\in[n], l\in[T].
\end{align}
\end{defin}
For convenience, we use ``MLDCSBM parameterized by $\{Z, \Theta, \{\tilde{B}_{l}\}^{T}_{l=1}\}$'' to denote the MLDCSBM in Definition \ref{DefinMLDCSBM}. When $T=1$, MLDCSBM degenerates to the popular DCSBM. When setting $\Theta=\sqrt{\rho}I$ for $\rho\in(0,1]$, the MLDCSBM model in Definition \ref{DefinMLDCSBM} reduces to the MLSBM model. The proposition below demonstrates that the MLDCSBM model is identifiable.
\begin{prop}\label{idMLDCSBM}
(Identifiability of MLDCSBM). The MLDCSBM model is identifiable: For valid $(Z,\Theta,\{\tilde{B}_{l}\}^{T}_{l=1})$ and $(\breve{Z},\breve{\Theta}, \{\breve{B}_{l}\}^{T}_{l=1})$, if $\Theta ZB_{l}Z'\Theta=\breve{\Theta}\breve{Z}\breve{B}_{l}\breve{Z}'\breve{\Theta}$ holds for all $l\in[T]$, then $Z$ and $\breve{Z}$ are equivalent up to a permutation of labels.
\end{prop}

Without confusion, under the MLDCSBM model, we also let $\Omega_{l}$ be the population adjacency matrix of $A_{l}$ for $l\in[T]$, $\mathcal{S}:=\sum_{l\in[T]}\Omega^{2}_{l}$ be the sum of the $T$ squared population adjacency matrices, and $\mathcal{L}_{\tau}:=\mathcal{D}^{-\frac{1}{2}}_{\tau}\mathcal{S}\mathcal{D}^{-\frac{1}{2}}_{\tau}$ be the population regularized Laplacian matrix. Under the MLDCSBM model, we have
\begin{align*}
\Omega_{l}=\Theta Z\tilde{B}_{l}Z'\Theta,\qquad l\in[T].
\end{align*}
We see that $\mathbb{E}(\sum_{j\in[n]}A_{l}(i,j))=\sum_{j\in[n]}\Omega_{l}(i,j)=\sum_{j\in[n]}\theta(i)\theta(j)Z(i,:)\tilde{B}Z'(j,:)=\theta(i)Z(i,:)\sum_{j\in[n]}\theta(j)\tilde{B}Z'(j,:)$ for $i\in[n], l\in[T]$ under the MLDCSBM, which implies that vertices from the same group may still have distinct expected degrees if their degree heterogeneity parameters are different.
\begin{rem}
It is worth noting that the MLDCSBM model imposes no constraints on the parameter $\theta(i)$ beyond the requirement that it lies within the interval $(0,1]$ for any node $i \in [n]$. This feature ensures that the MLDCSBM is capable of modeling complex real-world multi-layer networks. For instance, to describe networks with hub nodes, one may assign significantly smaller heterogeneity parameters to non-hub nodes compared to hub nodes. Furthermore, to capture the power-law distribution of node degrees observed in real networks, one can initially generate $n$ random values from a power-law distribution (say, let these values be stored in the vector $\xi$, with $\varsigma$ denoting the maximum element of $\xi$), and subsequently set $\theta(i) = \xi(i)/\varsigma$ to make each element of $\theta$ range in $(0,1]$. By adopting this setting of $\theta$, the MLDCSBM model can generate simulated multi-layer networks where the degree distribution of each layer closely approximates a power-law distribution. In general, due to the arbitrary nature of the settings of the heterogeneity parameter $\theta$, the MLDCSBM model possesses considerable flexibility in modeling the complex structures inherent in real-world networks. Furthermore, as MLDCSBM is an extension of MLSBM, it inherits MLSBM's advantage analyzed in Remark \ref{MLSBMADiS}.
\end{rem}
In the next subsection, we introduce a regularized debiased spectral clustering method to estimate nodes' label information under the MLDCSBM model.
\subsection{The DC-RDSoS algorithm}
The next lemma functions similarly to Lemma \ref{PopulationLtauMLSBM} and reveals the structure in  $\mathcal{L}_{\tau}$ under the MLDCSBM.
\begin{lem}\label{PopulationLtauMLDCSBM}
Under the MLDCSBM parameterized by $(Z,\Theta, \{\tilde{B}_{l}\}^{T}_{l=1})$, when the rank of the sum of squared connectivity matrices $\sum_{l\in[T]}\tilde{B}^{2}_{l}$ is $K$. Without confusion with the case under the MLSBM, we also let $U\Sigma U'$ be $\mathcal{L}_{\tau}$'s leading $K$ eigen-decomposition, where $U'U=I$. Let $U_{*}(i,:)=\frac{U(i,:)}{\|U(i,:)\|_{F}}$ for $i\in[n]$. Then, we have: (a) if $\ell(i)=\ell(\bar{i})$ for $i\in[n], \bar{i}\in[n]$, $U_{*}(i,:)=U_{*}(\bar{i},:)$; (b) $U_{*}=ZY$ with $Y$ satisfying $\|Y(\tilde{k},:)-Y(k,:)\|_{F}=\sqrt{2}$ for $\tilde{k}\neq k, \tilde{k}\in[K], k\in[K]$.
\end{lem}
\begin{figure}
\centering
\subfigure{\includegraphics[width=1\textwidth]{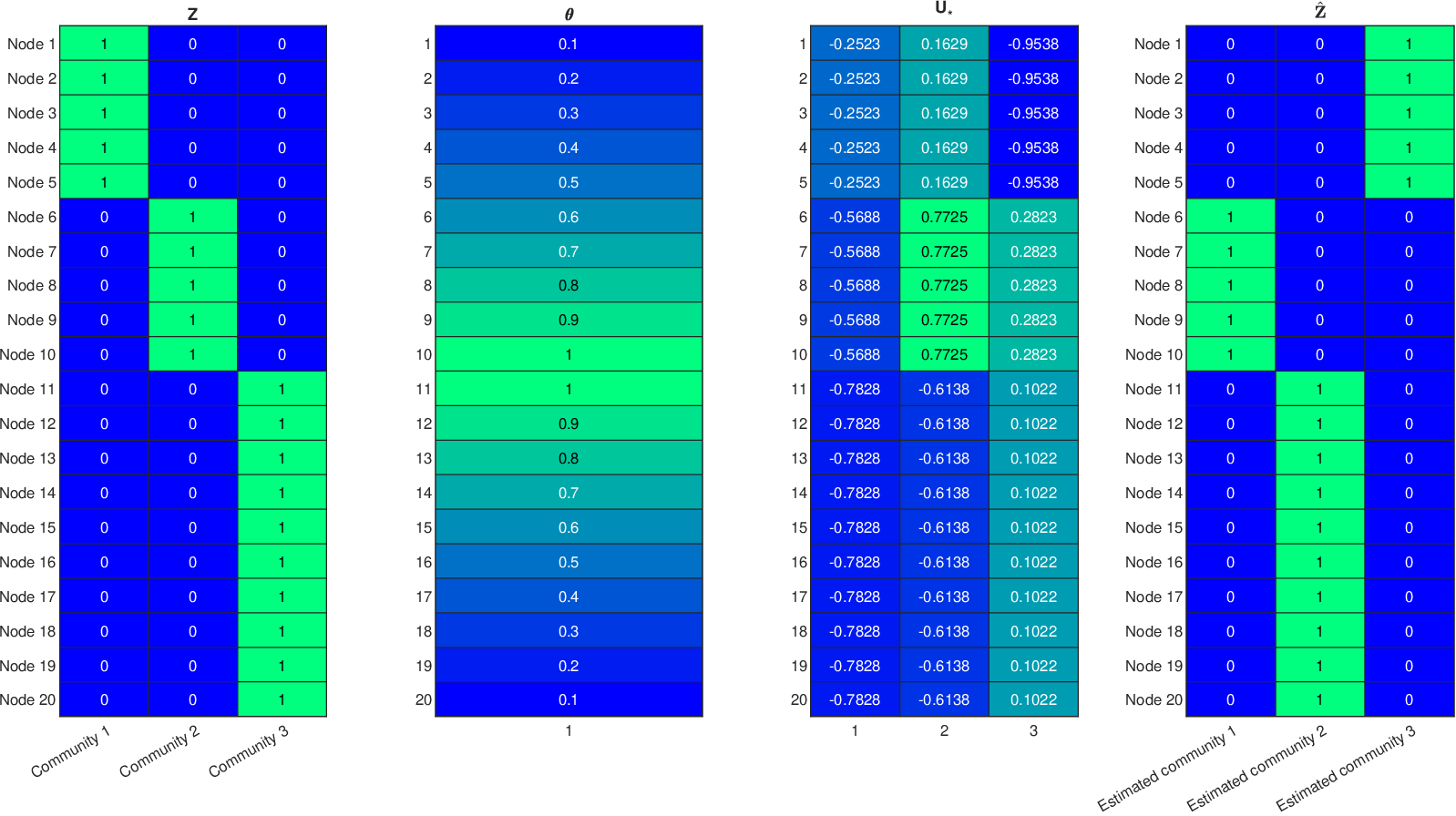}}
\caption{A toy example showing that running K-means clustering on $U_{*}$ exactly recovers $Z$ up to a label permutation. Here, $\hat{Z}$ is returned by running K-means to $U_{*}$ with 3 clusters. In this example, $\tilde{B}_{1}, \tilde{B_{2}}$, and $\tilde{B}_{3}$ are set the same as the $B_{1}, B_{2}$, and $B_{3}$ in Fig.~\ref{ZUZhatSimulated}, respectively.}
\label{thetaUstarZhatSimulated} 
\end{figure}

Lemma \ref{PopulationLtauMLDCSBM} says that there may exist more than $K$ distinct rows in the eigenvector matrix $U$ under the MLDCSBM due to the effect of the degree heterogeneity parameters $\Theta$. Therefore, applying K-means to $U$ may not exactly recover nodes' communities under the MLDCSBM. Instead, $U_{*}$, $U$'s row-normalized version, has exactly $K$ different rows, and employing K-means to it can perfectly recover nodes' communities, as demonstrated by a toy example in Fig.~\ref{thetaUstarZhatSimulated}. It is easy to check that $\|U_{*}(i,:)-U_{*}(j,:)\|=\sqrt{2}$ if nodes $i$ and $j$ are from different communities, and the estimated membership matrix $\hat{Z}$ equals to the true membership matrix $Z$ up to a label permutation. Again, given that $Z, \theta, \{\tilde{B}^{T}_{l=1}\}$ and $\mathcal{L}_{\tau}$ are provided in this example, one can compute the row-normalized version of $\mathcal{L}_{\tau}$'s leading $K$ eigenvector matrix $U$ to confirm the results presented in Fig.~\ref{thetaUstarZhatSimulated}. Define $S$ and $L_{\tau}$ the same as that of Algorithm \ref{alg:RDSoS}. Set $\hat{U}_{*}$ as $\hat{U}$'s row-normalized version  such that $\hat{U}_{*}(i,:)=\frac{\hat{U}(i,:)}{\|\hat{U}(i,:)\|_{F}}$ for $i\in[n]$. We see that $\hat{U}_{*}$ is a good approximation of $U_{*}$ and applying K-means to it should return good estimations of nodes' community partitions. Our second algorithm, called ``\textbf{D}egree-\textbf{c}orrected \textbf{r}egularized \textbf{d}ebiased \textbf{s}um \textbf{o}f \textbf{s}quared adjacency matrices" (DC-RDSoS, Algorithm \ref{alg:DCRDSoS}) summarizes the above analysis. Sure, our DC-RDSoS is also a regularized debiased spectral clustering method. The flowchart for DC-RDSoS is similar to that for RDSoS, thus we omit it here. Substituting $\mathcal{S}$ for $S$ in DC-RDSoS can exactly recover the true label vector $\ell$ up to a label permutation by Lemma \ref{PopulationLtauMLDCSBM}, which also supports the identifiability of MLDCSBM. The complexity cost of the DC-RDSoS algorithm is identical to that of RDSoS. Moreover, when we use $\tilde{S}$ to substitute $S$ in Algorithm \ref{alg:DCRDSoS}, we call the resulting method ``\textbf{D}egree-\textbf{c}orrected \textbf{r}egularized \textbf{s}um of \textbf{s}quared adjacency matrices" (DC-RSoS for short).
\begin{algorithm}
\caption{\textbf{DC-RDSoS}}
\label{alg:DCRDSoS}
\begin{algorithmic}[1]
\Require $\{A_{l}\}^{T}_{l=1}$, $K$, and $\tau$.
\Ensure $\hat{\ell}$.
\State Compute $\hat{U}$ using the steps 1-4 of Algorithm \ref{alg:RDSoS}.
\State Compute $\hat{U}_{*}$.
\State Run K-means to $\hat{U}_{*}$ with $K$ clusters to get $\hat{\ell}$.
\end{algorithmic}
\end{algorithm}
\subsection{Consistency of DC-RDSoS in MLDCSBM}
Our DC-RDSoS method also holds consistent estimation under the MLDCSBM model. The technical components of our theoretical analysis for DC-RDSoS are almost the same as those for RDSoS, and thus we omit them here for brevity. Set $\theta_{\mathrm{min}}=\mathrm{min}_{i\in[n]}\theta(i)$ and $\theta_{\mathrm{max}}=\mathrm{max}_{i\in[n]}\theta(i)$. Assumption \ref{Assum2} below functions similar to Assumption \ref{Assum1}.
\begin{assum}\label{Assum2}
 $\theta_{\mathrm{max}}\|\theta\|_{1}\|\theta\|^{2}_{F}T\geq\mathrm{log}(n+T)$.
\end{assum}
When the MLDCSBM reduces to the MLSBM by setting $\Theta=\sqrt{\rho}I$, we see that Assumption \ref{Assum2} degenerates to Assumption \ref{Assum1}.

The following lemma bounds $\|L_{\tau}-\mathcal{L}_{\tau}\|$ under the MLDCSBM and it shows that $L_{\tau}$ is close to its population version $\mathcal{L}_{\tau}$ in spectral norm. In proving Theorem \ref{MainMlSBM}, the bound of $\|L_{\tau}-\mathcal{L}_{\tau}\|$ under the MLSBM is also required. This bound can be immediately obtained from Lemma \ref{BoundLMLDCSBM} below by simply setting $\theta(i)=\sqrt{\rho}$ for $i\in[n]$.
\begin{lem}\label{BoundLMLDCSBM}
Under the MLDCSBM parameterized by $(Z,\Theta, \{\tilde{B}_{l}\}^{T}_{l=1})$. If Assumption \ref{Assum2} is satisfied, we have
\begin{align*}
\|L_{\tau}-\mathcal{L}_{\tau}\|=O(\frac{\sqrt{\theta_{\mathrm{max}}\|\theta\|_{1}\|\theta\|^{2}_{F}T\mathrm{log}(n+T)}}{\tau+\delta_{\mathrm{min}}})+O(\frac{\theta^{2}_{\mathrm{max}}\|\theta\|^{2}_{F}T}{\tau+\delta_{\mathrm{min}}})+O(\frac{\theta_{\mathrm{max}}\|\theta\|_{1}\|\theta\|^{2}_{F}T\mathrm{log}(n+T)}{(\tau+\delta_{\mathrm{min}})^{2}})+O(\frac{\theta^{4}_{\mathrm{max}}\|\theta\|^{4}_{F}T^{2}}{(\tau+\delta_{\mathrm{min}})^{2}}),
\end{align*}
 with probability at least $1-o(\frac{1}{n+T})$.
\end{lem}
Theorem \ref{MainMlDCSBM} quantifies DC-RDSoS's clustering performance under the MLDCSBM model.
\begin{thm}\label{MainMlDCSBM}
Under the MLDCSBM parameterized by $(Z,\Theta, \{\tilde{B}_{l}\}^{T}_{l=1})$, suppose that Assumptions \ref{Assum11} and \ref{Assum2} are satisfied, we have
\begin{align*}
\hat{f}_{DC-RDSoS}&=\frac{\theta^{2}_{\mathrm{max}}(\tau+\delta_{\mathrm{max}})^{3}K^{2}n_{\mathrm{max}}}{\theta^{10}_{\mathrm{min}}(\tau+\delta_{\mathrm{min}})n^{5}_{\mathrm{min}}T^{2}}(O(\frac{\theta_{\mathrm{max}}\|\theta\|_{1}\|\theta\|^{2}_{F}T\mathrm{log}(n+T)}{(\tau+\delta_{\mathrm{min}})^{2}})+O(\frac{\theta^{4}_{\mathrm{max}}\|\theta\|^{4}_{F}T^{2}}{(\tau+\delta_{\mathrm{min}})^{2}})+O(\frac{\theta^{2}_{\mathrm{max}}\|\theta\|^{2}_{1}\|\theta\|^{4}_{F}T^{2}\mathrm{log}^{2}(n+T)}{(\tau+\delta_{\mathrm{min}})^{4}})\\
&~~~+O(\frac{\theta^{8}_{\mathrm{max}}\|\theta\|^{8}_{F}T^{4}}{(\tau+\delta_{\mathrm{min}})^{4}})),
\end{align*}
 with probability at least $1-o(\frac{1}{n+T})$.
\end{thm}
Similar to the analysis after Theorem \ref{MainMlSBM}, a moderate value of $\tau$ is preferred for DC-RDSoS. Setting $\tau$ as its default value $\frac{\sum_{i\in[n]}D(i,i)}{10n}$ gives
\begin{align}\label{Simple2}
\hat{f}_{DC-RDSoS}&=\frac{\theta^{2}_{\mathrm{max}}K^{2}n_{\mathrm{max}}}{\theta^{10}_{\mathrm{min}}n^{5}_{\mathrm{min}}T^{2}}(O(\theta_{\mathrm{max}}\|\theta\|_{1}\|\theta\|^{2}_{F}T\mathrm{log}(n+T))+O(\theta^{4}_{\mathrm{max}}\|\theta\|^{4}_{F}T^{2})).
\end{align}

By setting $\Theta=\sqrt{\rho}I$, MLDCSBM becomes MLSBM and we see that the theoretical bound of $\hat{f}_{DC-RDSoS}$ in Equation (\ref{Simple2}) is consistent with that of Equation (\ref{Simple1}). The next corollary functions similar to Corollary \ref{CorSimpleForm}.
\begin{cor}\label{CorSimpleForm2}
If $K=O(1),\frac{n_{\mathrm{min}}}{n_{\mathrm{max}}}=O(1)$, and $\theta(i)=O(\sqrt{\rho})$ for any $\rho\in(0,1]$ for $i\in[n]$, Theorem \ref{MainMlDCSBM} can be simplified as
\begin{align*}
\hat{f}_{DC-RDSoS}=O(\frac{\mathrm{log}(n+T)}{\rho^{2}n^{2}T})+O(\frac{1}{n^{2}}).
\end{align*}
\end{cor}
\begin{rem}
In comparison to \citep{qingMLDCSBM}, several distinctions are noteworthy. Firstly, the focal points of the models diverge. The prior work concentrated on the MLDCSBM model, introducing a debiased spectral clustering method specifically tailored to it. Conversely, the current work expands its horizons by proposing two novel methods, RDSoS and DC-RDSoS, designed to accommodate MLSBM and MLDCSBM, respectively. This expansion enhances the flexibility and generality of the community detection methodology. Secondly, while the previous work did not consider the identifiability problem of the MLDCSBM model, the current work studies the identifiability of both MLSBM and MLDCSBM. Thirdly, this work introduces a novel regularized Laplacian matrix, extending the classical concept from single-layer to multi-layer networks. This regularization strategy mitigates the effects of noise and sparsity, potentially enhancing the performance of community detection, as evidenced by subsequent experimental results. Fourthly, the present work establishes theoretical consistency for both proposed methods under the MLSBM and MLDCSBM, demonstrating their robustness to the choice of the regularization parameter. Furthermore, this work proposes a novel metric in Section \ref{SecSoSModularity} to measure the quality of community detection in multi-layer networks. The numerical results presented in Section \ref{secSim} demonstrate that this metric is more reliable than an existing metric for evaluating community quality in multi-layer networks.
\end{rem}
\begin{rem}
Utilizing a proof analogous to that of Theorems \ref{MainMlSBM} and \ref{MainMlDCSBM}, we can establish the estimation consistencies of RSoS and DC-RSoS under the MLSBM and the MLDCSBM, respectively. Additionally, by adopting a similar theoretical analysis from \citep{qingMLDCSBM}, we can demonstrate that RDSoS outperforms RSoS under MLSBM, and DC-RDSoS outperforms DC-RSoS under MLDCSBM. For brevity, detailed proofs are omitted.
\end{rem}
\section{Estimating the number of communities in multi-layer networks}\label{SecSoSModularity}
In this section, we introduce an efficient methodology for estimating the number of communities $K$ in multi-layer networks. Typically, when $K$ is known, applying a community detection method (like those presented in this paper) to networks will yield a community partition. However, in real-world multi-layer networks, $K$ is often unknown. Therefore, determining $K$ is crucial for proceeding with community detection tasks. To estimate $K$, we first introduce a metric inspired by the well-known Newman-Girvan modularity \citep{newman2004finding} to assess the quality of community detection in multi-layer networks. Recall that $\tilde{S}=\sum_{l\in[T]}A^{2}_{l}$, let $d_{SoS}$ be an $n\times1$ vector, where $d_{SoS}(i)=\sum_{j\in[n]}\tilde{S}(i,j)$ for $i\in[n]$ and $SoS$ stands for sum of squared adjacency matrices. We then set $m_{SoS}=\sum_{i\in[n]}d_{SoS}(i)/2$. Given any community detection method $\mathcal{M}$ applied to a real multi-layer network with $T$ adjacency matrices $\{A_{l}\}^{T}_{l=1}$, and assuming there are $k$ communities, we denote the $n\times1$ estimated node label vector returned by $\mathcal{M}$ as $\hat{\ell}_{\mathcal{M},k}$. We define this metric in detail below:
\begin{align}\label{DefinQSoS}
Q_{SoS}(\mathcal{M},k)=\frac{1}{2m_{SoS}}\sum_{i\in[n],j\in[n]}(\tilde{S}(i,j)-\frac{d_{SoS}(i)d_{SoS}(j)}{2m_{SoS}})\mathbbm{1}(\hat{\ell}_{\mathcal{M},k}(i)=\hat{\ell}_{\mathcal{M},k}(j)),
\end{align}
where $\mathbbm{1}$ is the indicator function. We call the metric defined in Equation (\ref{DefinQSoS}) the sum of squared adjacency matrices modularity (SoS-modularity for short) and use $Q_{SoS}$ to denote the value of it. Notably, if we substitute the sum of squared adjacency matrices $\tilde{S}$ with the adjacency matrix of a single-layer network, $Q_{SoS}$ reduces to the Newman-Girvan modularity \citep{newman2004finding}. Analogous to the Newman-Girvan modularity, a higher $Q_{SoS}$ value is desirable, indicating a better community partition for the multi-layer network.

After defining our SoS-modularity metric to assess the quality of community partitions in multi-layer networks, we present our methodology for estimating the number of communities $K$. For real-world multi-layer networks where the true $K$ is unknown, we assume $K$ lies within the range $\{1, 2, \ldots, K_{C}\}$. In this paper, based on the observation that $K$ is typically not excessively large in real multi-layer networks, we set $K_{C}$ to 20. For a given method $\mathcal{M}$, we compute $K_{C}$ SoS-modularity values, denoted as $\{Q_{SoS}(\mathcal{M}, k)\}_{k=1}^{K_{C}}$. Following the strategy in \citep{nepusz2008fuzzy,qing2024applications}, we estimate $K$ by maximizing the SoS-modularity, which we denote as $K_{SoS,\mathcal{M}} = \arg\max_{k \in [K_{C}]} Q_{SoS}(\mathcal{M}, k)$ for method $\mathcal{M}$. Let $Q_{SoS,\mathcal{M}} = Q_{SoS}(\mathcal{M}, K_{SoS,\mathcal{M}})$ represent the highest SoS-modularity achieved by method $\mathcal{M}$. We consider method $\mathcal{M}_{1}$ to provide a better community partition than method $\mathcal{M}_{2}$ if $Q_{SoS,\mathcal{M}_{1}} > Q_{SoS,\mathcal{M}_{2}}$.

We note that \cite{paul2021null} introduced a metric known as multi-normalized average modularity (MNavrg-modularity for brevity) to evaluate the quality of community partitions in multi-layer networks. This metric can be understood as the average of the Newman-Girvan modularity across all layers. For completeness, we provide a brief description of this metric here.
\begin{align*}
Q_{MNavrg}(\mathcal{M},k)=\frac{1}{T}\sum_{l\in[T]}\sum_{i\in[n], j\in[n]}\frac{1}{2e_{l}}(A_{l}(i,j)-\frac{d_{l}(i)d_{l}(j)}{2e_{l}})\mathbbm{1}(\hat{\ell}_{\mathcal{M},k}(i)=\hat{\ell}_{\mathcal{M},k}(j)),
\end{align*}
where $e_{l}=\sum_{i\in[n]}d_{l}(i)/2$ for $l\in[T]$. The value $Q_{MNavrg}(\mathcal{M},k)$ lies in the range $[0,1]$. Similar to the Newman-Girvan modularity, a higher value of the MNavrg-modularity indicates a better quality of the estimated communities. For any community detection method $\mathcal{M}$, we define $K_{MNavrg,\mathcal{M}} = \arg\max_{k \in [K_{C}]} Q_{MNavrg}(\mathcal{M}, k)$ as the number of communities estimated by the method $\mathcal{M}$ through maximizing the MNavrg-modularity, and we let $Q_{MNavrg,\mathcal{M}} = Q_{MNavrg}(\mathcal{M}, K_{MNavrg,\mathcal{M}})$ denote the highest MNavrg-modularity achieved by the method $\mathcal{M}$.

Given that the true community structure of real-world multi-layer networks is often unknown, and considering that there are two metrics available—our proposed SoS-modularity and the MNavrg-modularity introduced by \citep{paul2021null}, it is natural to ask which metric is more preferable and more believable for assessing the quality of community detection in multi-layer networks. To address this question, we will compare the accuracy rates of determining the number of communities $K$ based on these two metrics in computer-generated multi-layer networks where the true $K$ is known in the next section.
\section{Simulations}\label{secSim}
This section conducts simulation studies to investigate the performance of our RDSoS, DC-RDSoS, RSoS, and DC-RSoS by comparing them with the following methods:
\begin{itemize}
  \item \textbf{SoS-Debias}: this method is proposed by \citep{lei2023bias} and it is designed by substituting $S$ for $L_{\tau}$ in Algorithm \ref{alg:RDSoS}, i.e., it does not consider regularized Laplacian matrix. Estimation consistency of SoS-Debias under the MLSBM is provided in \citep{lei2023bias}.
  \item \textbf{NDSoSA}: this algorithm is introduced in \citep{qingMLDCSBM} and it is designed by substituting $S$ for $L_{\tau}$ in Algorithm \ref{alg:DCRDSoS}. Its theoretical convergence is analyzed in \citep{qingMLDCSBM} under the MLDCSBM.
  \item \textbf{MASE}: it is a spectral embedding algorithm proposed by \citep{arroyo2021inference} to estimate nodes' communities.
  \item \textbf{RSum} and \textbf{DC-RSum}: Both methods are derived by substituting $A_{\mathrm{sum}}\equiv \sum_{l \in [T]} A_{l}$ for $S$ in Algorithms \ref{alg:RDSoS} and \ref{alg:DCRDSoS}, respectively.
\end{itemize}
\begin{rem}
In this work, we do not consider other baseline methods mentioned in \citep{arroyo2021inference,lei2023bias,qingMLDCSBM} for comparison, as the primary algorithms proposed in these works typically exhibit superior performance compared to these baseline methods, as evidenced by their numerical studies.
\end{rem}

For simulated datasets with knowing the true node community information $\ell$, the performance of the above methods is evaluated using the following measures: Clustering error \citep{joseph2016impact}, Hamming error \citep{SCORE}, ARI \citep{hubert1985comparing}, and NMI \citep{strehl2002cluster}. For Clustering error and Hamming error, lower values indicate better performance, while for ARI and NMI, higher values indicate better performance. To evaluate the accuracy of all previously mentioned methods in estimating $K$ through the maximization of $Q_{SoS}$ (or $Q_{MNavrg}$), we adopt the Accuracy rate employed in \citep{qing2024finding}. This metric is the ratio of successful estimations of $K$ to the total number of independent trials, expressed as a value within the interval $[0,1]$. A higher Accuracy rate signifies superior precision in determining $K$.

For all simulation studies, we set $K=3$ and generate three communities with imbalanced sizes: $n_{1}=\frac{n}{2}$, $n_{2}=\frac{n}{5}$, and $n_{3}=\frac{3n}{10}$, where $n$ is a multiple of 10. First, let each entry of $\tilde{B}_{l}$ be generated from Unif(0,1). Then update $\tilde{B}_{l}$ by letting it be $\frac{\tilde{B}_{l}+\tilde{B}'_{l}}{2}$ to make it symmetric for $l\in[T]$. For the MLDCSBM, unless specified, we set $\theta(i)=\sqrt{\rho}\frac{\ell(i)}{K}\mathrm{rand}(1)$ for $i\in[n]$, where $\mathrm{rand}(1)$ is a random value generated from $\mathrm{Unif}(0,1)$. Thus, in simulated multi-layer networks generated from the MLDCSBM, each node may have a different heterogeneity parameter. Each simulation study considers two cases: the MLSBM case and the MLDCSBM case, where $\Omega_{l}=\rho Z\tilde{B}_{l}Z'$ under the MLSBM case and $\Omega_{l}=\rho \Theta Z\tilde{B}_{l}Z'\Theta$ under the MLDCSBM case for all $l\in[T]$. For each simulation study, $n$, $T$, and $\rho$ are set independently. Finally, 50 independent replicates are generated in every simulation setting, and we report the mean outcomes across these 50 repetitions for each metric. In addition, all experiments presented in this paper were conducted using MATLAB R2021b on a standard personal computer, specifically a Thinkpad X1 Carbon Gen 8, equipped with a 64-bit Windows-10 operating system, a 1.61 GHz Intel(R) Core(TM) i7-10710U CPU, and 16 GB of RAM.

\begin{figure}
\centering
\resizebox{\columnwidth}{!}{
\subfigure[MLSBM case]{\includegraphics[width=0.2\textwidth]{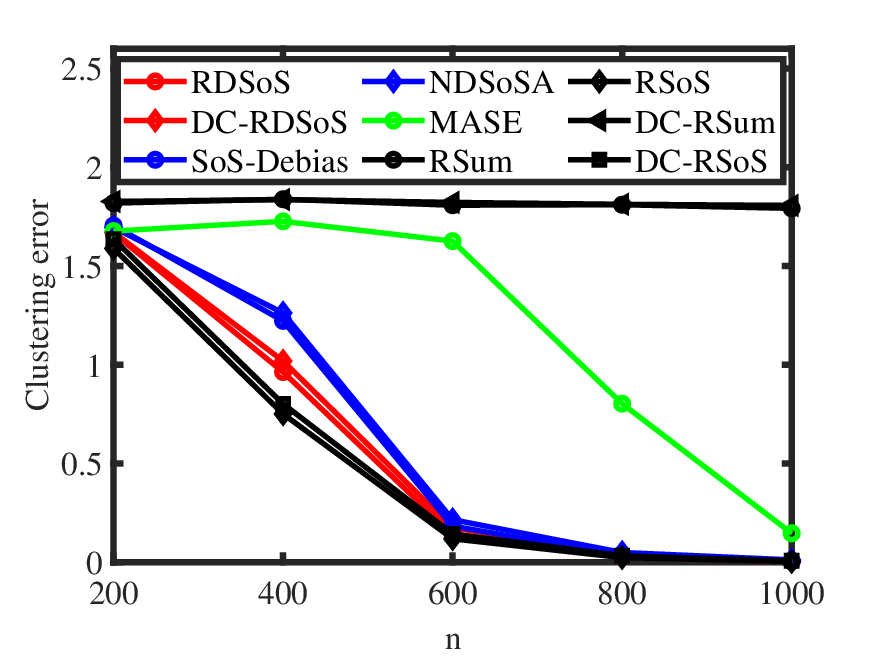}}
\subfigure[MLSBM case]{\includegraphics[width=0.2\textwidth]{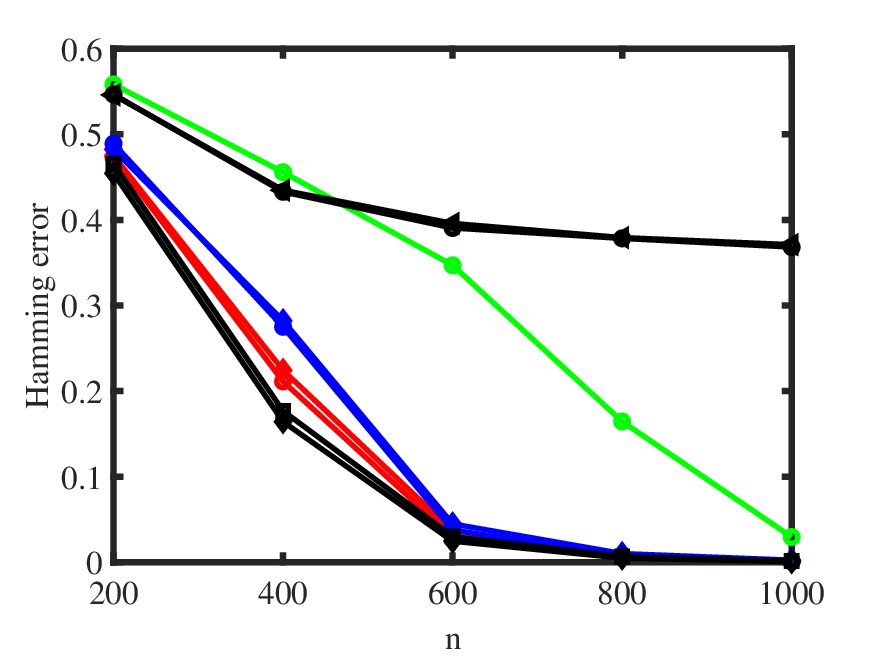}}
\subfigure[MLSBM case]{\includegraphics[width=0.2\textwidth]{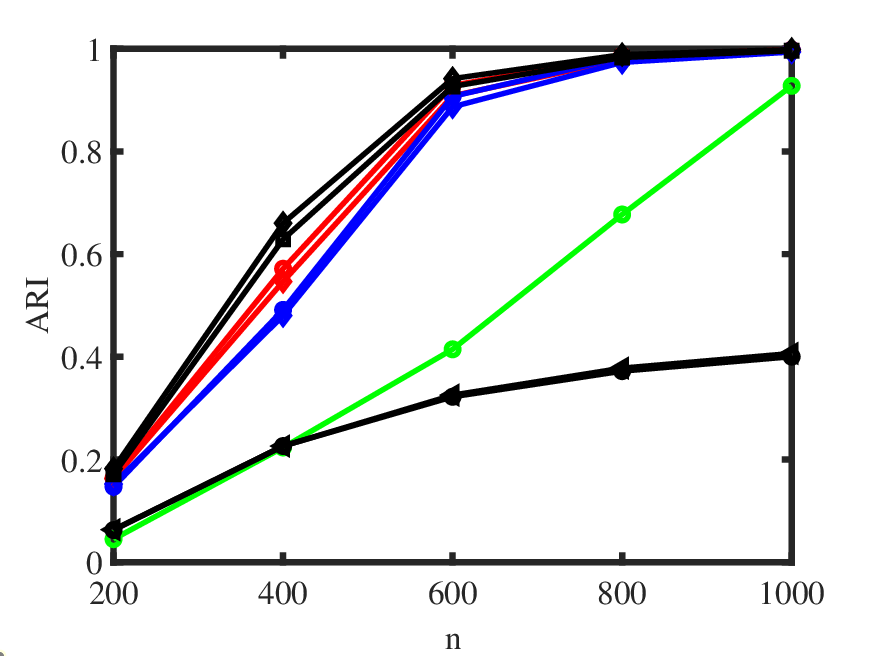}}
\subfigure[MLSBM case]{\includegraphics[width=0.2\textwidth]{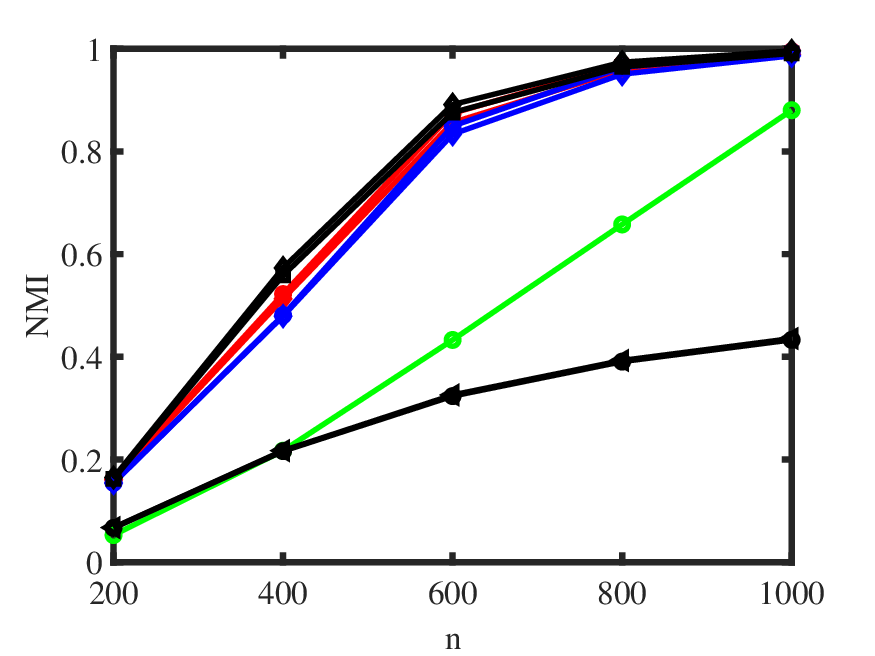}}
\subfigure[MLSBM case]{\includegraphics[width=0.2\textwidth]{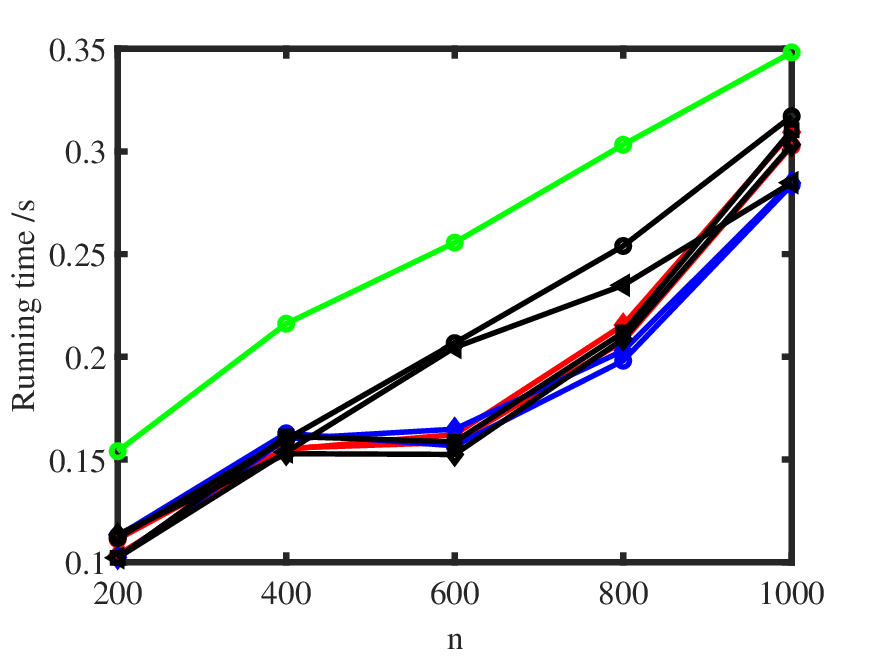}}
}
\resizebox{\columnwidth}{!}{
\subfigure[MLDCSBM case]{\includegraphics[width=0.2\textwidth]{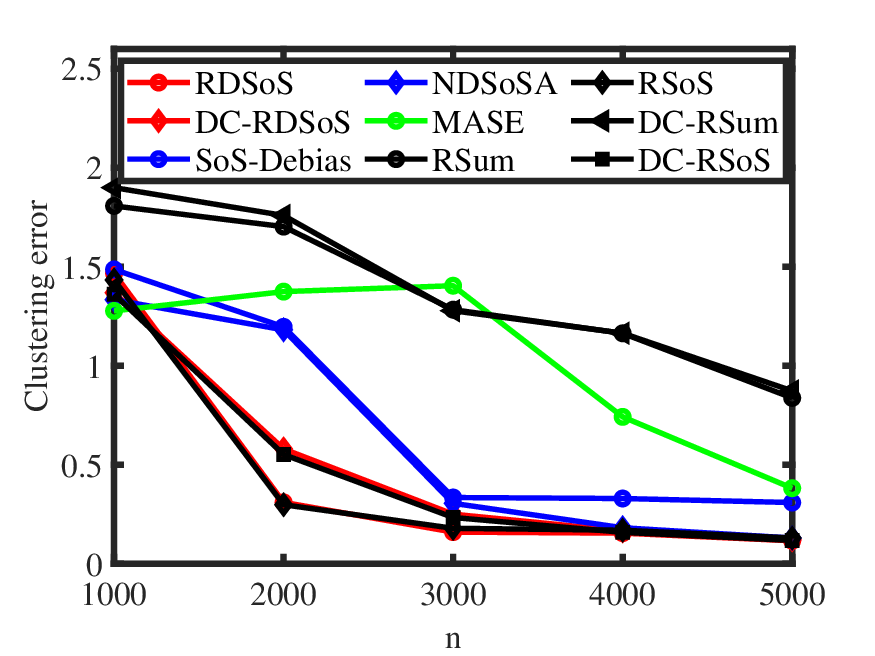}}
\subfigure[MLDCSBM case]{\includegraphics[width=0.2\textwidth]{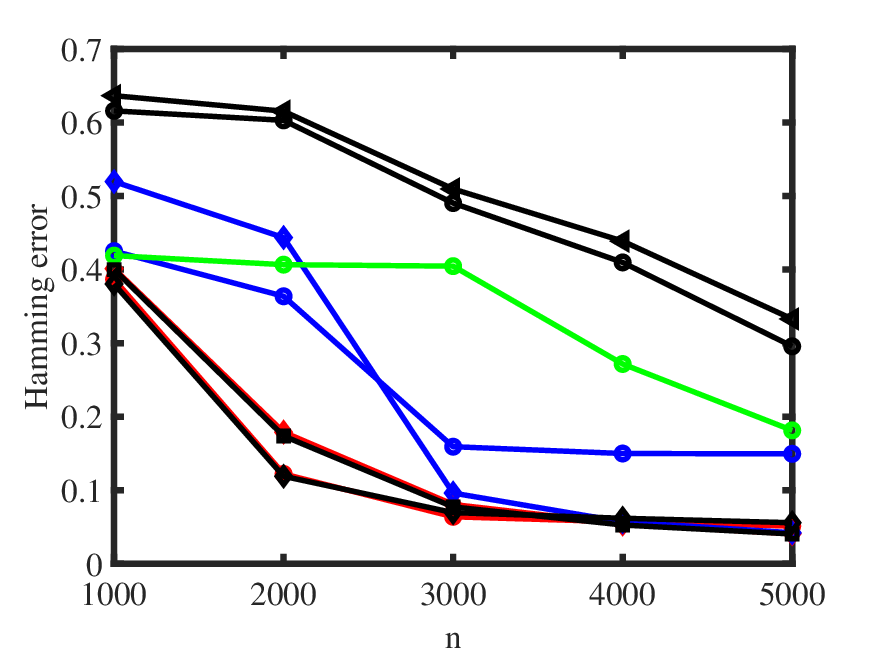}}
\subfigure[MLDCSBM case]{\includegraphics[width=0.2\textwidth]{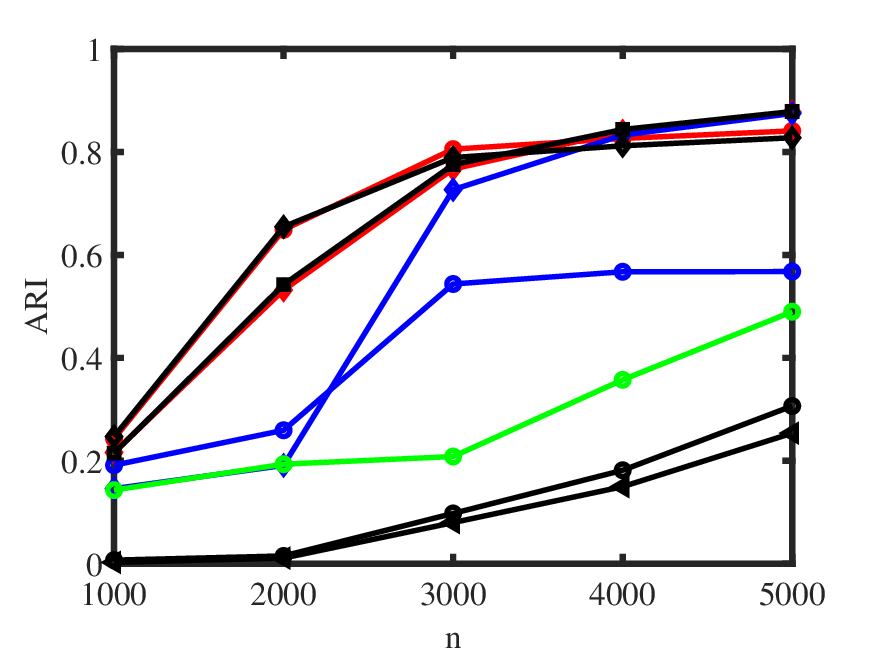}}
\subfigure[MLDCSBM case]{\includegraphics[width=0.2\textwidth]{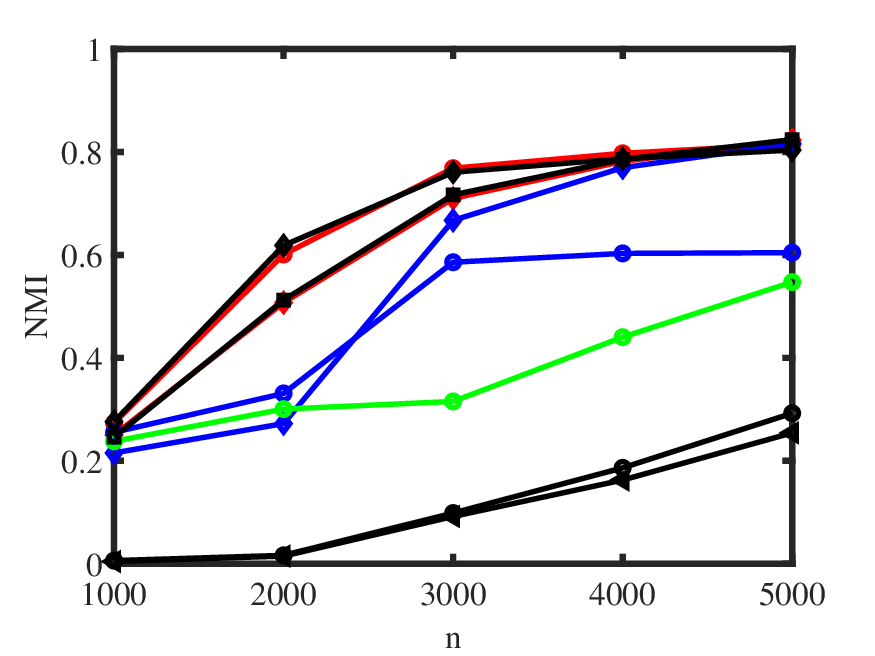}}
\subfigure[MLDCSBM case]{\includegraphics[width=0.2\textwidth]{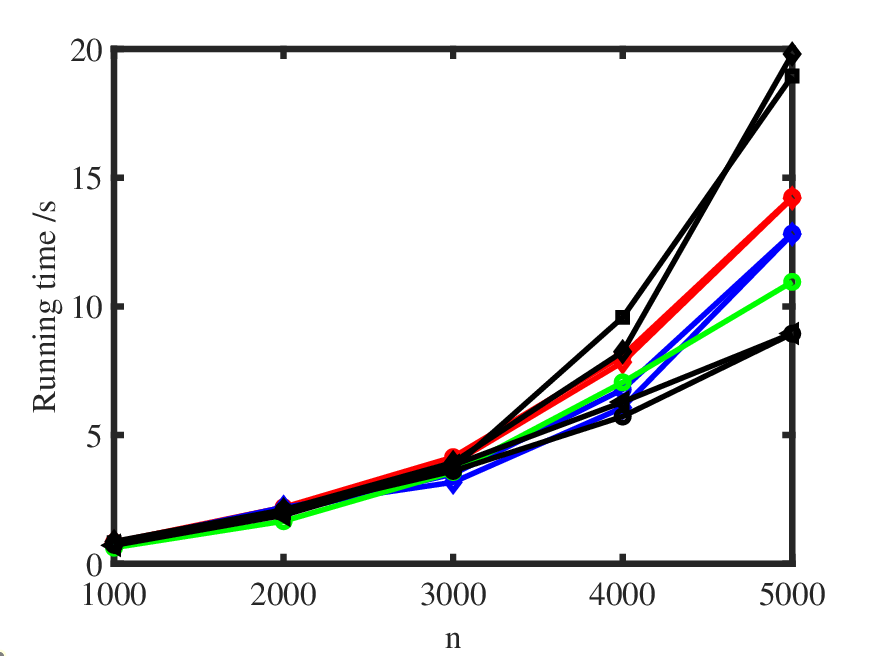}}
}
\caption{Performances of different methods for community detection in Simulation 1. The results indicate that as $n$ increases, all methods, except for MASE, RSum, and DC-RSum, exhibit improved performance. Notably, the proposed methods, RDSoS, DC-RDSoS, RSoS, and DC-RSoS, surpass their competitors. Furthermore, in the MLSBM case, MASE is the slowest method among all.}
\label{Ex1} 
\end{figure}

\begin{figure}
\centering
\resizebox{\columnwidth}{!}{
\subfigure[MLSBM case]{\includegraphics[width=0.2\textwidth]{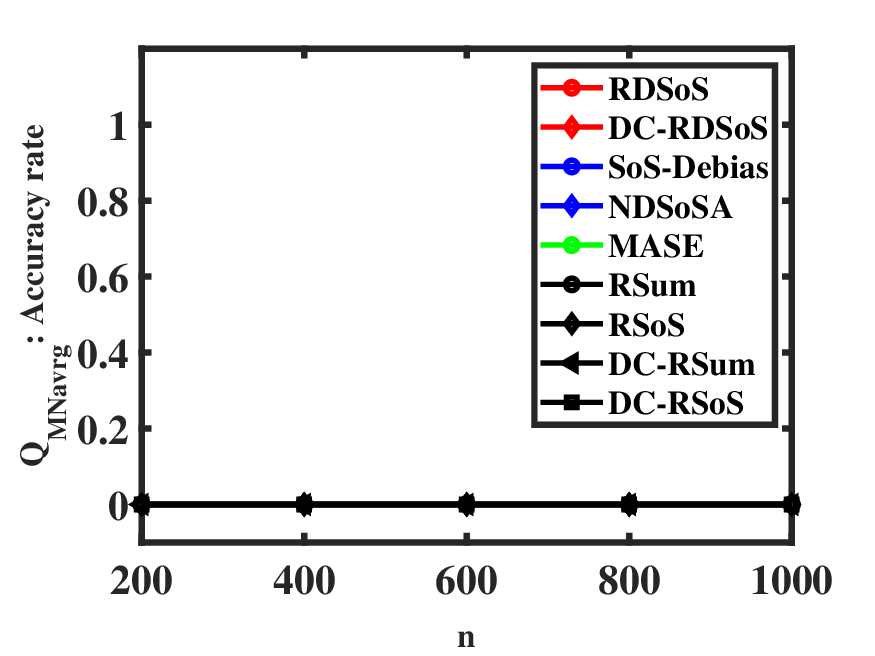}}
\subfigure[MLSBM case]{\includegraphics[width=0.2\textwidth]{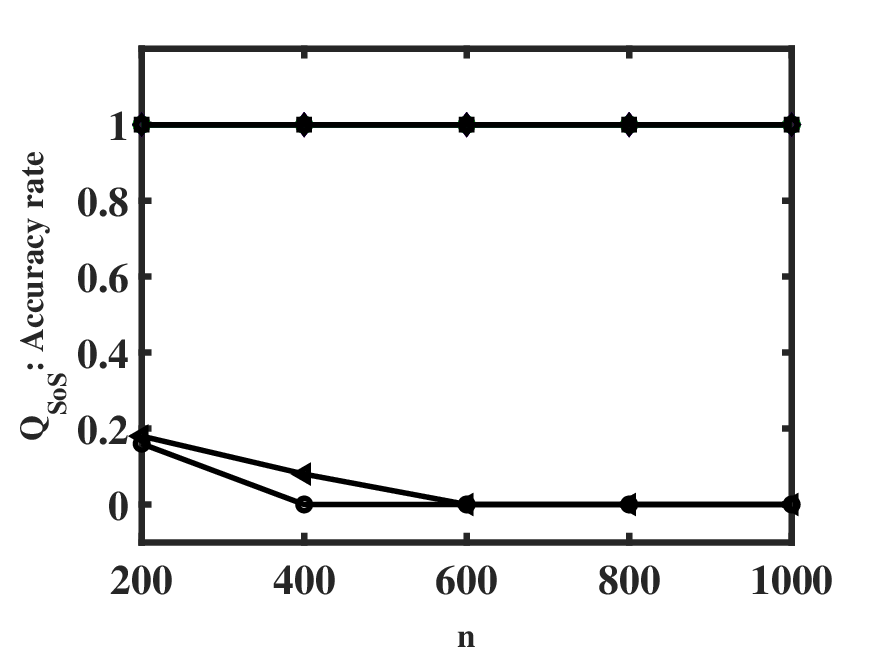}}
\subfigure[MLDCSBM case]{\includegraphics[width=0.2\textwidth]{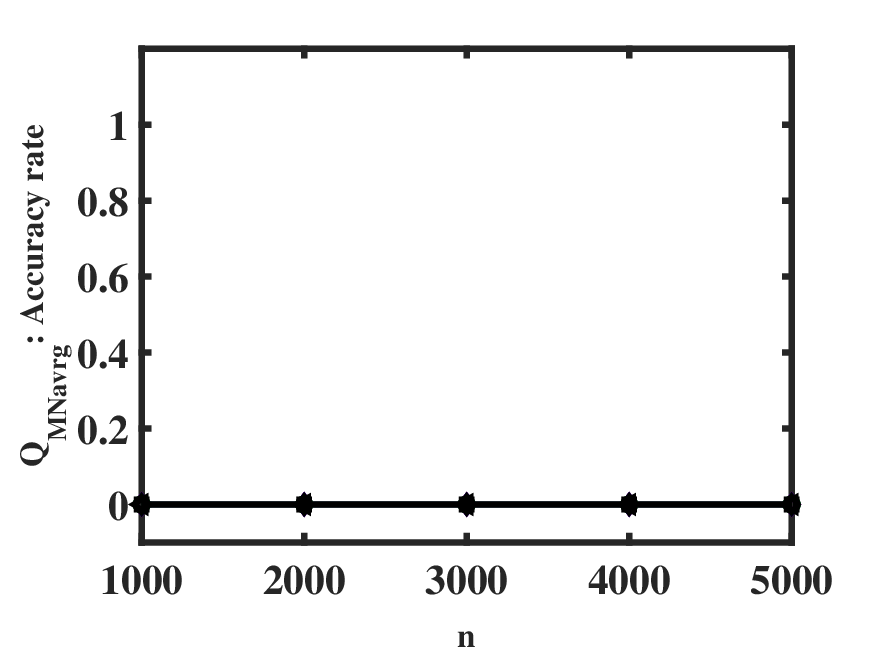}}
\subfigure[MLDCSBM case]{\includegraphics[width=0.2\textwidth]{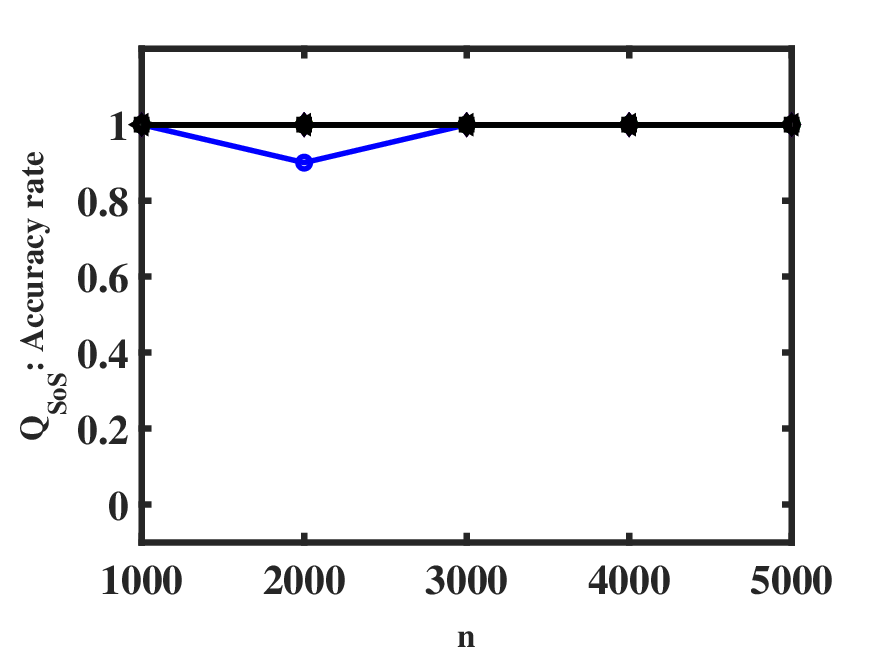}}
}
\caption{Accuracy rates of different methods for estimating $K$ in Simulation 1. It is noteworthy that estimating $K$ by maximizing the MNavrg-modularity fails to determine the value of $K$ for all methods. Except for RSum and DC-RSum, the remaining seven methods successfully estimate $K$ by maximizing our proposed SoS-modularity.}
\label{Ex1K} 
\end{figure}

\textbf{Simulation study 1: varying the number of nodes $n$.} For the MLSBM case, we set $\rho=0.04$, $T=10$, and vary $n$ in $\{200, 400, \ldots, 1000\}$, where $\rho$ is set as $0.25$ when we compare the Accuracy rates of different methods in estimating $K$. For the MLDCSBM case, we set $\rho=0.16$, $T=5$, and vary $n$ in $\{1000, 2000, \ldots, 5000\}$, where $\rho$ and $T$ are set as $1$ and $20$ for the task of estimating $K$, respectively. Results shown in Fig.~\ref{Ex1} say that for community detection, (a) all algorithms behave better when $n$ grows, except for MASE, RSum, and DC-RSum, whose Clustering Error and Hamming Error almost do not decrease in the MLDCSBM case; (b) our RDSoS, DC-RDSoS, RSoS, and DC-RSoS slightly outperform SoS-Debias and NDSoSA, and these methods outperform MASE, RSum, and DC-RSum in the MLSBM case; (c) our RDSoS, DC-RDSoS, RSoS, and DC-RSoS significantly outperform their competitors in the MLDCSBM case; (d) MASE runs slowest in the MLSBM case while RSum, DC-RSum, and MASE run slightly faster than the other six methods in the MLDCSBM case. Fig.~\ref{Ex1K} shows the Accuracy rates of different methods in estimating $K$. We observe that estimating $K$ via maximizing the MNavrg-modularity fails to determine the true $K$ for all methods. For comparison, except RSum and DC-RSum, the other seven methods accurately estimate $K$ through maximizing our SoS-modularity, which implies that our SoS-modularity is more believable than the MNavrg-modularity in assessing the quality of community partitions in multi-layer networks.

\begin{figure}
\centering
\resizebox{\columnwidth}{!}{
\subfigure[MLSBM case]{\includegraphics[width=0.2\textwidth]{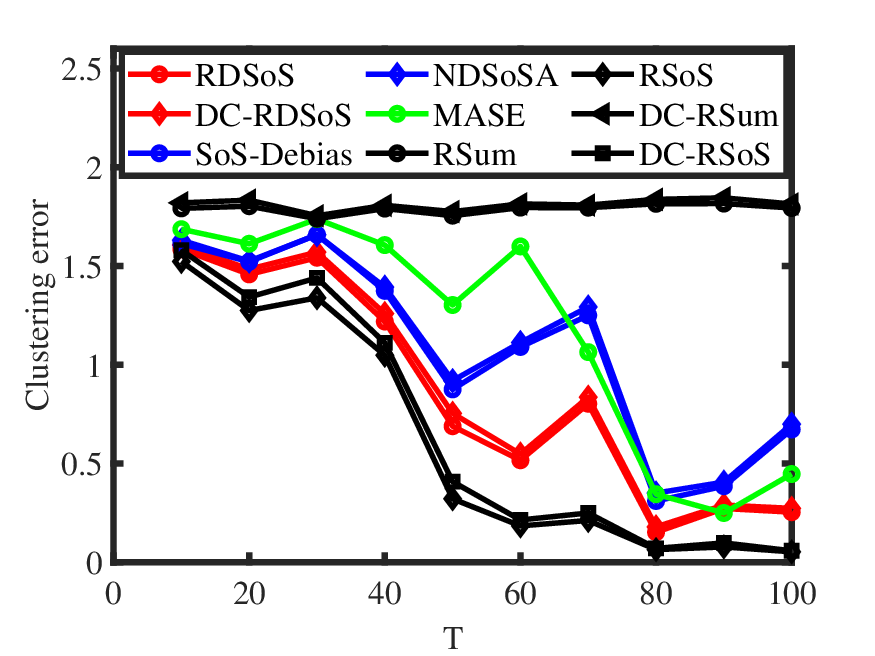}}
\subfigure[MLSBM case]{\includegraphics[width=0.2\textwidth]{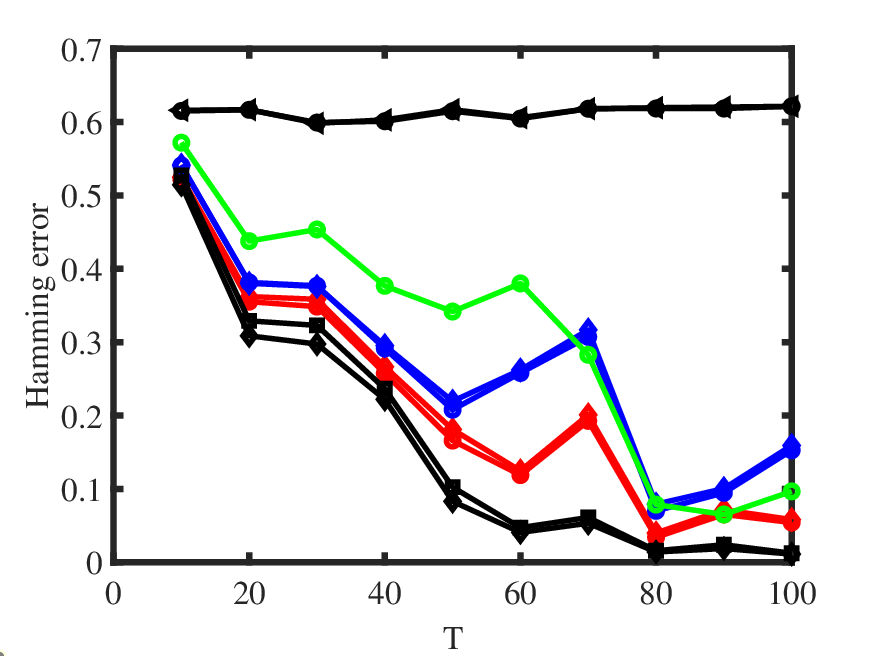}}
\subfigure[MLSBM case]{\includegraphics[width=0.2\textwidth]{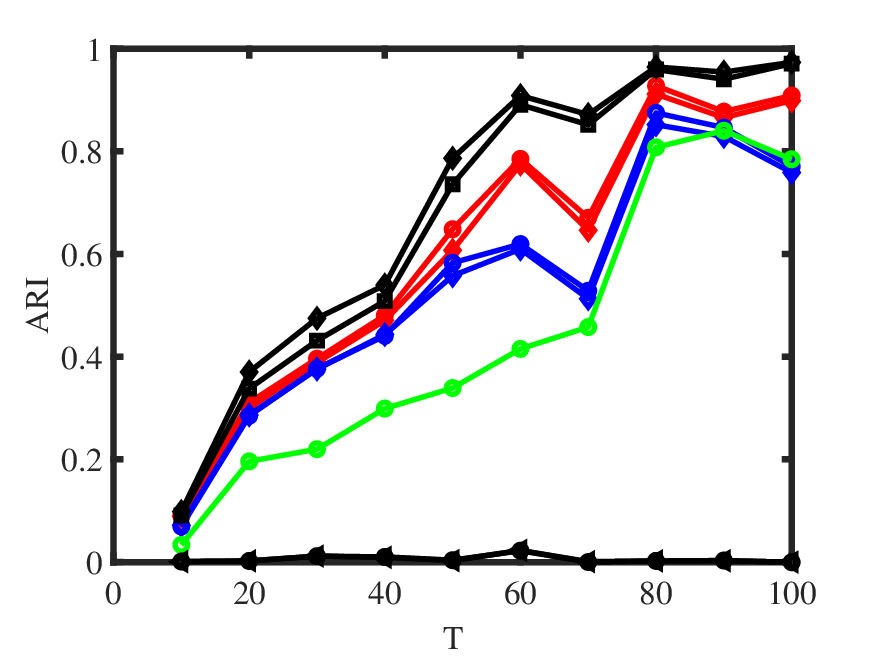}}
\subfigure[MLSBM case]{\includegraphics[width=0.2\textwidth]{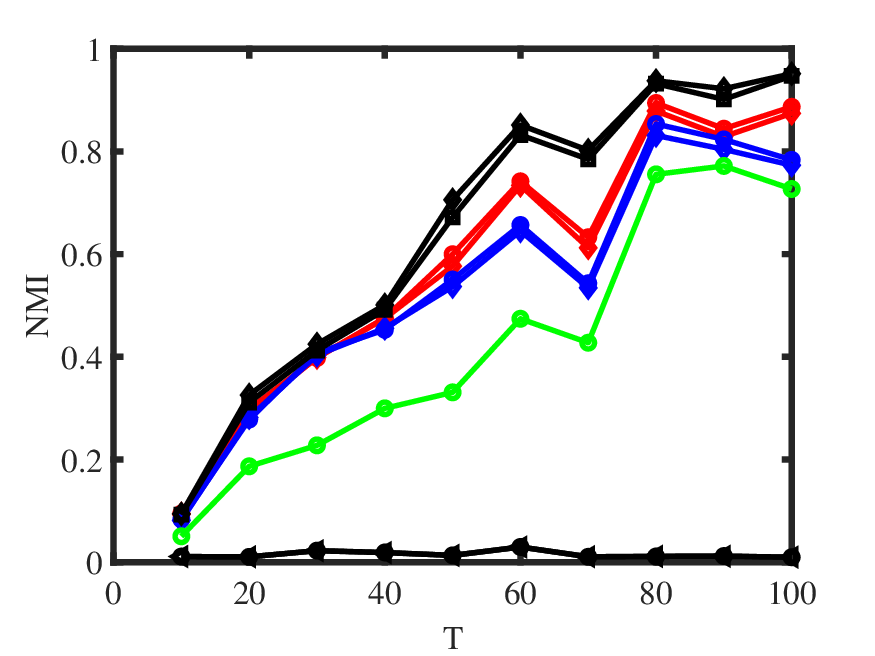}}
\subfigure[MLSBM case]{\includegraphics[width=0.2\textwidth]{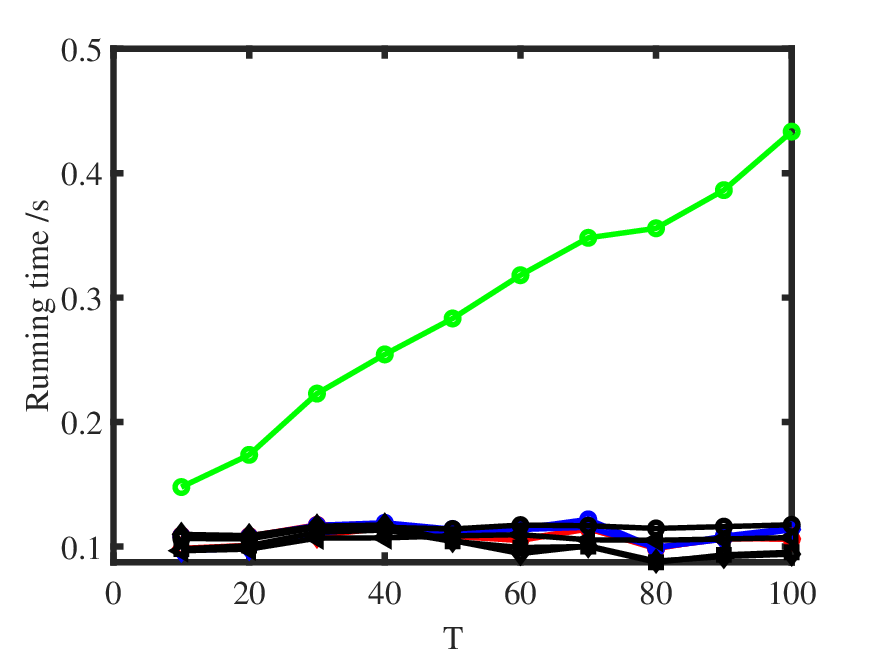}}
}
\resizebox{\columnwidth}{!}{
\subfigure[MLDCSBM case]{\includegraphics[width=0.2\textwidth]{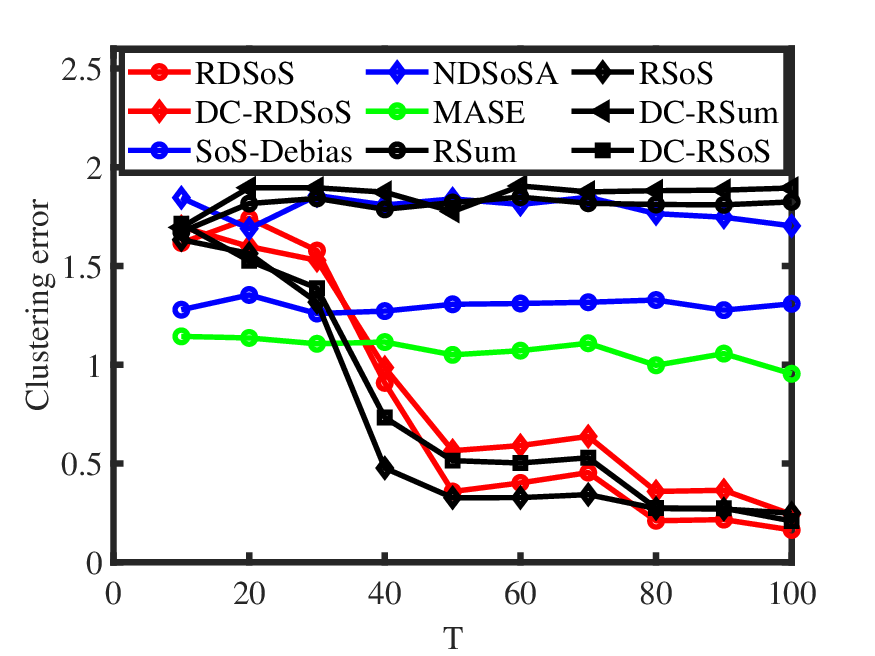}}
\subfigure[MLDCSBM case]{\includegraphics[width=0.2\textwidth]{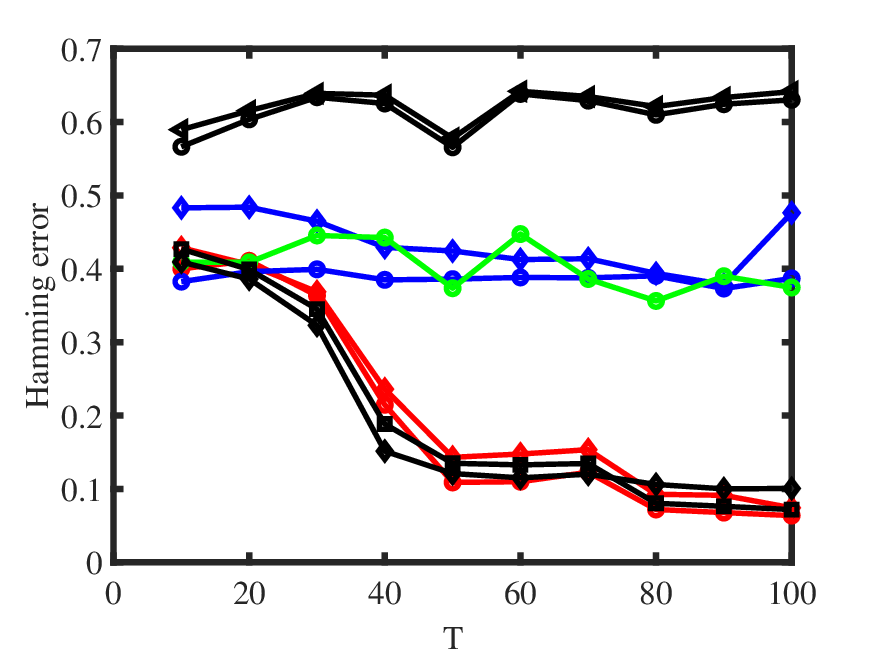}}
\subfigure[MLDCSBM case]{\includegraphics[width=0.2\textwidth]{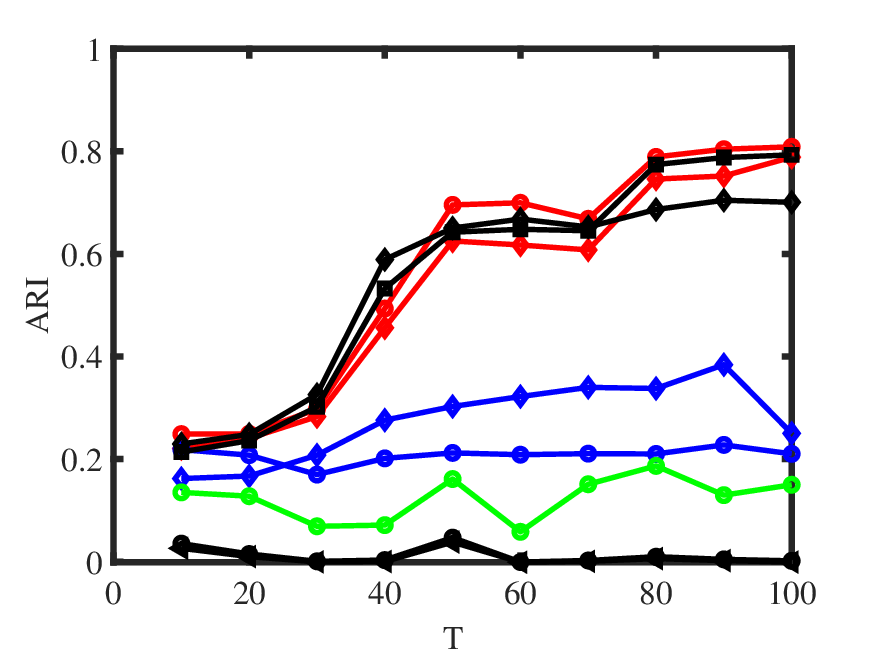}}
\subfigure[MLDCSBM case]{\includegraphics[width=0.2\textwidth]{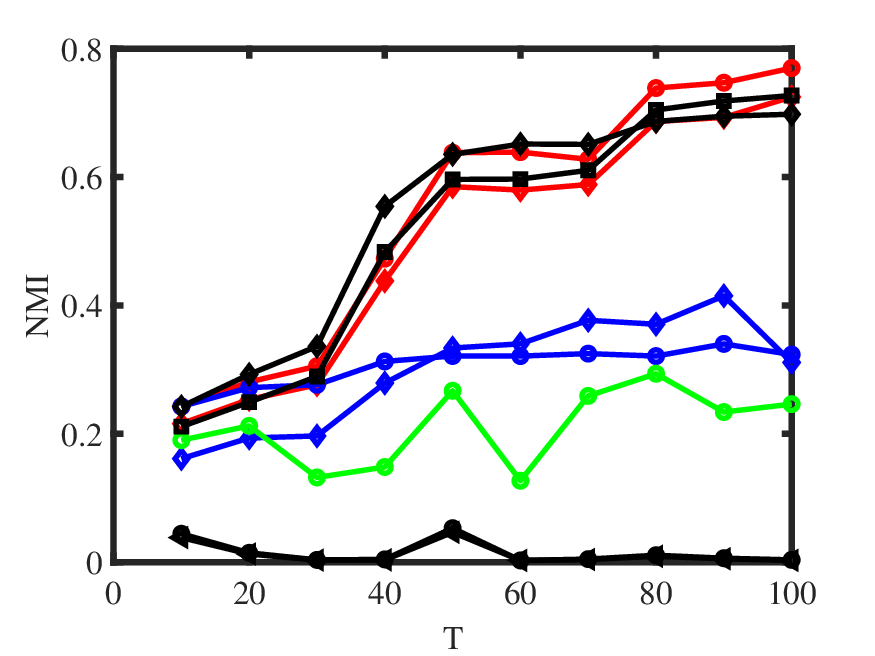}}
\subfigure[MLDCSBM case]{\includegraphics[width=0.2\textwidth]{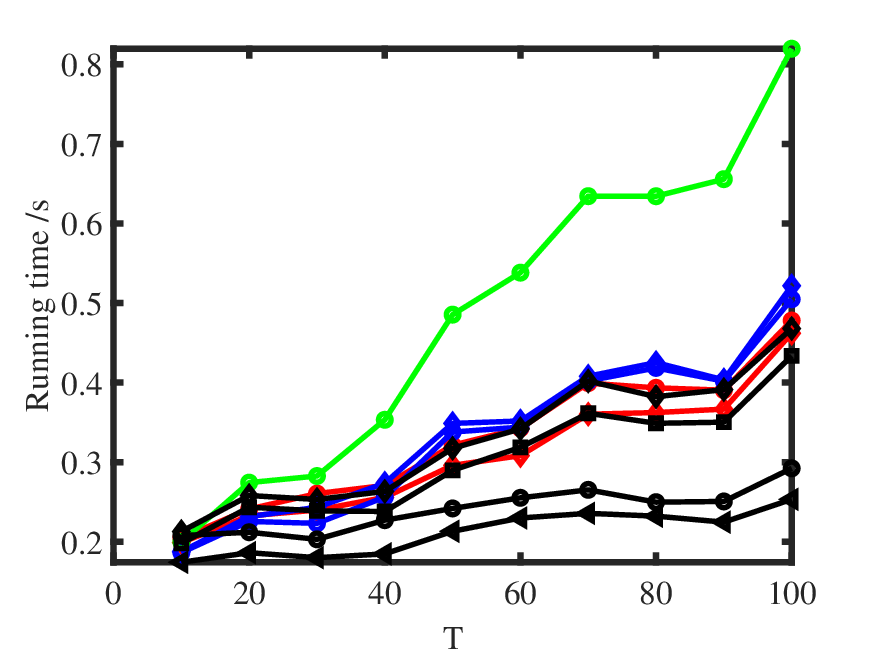}}
}
\caption{Performances of different methods for community detection in Simulation 2. The results demonstrate that as
$T$ grows, all methods except RSum and DC-RSum exhibit improved performance. The proposed methods, RDSoS, DC-RDSoS, RSoS, and DC-RSoS, consistently outperform their competitors, particularly in the MLDCSBM case. The running time of MASE is notably slower compared to the other methods, while the remaining methods exhibit similar computational efficiency.}
\label{Ex2} 
\end{figure}

\begin{figure}
\centering
\resizebox{\columnwidth}{!}{
\subfigure[MLSBM case]{\includegraphics[width=0.2\textwidth]{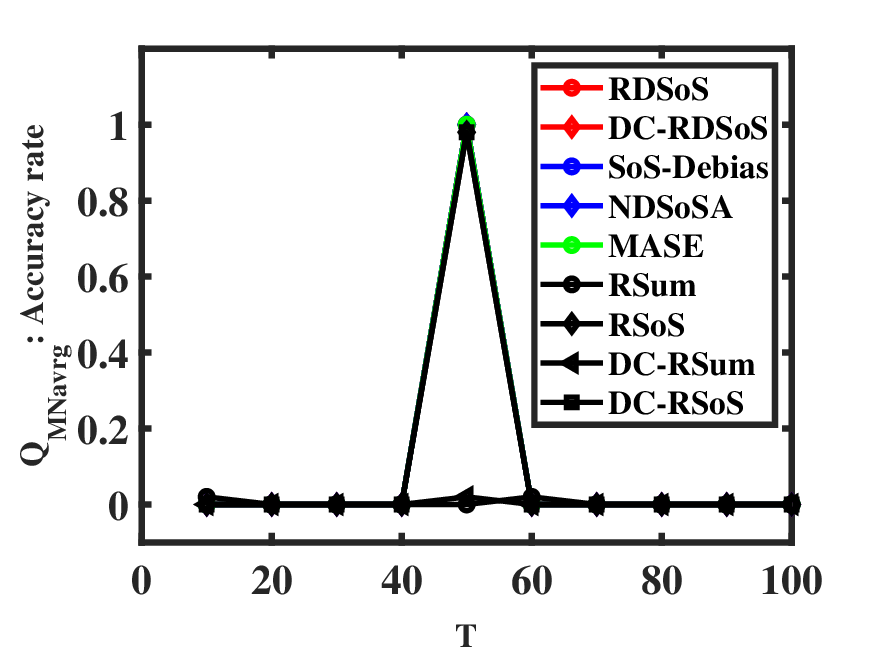}}
\subfigure[MLSBM case]{\includegraphics[width=0.2\textwidth]{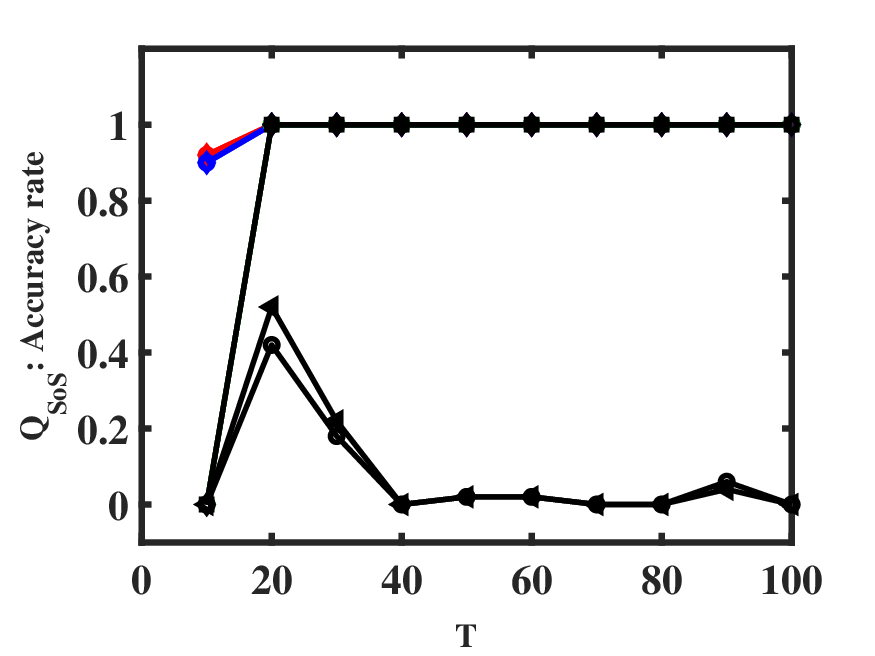}}
\subfigure[MLDCSBM case]{\includegraphics[width=0.2\textwidth]{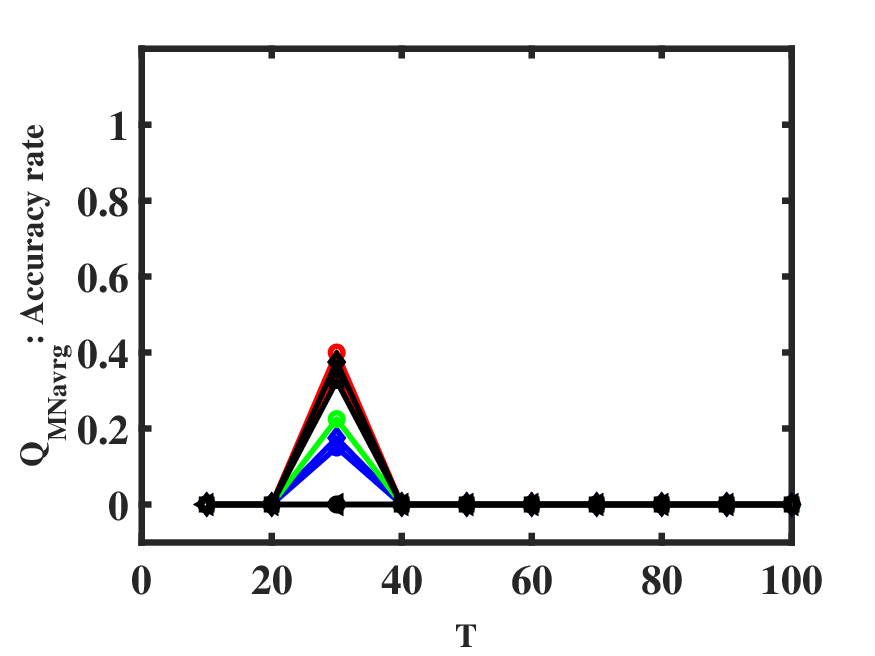}}
\subfigure[MLDCSBM case]{\includegraphics[width=0.2\textwidth]{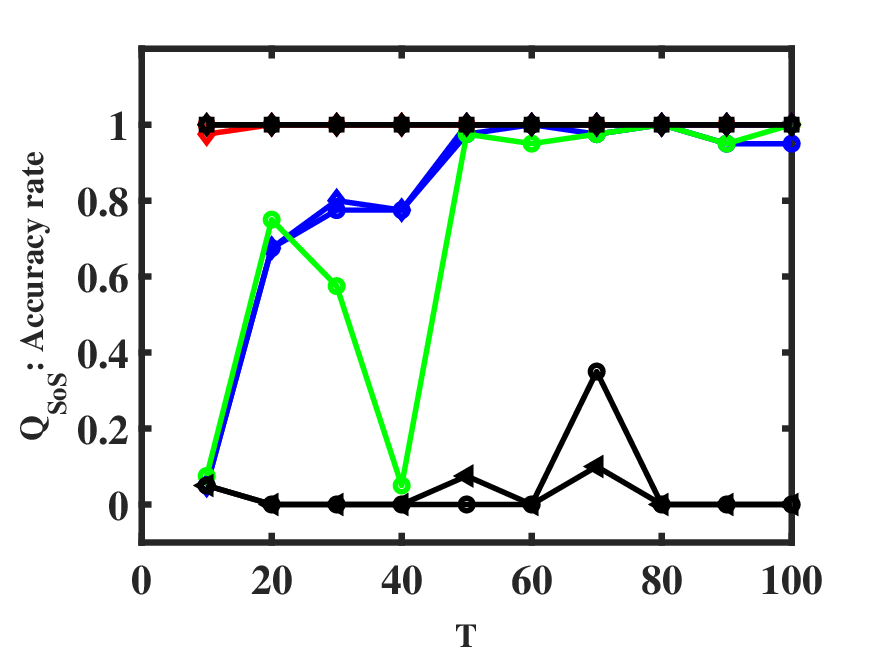}}
}
\caption{Accuracy rates of different methods for estimating $K$ in Simulation 2. Methods based on the MNavrg-modularity fails to find the true $K$. Except for RSum and DC-RSum, the remaining methods that estimate $K$ by maximizing the SoS-modularity exhibit high accuracy rates.}
\label{Ex2K} 
\end{figure}

\textbf{Simulation study 2: varying the number of layers $T$.} For the MLSBM case, we set $n=200$ and $\rho=0.04$, where $\rho$ is 0.16 for estimating $K$. For the MLDCSBM case, we set $n=500$ and $\rho=0.16$, where $\rho=0.6$ when estimating $K$. In both cases, we vary $T$ in $\{10, 20, \ldots, 100\}$. From Fig.~\ref{Ex2}, we observe that (a) as the number of layers grows, all approaches except RSum and DC-RSum enjoy better performance; (b) the proposed RDSoS, DC-RDSoS, RSoS, and DC-RSoS methods outperform their competitors in this simulation; (c) MASE runs slowest while the other eight methods enjoy similar running time in this simulation. The results presented in Fig.~\ref{Ex2K} reveal that all methods fail to identify the true $K$ via maximizing the MNavrg-modularity. In contrast, when estimating $K$ by maximizing our SoS-modularity, our proposed methods RDSoS, DC-RDSoS, RSoS, and DC-RSoS estimate $K$ with high accuracy. These findings suggest that our SoS-modularity serves as an effective metric for assessing the quality of communities in multi-layer networks.

\begin{figure}
\centering
\resizebox{\columnwidth}{!}{
\subfigure[MLSBM case]{\includegraphics[width=0.2\textwidth]{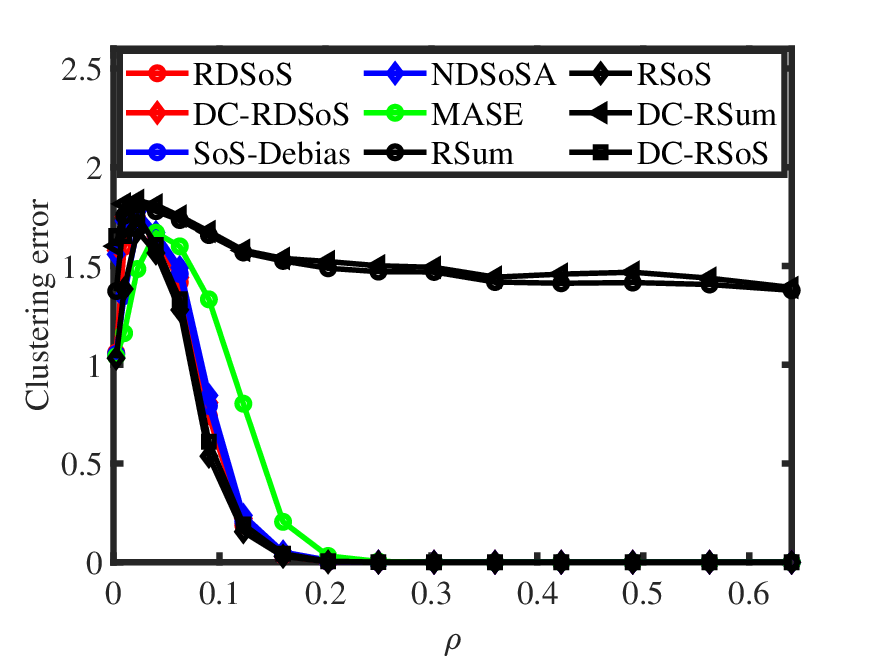}}
\subfigure[MLSBM case]{\includegraphics[width=0.2\textwidth]{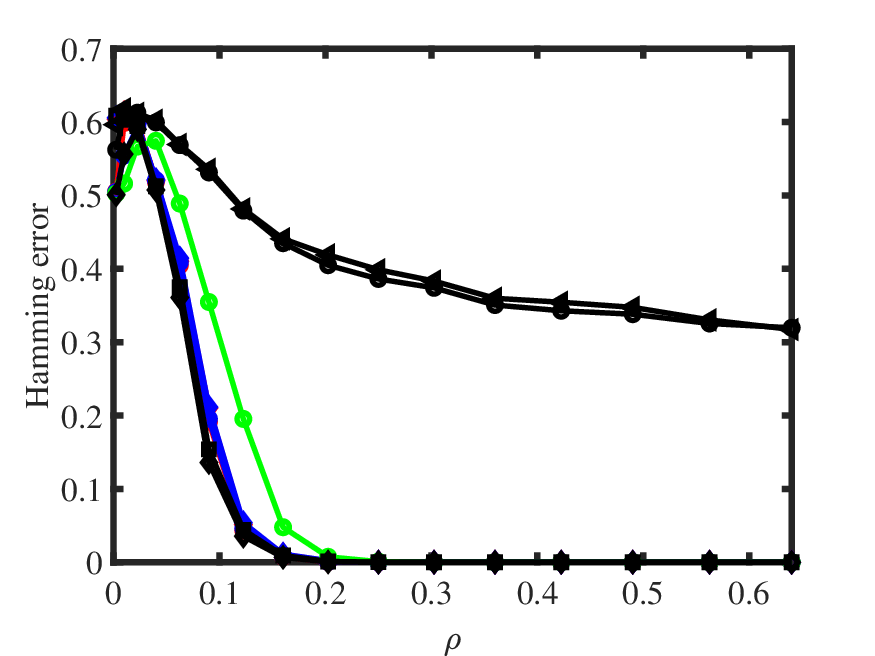}}
\subfigure[MLSBM case]{\includegraphics[width=0.2\textwidth]{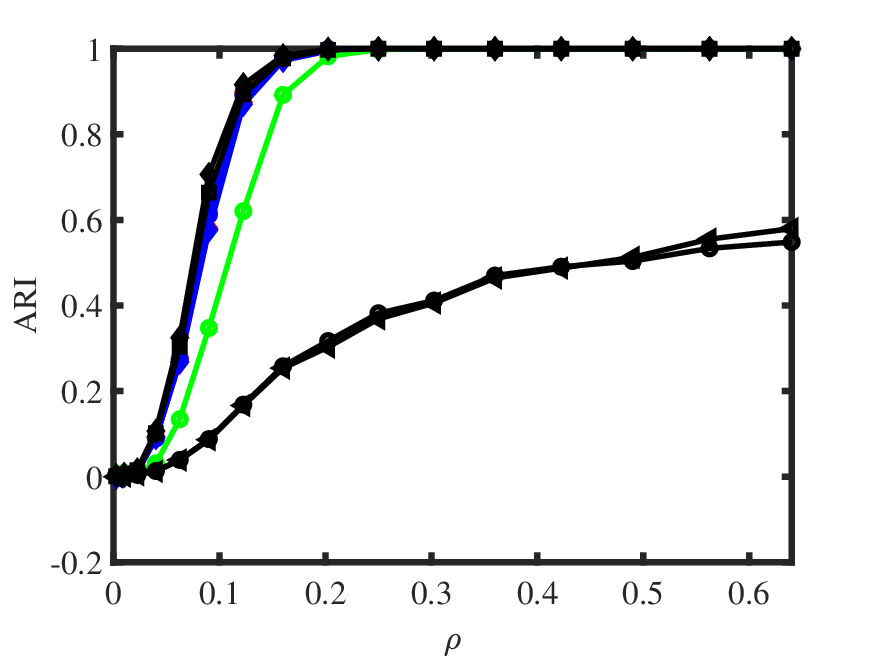}}
\subfigure[MLSBM case]{\includegraphics[width=0.2\textwidth]{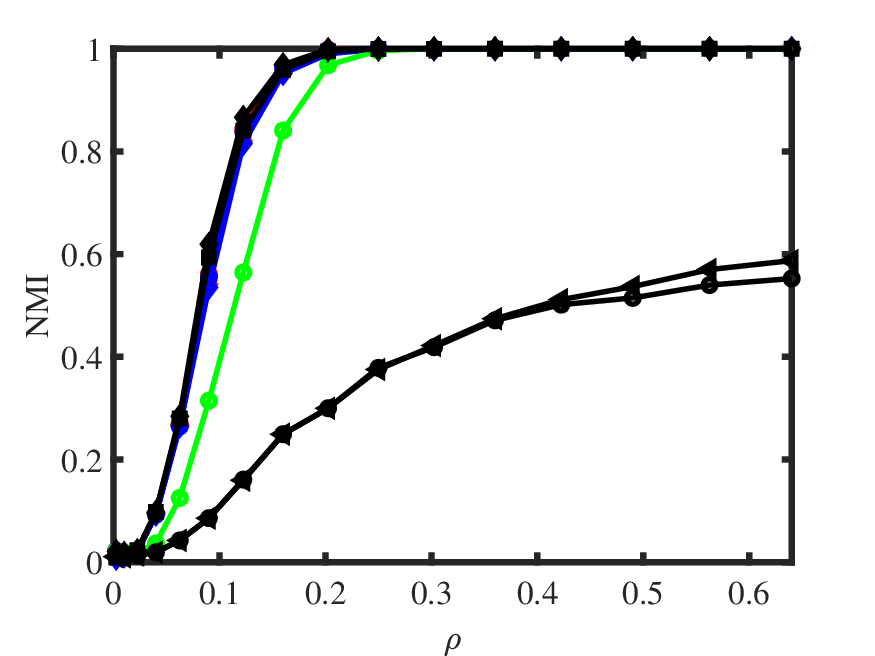}}
\subfigure[MLSBM case]{\includegraphics[width=0.2\textwidth]{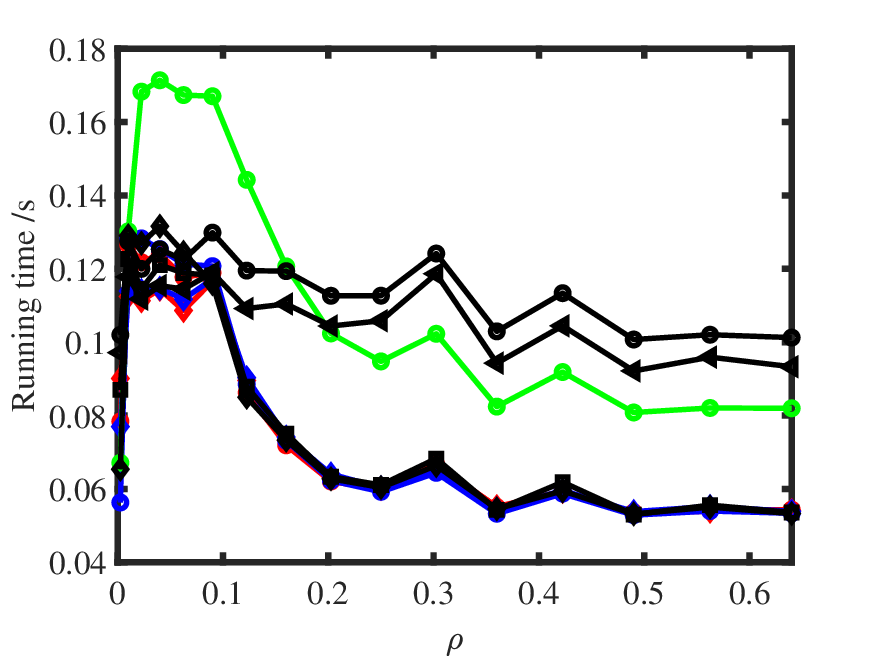}}
}
\resizebox{\columnwidth}{!}{
\subfigure[MLDCSBM case]{\includegraphics[width=0.2\textwidth]{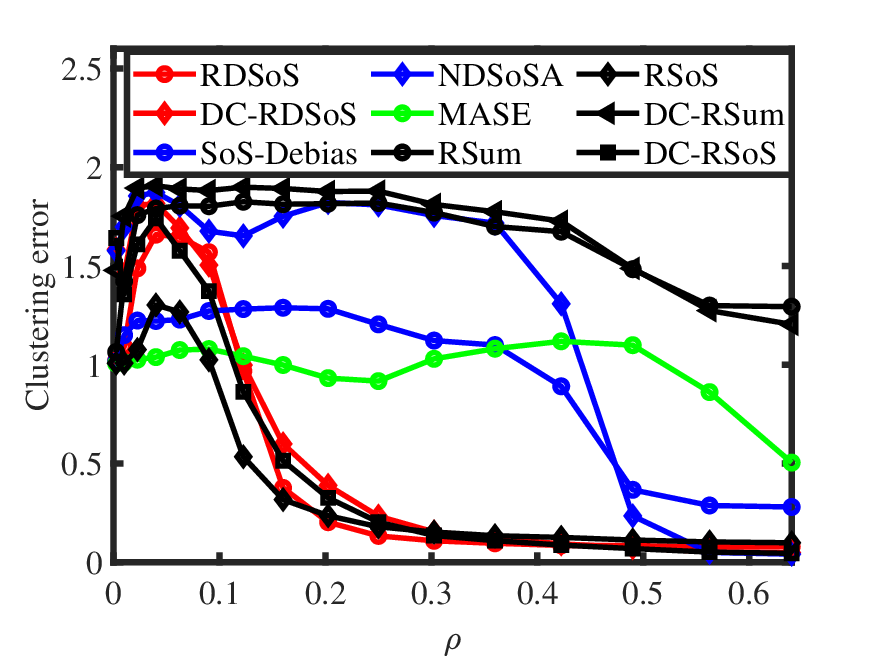}}
\subfigure[MLDCSBM case]{\includegraphics[width=0.2\textwidth]{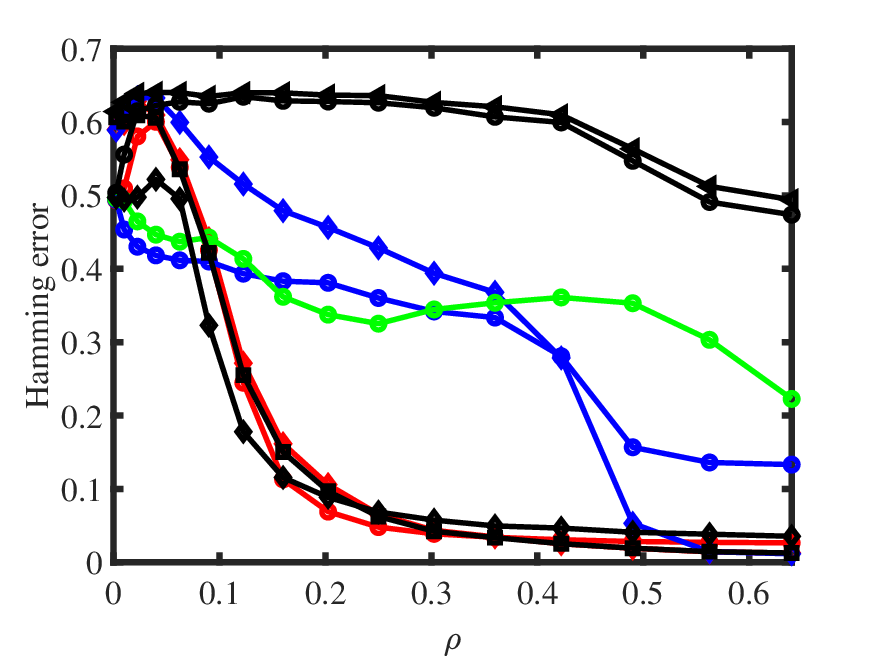}}
\subfigure[MLDCSBM case]{\includegraphics[width=0.2\textwidth]{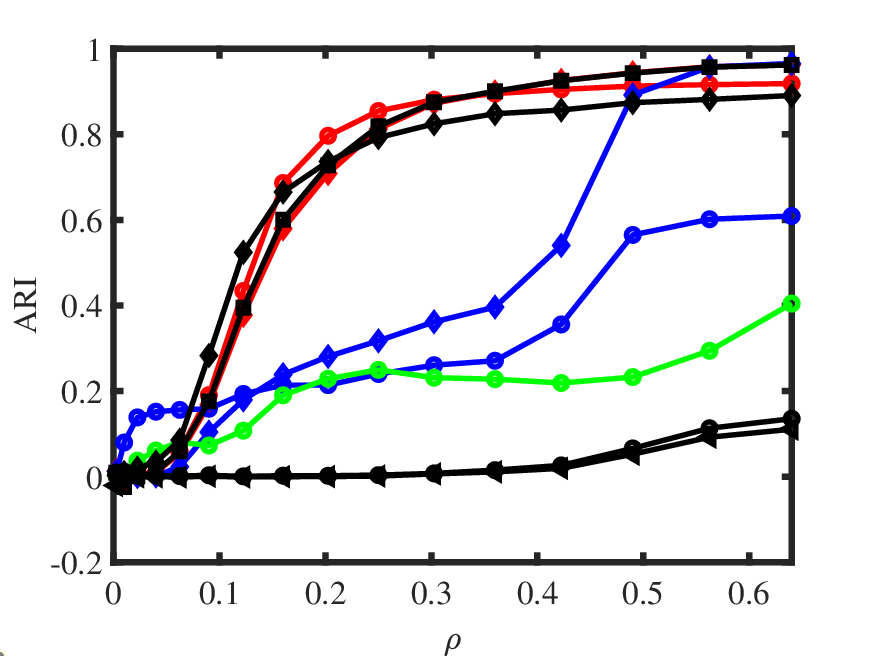}}
\subfigure[MLDCSBM case]{\includegraphics[width=0.2\textwidth]{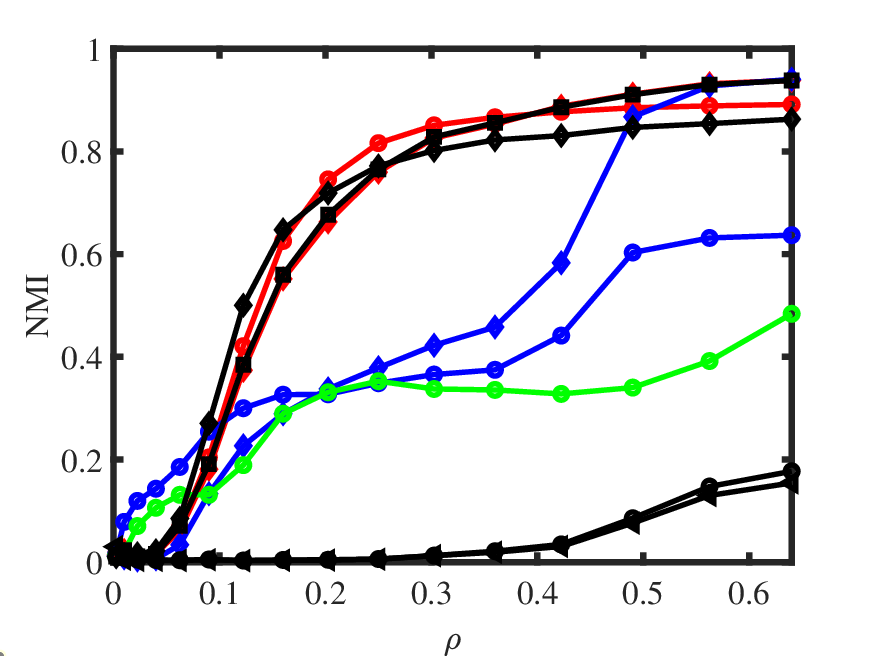}}
\subfigure[MLDCSBM case]{\includegraphics[width=0.2\textwidth]{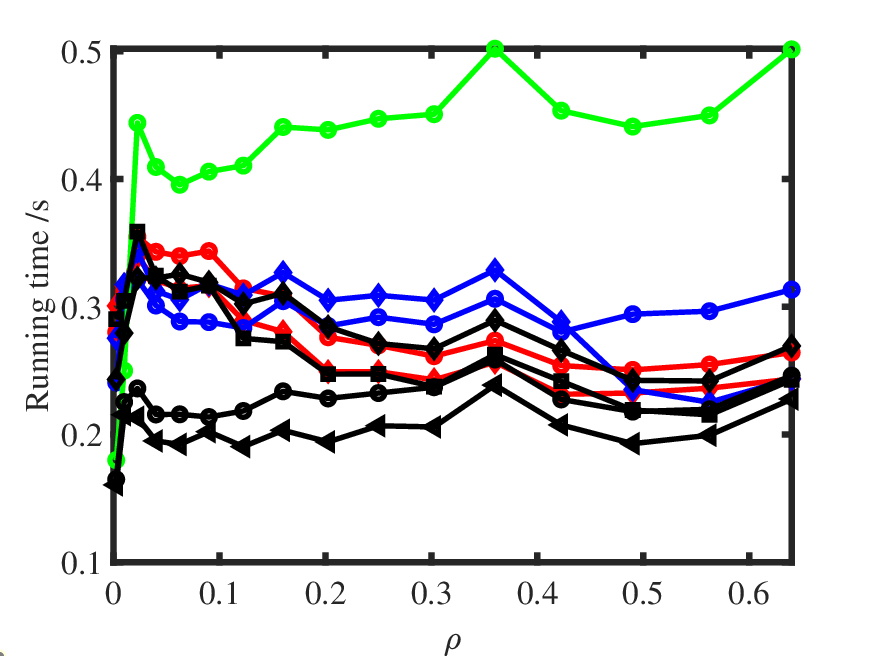}}
}
\caption{Performances of different methods for community detection in Simulation 3. The results indicate that as $\rho$ increases, all methods, except for RSum, and DC-RSum, exhibit improved performance. All methods perform similar in the MLSBM case and our methods perform best in the MLDCSBM case.}
\label{Ex3} 
\end{figure}

\begin{figure}
\centering
\resizebox{\columnwidth}{!}{
\subfigure[MLSBM case]{\includegraphics[width=0.2\textwidth]{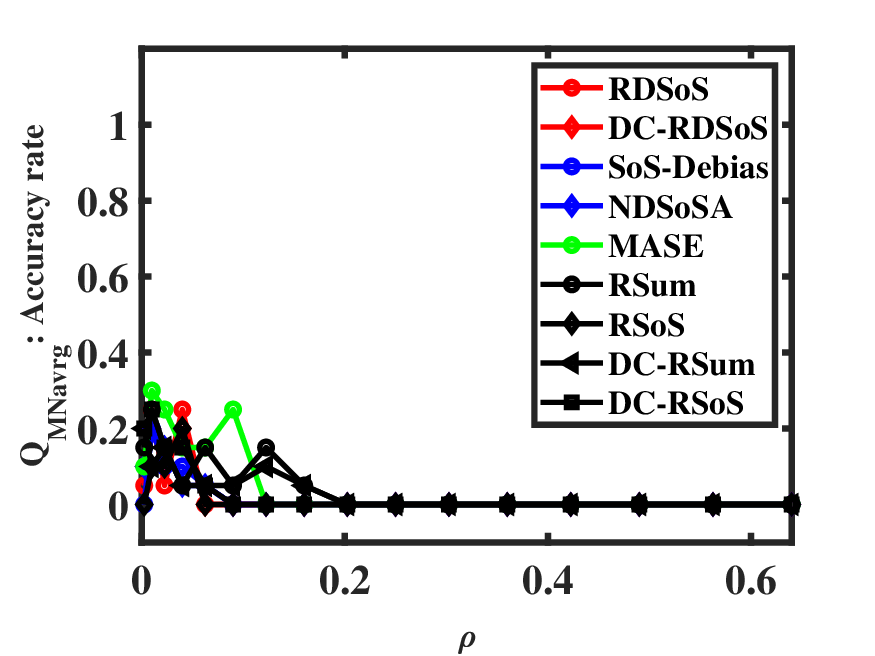}}
\subfigure[MLSBM case]{\includegraphics[width=0.2\textwidth]{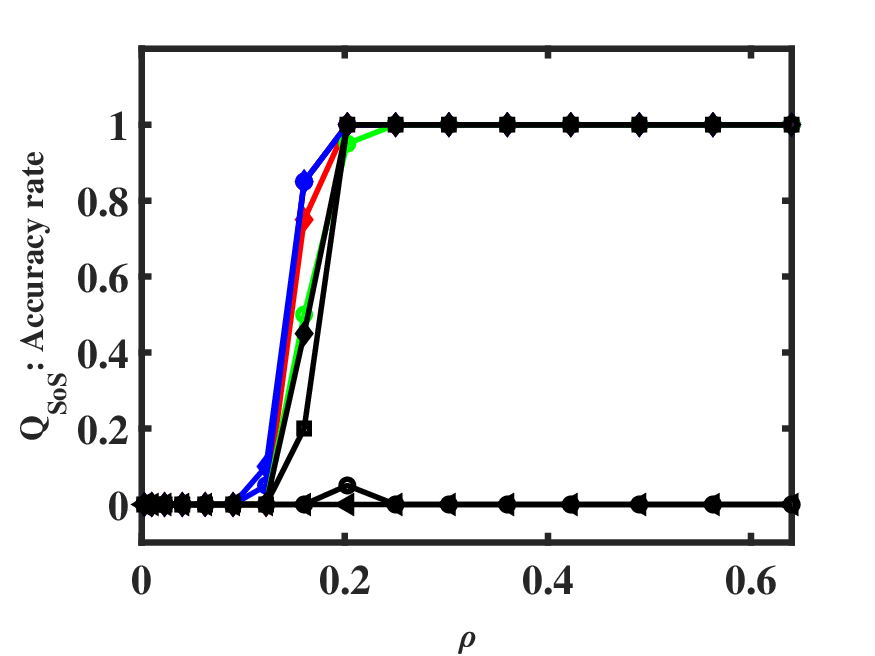}}
\subfigure[MLDCSBM case]{\includegraphics[width=0.2\textwidth]{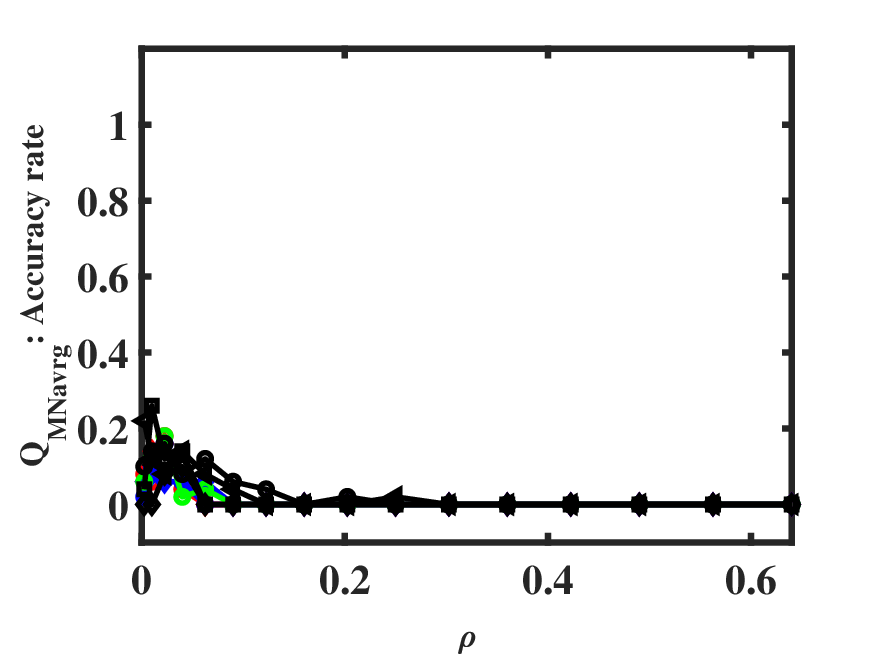}}
\subfigure[MLDCSBM case]{\includegraphics[width=0.2\textwidth]{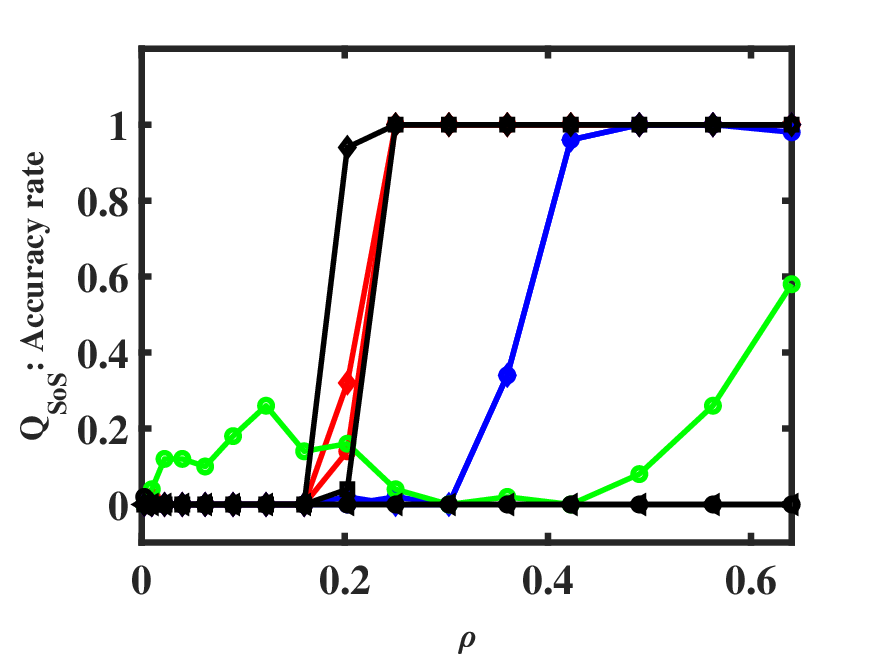}}
}
\caption{Accuracy rates of different methods for estimating $K$ in Simulation 3. Methods relying on the MNavrg-modularity fail to identify the true $K$. Except for RSum and DC-RSum, the remaining methods that estimate $K$ by maximizing the SoS-modularity generally demonstrate higher accuracy rates as the sparsity parameter $\rho$ increases.}
\label{Ex3K} 
\end{figure}

\textbf{Simulation study 3: varying the sparsity parameter $\rho$.} For the MLSBM case, we set $n=200$ and $T=10$. For the MLDCSBM case, we set $n=500$ and $T=50$. In both cases, we vary $\rho$ in $\{0.05^2, 0.1^2, 0.15^2, \ldots, 0.8^2\}$. According to Fig.~\ref{Ex3}, we observe that (a) each method except RSum and DC-RSum behaves better in the task of estimating $\ell$ when we increase the sparsity parameter $\rho$; (b) our RDSoS, DC-RDSoS, RSoS, and DC-RSoS have similar performances to SoS-Debias and NDSoSA in the MLSBM case; (c) our proposed methods significantly outperform SoS-Debias, NDSoSA, and MASE in the MLDCSBM case; (d) all methods run fast in this simulation. The results displayed in Fig.~\ref{Ex3K} suggest that our SoS-modularity outperforms its competitor, MNavrg-modularity, in assessing the quality of community detection in multi-layer networks. Furthermore, except RSum and DC-RSum, all methods estimate $K$ more precisely as the sparsity parameter $\rho$ increases when determining $K$ via maximizing our SoS-modularity.

\begin{figure}
\centering
\resizebox{\columnwidth}{!}{
\subfigure[MLSBM case]{\includegraphics[width=0.2\textwidth]{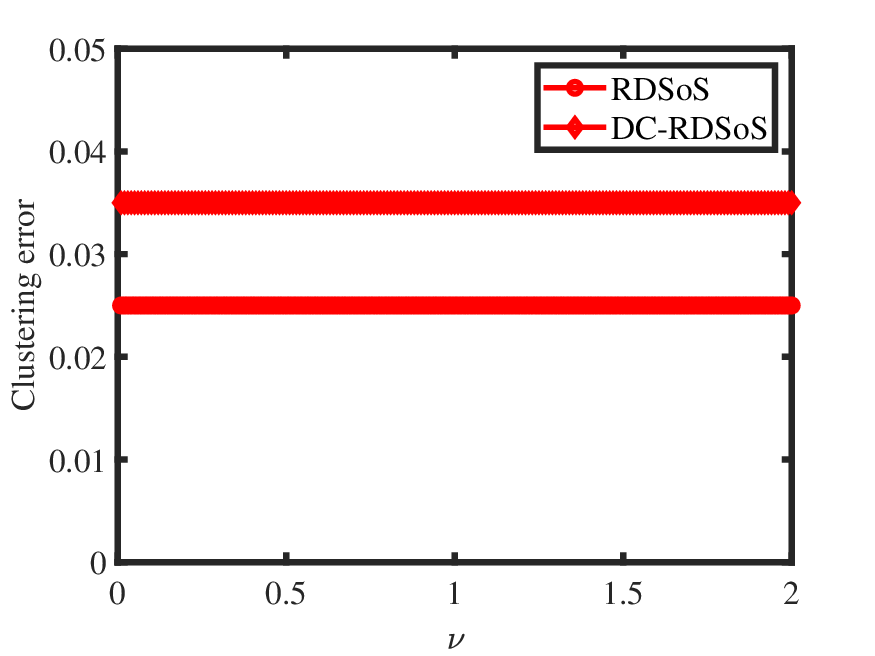}}
\subfigure[MLSBM case]{\includegraphics[width=0.2\textwidth]{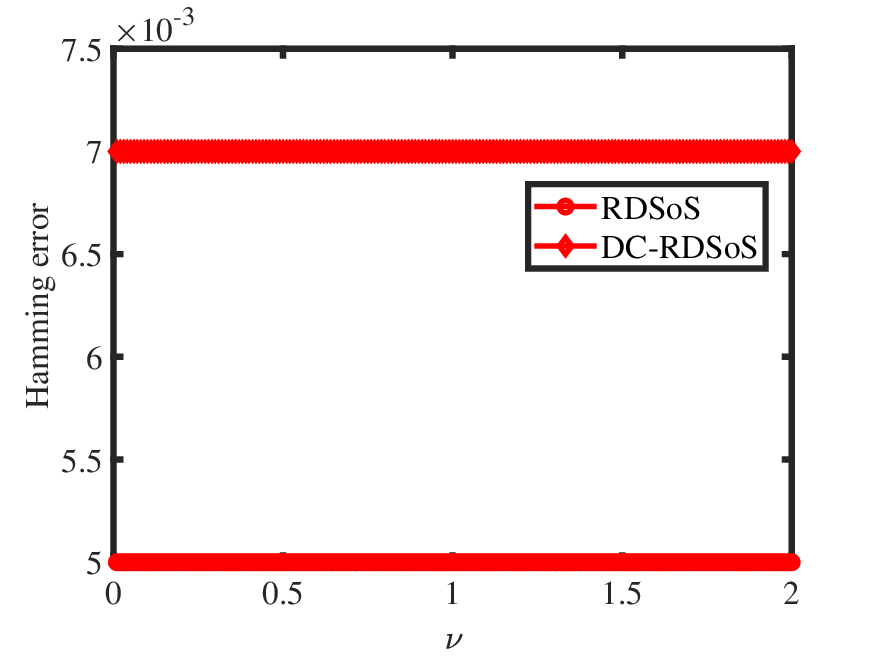}}
\subfigure[MLSBM case]{\includegraphics[width=0.2\textwidth]{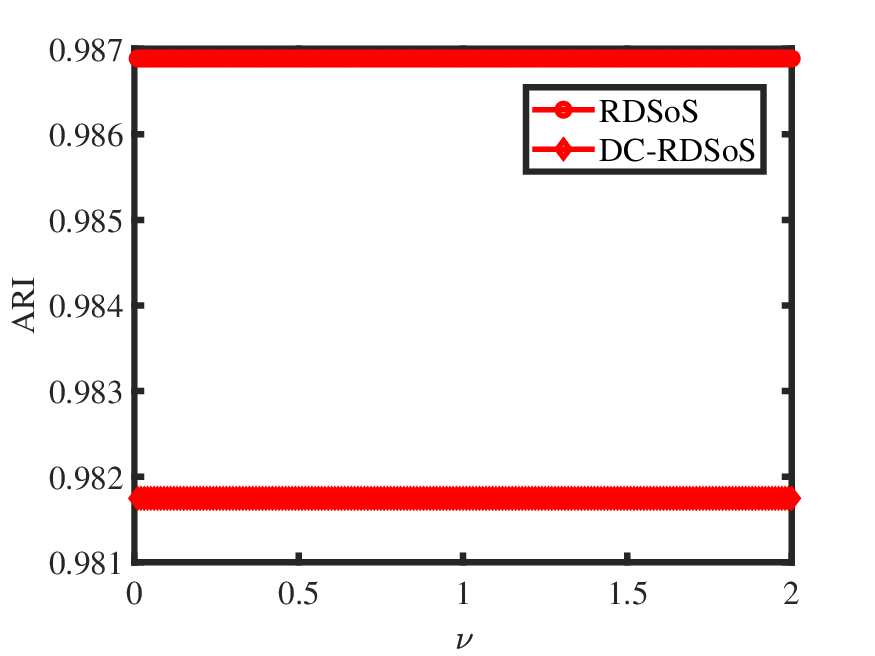}}
\subfigure[MLSBM case]{\includegraphics[width=0.2\textwidth]{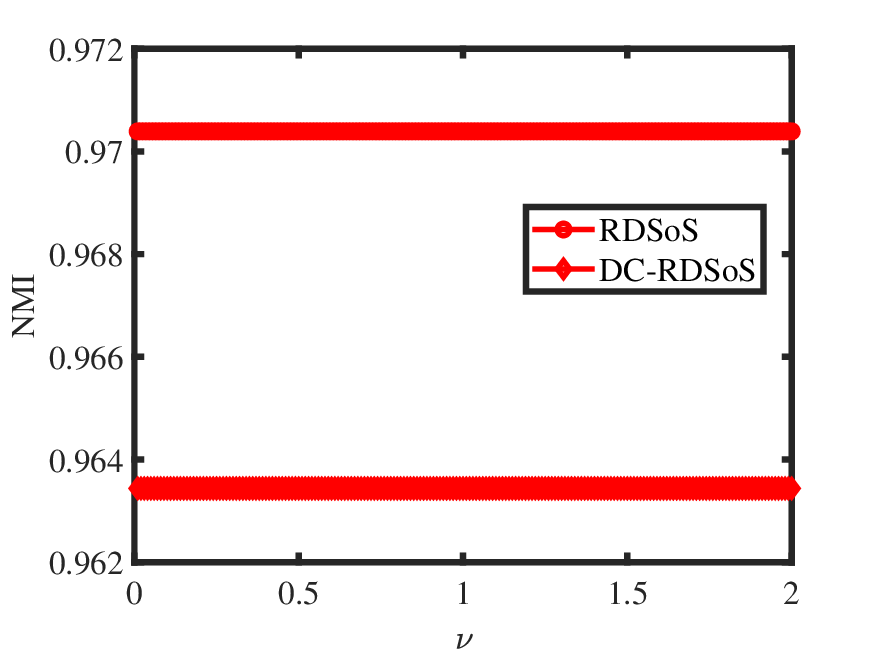}}
}
\resizebox{\columnwidth}{!}{
\subfigure[MLDCSBM case (a)]{\includegraphics[width=0.2\textwidth]{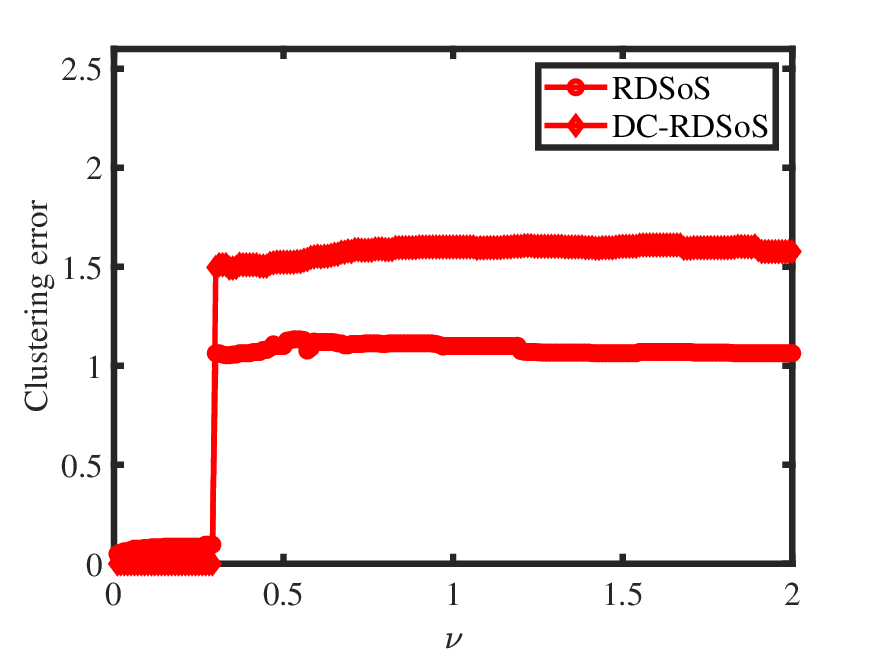}}
\subfigure[MLDCSBM case (a)]{\includegraphics[width=0.2\textwidth]{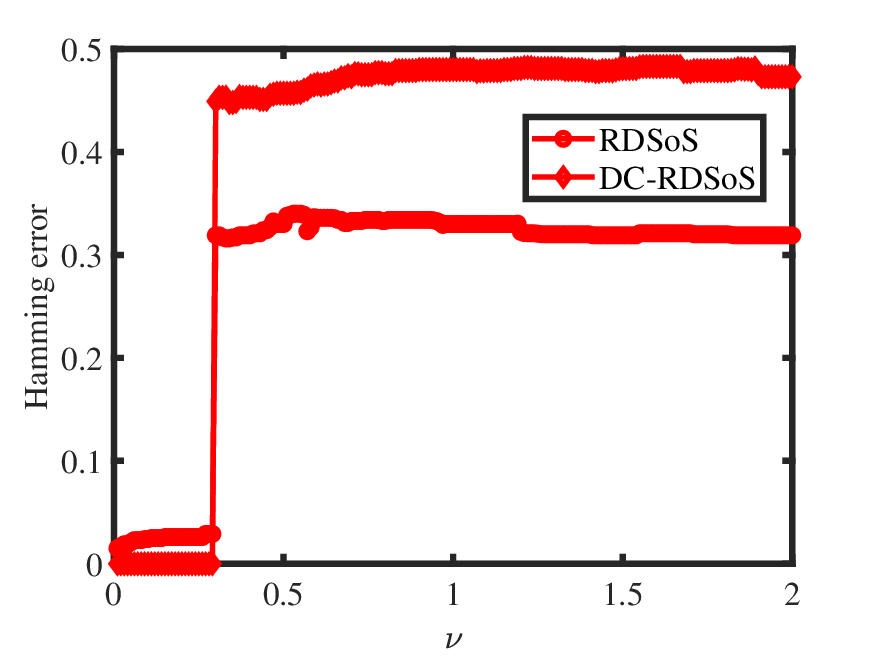}}
\subfigure[MLDCSBM case (a)]{\includegraphics[width=0.2\textwidth]{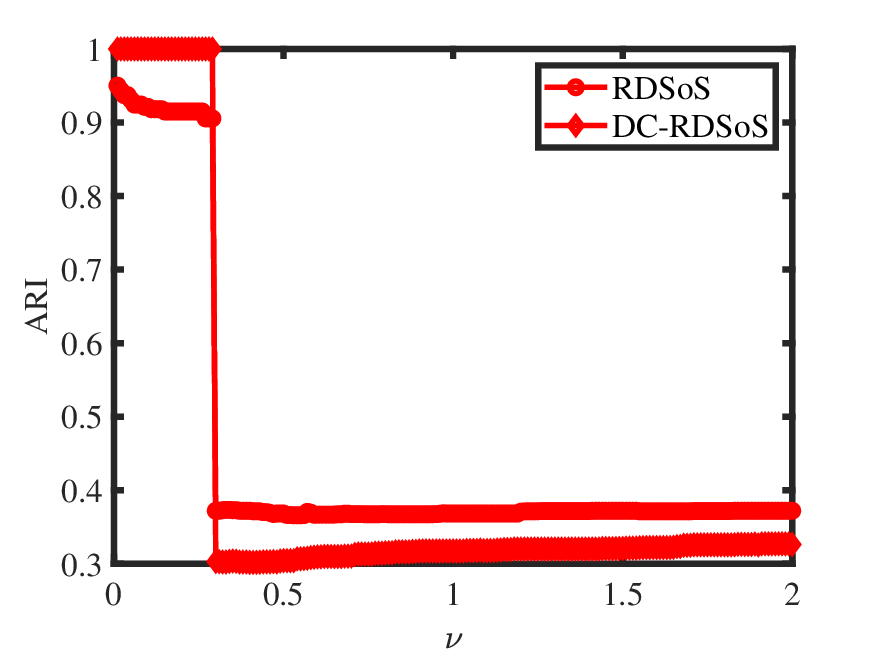}}
\subfigure[MLDCSBM case (a)]{\includegraphics[width=0.2\textwidth]{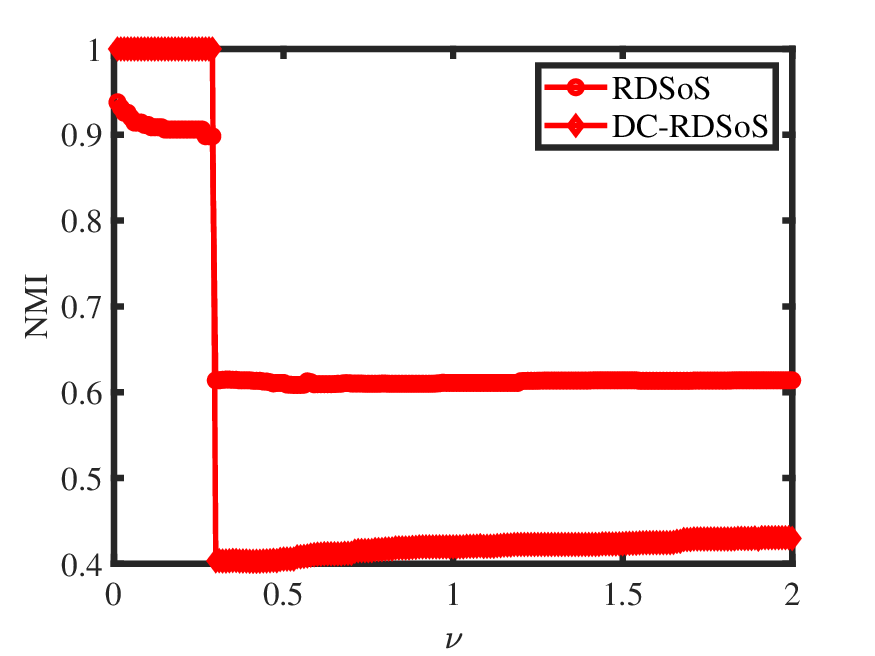}}
}
\resizebox{\columnwidth}{!}{
\subfigure[MLDCSBM case (b)]{\includegraphics[width=0.2\textwidth]{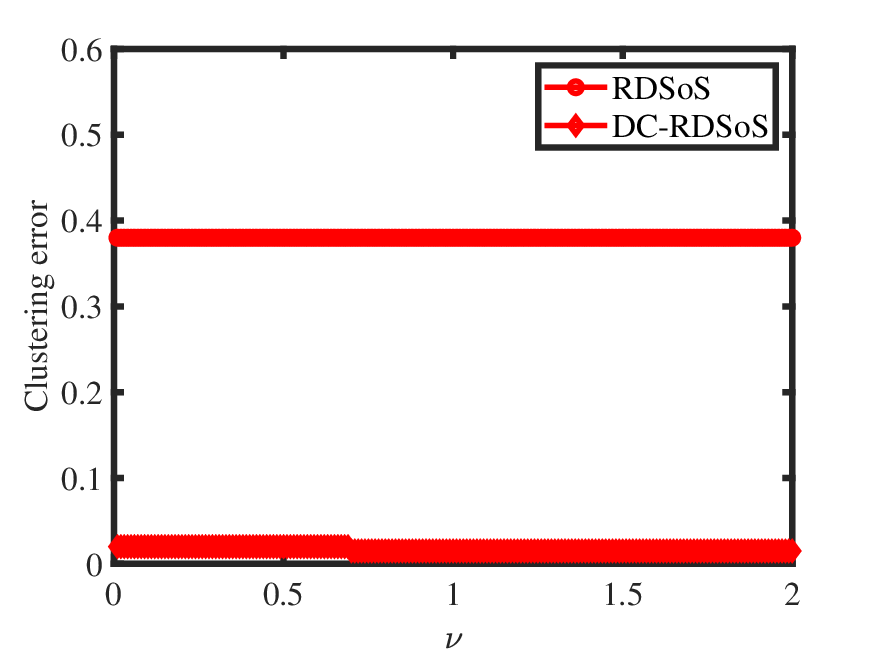}}
\subfigure[MLDCSBM case (b)]{\includegraphics[width=0.2\textwidth]{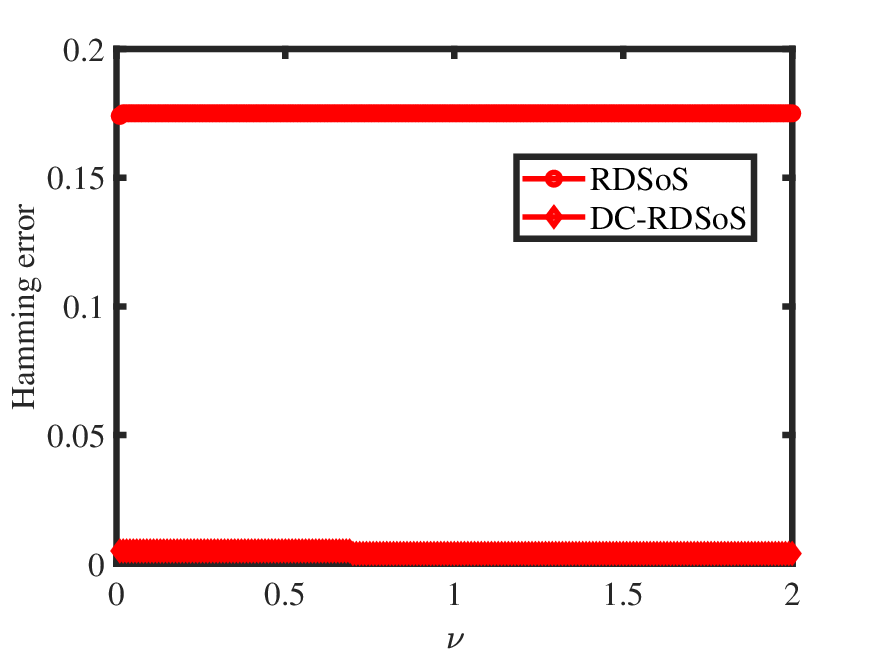}}
\subfigure[MLDCSBM case (b)]{\includegraphics[width=0.2\textwidth]{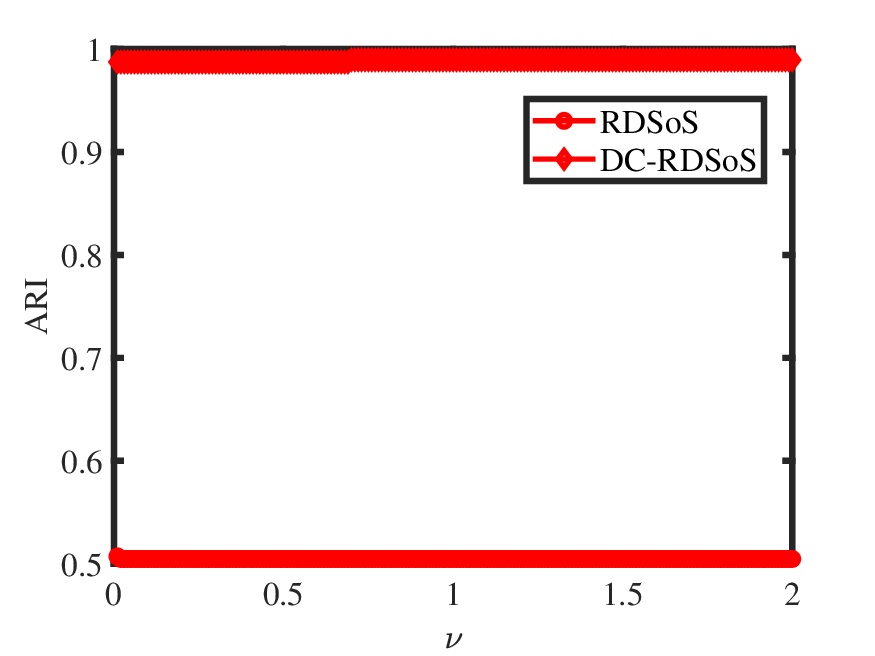}}
\subfigure[MLDCSBM case (b)]{\includegraphics[width=0.2\textwidth]{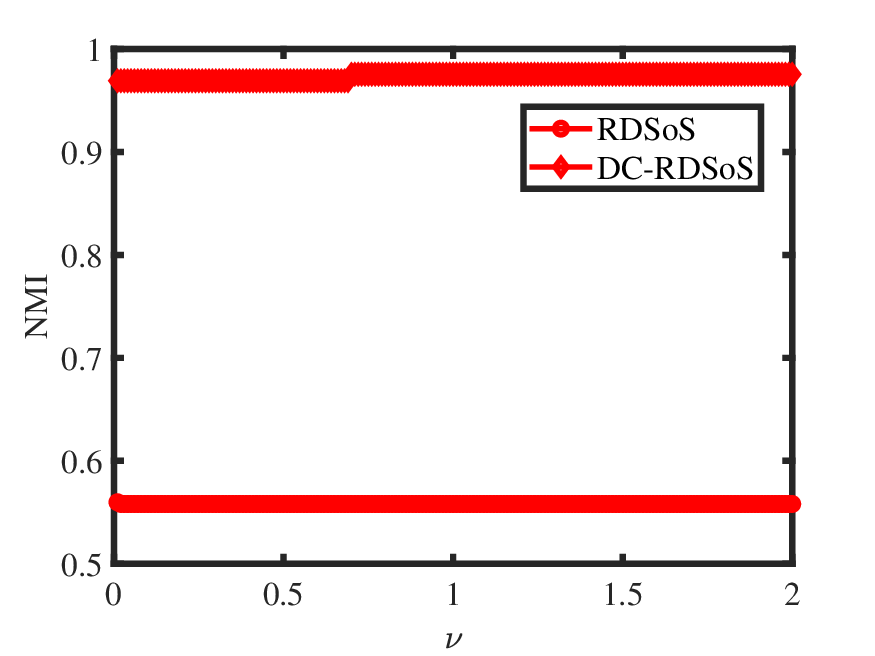}}
}
\caption{Performances of RDSoS and DC-RDSoS in Simulation 4.}
\label{Ex4} 
\end{figure}

\textbf{Simulation study 4: varying the regularizer $\tau$}. For this simulation, we set $n=1000$ and $T=20$. To investigate the impact of the regularizer $\tau$ on the performance of RDSoS and DC-RDSoS, we define $\tau$ as $\tau=\nu\frac{\sum_{i\in[n]}D(i,i)}{n}$ and allow $\nu$ to vary within the set $\{0.01, 0.02,\ldots,2\}$. For the MLSBM case, we set $\rho=0.02$. For the MLDCSBM case, we consider two scenarios: MLDCSBM cases (a) and (b). In MLDCSBM case (a), we set $\rho=1$ and define $\theta(i)$ as follows: $\theta(i)=0.9+\mathrm{rand}(1)/10$ if $\ell(i)=1$, $\theta(i)=0.4+\mathrm{rand}(1)/5$ if $\ell(i)=2$, and $\theta(i)=\mathrm{rand}(1)/10$ if $\ell(i)=3$ for $i\in[n]$. Given that $\mathrm{rand}(1)$ falls within the interval $(0,1)$, the degree heterogeneity parameter for nodes in the first community ranges between $0.9$ and $1$, for nodes in the second community it ranges between $0.4$ and $0.6$, and for nodes in the third community it ranges between $0$ and $0.1$. This setup ensures that the degree varies significantly across nodes belonging to different communities. In MLDCSBM case (b), we also set $\rho=1$ but allow $\theta(i)$ to be randomly selected from the set $\{0.9+\mathrm{rand}(1)/10, 0.4+\mathrm{rand}(1)/5, \mathrm{rand}(1)/10\}$ with equal probability for any node $i\in[n]$. Such a setting ensures that the degree variability of a node is primarily determined by itself rather than the community it belongs to. Additionally, this setting guarantees significant degree variation for the majority of nodes. Fig.~\ref{Ex4} presents the numerical results. Our observations indicate that (a) both RDSoS and DC-RDSoS exhibit insensitivity to the selection of $\tau$ in the MLSBM scenario; (b) in cases where the degree variability of a node is strongly influenced by the community it belongs to (i.e., MLDCSBM case (a)), both methods demonstrate robustness for large values of $\tau$, with a preference for moderate $\tau$ values; and (c) when the degree variability of a node is minimally affected by its community affiliation (i.e., MLDCSBM case (b)), both methods maintain robustness across various $\tau$ values.
\section{Real-world multi-layer networks}
\begin{figure}
\centering
\resizebox{\columnwidth}{!}{
\subfigure[LLF]{\includegraphics[width=0.5\textwidth]{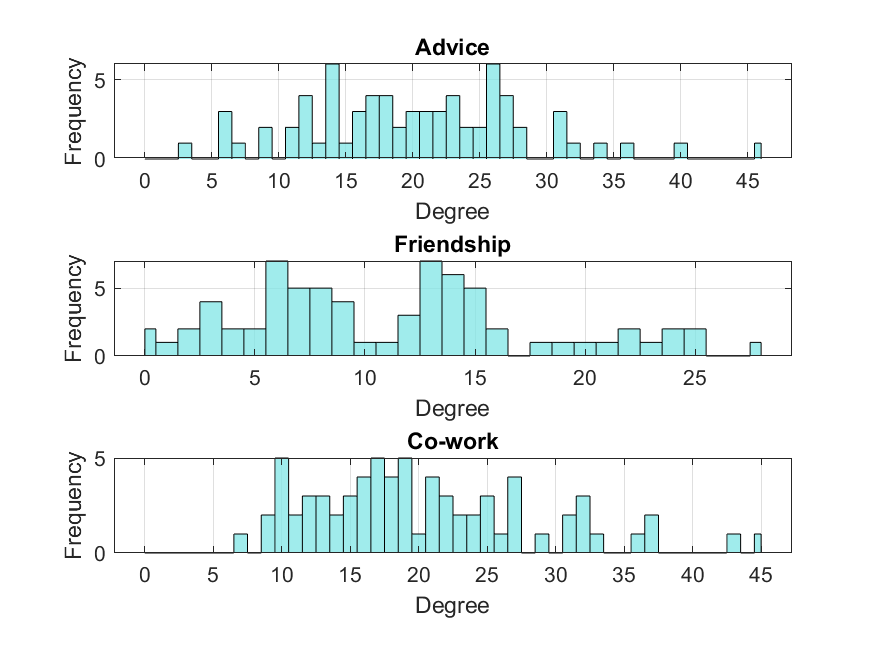}}
\subfigure[C.E]{\includegraphics[width=0.5\textwidth]{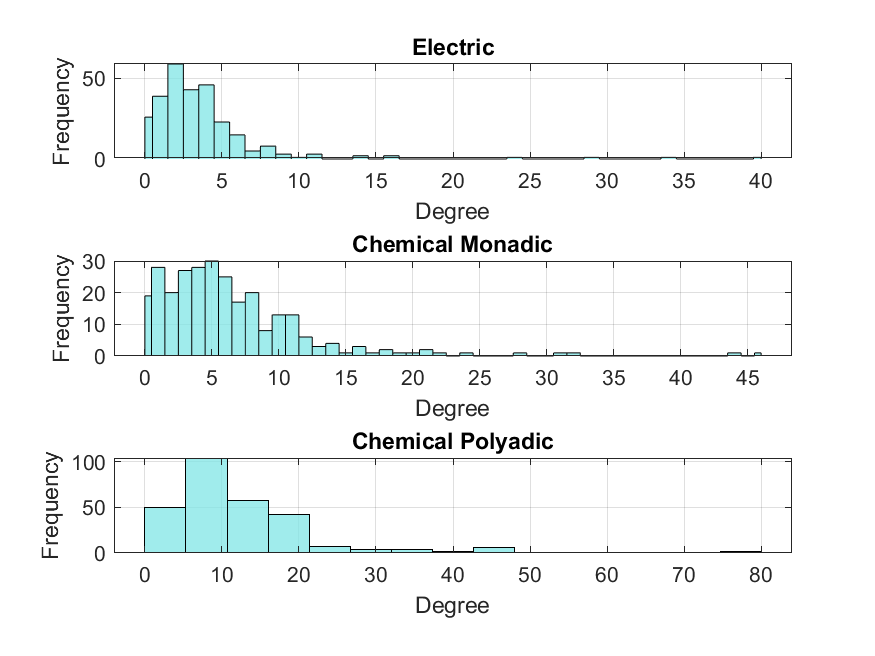}}
}
\resizebox{\columnwidth}{!}{
\subfigure[CS-A]{\includegraphics[width=0.5\textwidth]{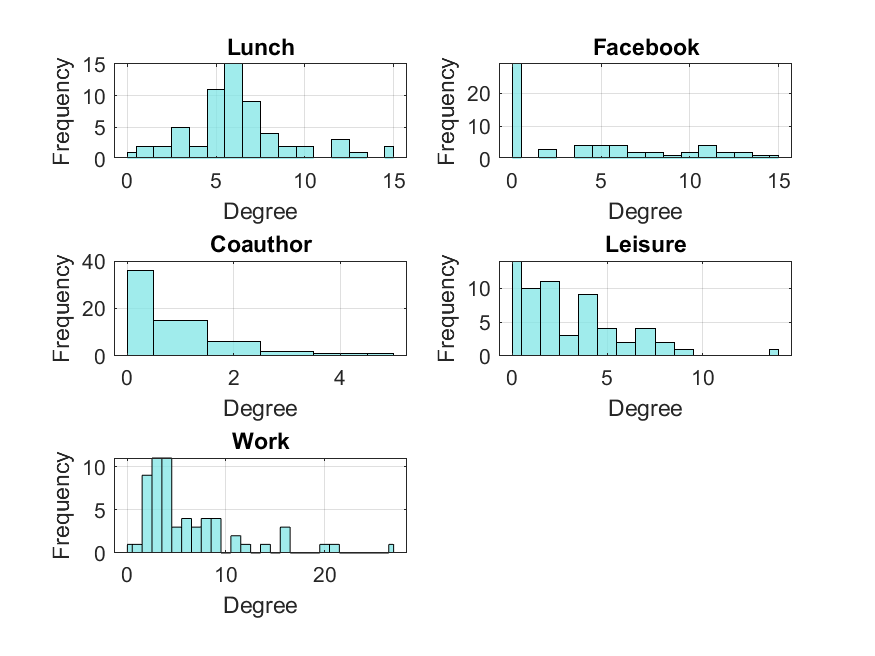}}
\subfigure[FAO-t]{\includegraphics[width=0.5\textwidth]{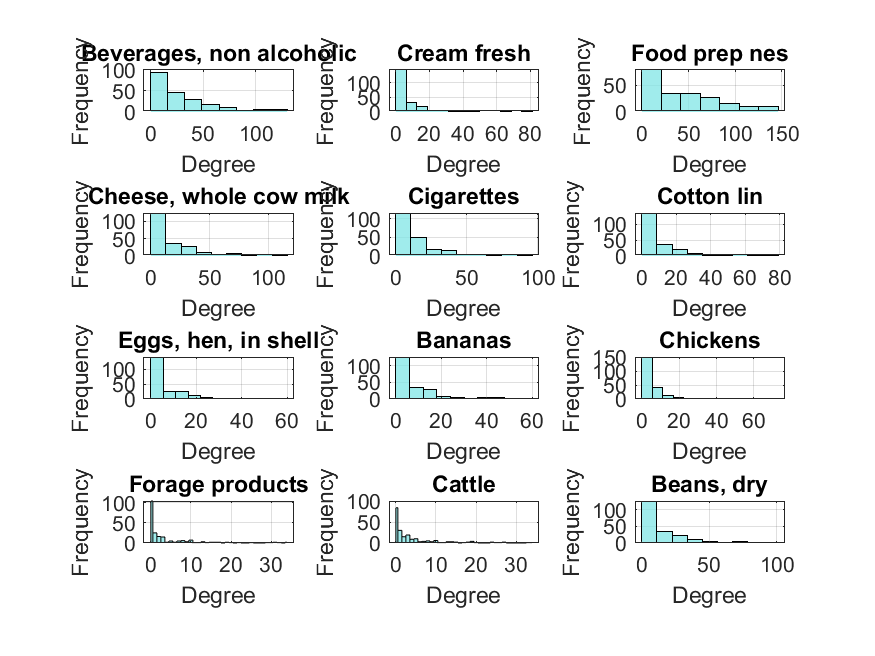}}
}
\caption{Degree distribution across layers for the four real multi-layer networks listed in Table \ref{realdata}.}
\label{DegreeRealData} 
\end{figure}

In this section, we assess the effectiveness of the proposed methods and their competitors in real-world multi-layer networks. We consider four real datasets, with their main characteristics summarized in Table \ref{realdata}. Fig.~\ref{DegreeRealData} presents the degree distributions for each layer within these datasets, highlighting significant variations in node degrees. Given that ground truth community labels and the number of communities are often unknown in real-world data, the five metrics used in Section \ref{secSim} are not applicable. Consequently, we employ our SoS-modularity and its competitor, MNavrg-modularity, to measure the quality of community partitions in real data when true node labels are unavailable. 

\begin{table}[h!]
\footnotesize
	\centering
	\caption{The four multi-layer networks analyzed in this work.}
	\label{realdata}
	\resizebox{\columnwidth}{!}{
	\begin{tabular}{cccccccccccc}
\hline
&Source&Node meaning&Edge meaning&Layer meaning&$n$&$T$\\
\hline
Lazega Law Firm (LLF)&\cite{snijders2006new}&Partners and associates&Communication &Advice/Friendship/Co-work&71&3\\
C.Elegans (C.E)&\cite{chen2006wiring}&Caenorhabditis elegans&Connectome&Electric/Chemical Monadic/Chemical Polyadic&279&3\\
CS-Aarhus (CS-A)&\cite{magnani2013combinatorial}&Employees&Relationship&Lunch/Facebook/Coauthor/Leisure/Work&61&5\\
FAO-trade (FAO-t)&\cite{de2015structural}&Countries&Import/export relationship&Food products&214&364\\
\hline
\end{tabular}}
\end{table}

\begin{figure}
\centering
\resizebox{\columnwidth}{!}{
\subfigure[LLF]{\includegraphics[width=0.2\textwidth]{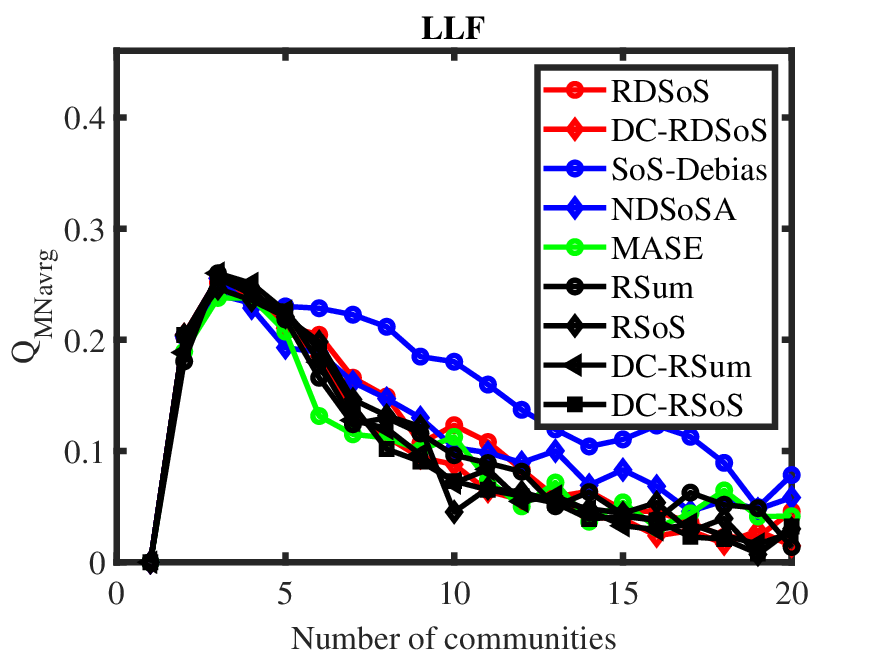}}
\subfigure[C.E]{\includegraphics[width=0.2\textwidth]{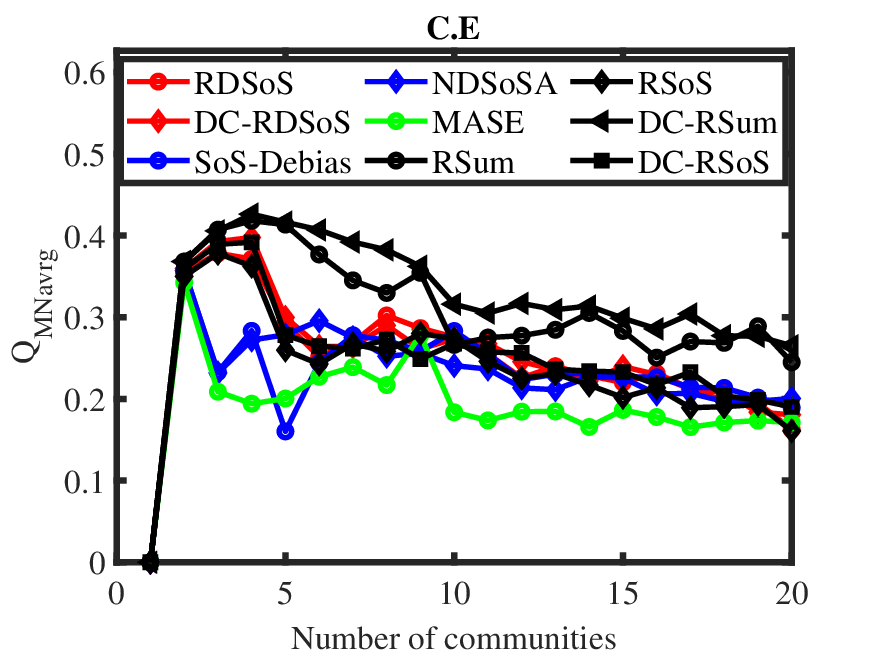}}
\subfigure[CS-A]{\includegraphics[width=0.2\textwidth]{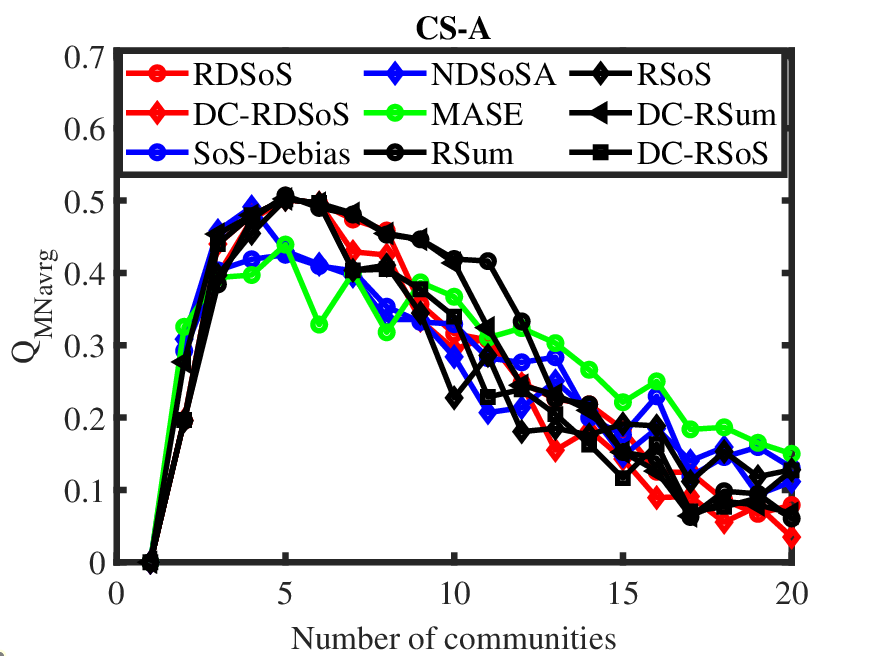}}
\subfigure[FAO-t]{\includegraphics[width=0.2\textwidth]{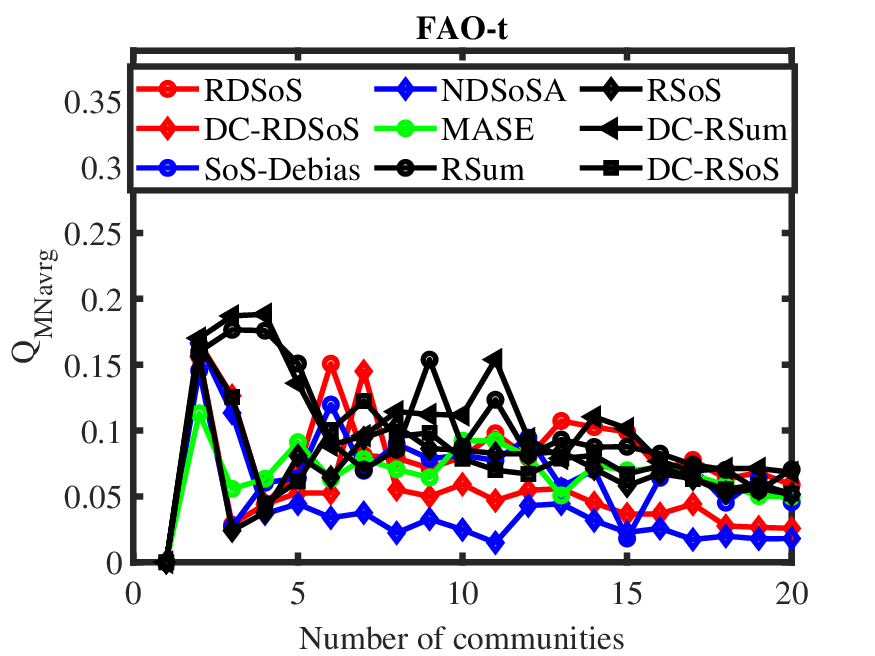}}
}
\caption{$Q_{MNavrg}$ against the possible number of communities $k$ for the four real multi-layer networks listed in Table \ref{realdata}.}
\label{Q} 
\end{figure}

\begin{figure}
\centering
\resizebox{\columnwidth}{!}{
\subfigure[LLF]{\includegraphics[width=0.2\textwidth]{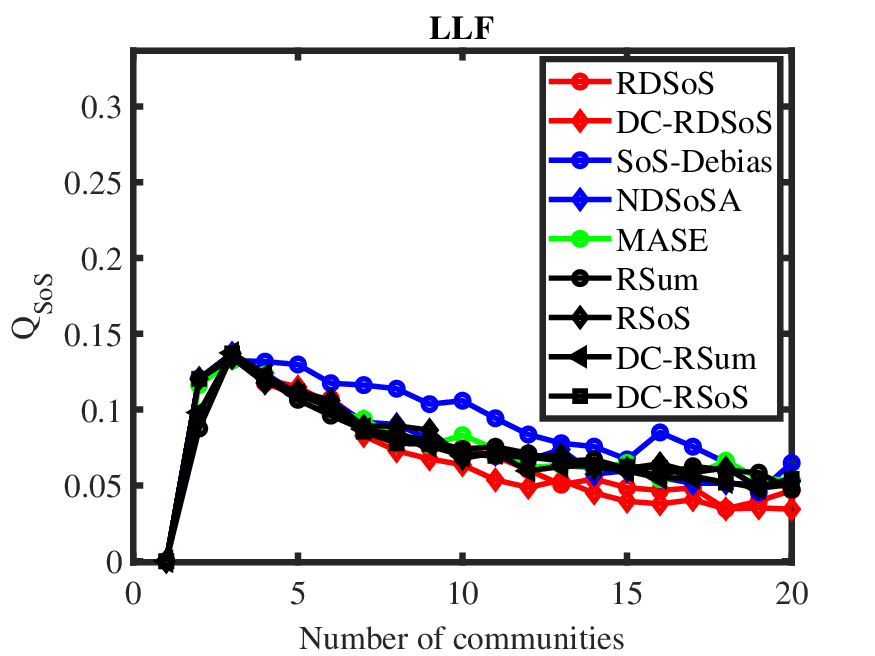}}
\subfigure[C.E]{\includegraphics[width=0.2\textwidth]{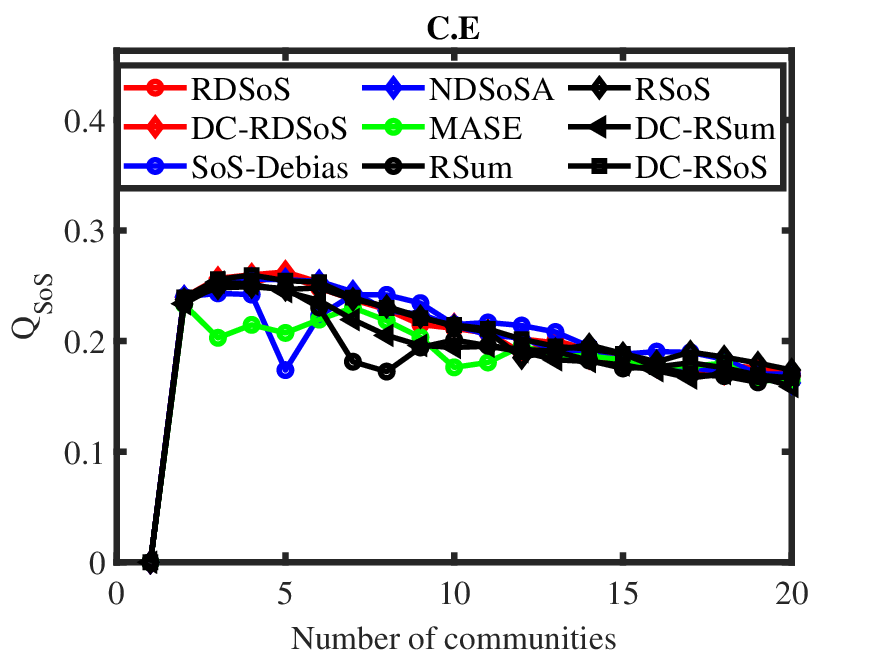}}
\subfigure[CS-A]{\includegraphics[width=0.2\textwidth]{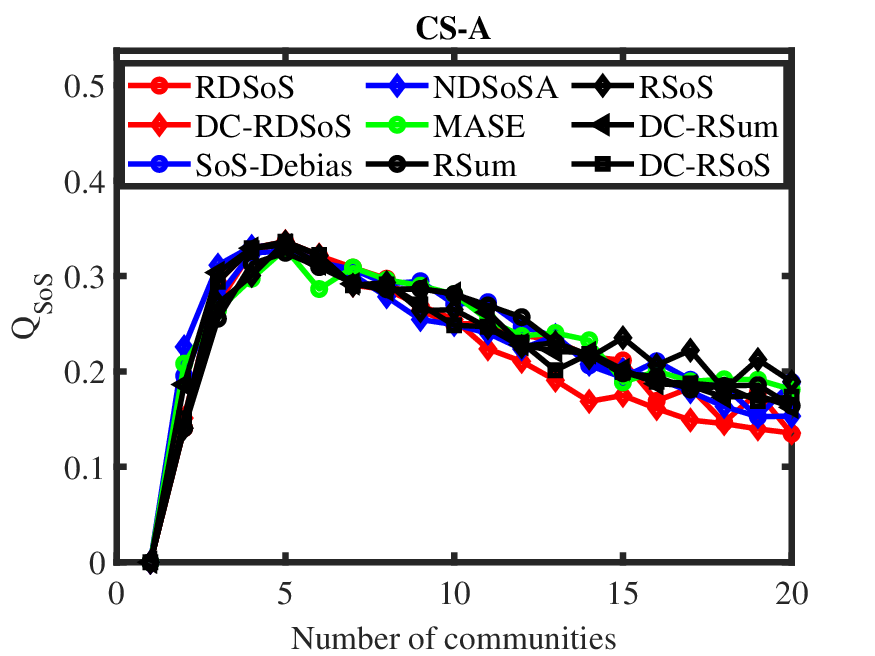}}
\subfigure[FAO-t]{\includegraphics[width=0.2\textwidth]{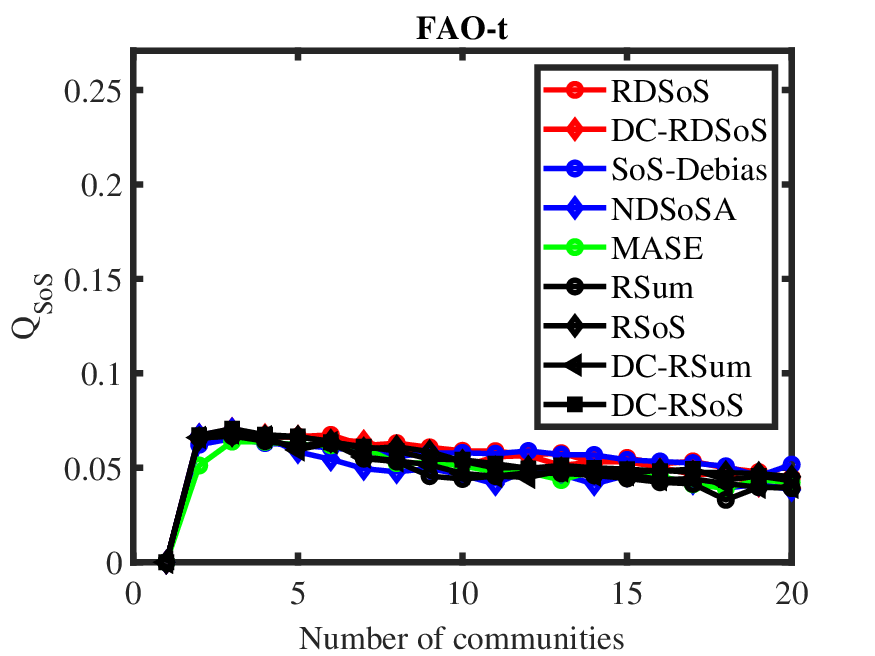}}
}
\caption{$Q_{SoS}$ against the possible number of communities $k$ for the four real multi-layer networks listed in Table \ref{realdata}.}
\label{QSoS} 
\end{figure}

\begin{table}[h!]
\footnotesize
	\centering
	\caption{$(K_{MNavrg,\mathcal{M}}, Q_{MNavrg,\mathcal{M}})$ for the four real multi-layer networks, with the largest $Q_{MNavrg,\mathcal{M}}$ in bold and the second-largest underlined.}
	\label{realdataQ}
	\begin{tabular}{cccccccccccc}
\hline
Dataset&RDSoS&DC-RDSoS&SoS-Debias&NDSoSA&MASE&RSum&RSoS&DC-RSum&DC-RSoS\\
\hline
LLF&(3,0.2515)&(3,\underline{0.2553})&(3,0.2406)&(3,\underline{0.2553})&(3,0.2377)&(3,\textbf{0.2599})&(3,0.2463)&(3,\textbf{0.2599})&(3,\underline{0.2553})\\
C.E&(3,0.3783)&(4,0.3976)&(2,0.3570)&(2,0.3570)&(2,0.3426)&(4,\underline{0.4181})&(3,0.3782)&(4,\textbf{0.4263})&(4,0.3917)\\
CS-A&(5,0.5015)&(5,0.5018)&(5,0.4252)&(4,0.4913)&(5,0.4392)&(5,\textbf{0.5071})&(5,0.5015)&(5,\underline{0.5023})&(5,0.5018)\\
FAO-t&(2,0.1563)&(2,0.1668)&(2,0.1456)&(2,0.1658)&(2,0.1131)&(3,\underline{0.1764})&(2,0.1563)&(4,\textbf{0.1883})&(2,0.1658)\\
\hline
\end{tabular}
\end{table}

\begin{table}[h!]
\footnotesize
	\centering
	\caption{$(K_{SoS,\mathcal{M}}, Q_{SoS,\mathcal{M}})$ for the four real multi-layer networks, with the largest $Q_{SoS,\mathcal{M}}$ in bold and the second-largest underlined.}
	\label{realdataQSoS}
	\begin{tabular}{cccccccccccc}
\hline
Dataset&RDSoS&DC-RDSoS&SoS-Debias&NDSoSA&MASE&RSum&RSoS&DC-RSum&DC-RSoS\\
\hline
LLF&(3,0.1359)&(3,\underline{0.1370})&(3,0.1325)&(3,\underline{0.1370})&(3,0.1324)&(3,0.1365)&(3,0.1342)&(3,\textbf{0.1374})&(3,0.1370)\\
C.E&(5,\underline{0.2600})&(5,\textbf{0.2625})&(3,0.2432)&(4,0.2560)&(2,0.2331)&(3,0.2505)&(4,0.2492)&(3,0.2533)&(4,0.2597)\\
CS-A&(5,\underline{0.3357})&(5,\textbf{0.3361})&(5,0.3262)&(5,0.3314)&(5,0.3286)&(5,0.3244)&(5,\underline{0.3357})&(5,0.3327)&(5,\textbf{0.3361})\\
FAO-t&(6,0.0674)&(3,\underline{0.0703})&(3,0.0654)&(3,0.0702)&(3,0.0638)&(3,0.0668)&(3,0.0673)&(3,0.0685)&(3,\textbf{0.0707})\\
\hline
\end{tabular}
\end{table}

Fig.~\ref{Q} (or Fig.~\ref{QSoS}) shows how $Q_{MNavrg}$ (or $Q_{SoS}$) varies with the number of communities for each network. From these results, we can find the $k$ that gives the highest $Q_{MNavrg}$ (or $Q_{SoS}$) for each method and dataset. Tables~\ref{realdataQ} and~\ref{realdataQSoS} list the specific $(K_{MNavrg,\mathcal{M}}, Q_{MNavrg,\mathcal{M}})$ and $(K_{SoS,\mathcal{M}}, Q_{SoS,\mathcal{M}})$ values for each method across all real datasets. Our observations are as follows:
\begin{itemize}
\item The community structure of CS-A is more pronounced than that of LLF, C.E, and FAO-t, as evidenced by the highest maximum MNavrg-modularity and SoS-modularity values for CS-A in Tables \ref{realdataQ} and \ref{realdataQSoS} across all methods. By contrast, FAO-t's community structure is the least pronounced, with its maximum MNavrg-modularity and SoS-modularity values being lower than the minimum values observed for the other datasets across all methods.
\item In general, the six methods, namely RDSoS, DC-RDSoS, RSum, RSoS, DC-RSum, and DC-RSoS, which are formulated based on an application of our regularized Laplacian matrix $L_{\tau}$, yield higher values of MNavrg-modularity and SoS-modularity compared to SoS-Debias, NDSoSA, and MASE, thereby highlighting the advantages of regularization.
\item DC-RSum yields the highest $Q_{MNavrg}$ values across all datasets when utilizing the MNavrg-modularity. For every network, the $Q_{MNavrg}$ values of RSum and DC-RSum surpass those of RDSoS (or RSoS) and DC-RDSoS (or DC-RSoS), respectively. Nevertheless, these findings are at odds with the results obtained from all simulation experiments presented in Section \ref{secSim} which demonstrate that RDSoS, RSoS, DC-RDSoS, and DC-RSoS almost always outperform RSum and DC-RSum in community detection. This further confirms that the MNavrg-modularity is not a reliable measure for assessing the quality of community partitions in multi-layer networks.
  \item Our RDSoS, DC-RDSoS, RSoS, and DC-RSoS methods exhibit competitive performance, often matching or exceeding the best results of their competitors when using our SoS-modularity. Specifically, DC-RDSoS outperforms the other methods for C.E and CS-A. DC-RSoS achieves the best community partitions for FAO-t. More significantly, our RDSoS (or RSoS) and DC-RDSoS (or DC-RSoS) typically surpass RSum and DC-RSum, respectively, when utilizing our SoS-modularity. These outcomes align with our numerical results presented in Section \ref{secSim}, thereby reinforcing the credibility and effectiveness of our proposed SoS-modularity as a measure for assessing the quality of community detection in multi-layer networks.
  \item Considering that our SoS-modularity offers a more reliable assessment of community quality in multi-layer networks, we now determine the number of communities $K$ by maximizing the $Q_{SoS}$ value. According to Table \ref{realdataQSoS}, our findings are as follows: For the LLF network, all methods consistently identify 3 communities. In the case of C.E and CS-A networks, $K$ should be set to 5, as DC-RDSoS attains the highest SoS-modularity values at $k=5$ for both. Meanwhile, for the CS-A network, all methods identify 5 communities. Likewise, for the FAO-t network, $K$ should be set to 3, with all methods except RDSoS agreeing that the FAO-t network consists of 3 communities.
\end{itemize}

\begin{figure}
\centering
\resizebox{\columnwidth}{!}{
\subfigure[LLF]{\includegraphics[width=0.2\textwidth]{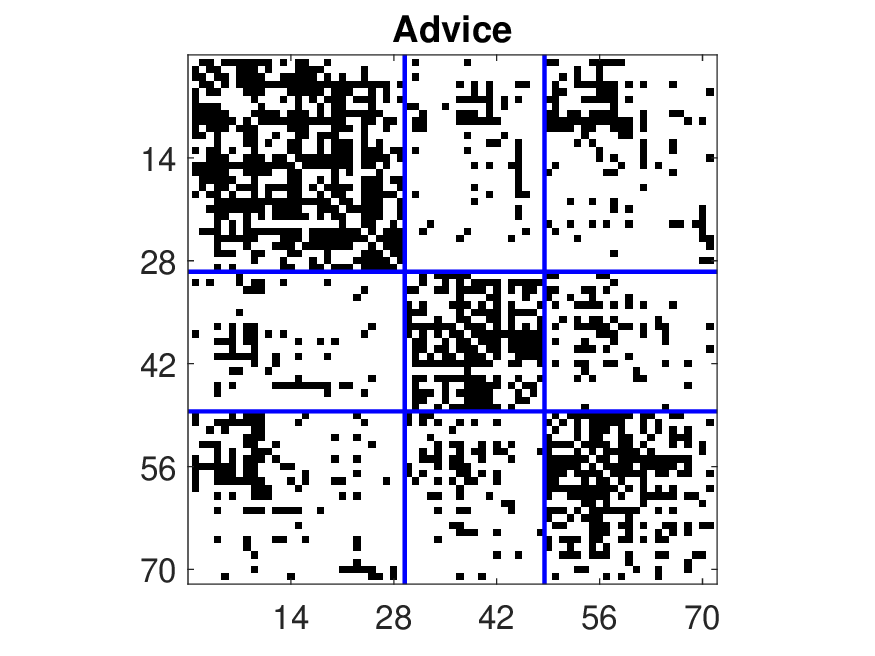}}
\subfigure[LLF]{\includegraphics[width=0.2\textwidth]{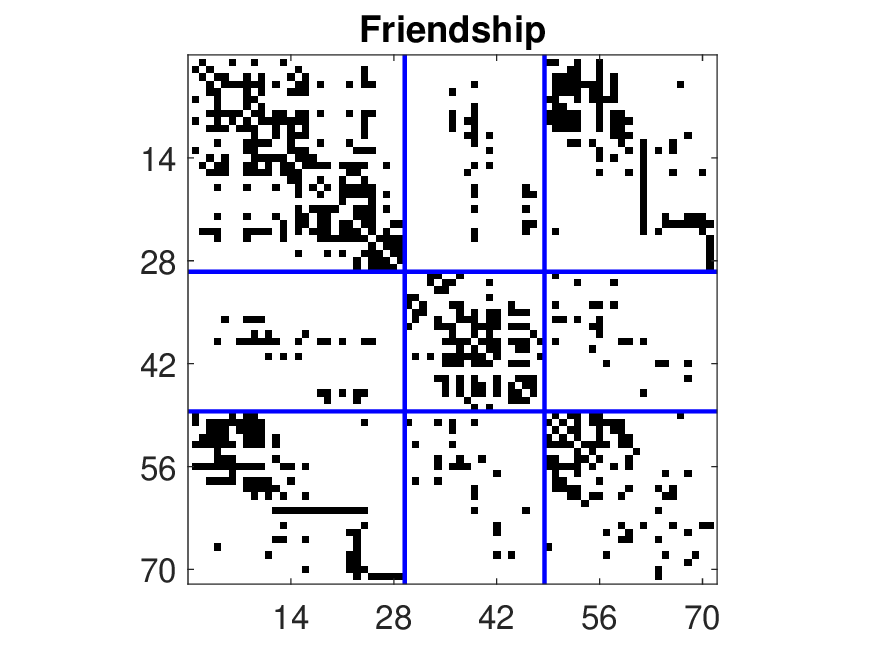}}
\subfigure[LLF]{\includegraphics[width=0.2\textwidth]{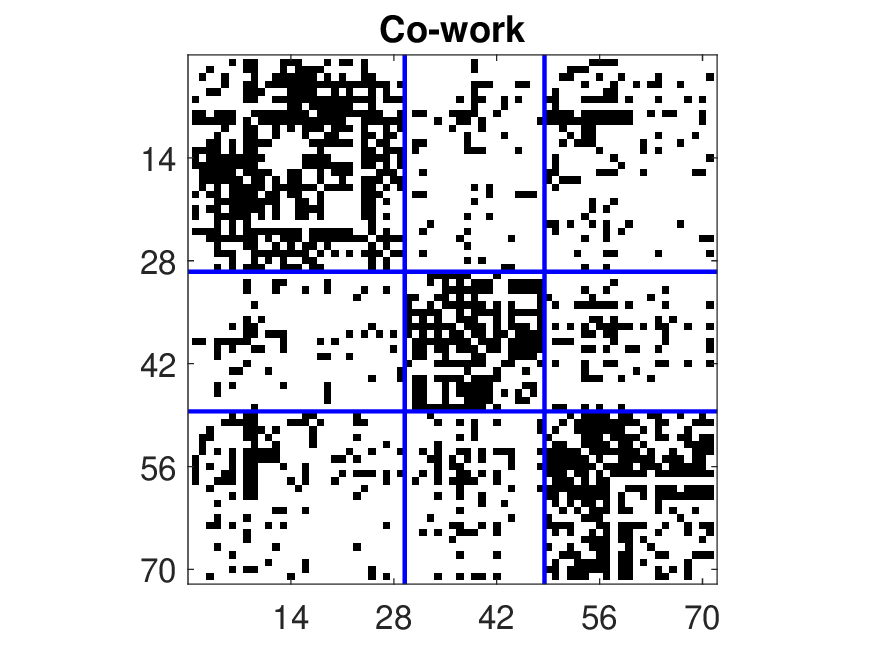}}
}
\resizebox{\columnwidth}{!}{
\subfigure[C.E]{\includegraphics[width=0.2\textwidth]{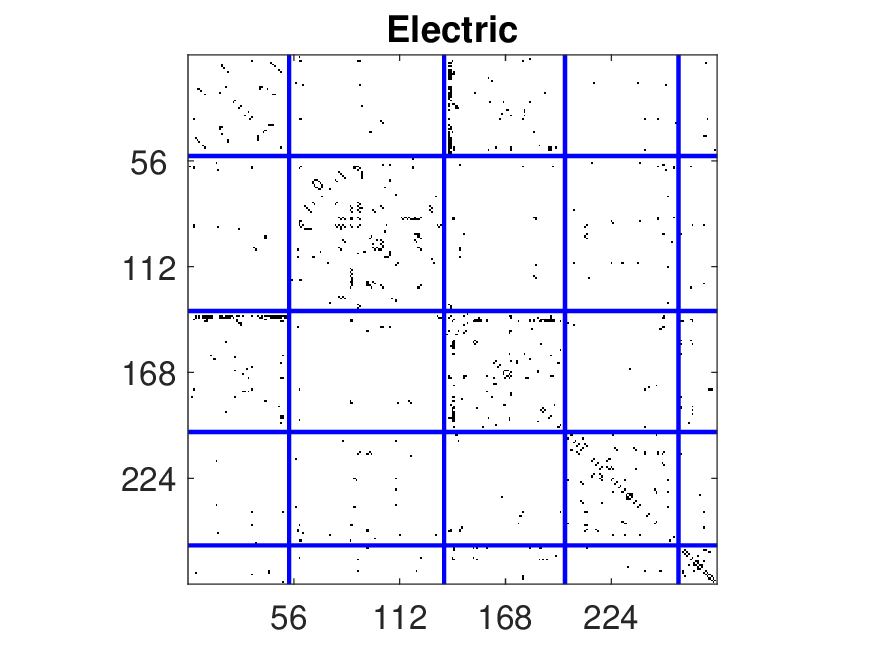}}
\subfigure[C.E]{\includegraphics[width=0.2\textwidth]{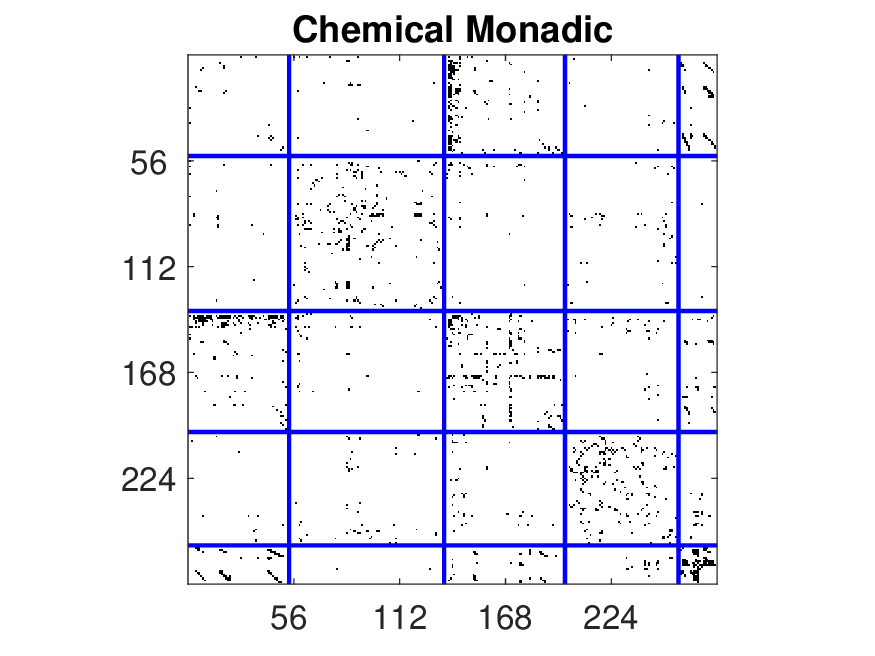}}
\subfigure[C.E]{\includegraphics[width=0.2\textwidth]{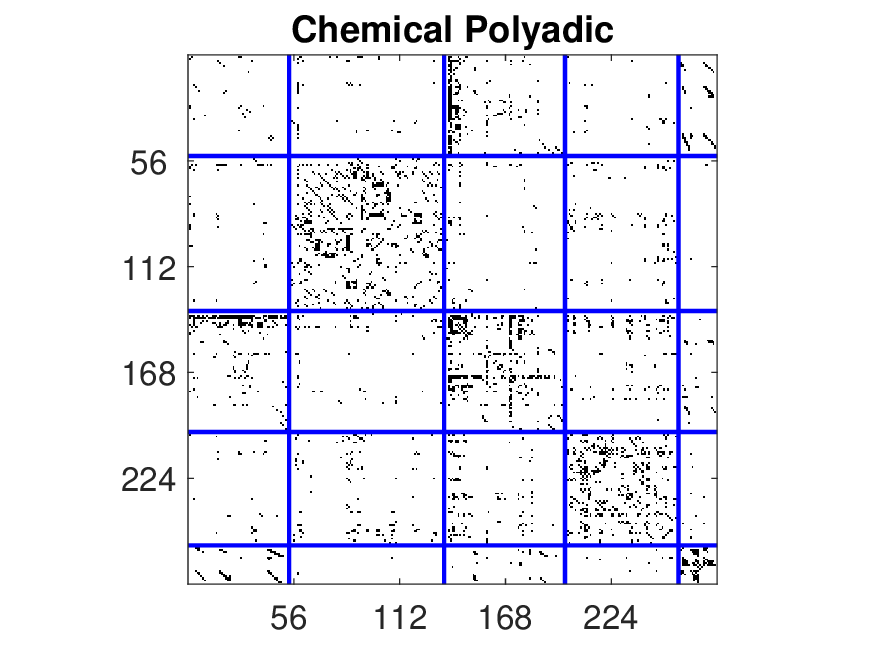}}
}
\resizebox{\columnwidth}{!}{
\subfigure[CS-A]{\includegraphics[width=0.2\textwidth]{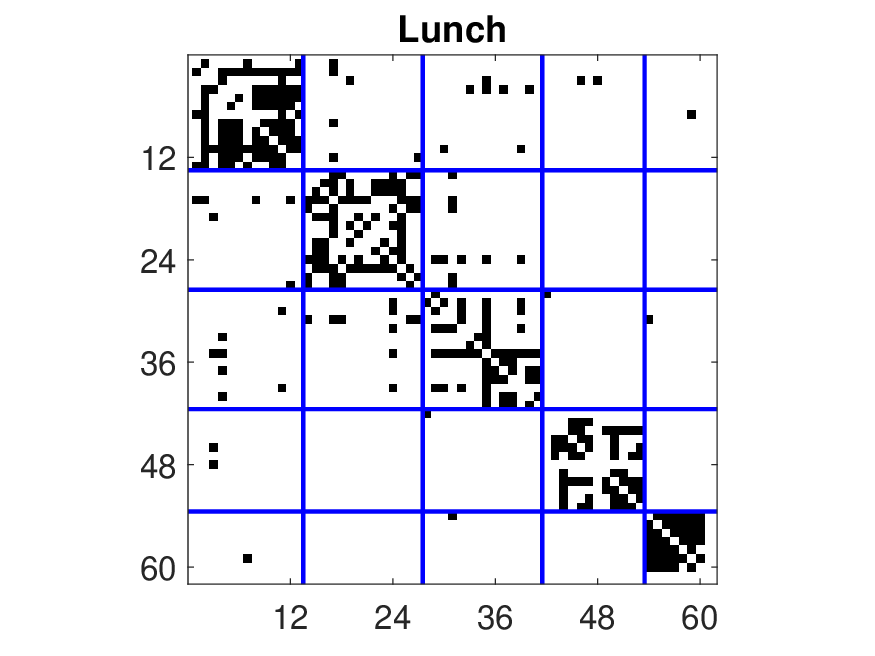}}
\subfigure[CS-A]{\includegraphics[width=0.2\textwidth]{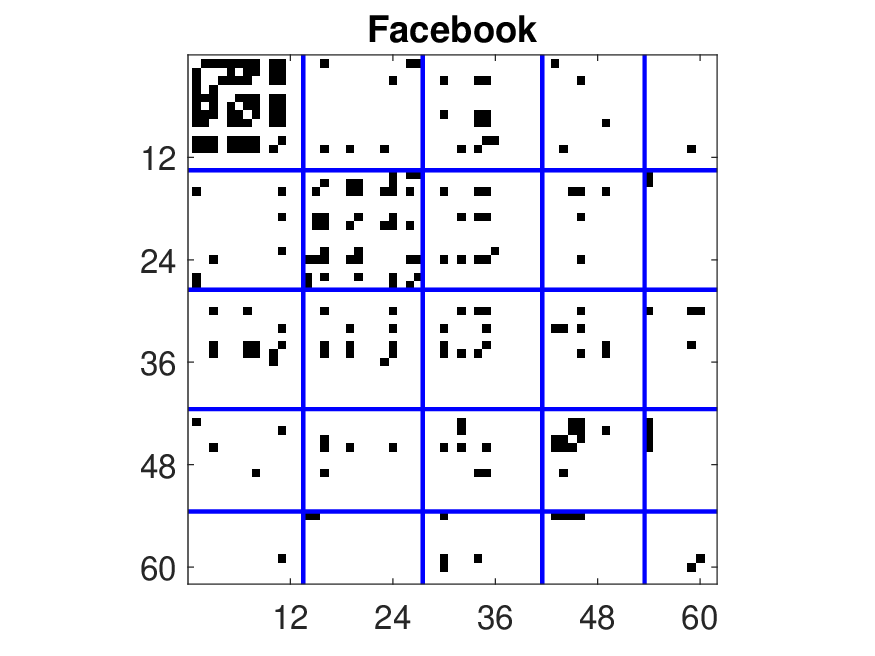}}
\subfigure[CS-A]{\includegraphics[width=0.2\textwidth]{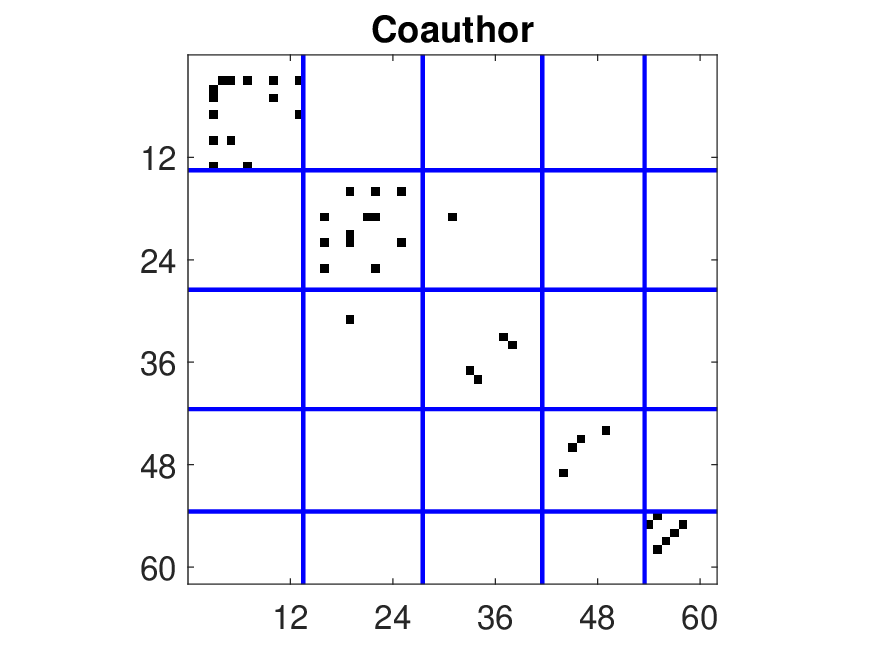}}
\subfigure[CS-A]{\includegraphics[width=0.2\textwidth]{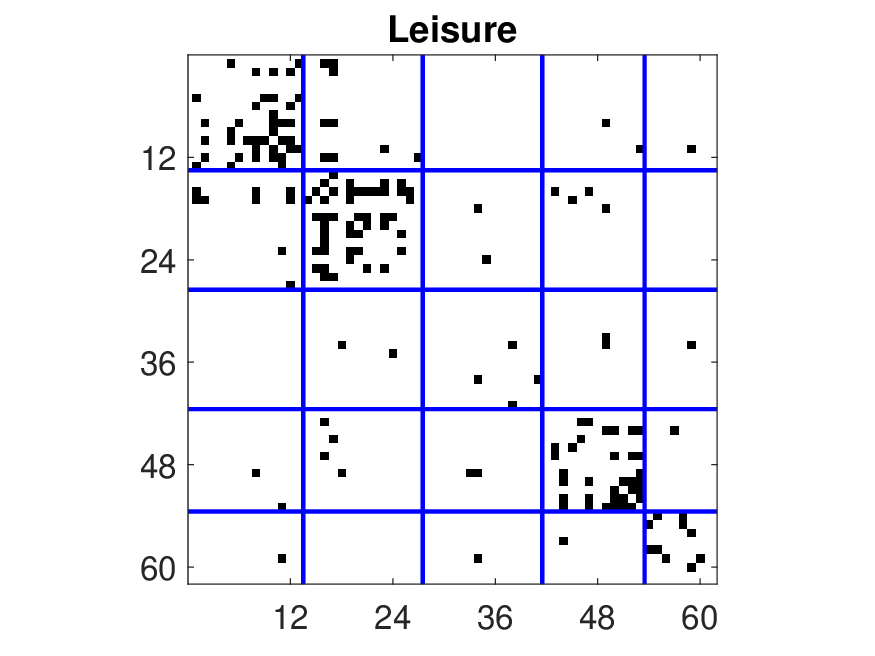}}
\subfigure[CS-A]{\includegraphics[width=0.2\textwidth]{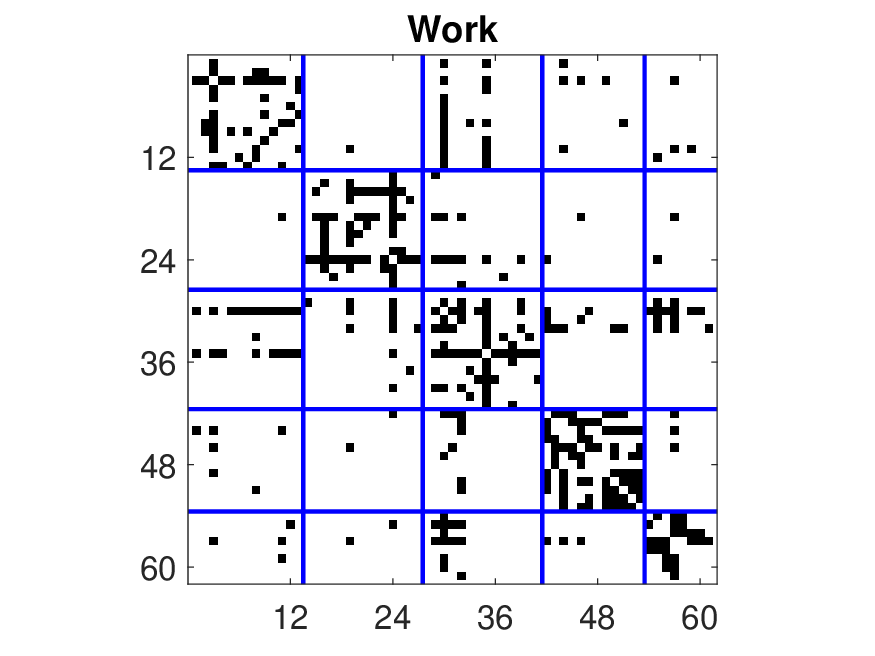}}
}
\resizebox{\columnwidth}{!}{
\subfigure[FAO-t]{\includegraphics[width=0.2\textwidth]{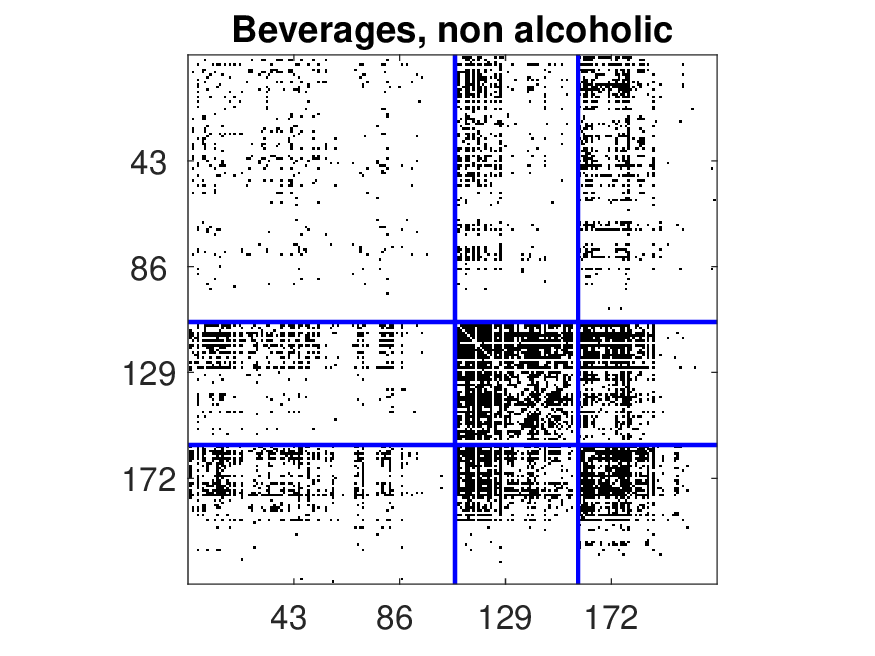}}
\subfigure[FAO-t]{\includegraphics[width=0.2\textwidth]{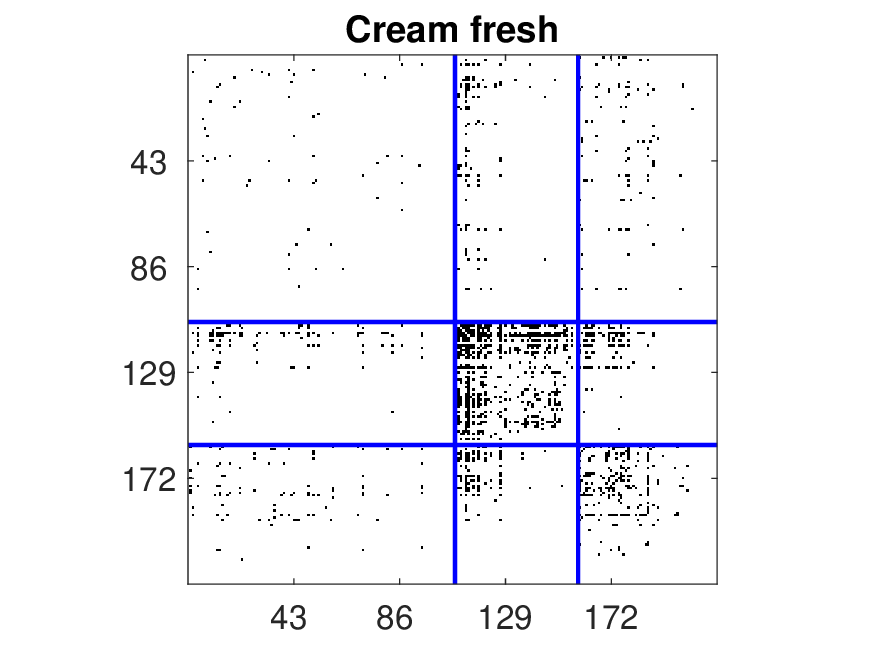}}
\subfigure[FAO-t]{\includegraphics[width=0.2\textwidth]{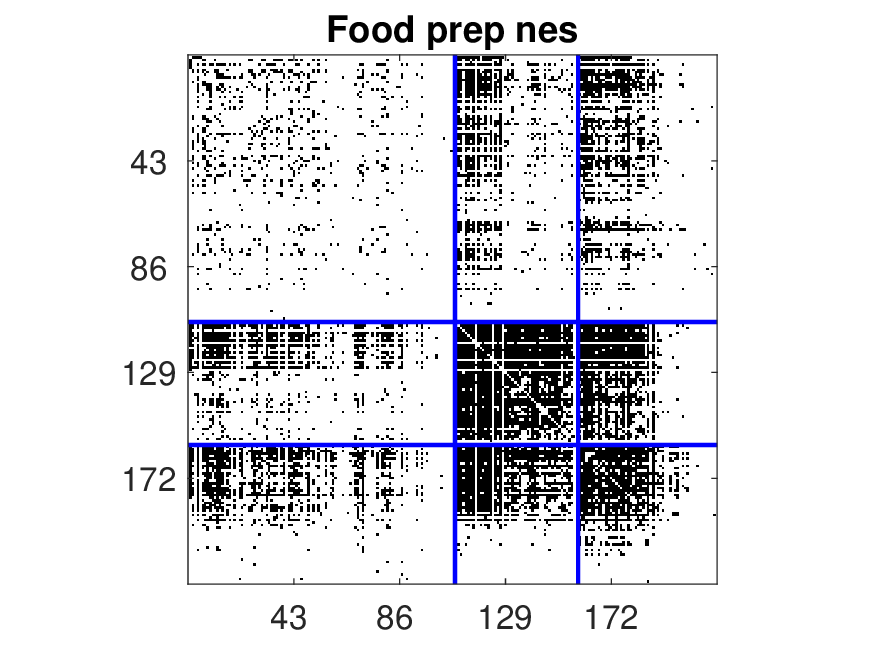}}
\subfigure[FAO-t]{\includegraphics[width=0.2\textwidth]{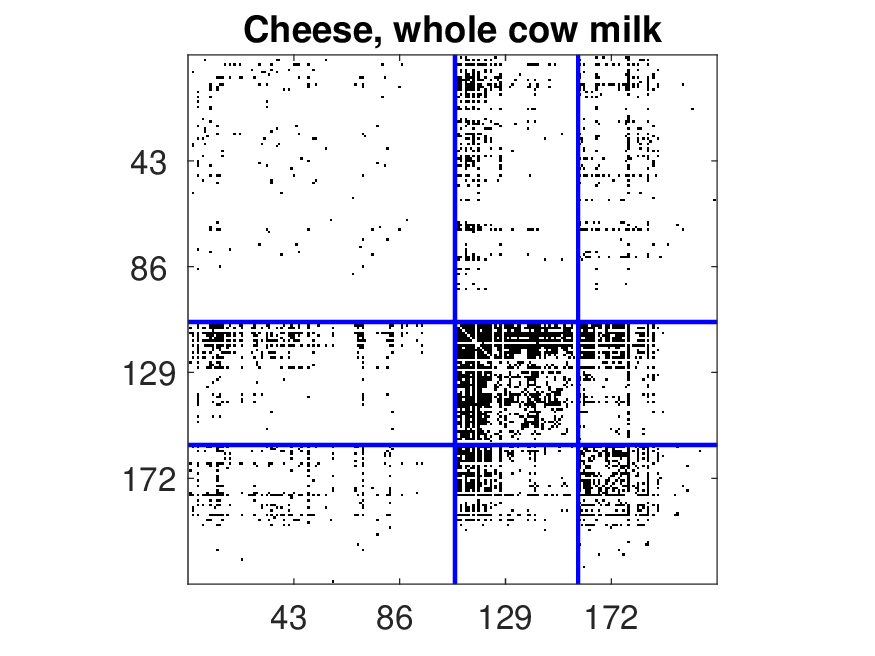}}
\subfigure[FAO-t]{\includegraphics[width=0.2\textwidth]{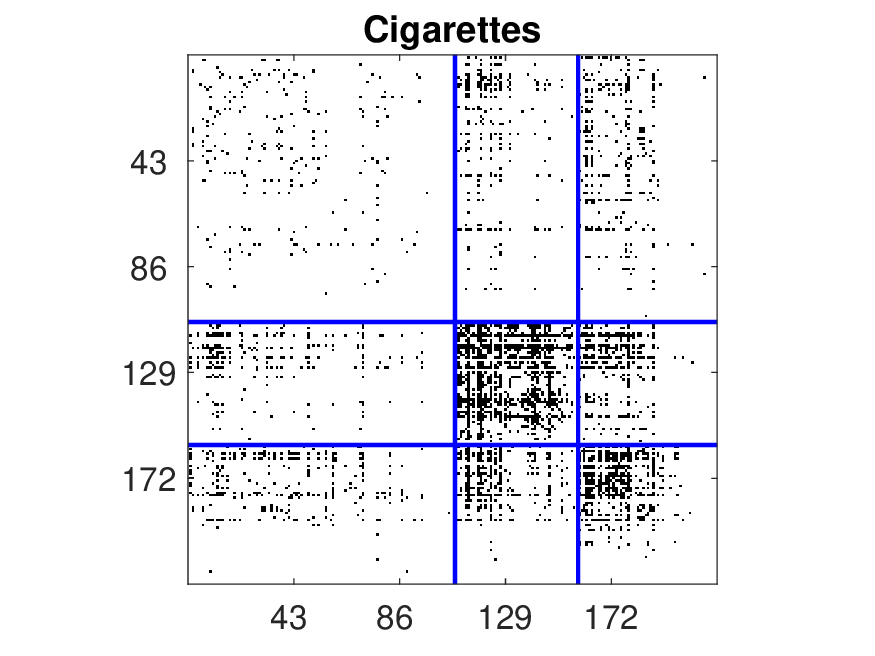}}
\subfigure[FAO-t]{\includegraphics[width=0.2\textwidth]{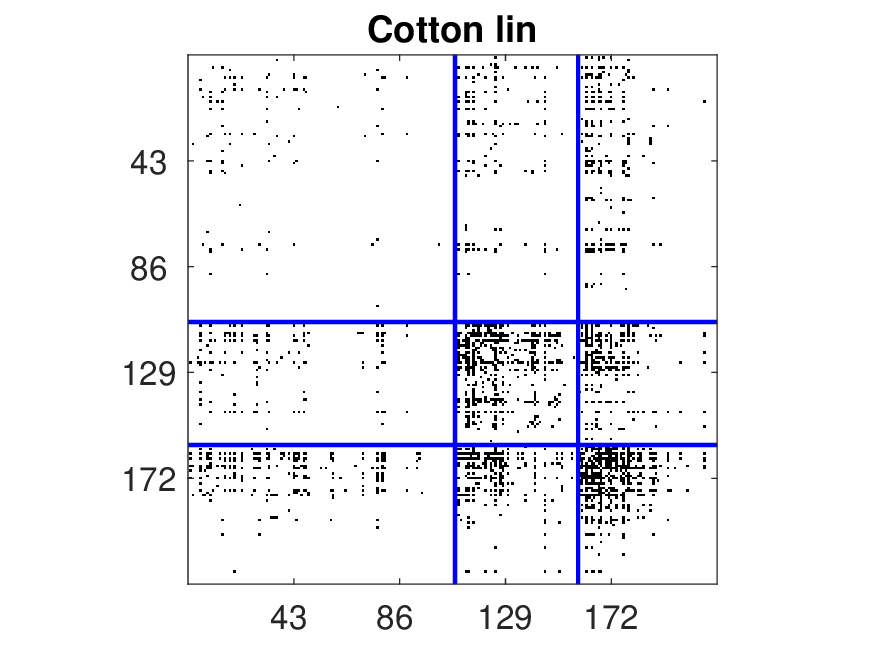}}
}
\resizebox{\columnwidth}{!}{
\subfigure[FAO-t]{\includegraphics[width=0.2\textwidth]{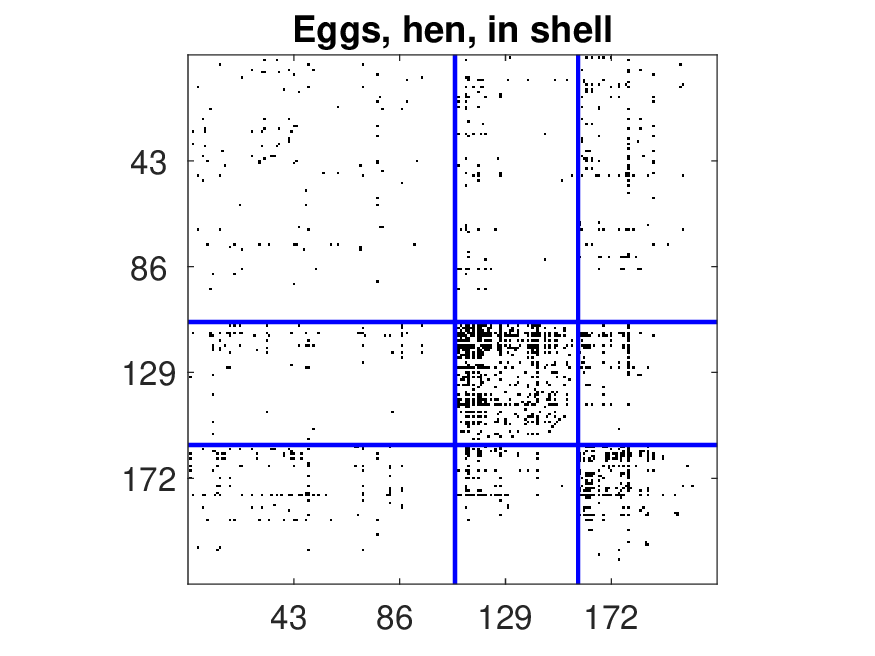}}
\subfigure[FAO-t]{\includegraphics[width=0.2\textwidth]{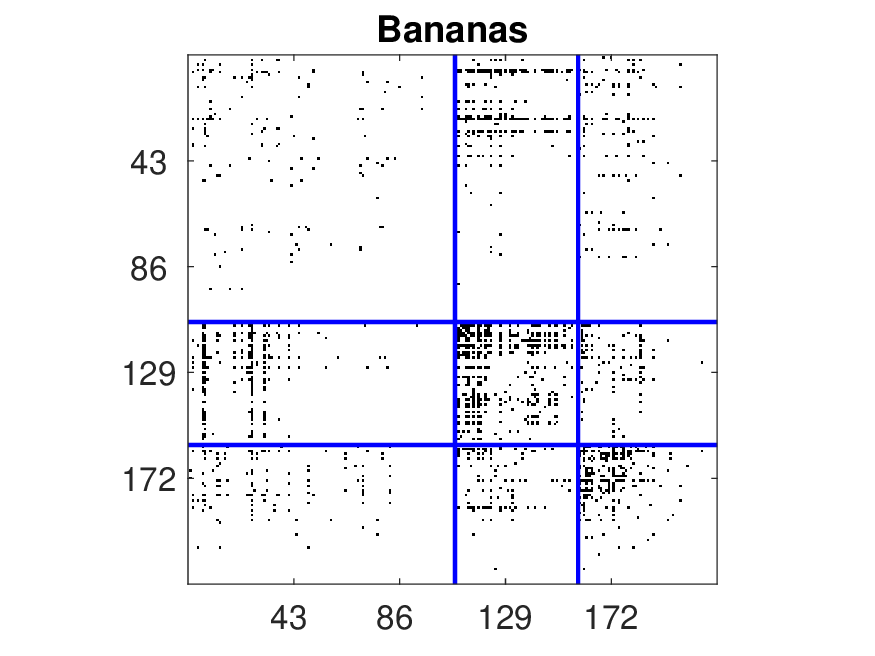}}
\subfigure[FAO-t]{\includegraphics[width=0.2\textwidth]{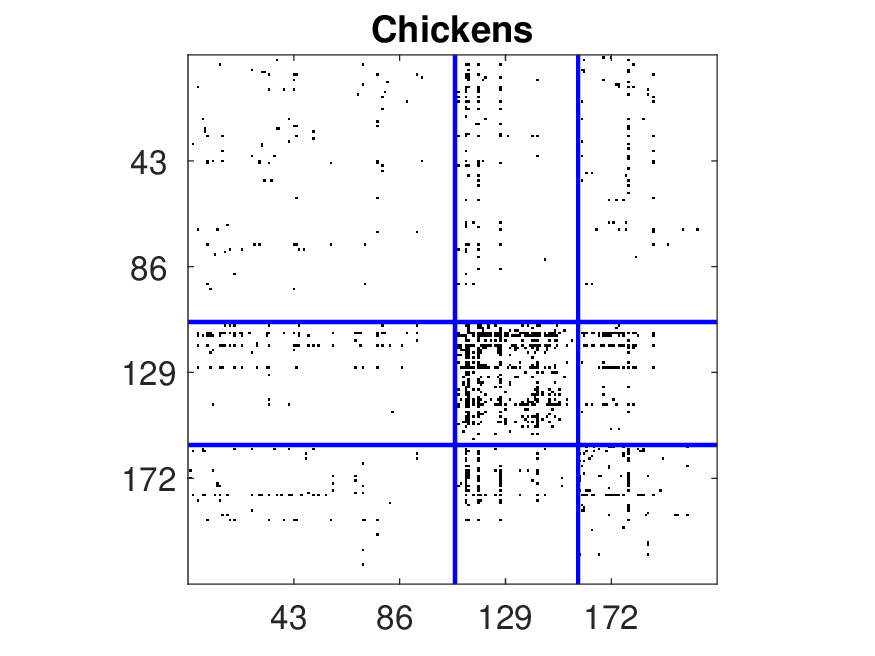}}
\subfigure[FAO-t]{\includegraphics[width=0.2\textwidth]{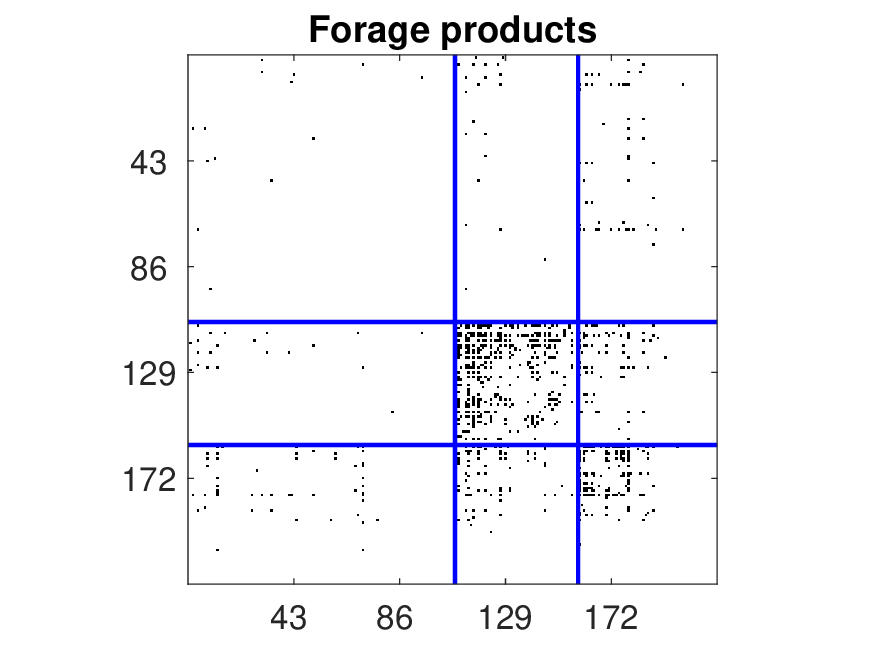}}
\subfigure[FAO-t]{\includegraphics[width=0.2\textwidth]{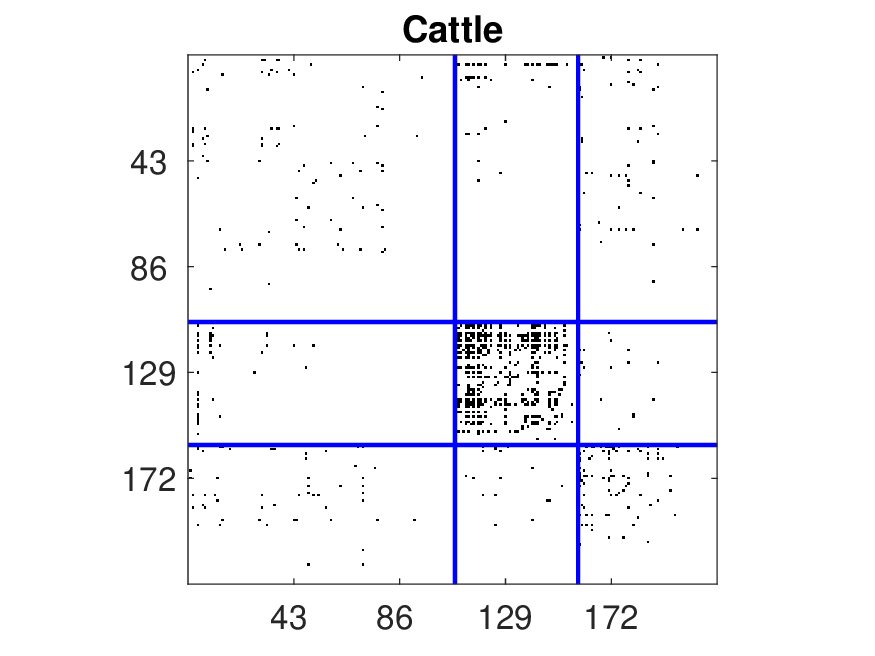}}
\subfigure[FAO-t]{\includegraphics[width=0.2\textwidth]{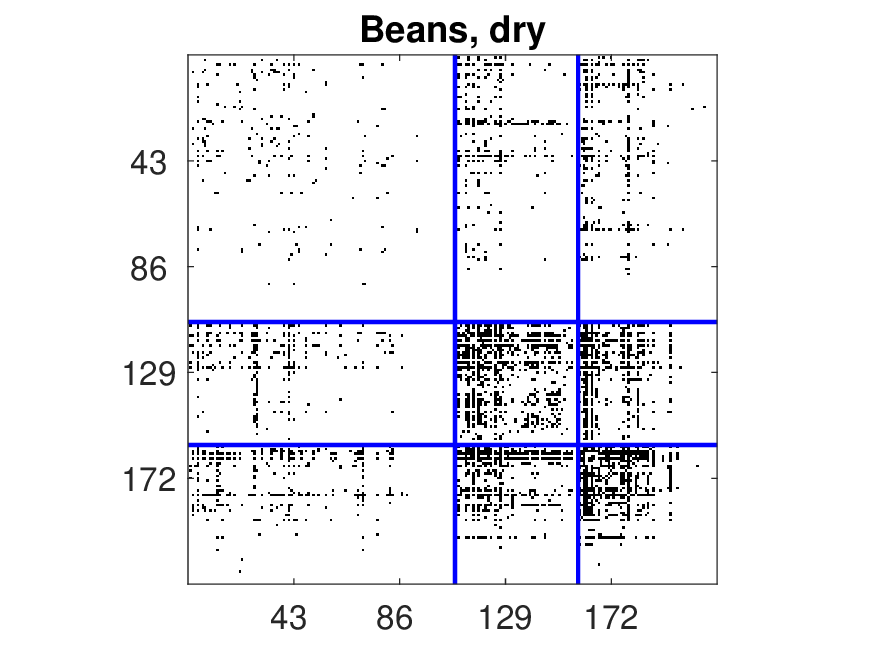}}
}
\caption{Panels (a)-(c) display the LLF multi-layer network. Panels (d)-(f) display the C.E multi-layer network. Panels (g)-(k) display the CS-A multi-layer network. Panels (l)-(w) display twelve products of FAO-t multi-layer network. The nodes in LLF, C.E, CS-A, and FAO-t have been reordered based on the 3, 5, 5, and 3 communities detected by DC-RDSoS, respectively. Blue lines represent community partitions. It is clear that except for nodes belonging to the first estimated community of C.E and FAO-t, nodes within the same community typically have more connections compared to nodes across different communities.}
\label{SortA} 
\end{figure}

\begin{figure}
\centering
\resizebox{\columnwidth}{!}{
\subfigure[LLF]{\includegraphics[width=0.2\textwidth]{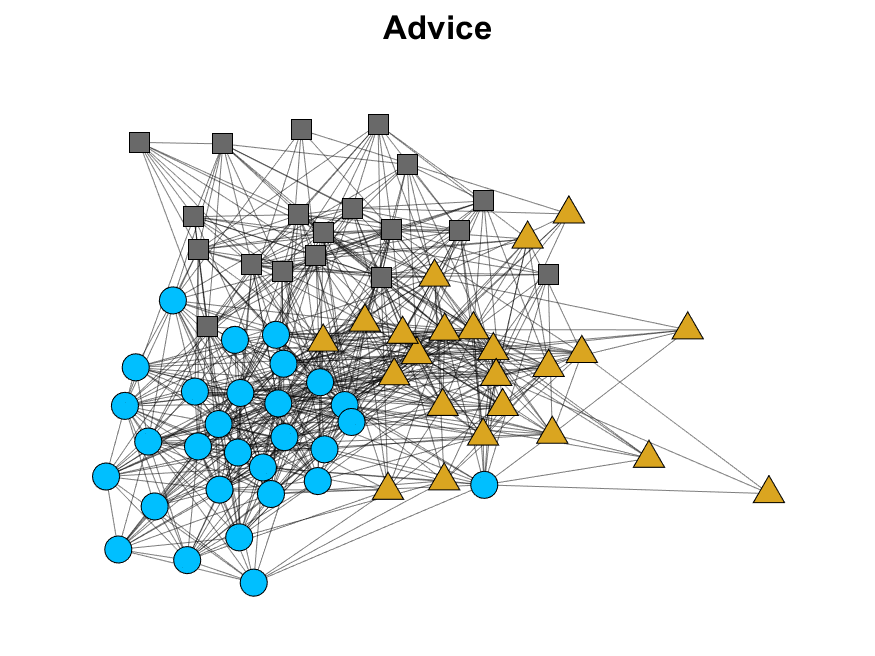}}
\subfigure[LLF]{\includegraphics[width=0.2\textwidth]{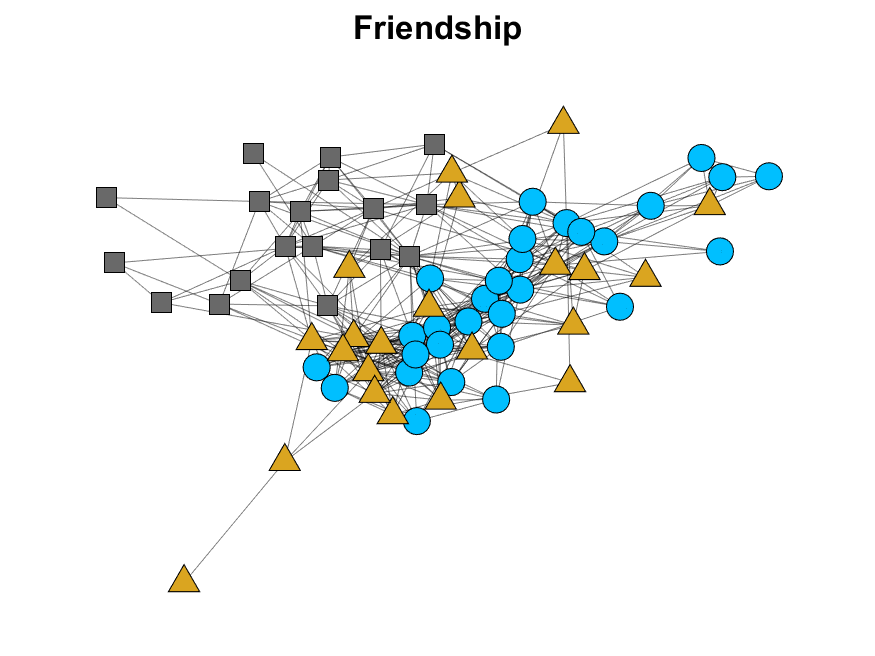}}
\subfigure[LLF]{\includegraphics[width=0.2\textwidth]{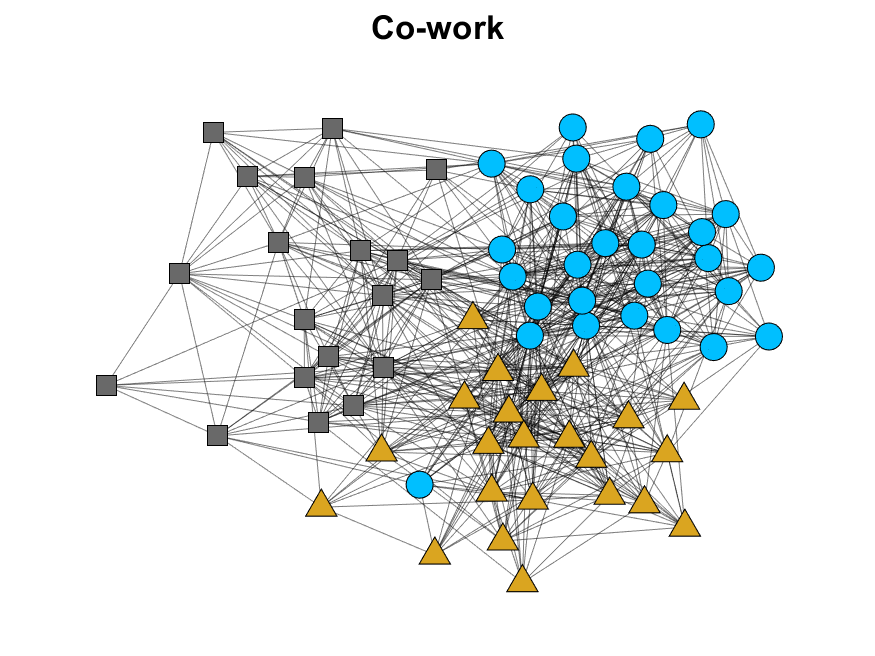}}
}
\resizebox{\columnwidth}{!}{
\subfigure[C.E]{\includegraphics[width=0.2\textwidth]{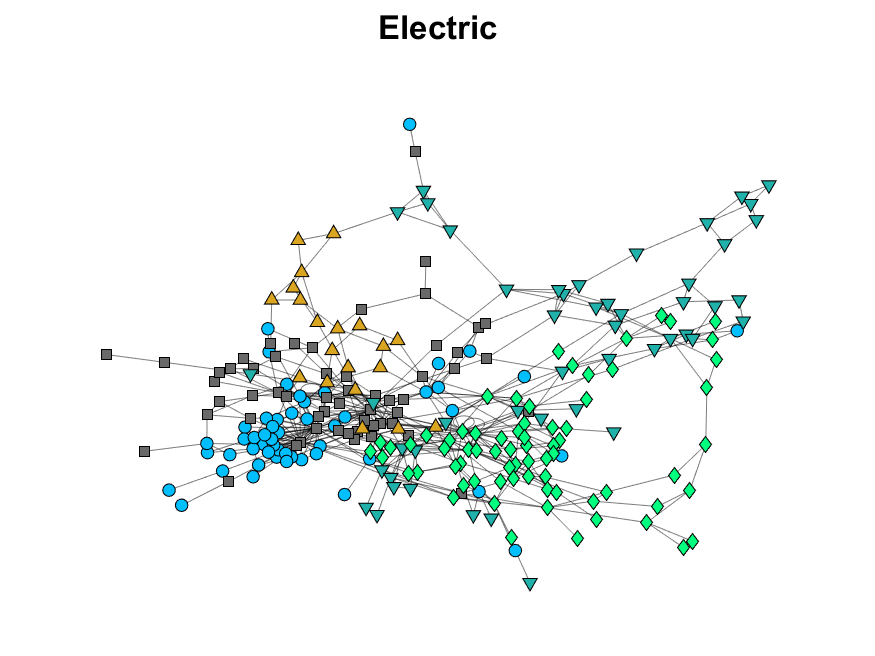}}
\subfigure[C.E]{\includegraphics[width=0.2\textwidth]{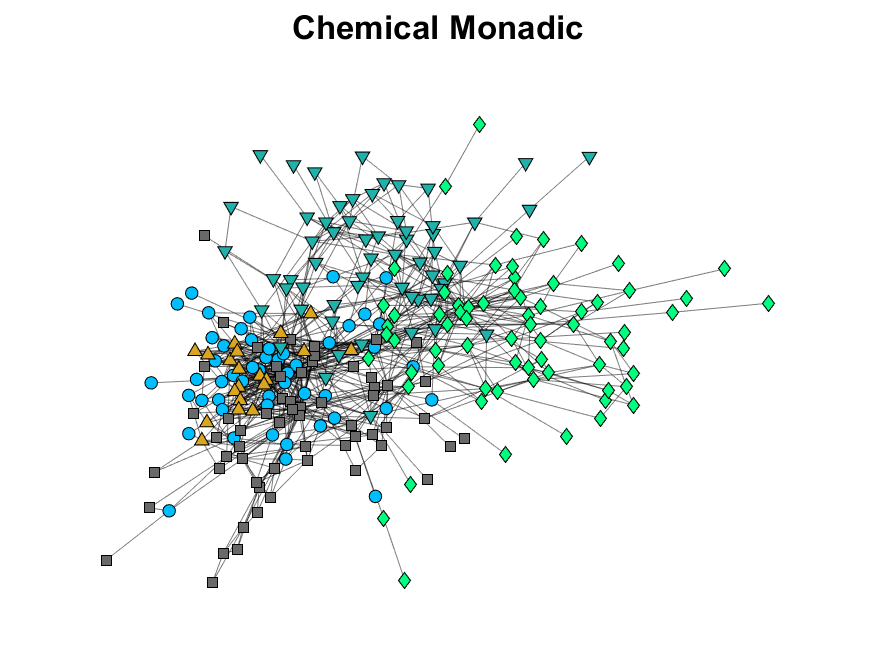}}
\subfigure[C.E]{\includegraphics[width=0.2\textwidth]{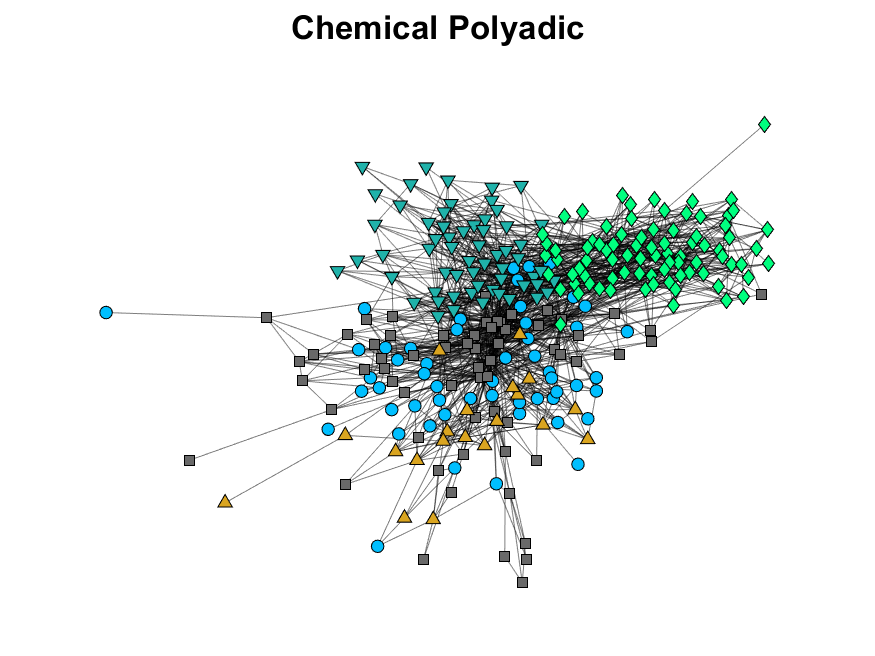}}
}
\resizebox{\columnwidth}{!}{
\subfigure[CS-A]{\includegraphics[width=0.2\textwidth]{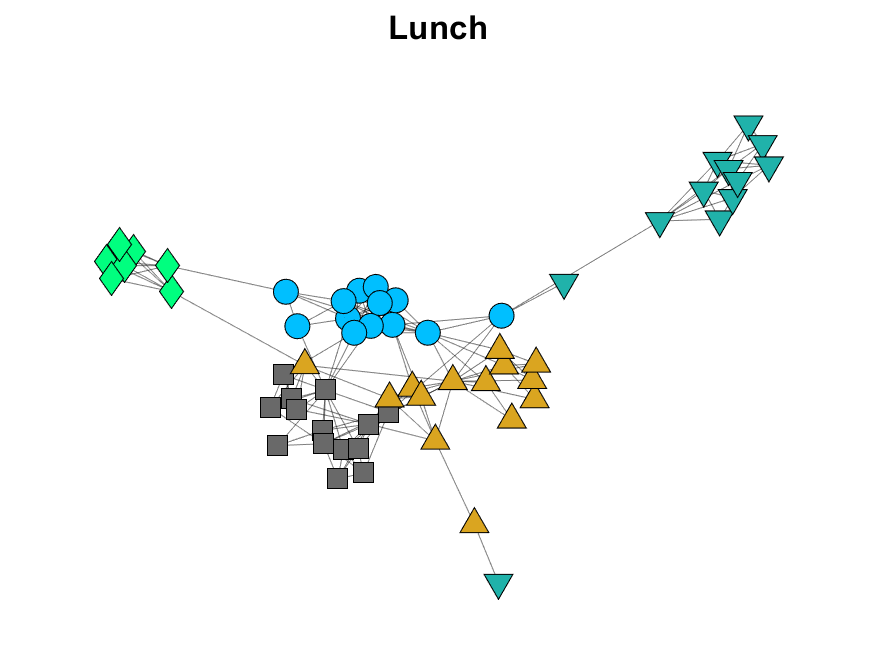}}
\subfigure[CS-A]{\includegraphics[width=0.2\textwidth]{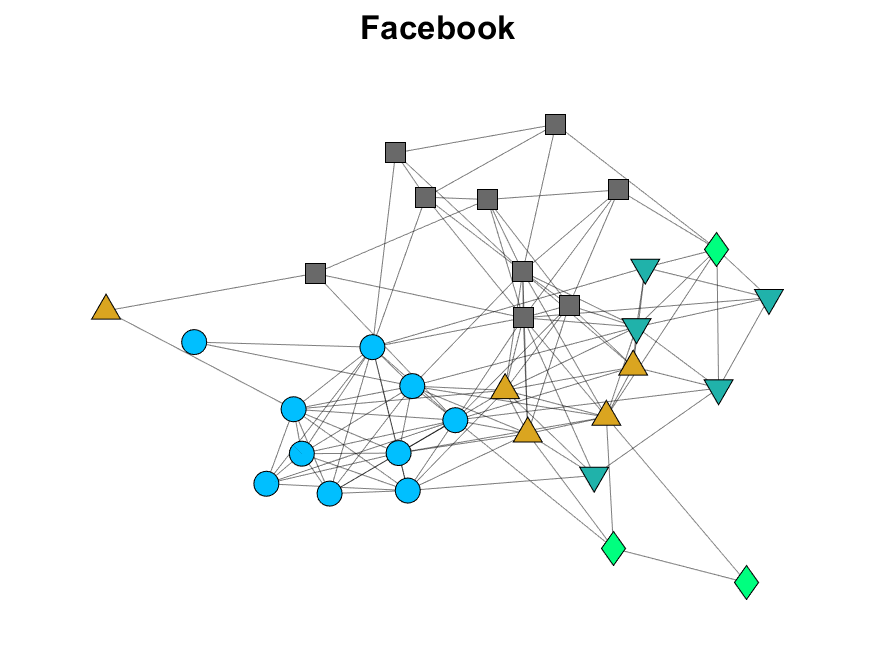}}
\subfigure[CS-A]{\includegraphics[width=0.2\textwidth]{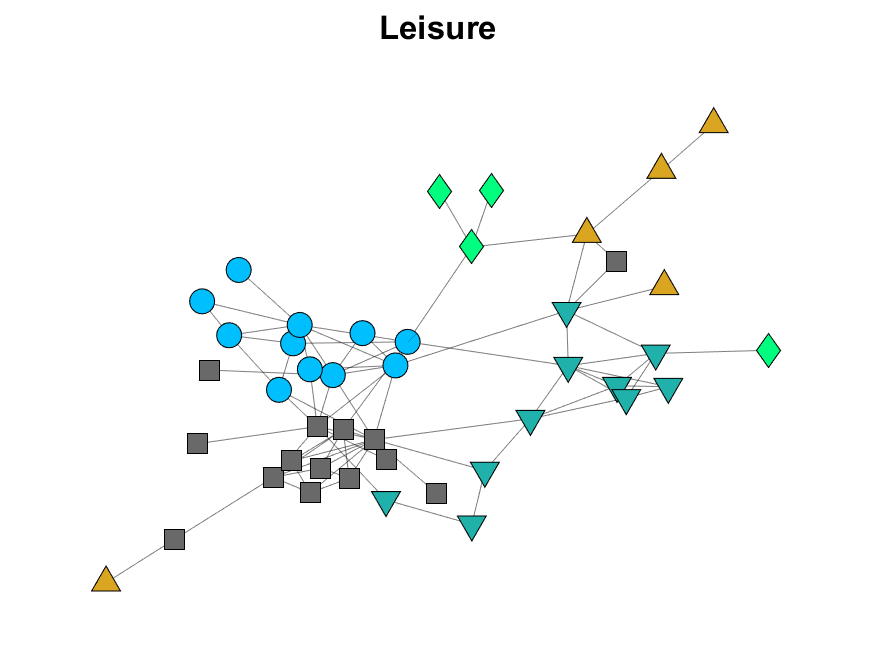}}
\subfigure[CS-A]{\includegraphics[width=0.2\textwidth]{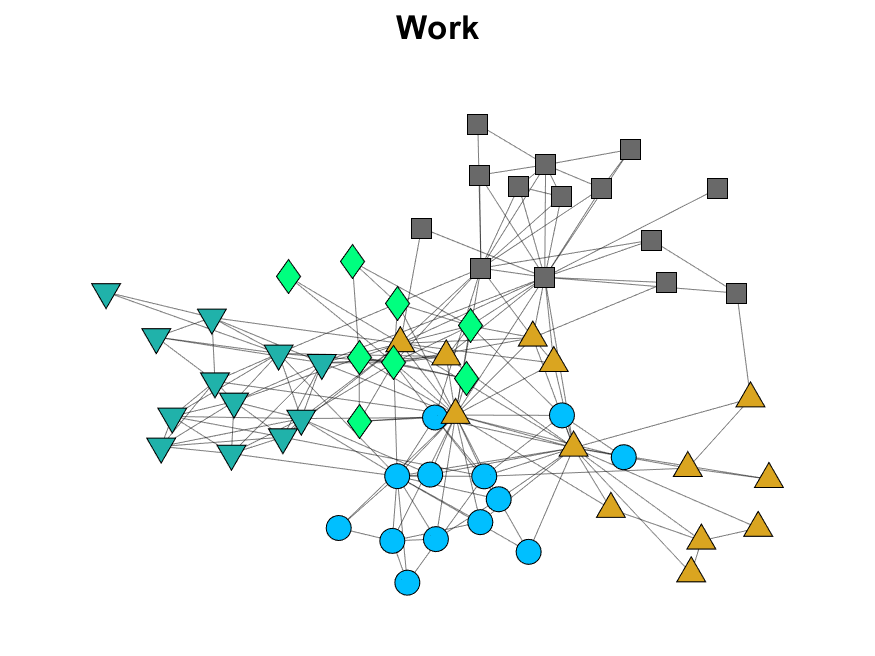}}
}
\resizebox{\columnwidth}{!}{
\subfigure[FAO-t]{\includegraphics[width=0.2\textwidth]{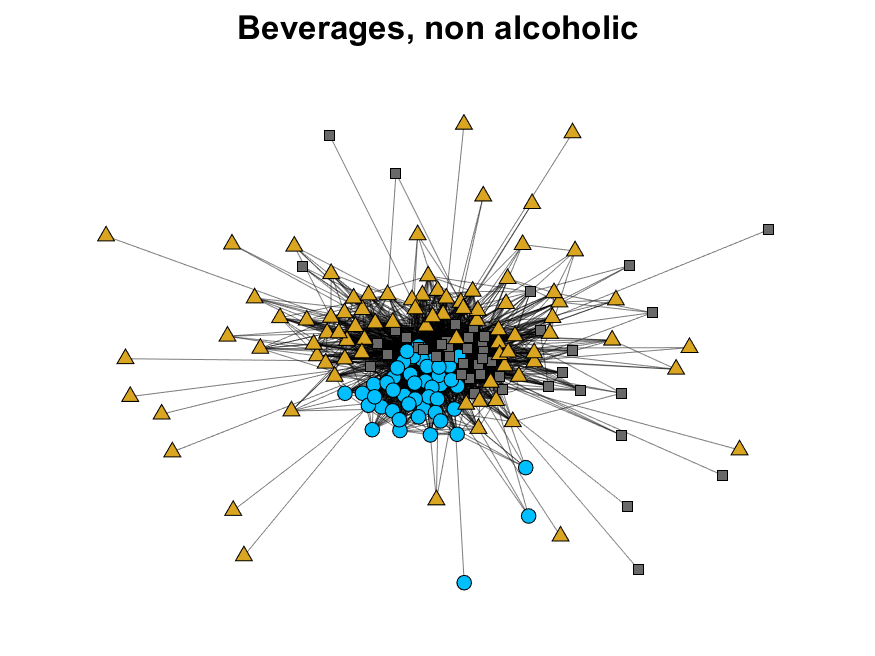}}
\subfigure[FAO-t]{\includegraphics[width=0.2\textwidth]{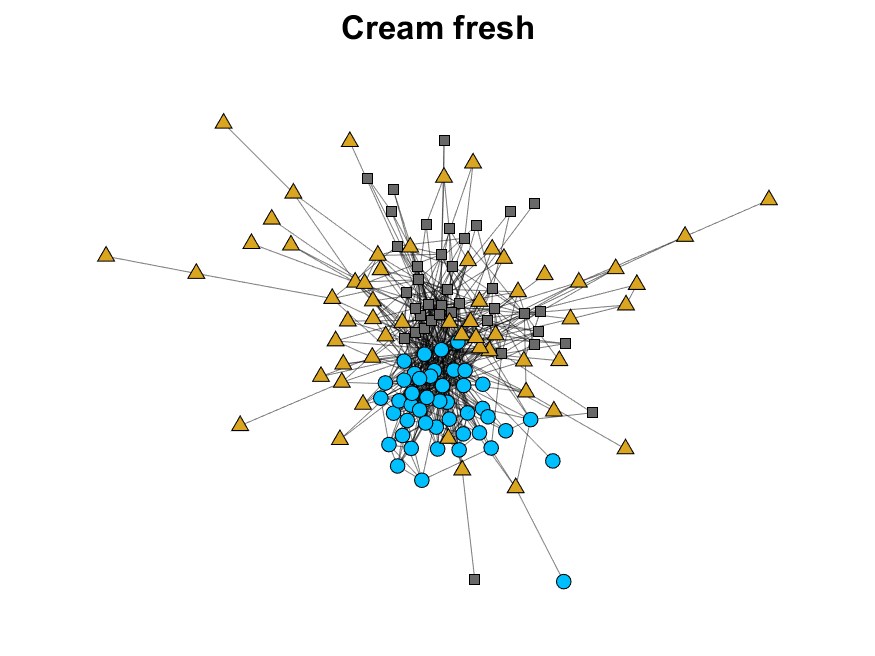}}
\subfigure[FAO-t]{\includegraphics[width=0.2\textwidth]{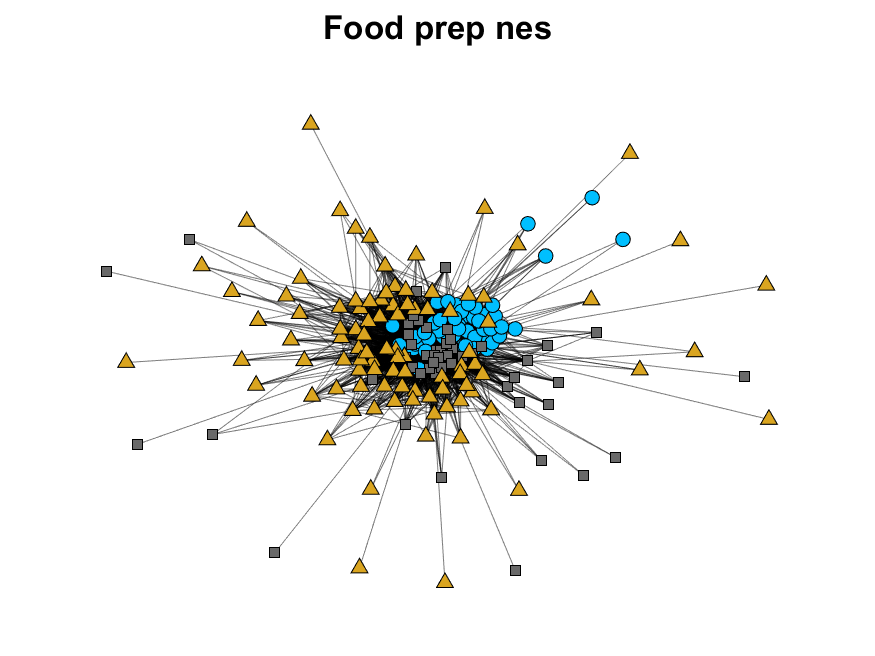}}
\subfigure[FAO-t]{\includegraphics[width=0.2\textwidth]{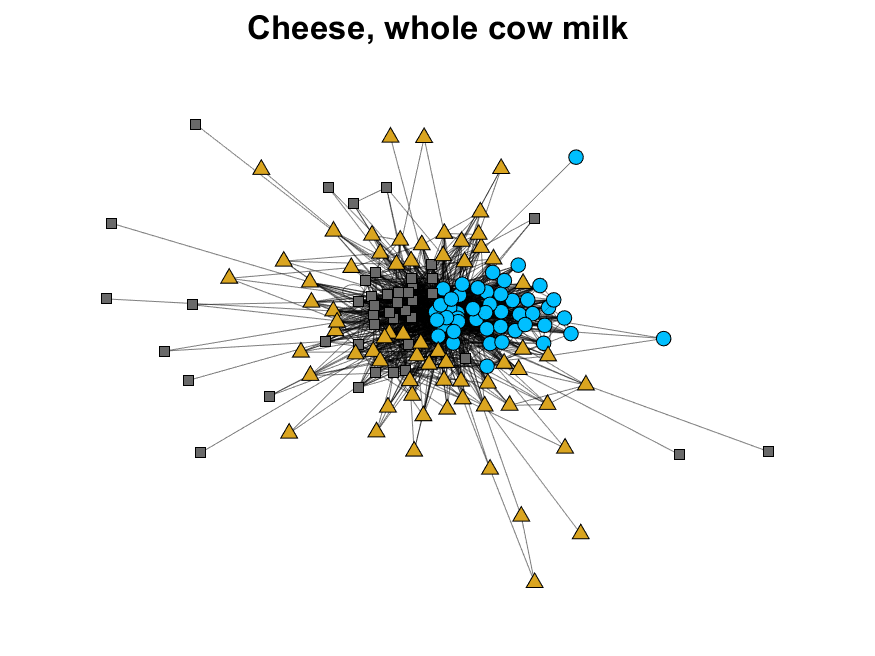}}
\subfigure[FAO-t]{\includegraphics[width=0.2\textwidth]{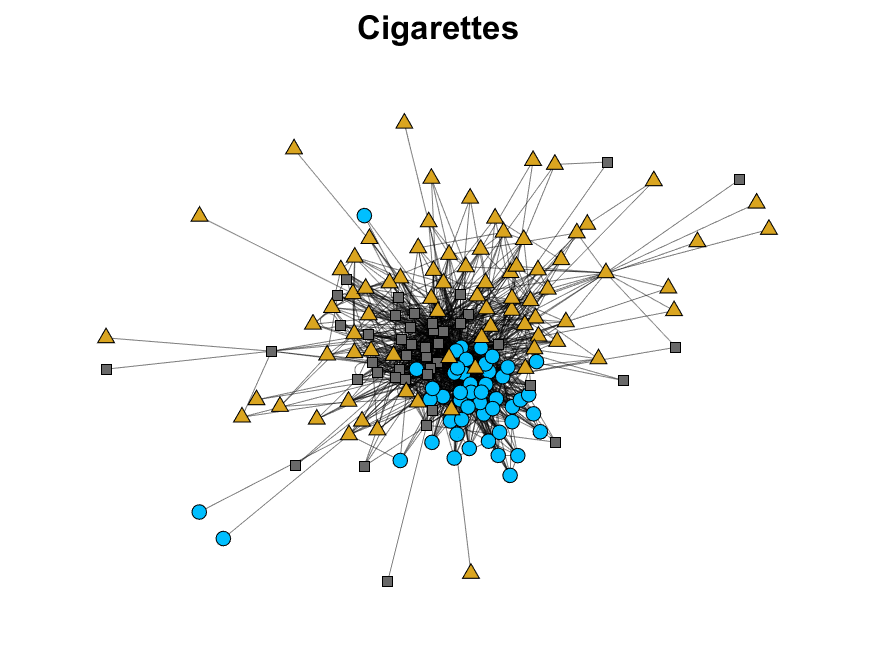}}
\subfigure[FAO-t]{\includegraphics[width=0.2\textwidth]{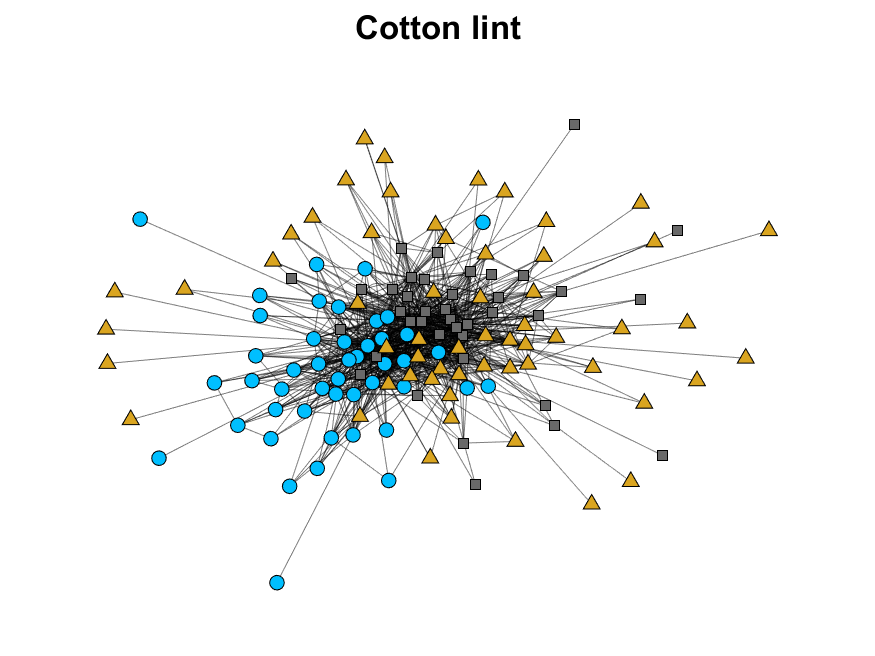}}
}
\resizebox{\columnwidth}{!}{
\subfigure[FAO-t]{\includegraphics[width=0.2\textwidth]{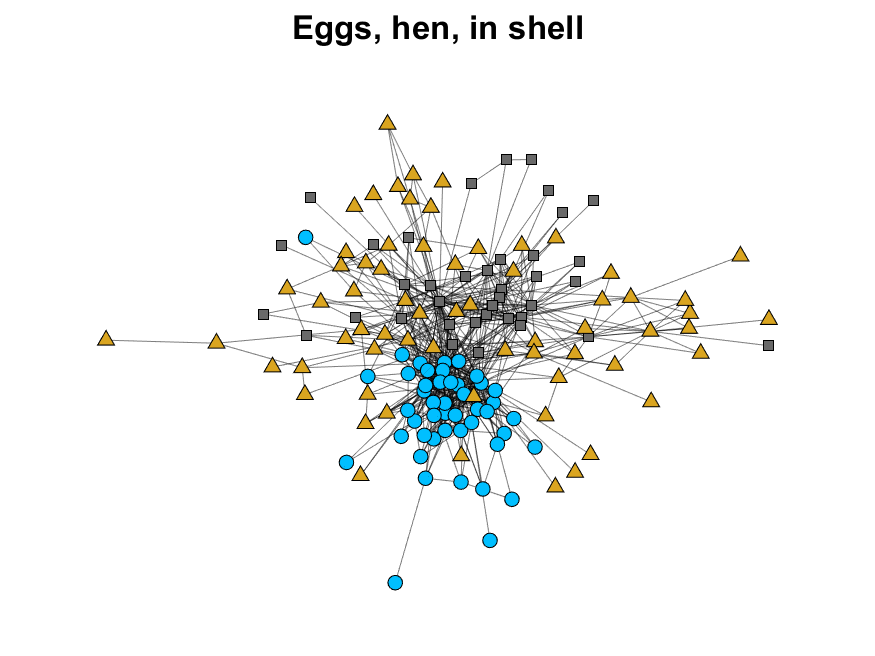}}
\subfigure[FAO-t]{\includegraphics[width=0.2\textwidth]{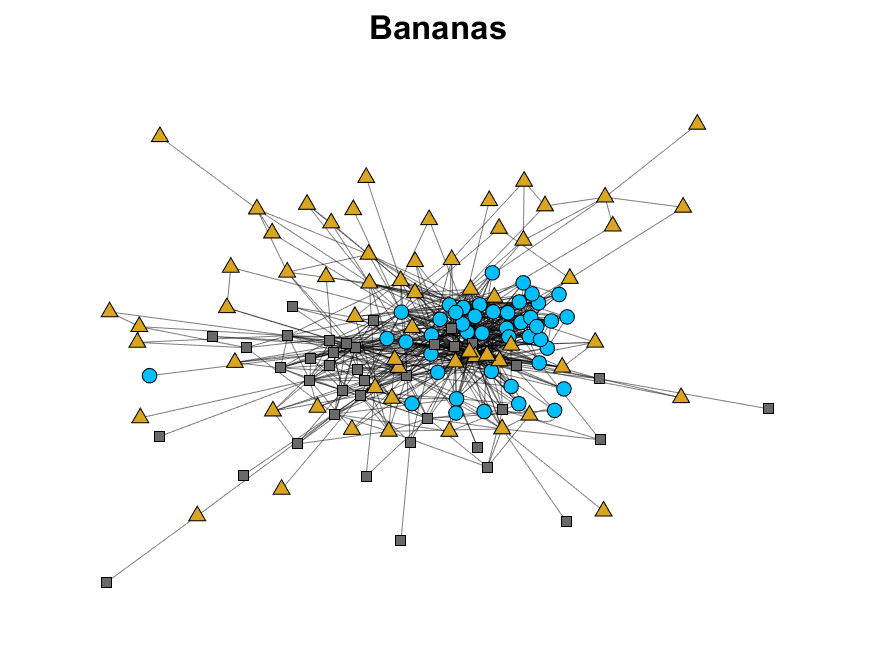}}
\subfigure[FAO-t]{\includegraphics[width=0.2\textwidth]{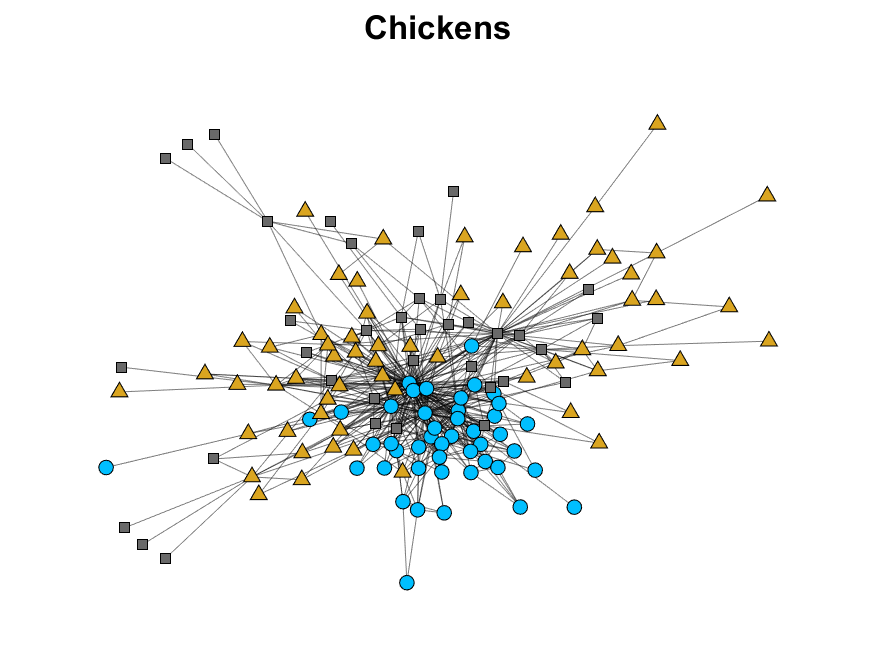}}
\subfigure[FAO-t]{\includegraphics[width=0.2\textwidth]{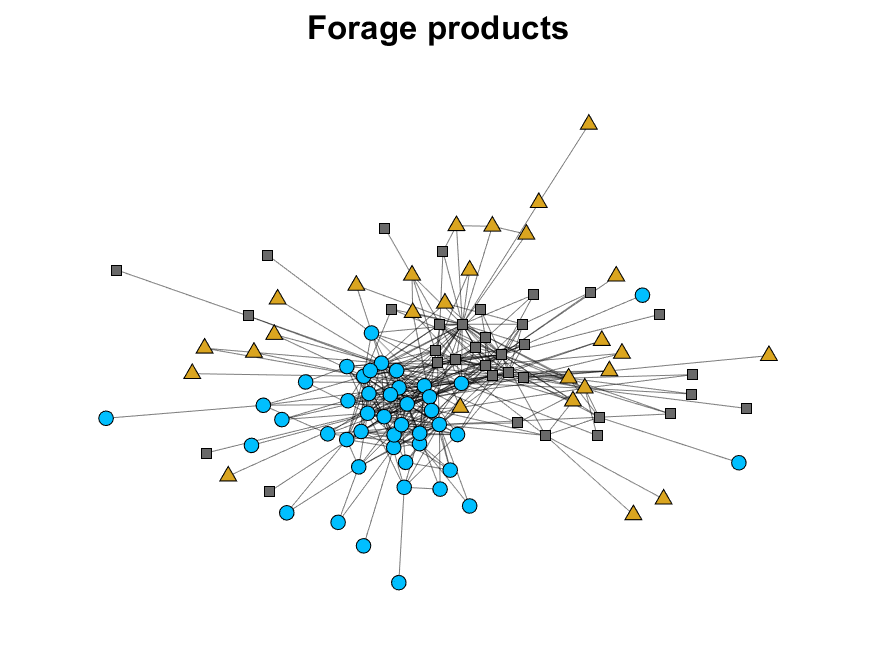}}
\subfigure[FAO-t]{\includegraphics[width=0.2\textwidth]{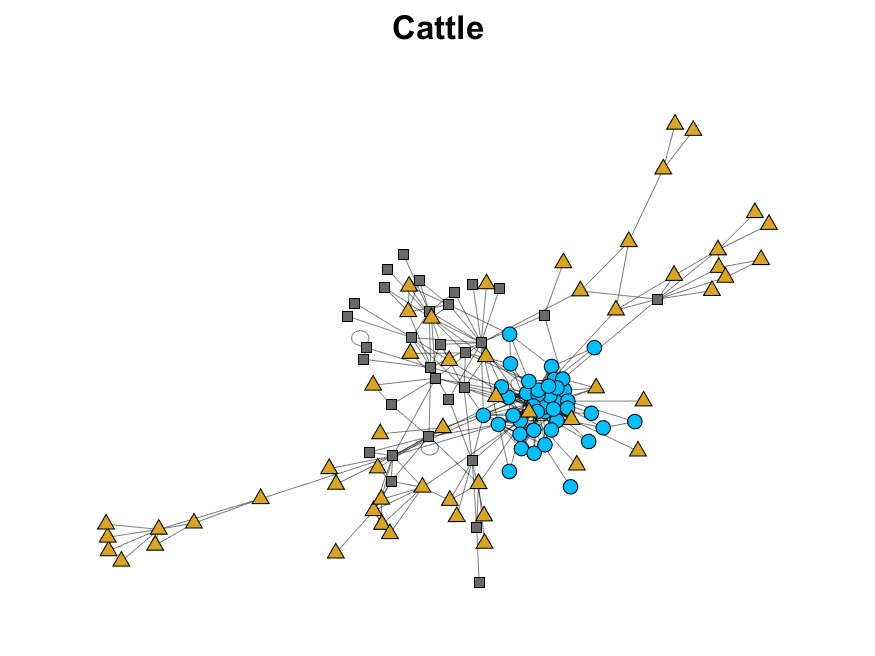}}
\subfigure[FAO-t]{\includegraphics[width=0.2\textwidth]{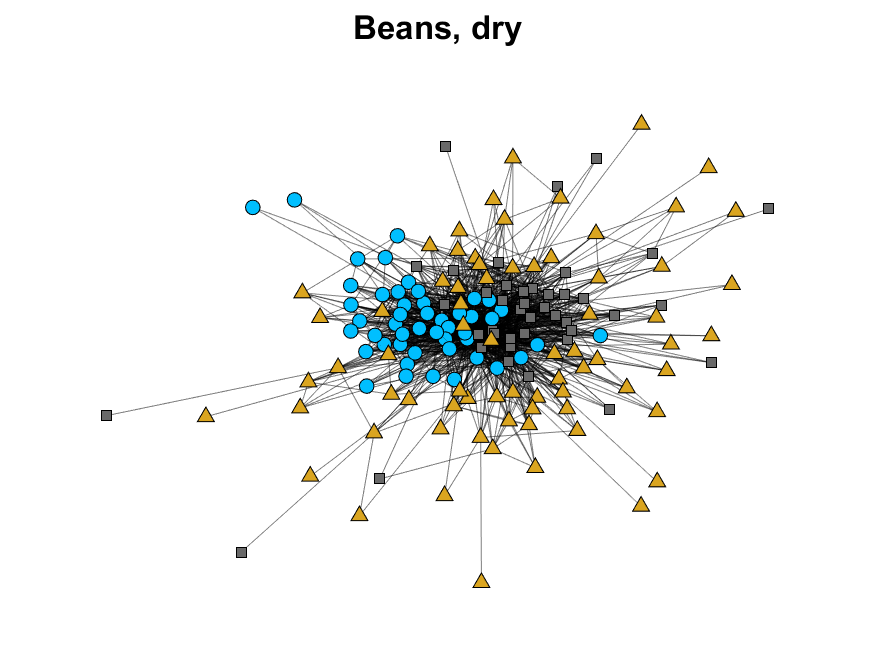}}
}
\caption{Visualization of the estimated community structure of the four real datasets. Only the largest connected graph for each network is displayed. The estimated communities for each multi-layer network are the same as those presented in Fig.~\ref{SortA}. We do not show the Coauthor layer of CS-A because it is too sparse. Colors (shapes) indicate estimated communities. Similar to Fig.~\ref{SortA}, in general, we observe that nodes within the same community tend to connect more than nodes across communities.}
\label{EstimatedC} 
\end{figure}

\begin{figure}
\centering
\resizebox{\columnwidth}{!}{
\subfigure[LLF]{\includegraphics[width=0.2\textwidth]{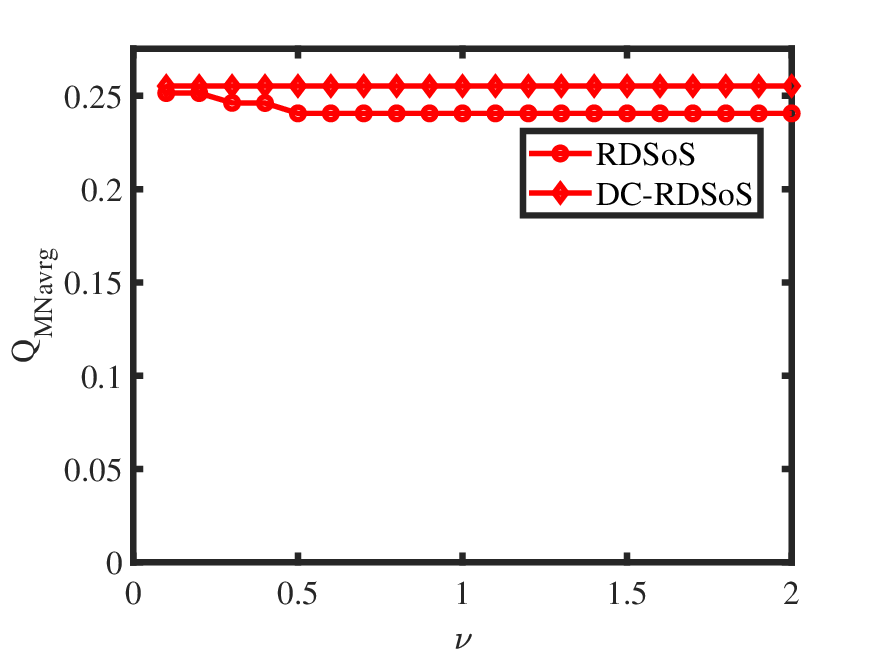}}
\subfigure[C.E]{\includegraphics[width=0.2\textwidth]{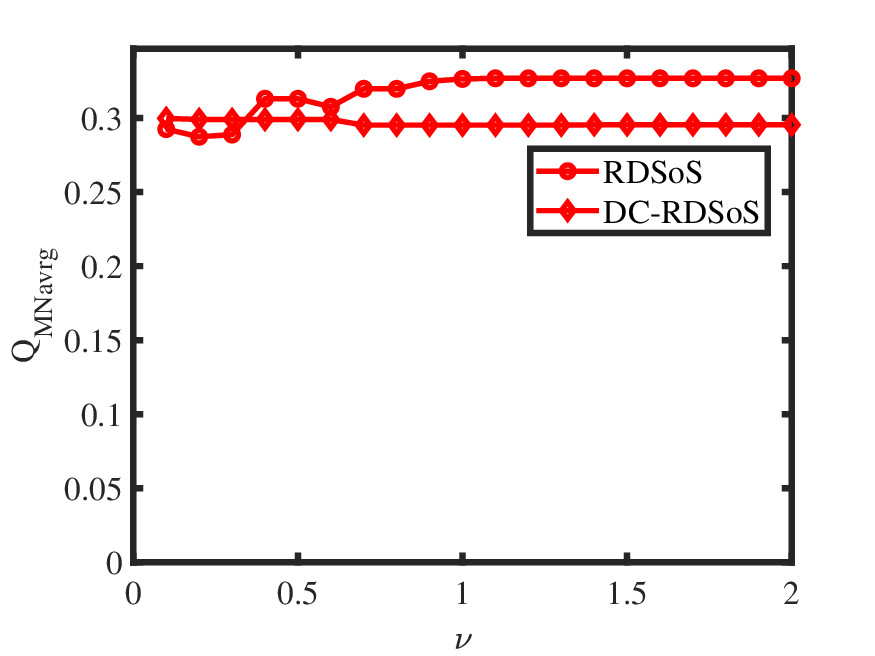}}
\subfigure[CS-A]{\includegraphics[width=0.2\textwidth]{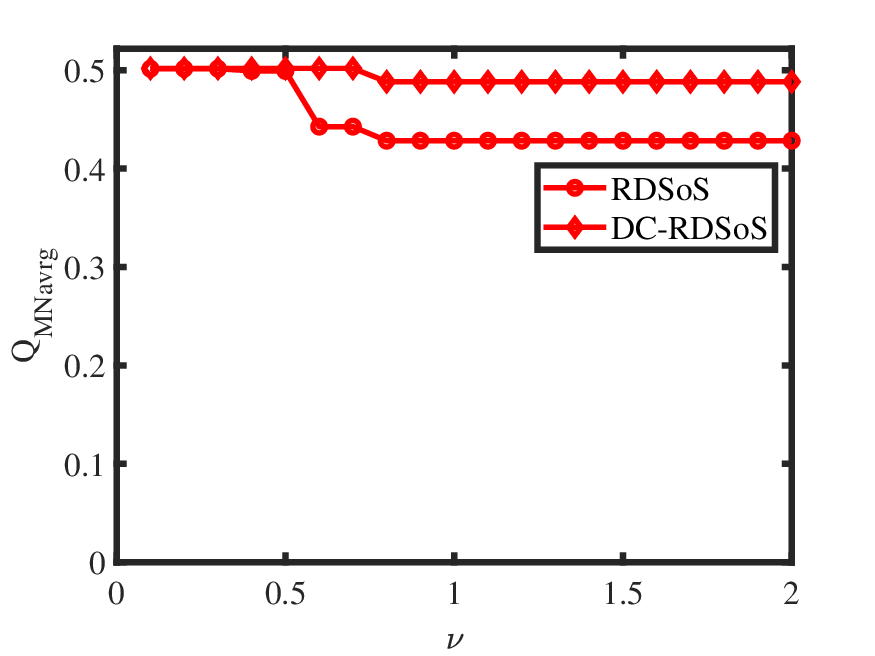}}
\subfigure[FAO-t]{\includegraphics[width=0.2\textwidth]{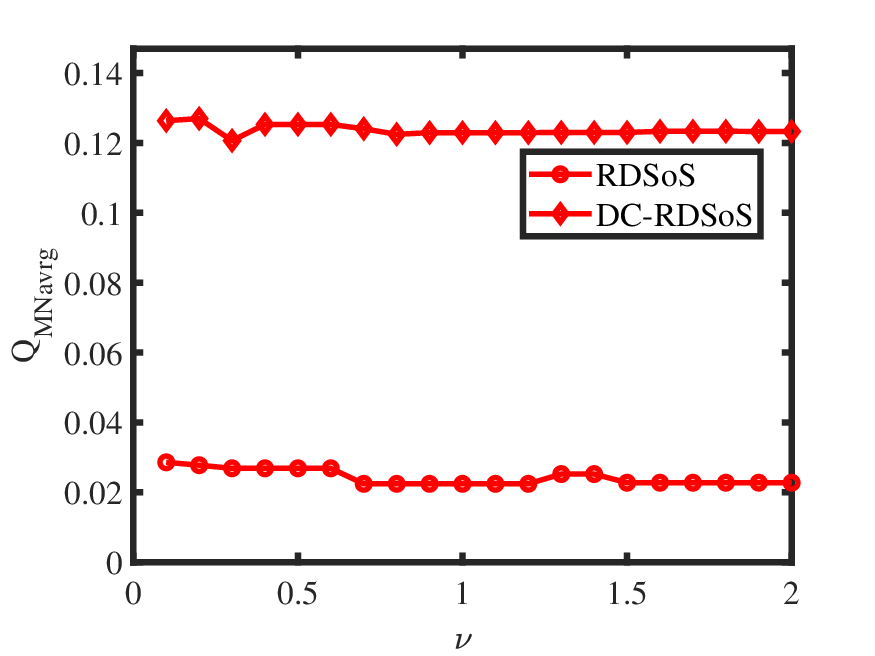}}
}
\resizebox{\columnwidth}{!}{
\subfigure[LLF]{\includegraphics[width=0.2\textwidth]{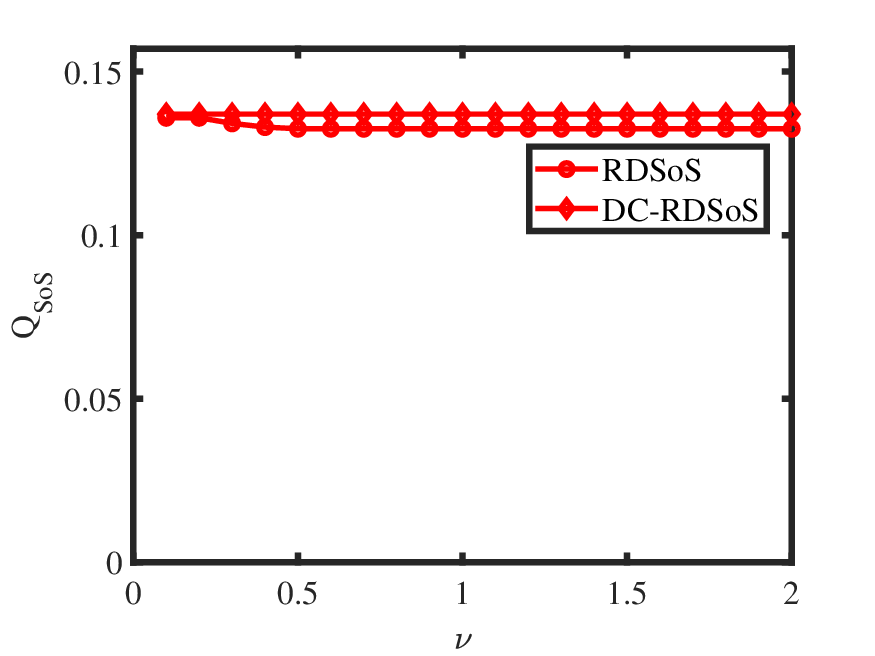}}
\subfigure[C.E]{\includegraphics[width=0.2\textwidth]{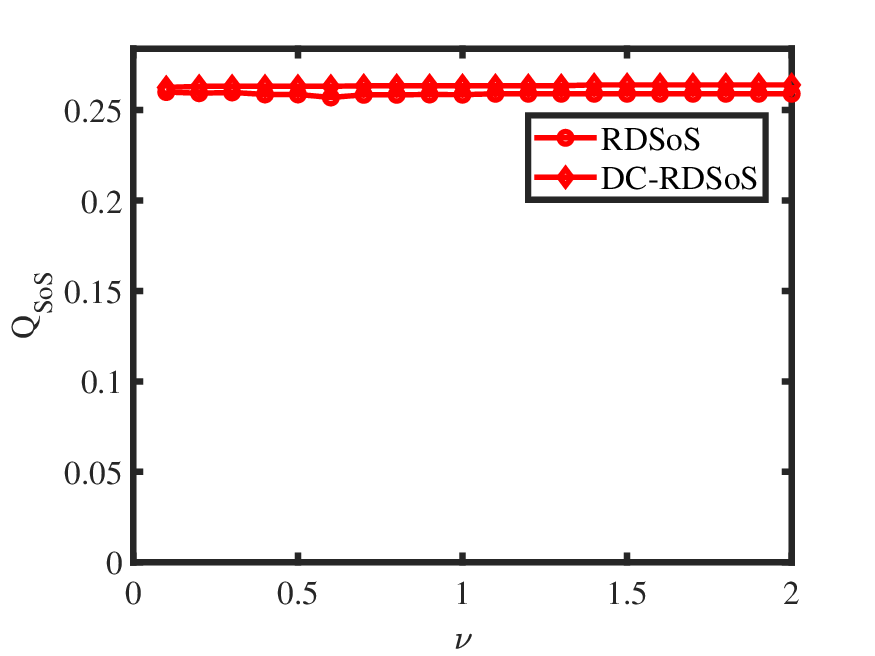}}
\subfigure[CS-A]{\includegraphics[width=0.2\textwidth]{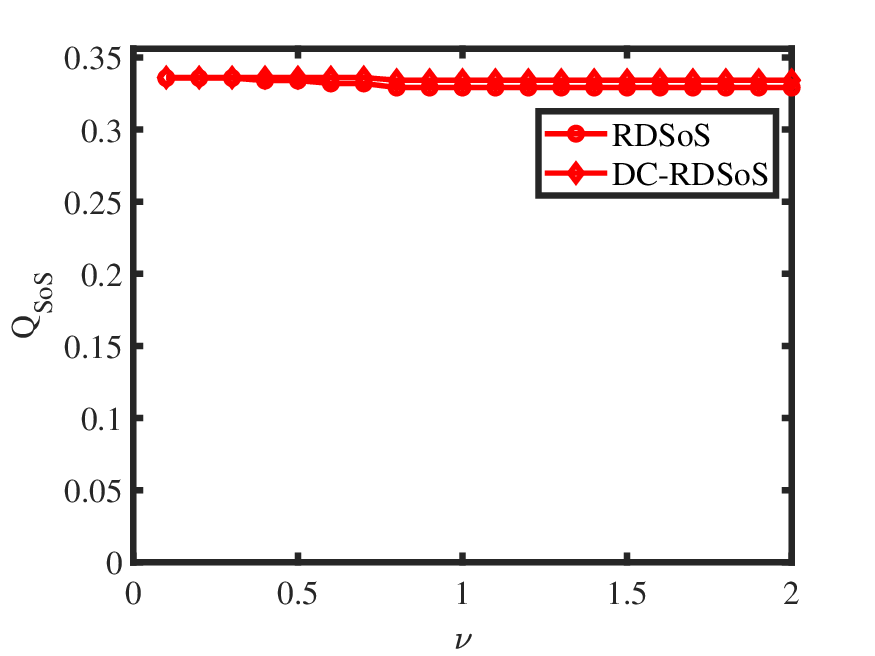}}
\subfigure[FAO-t]{\includegraphics[width=0.2\textwidth]{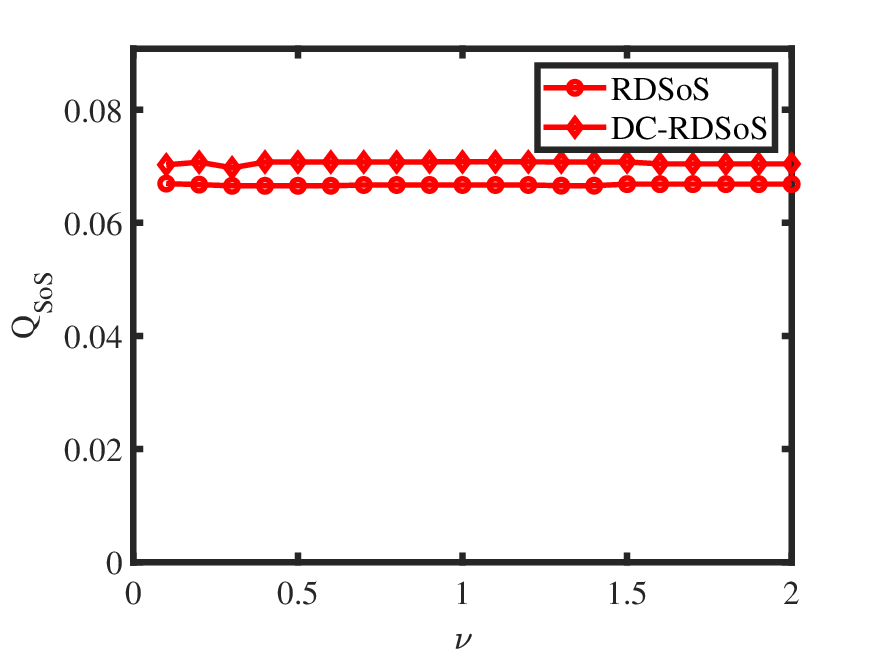}}
}
\caption{The modularity value against $\nu$ for the four real multi-layer networks listed in Table \ref{realdata}, where we let $\nu$ range in $\{0.1, 0.2, \ldots,2\}$. For the computation of modularity, we set $K$ as 3, 5, 5, and 3 for LLF, C.E, CS-A, and FAO-t, respectively.}
\label{TauQQs} 
\end{figure}

In Fig.~\ref{SortA}, we present the adjacency matrices of these real-world multi-layer networks, where the nodes are rearranged based on the estimated community partitions. From this figure, it is evident that, except nodes belonging to the first estimated community of C.E and FAO-t, nodes within the same community (indicated by the square section in the diagonal part of each adjacency matrix) generally exhibit a higher number of connections compared to nodes across different communities, suggesting that these multi-layer networks are assortative \citep{newman2002assortative}. For visualization purposes, we present the estimated community structure for each multi-layer network in Fig.~\ref{EstimatedC}. Similar to Simulation study 4, we also investigate the influence of the regularizer $\tau$ on the performance of RDSoS and DC-RDSoS across the four real-world multi-layer networks. The results presented in Fig.~\ref{TauQQs} suggest that both methods exhibit robustness to the selection of $\tau$.

To further analyze the detected communities (say the estimated community $k$ for $k\in[K]$), we define three indexes: mean degree across layers (mean-degree for short), mean in-degree across layers (mean-in-degree for short), and mean out-degree across layers (mean-out-degree for short). These indices are defined as follows:
\begin{align*}
&\mathrm{mean}-\mathrm{degree}=\frac{\sum_{l\in[T]}\sum_{i\in[n]}d_{l}(i)\mathbbm{1}(\hat{\ell}(i)=k)}{Tn_{k}},\mathrm{mean-in-degree}=\frac{\sum_{l\in[T]}\sum_{i,j: \hat{\ell}(i)=k, \hat{\ell}(j)=k}A_{l}(i,j)}{Tn_{k}},\\
&\mathrm{mean-out-degree}=\sum_{l\in[T]}\frac{\sum_{j:\hat{\ell}(j)\neq k}\sum_{i:\hat{\ell}(i)=k}A_{l}(i,j)}{T|j\in[n]:\hat{\ell}(j)\neq k, A_{l}(i,j)=1\mathrm{~for~at~least~one~node~}i\mathrm{~satisfying~}\hat{\ell}(i)=k|}.
\end{align*}

For $k\in[K]$, the index mean-degree represents the average degree of nodes that belong to the $k$-th estimated community. The mean-in-degree index quantifies the average degree of nodes' connections exclusively within the k-th estimated community. Lastly, the mean-out-degree index measures the average degree of nodes that are not part of the k-th estimated community but are connected to at least one node within it. These three indices are considered to offer a comprehensive perspective on the varying degrees within the estimated community across all layers. Tables \ref{CommunityLLFbasicInformation}-\ref{CommunityFAObasicInformation} record the three indexes for data analyzed in this paper. We have the following observations:
\begin{itemize}
  \item The size of communities varies significantly across the four networks, with some networks having more balanced community sizes (e.g., LLF) and others having more imbalanced sizes (e.g., C.E and FAO-t).
  \item In general, except for Community 1 of C.E and FAO-t, the mean-in-degree within communities is consistently higher than the mean-out-degree for all the other detected communities, indicating that nodes tend to connect more within their communities than with nodes in other communities, i.e., these detected communities are relatively well-separated in these networks and these networks are assortative \citep{newman2002assortative}.
  \item The mean degree across layers varies considerably within and across communities in all networks, reflecting the heterogeneity in node connectivity patterns.
  \item For LLF, the mean-degree across layers within each community is relatively high, with Community 1 having the highest mean-degree (19.4713) and Community 2 having the lowest (14.8286). The mean-in-degree is higher than the mean-out-degree in all communities, suggesting that nodes within the same community tend to connect more than in other communities. This is consistent with the connection patterns in panels (a)-(c) of Fig.~\ref{SortA}. Meanwhile, the LLF network exhibits a higher level of inter-community connectivity than the other networks. Similar observations can be found for the CS-A network.
  \item For C.E, the mean-in-degree for Communities 2-5 is higher than the mean-out-degree, indicating a higher density of connections within these four communities than between communities. However, the mean-in-degree of Community 1 is smaller than the mean-out-degree, suggesting that nodes in Community 1 tend to have fewer connections than across communities. These observations are consistent with panels (d)-(f) in Fig.~\ref{SortA}. Meanwhile, Community 5, which has the smallest size, exhibits a relatively high mean-degree (8.7169), suggesting the presence of hub nodes within this community. A similar analysis holds for the FAO-t network, where the mean-in-degree is smaller than the mean-out-degree for its Community 1 and there exists some hub nodes because its Community 2 with the smallest size exhibits a much higher mean-degree (14.0703) compared to the other communities.
\end{itemize}

\begin{table}[h!]
\footnotesize
	\centering
	\caption{Size, mean-degree, mean-in-degree, and mean-out-degree in each  community detected by our DC-RDSoS for the LLF network, where size denotes the number of nodes in the estimated community.}
	\label{CommunityLLFbasicInformation}
	\begin{tabular}{cccccccccccc}
\hline
Cluster&Size&Mean-degree&Mean-in-degree&Mean-out-degree\\
\hline
Community 1&29&19.4713&13.5172&5.1194\\
Community 2&19&14.8286&9.0526&2.7114\\
Community 3&23&16.5942&8.9275&4.3732\\
\hline
\end{tabular}
\end{table}

\begin{table}[h!]
\footnotesize
	\centering
	\caption{Size, mean-degree, mean-in-degree, and mean-out-degree in each community detected by our DC-RDSoS for the C.E network.}
	\label{CommunityCEbasicInformation}
	\begin{tabular}{cccccccccccc}
\hline
Cluster&Size&Mean-degree&Mean-in-degree&Mean-out-degree\\
\hline
Community 1&53&5.4340&0.6604&3.2227\\
Community 2&82&6.8984&5.0976&2.3330\\
Community 3&64&9.4688&4.6250&2.7213\\
Community 4&60&7.2833&4.5778&2.2222\\
Community 5&20&8.7169&3.9333&1.9253\\
\hline
\end{tabular}
\end{table}

\begin{table}[h!]
\footnotesize
	\centering
	\caption{Size, mean-degree, mean-in-degree, and mean-out-degree in each community detected by our DC-RDSoS for the CS-A network.}
	\label{CommunityCSAbasicInformation}
	\begin{tabular}{cccccccccccc}
\hline
Cluster&Size&Mean-degree&Mean-in-degree&Mean-out-degree\\
\hline
Community 1&13&5.2462&3.9692&1.4732\\
Community 2&14&4.4571&3.4000&1.4636\\
Community 3&14&3.4429&1.8286&1.5688\\
Community 4&12&3.7667&2.9000&1.2583\\
Community 5&8&3&2.0500&1.0111\\
\hline
\end{tabular}
\end{table}

\begin{table}[h!]
\footnotesize
	\centering
	\caption{Size, mean-degree, mean-in-degree, and mean-out-degree in each community detected by our DC-RDSoS for the FAO-t network.}
	\label{CommunityFAObasicInformation}
	\begin{tabular}{cccccccccccc}
\hline
Cluster&Size&Mean-degree&Mean-in-degree&Mean-out-degree\\
\hline
Community 1&108&2.5600&0.5604&4.4740\\
Community 2&50&14.0703&9.1748&4.1265\\
Community 3&56&8.8647&3.9924&3.7042\\
\hline
\end{tabular}
\end{table}

\begin{figure}
\centering
\resizebox{\columnwidth}{!}{
{\includegraphics[width=0.2\textwidth]{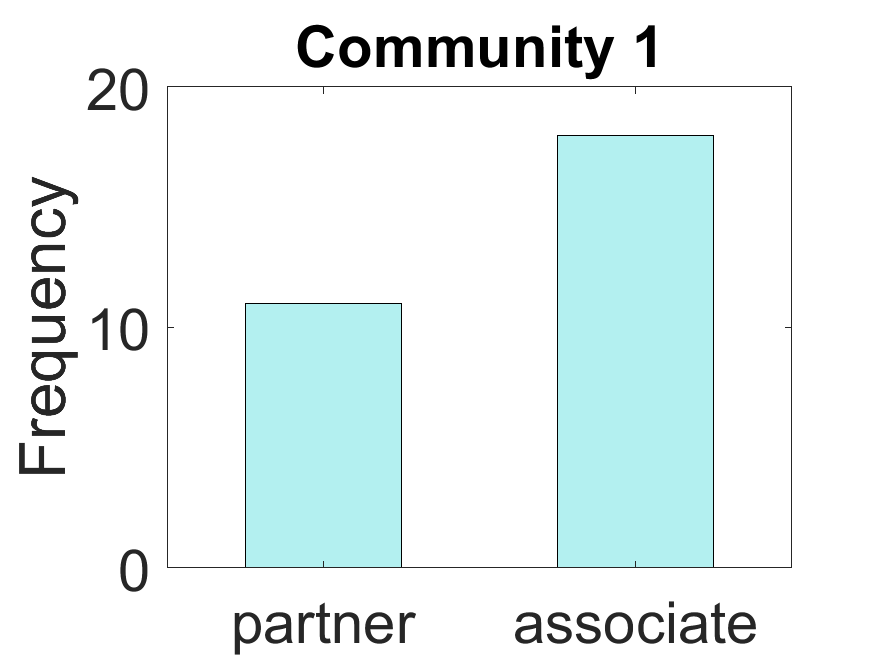}}
{\includegraphics[width=0.2\textwidth]{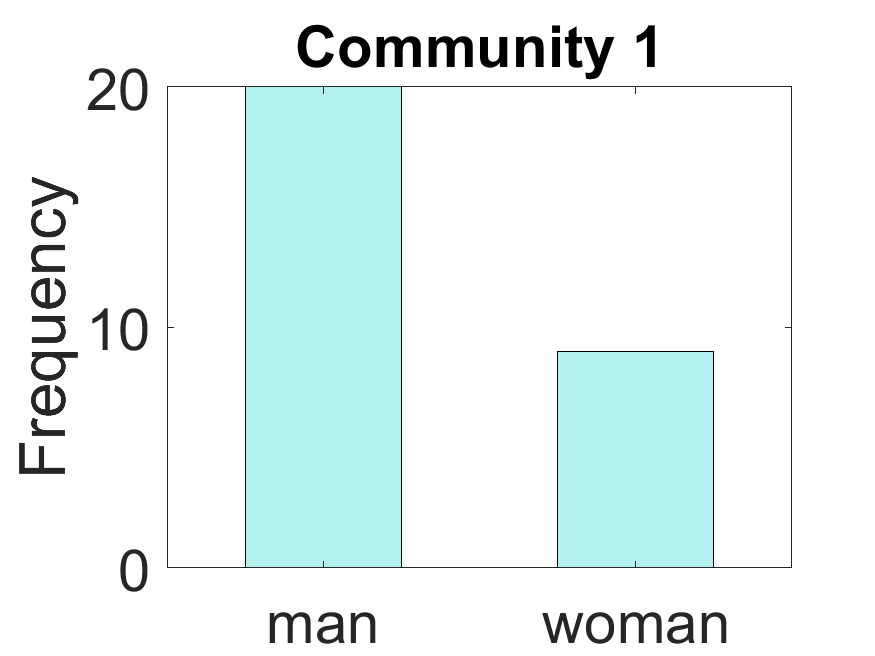}}
{\includegraphics[width=0.2\textwidth]{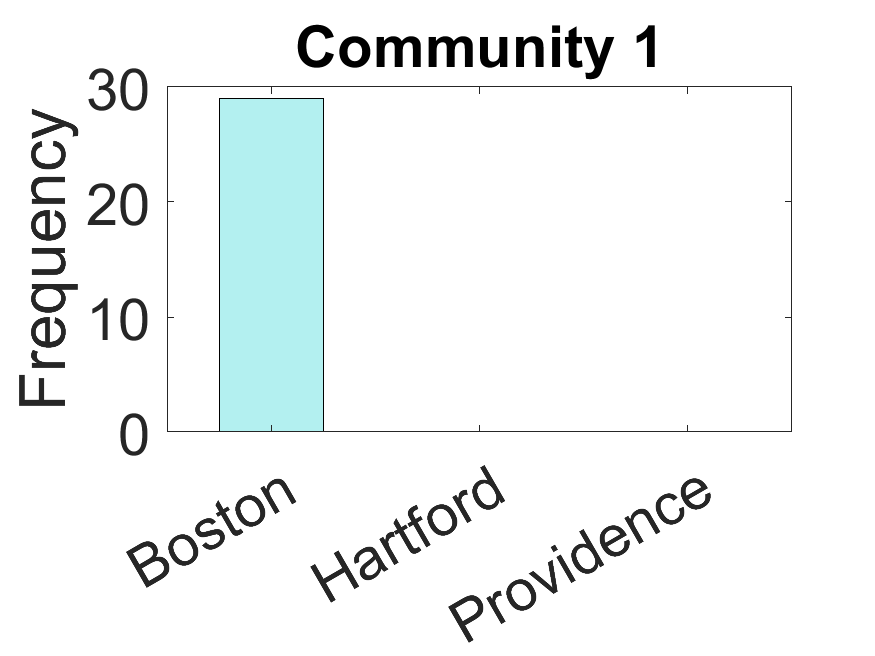}}
{\includegraphics[width=0.2\textwidth]{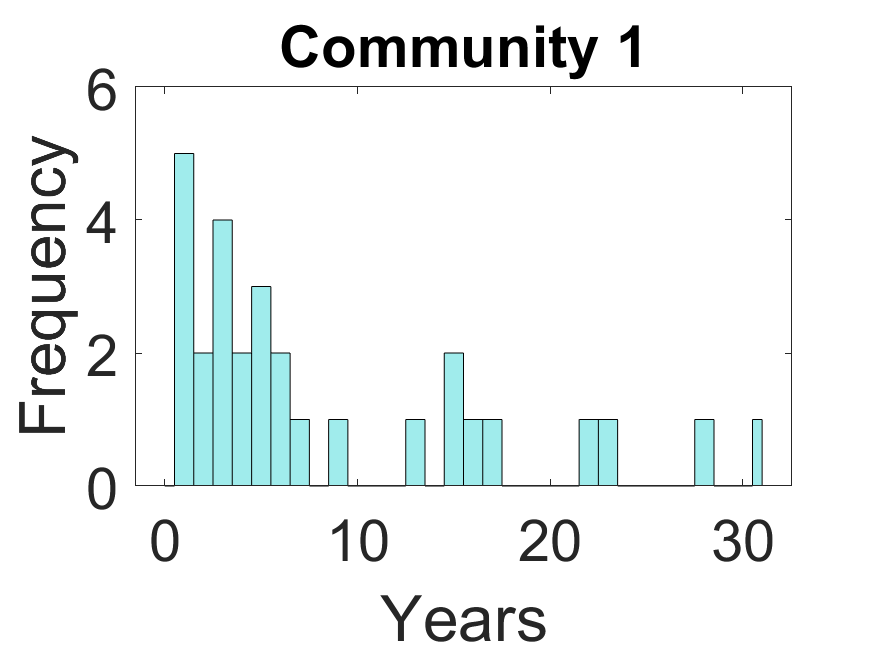}}
}
\resizebox{\columnwidth}{!}{
{\includegraphics[width=0.2\textwidth]{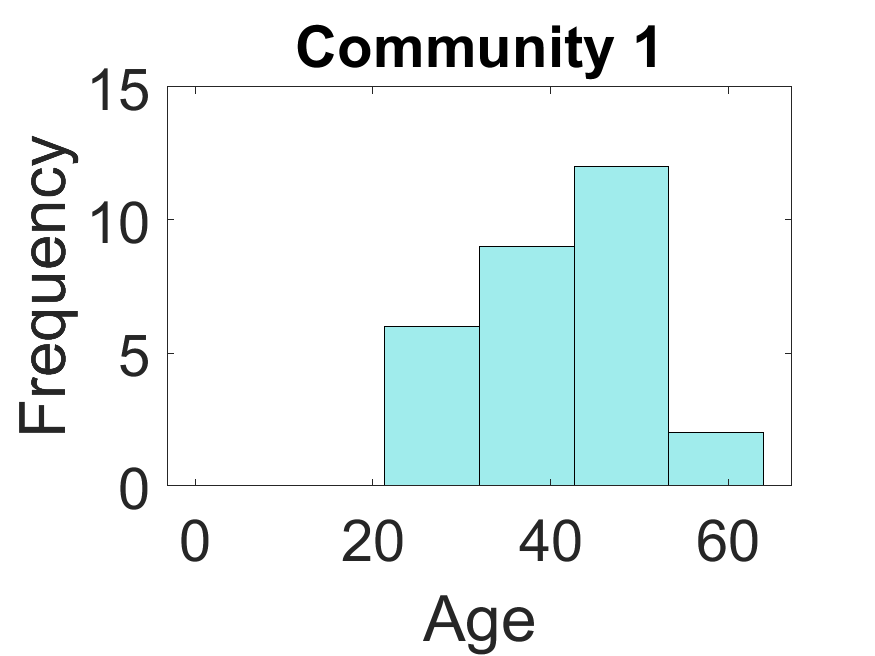}}
{\includegraphics[width=0.2\textwidth]{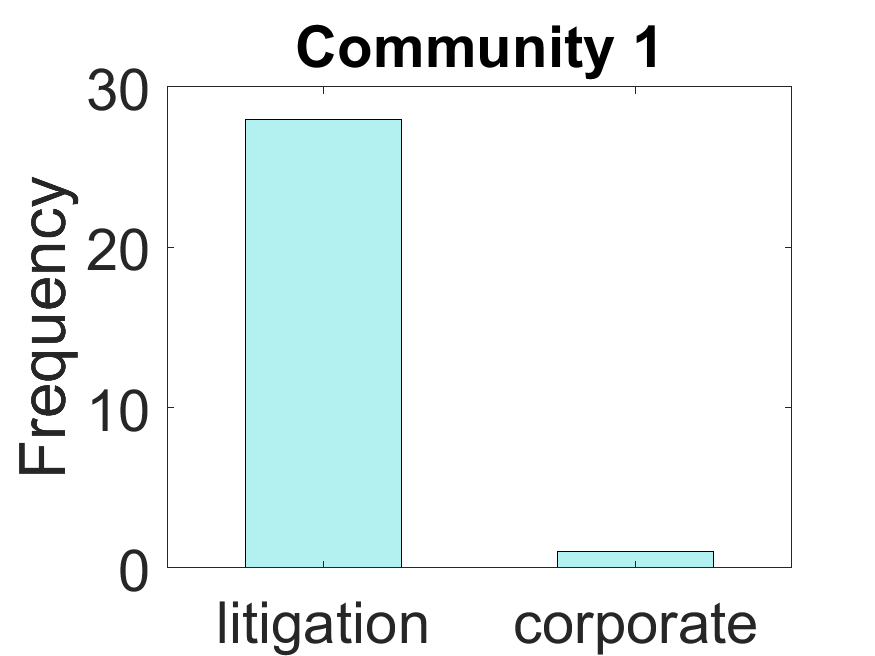}}
{\includegraphics[width=0.2\textwidth]{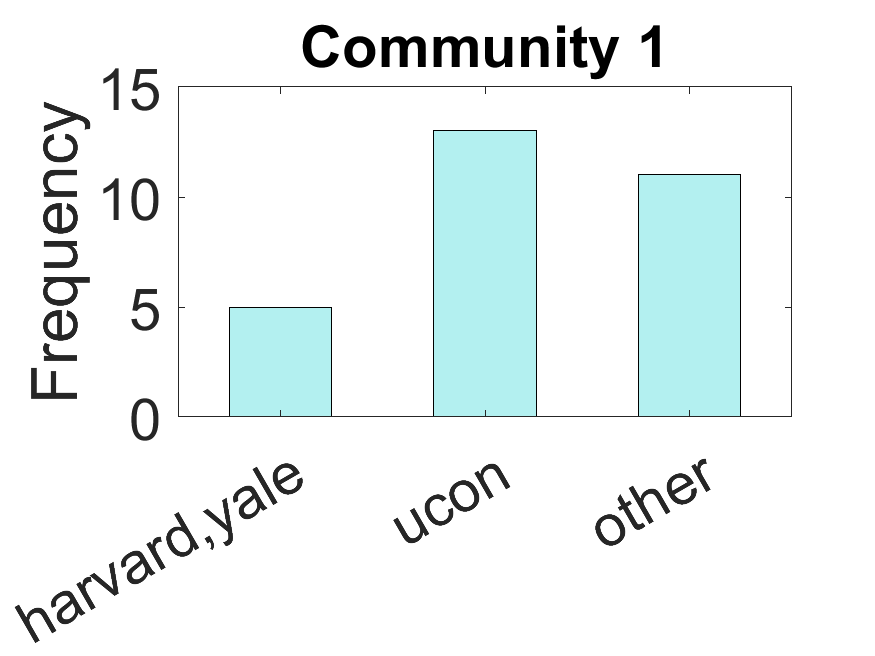}}
{\includegraphics[width=0.2\textwidth]{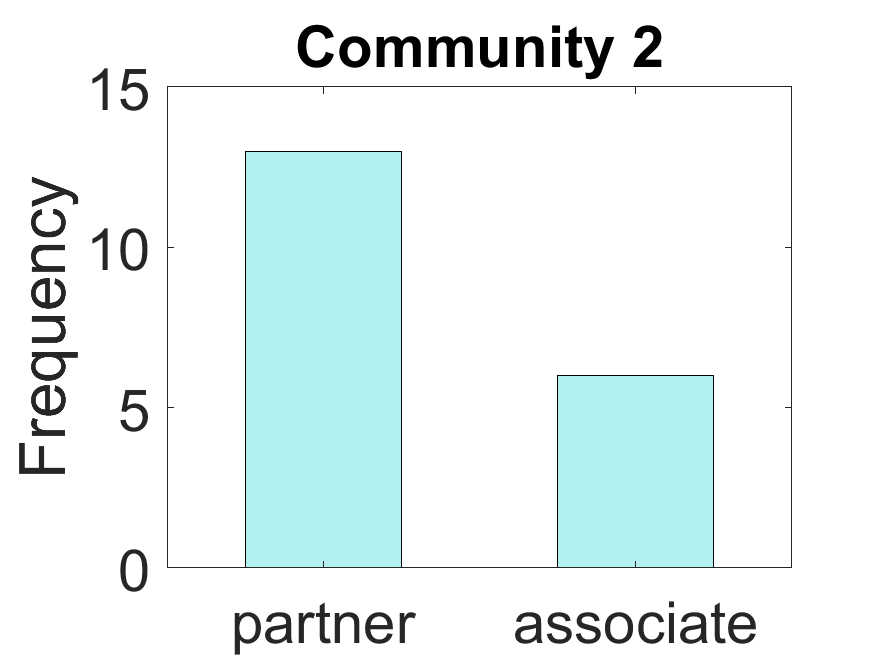}}
}
\resizebox{\columnwidth}{!}{
{\includegraphics[width=0.2\textwidth]{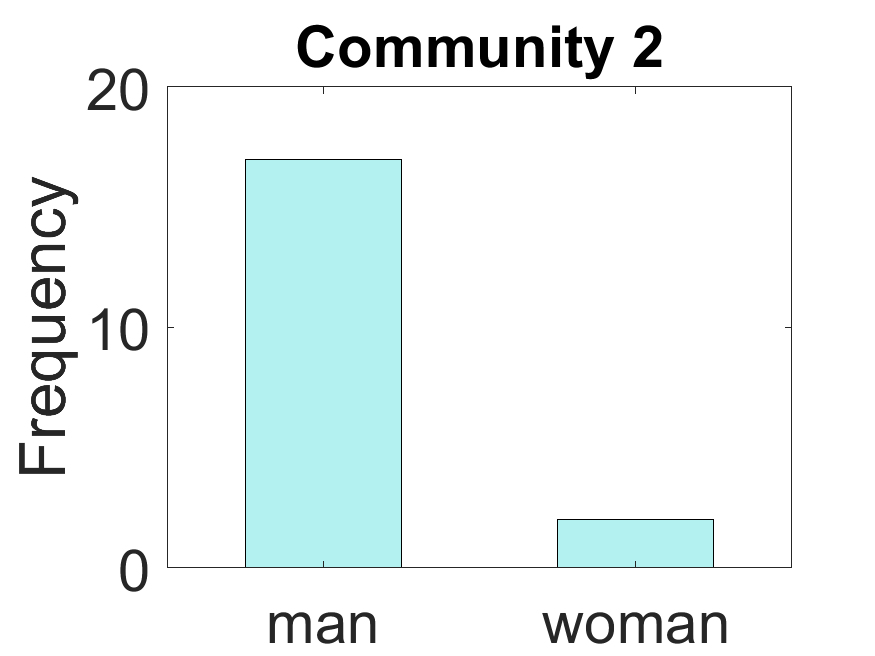}}
{\includegraphics[width=0.2\textwidth]{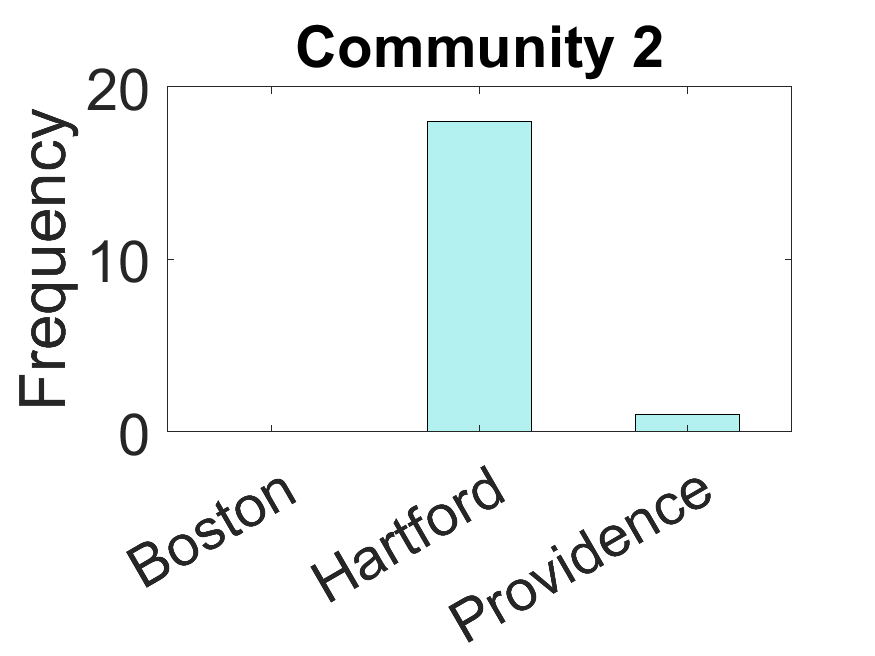}}
{\includegraphics[width=0.2\textwidth]{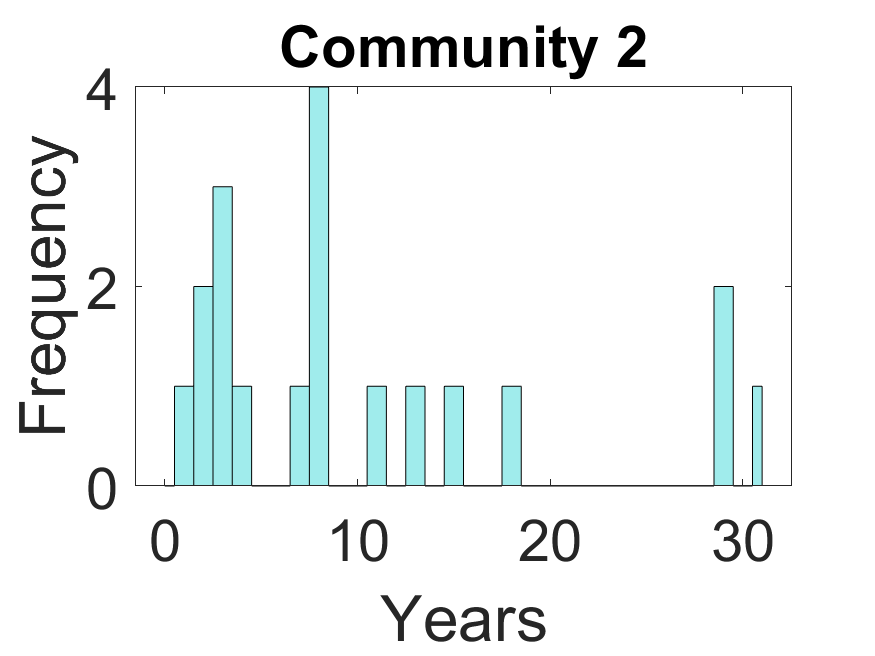}}
{\includegraphics[width=0.2\textwidth]{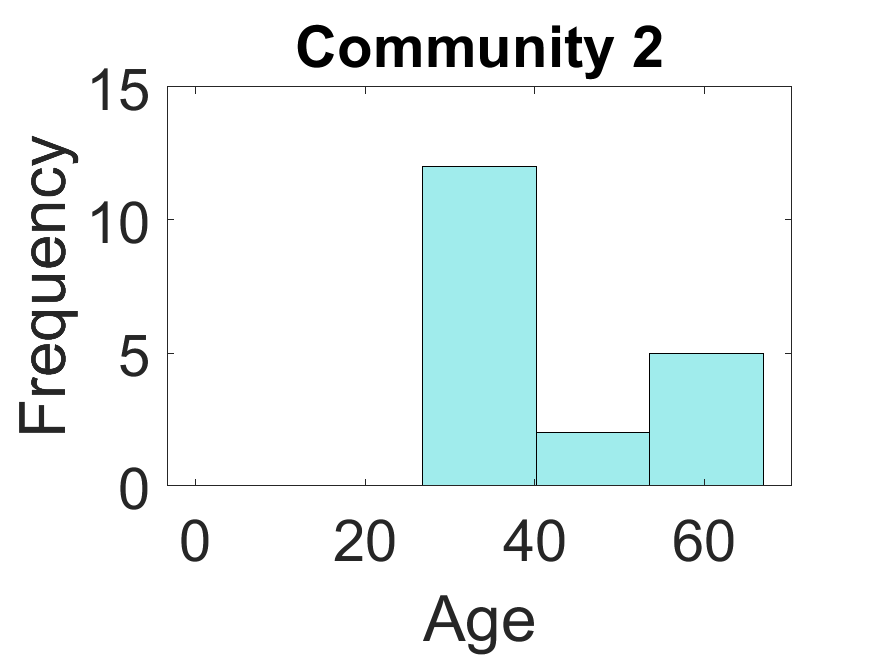}}
}
\resizebox{\columnwidth}{!}{
{\includegraphics[width=0.2\textwidth]{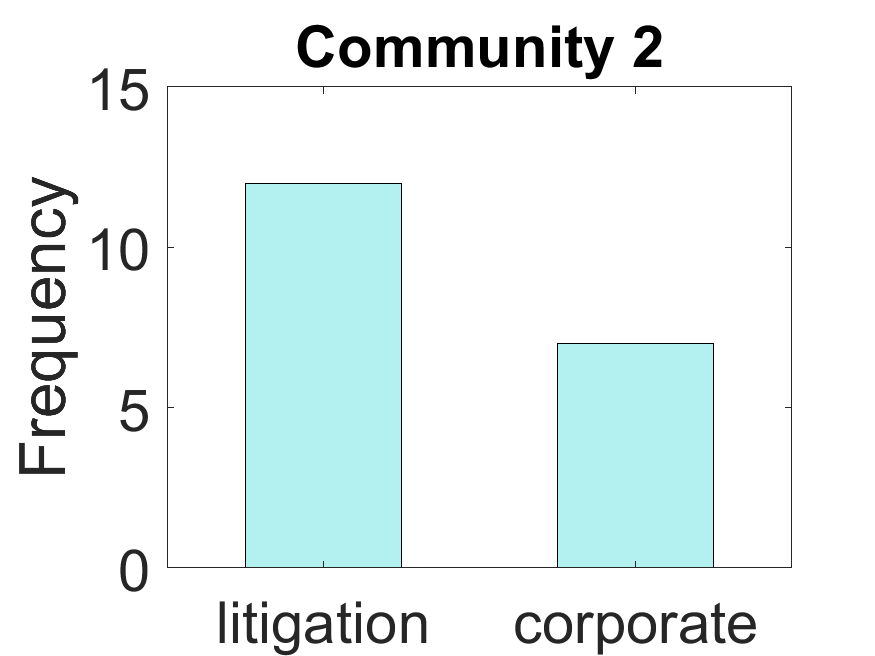}}
{\includegraphics[width=0.2\textwidth]{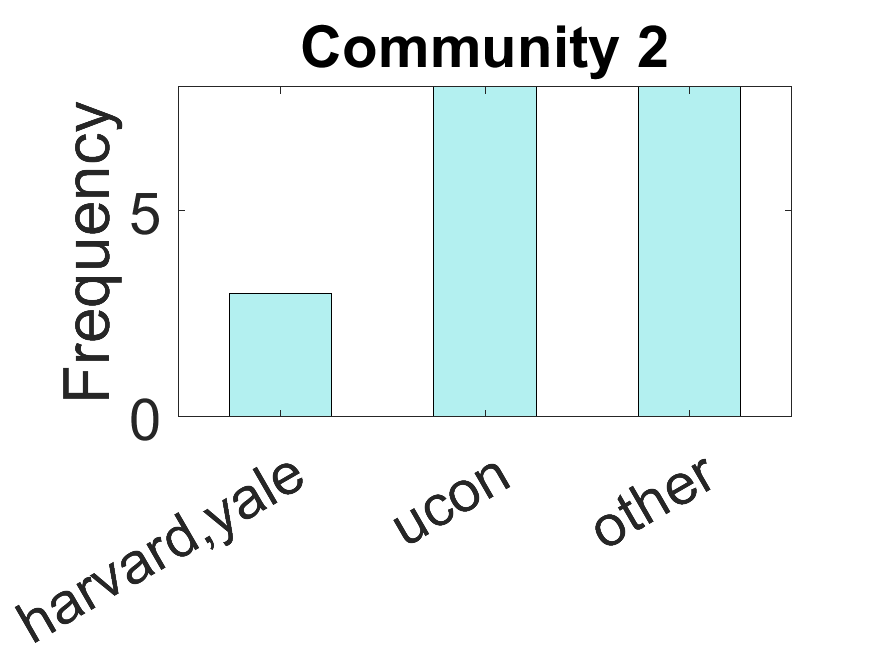}}
{\includegraphics[width=0.2\textwidth]{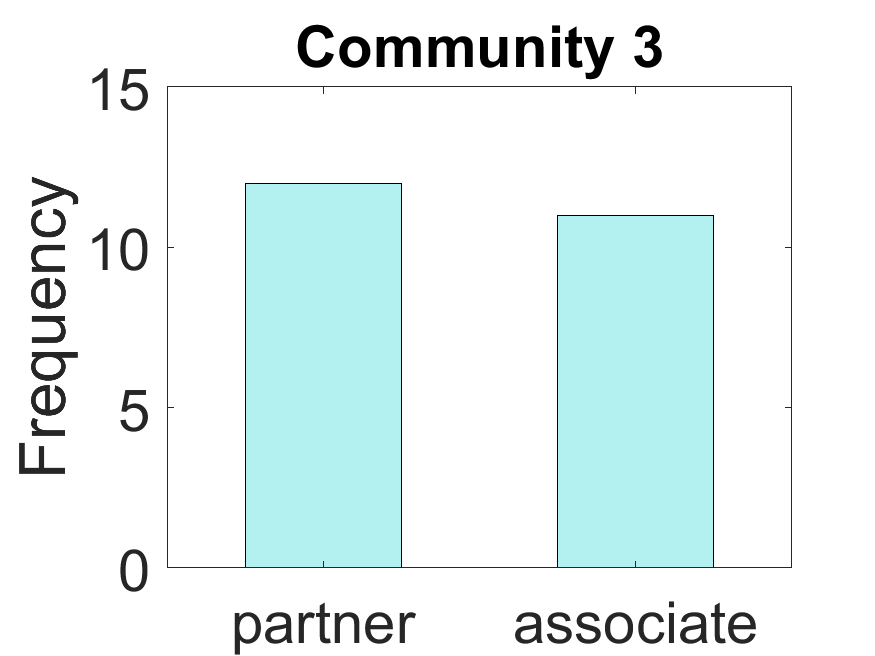}}
{\includegraphics[width=0.2\textwidth]{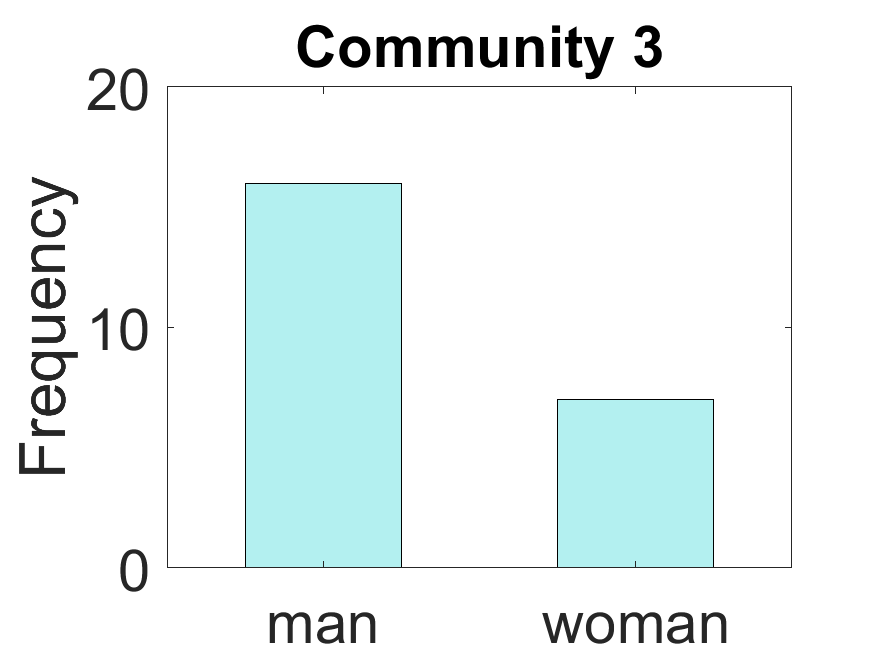}}
}
\resizebox{\columnwidth}{!}{
{\includegraphics[width=0.2\textwidth]{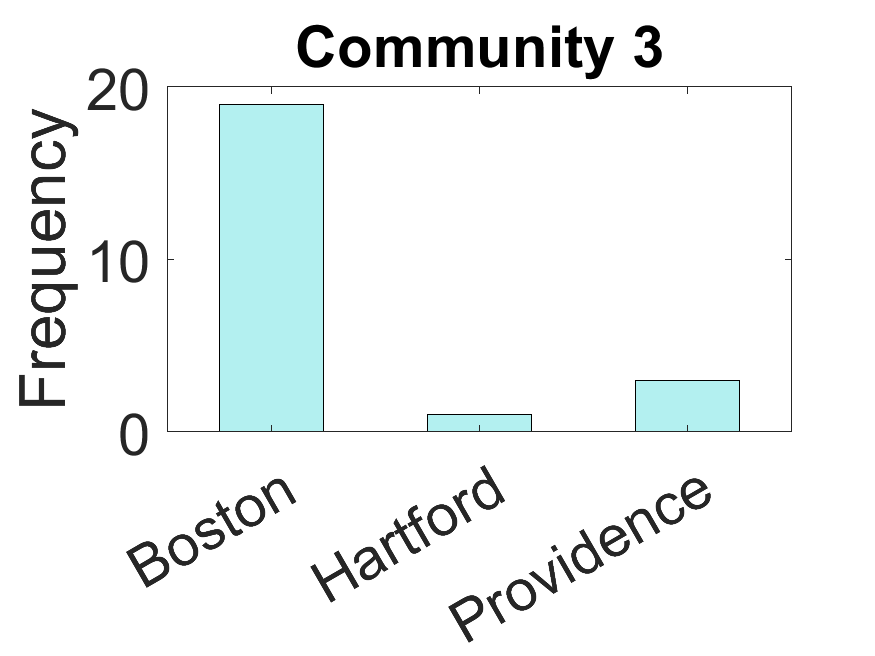}}
{\includegraphics[width=0.2\textwidth]{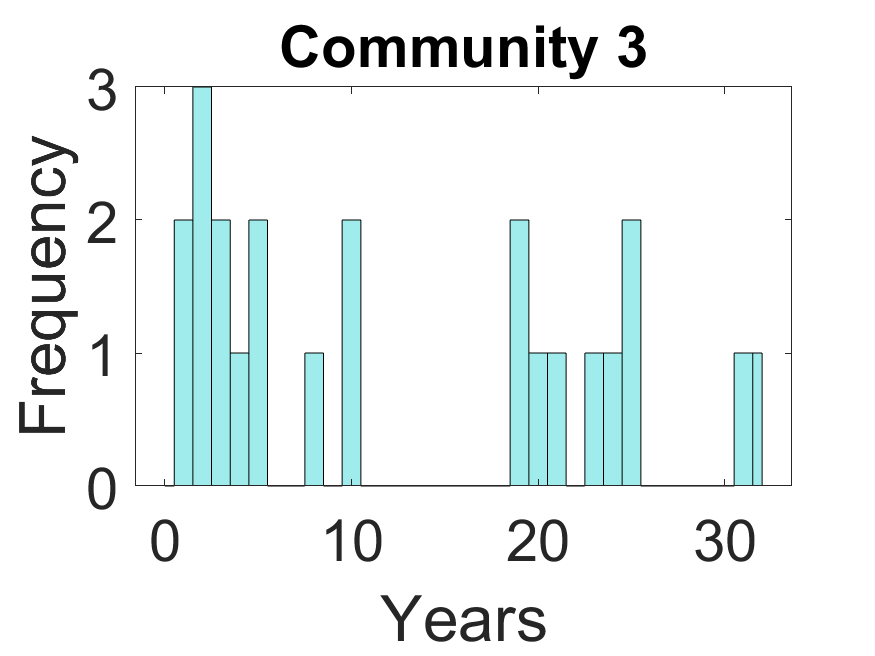}}
{\includegraphics[width=0.2\textwidth]{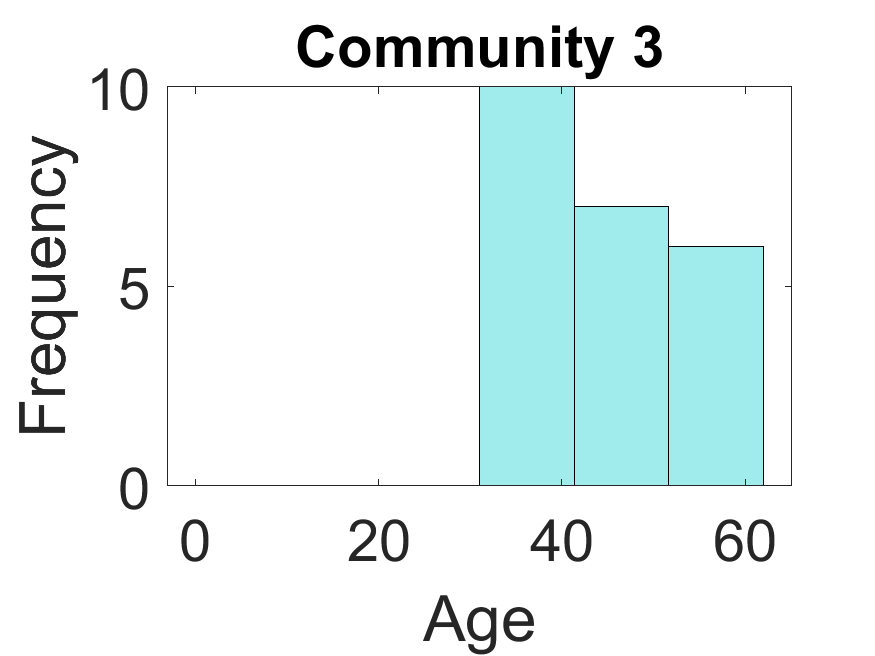}}
{\includegraphics[width=0.2\textwidth]{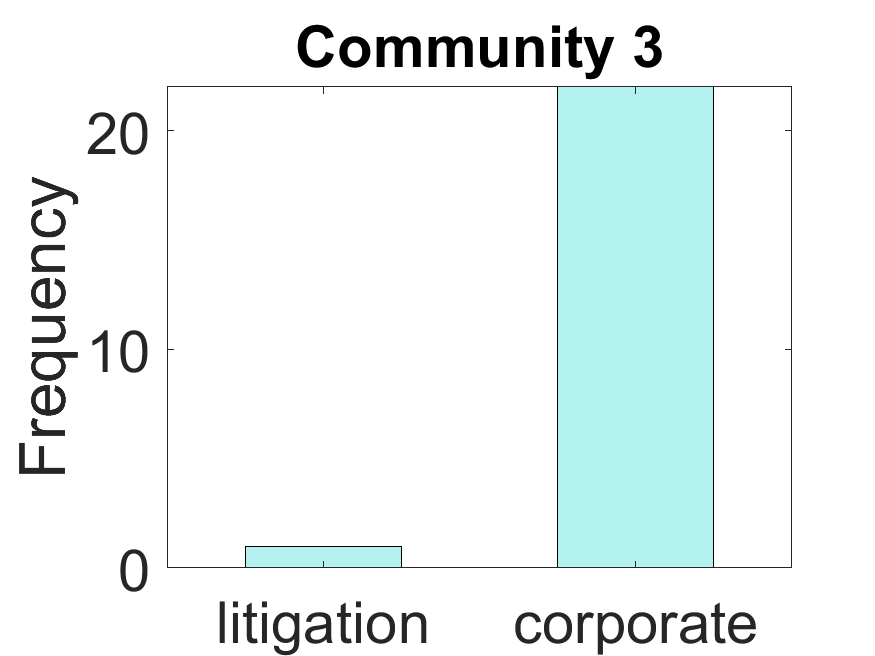}}
}
{\includegraphics[width=0.25\textwidth]{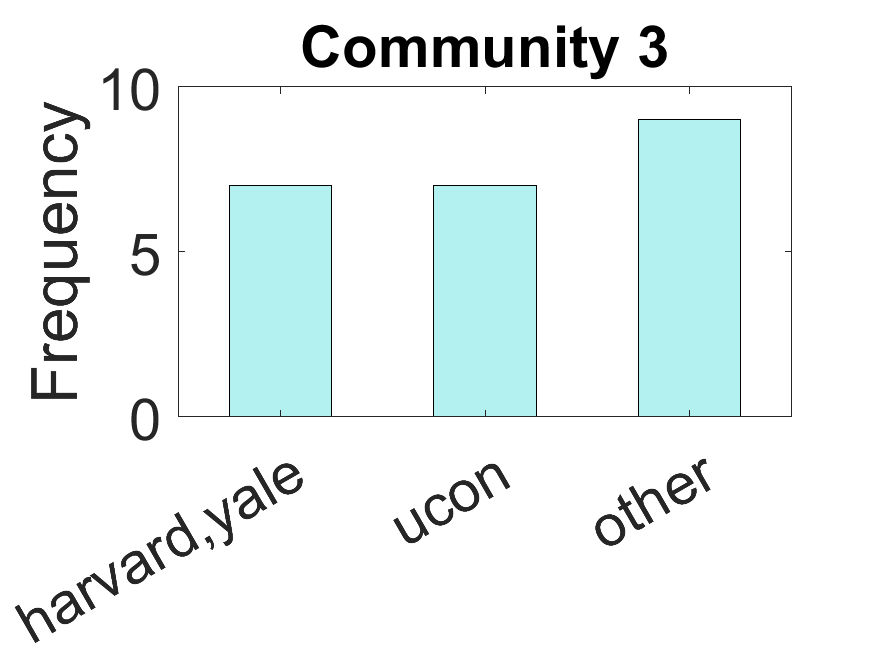}}
\caption{Attribute distribution within each detected community in the LLF network.}
\label{DegreeCommunityLLF} 
\end{figure}

Since the LLF datasets encompass multiple attributes, including status (partner and associate), gender (man and woman), office (Boston, Hartford, and Providence), years with the firm, age, practice (litigation and corporate), and law school (harvard$\&$yale, ucon, and other), we proceed to conduct a more in-depth analysis of the detected communities returned by our DC-RDSoS algorithm within this datasets. Fig.~\ref{DegreeCommunityLLF} showcases the distributions of various attributes across the detected communities for the LLF network. A closer examination reveals insightful patterns regarding the community composition and characteristics:
\begin{itemize}
  \item \textbf{Status}: The majority of nodes in Community 1 hold the status of associate, indicating that this community primarily comprises junior members of the law firm. In contrast, Community 2 and Community 3 are dominated by partners, suggesting that these communities encompass more senior members.
  \item \textbf{Gender}: Across all communities, the number of male nodes outnumbers the female nodes, reflecting a potential gender imbalance within the law firm network. However, no clear trend emerges in the gender distribution among the different communities.
  \item \textbf{Office location}: There is a notable geographic segregation in the office locations. Almost all nodes in Community 1 and Community 3 are based in Boston, whereas most nodes in Community 2 locate in Hartford. This suggests that office location might be a significant factor influencing community formation within the LLF network.
  \item \textbf{Years with the firm}: Each community exhibits a similar range of years with the firm, indicating that tenure does not significantly differ among the community members. This suggests that community structure is not primarily driven by the length of service within the firm.
  \item \textbf{Age}: Nodes in Community 1 tend to be younger than those in Community 2 and Community 3. This age difference aligns with the status distribution, where Community 1 comprises more junior associates and the other communities include more senior partners.
  \item \textbf{Practice area}: Community 1 predominantly focuses on litigation, while Community 3 specializes in corporate practice. Community 2 shows a slight bias towards litigation but also includes a significant number of corporate practitioners, suggesting a more mixed practice area. This specialization in different legal fields contributes to the structure of these communities.
  \item \textbf{Law school}: Most nodes in Community 1 and Community 2 graduated from UConn or other law schools, indicating a potential shared educational background. In contrast, Community 3 displays a more diverse law school distribution, with members from various institutions. This diversity might reflect broader networking and professional connections within this community.
\end{itemize}

Overall, the analysis of Fig.~\ref{DegreeCommunityLLF}  highlights that the detected communities in the LLF network are shaped by a combination of factors, including professional status, office location, practice area, and educational background. These attributes not only distinguish the communities but also provide insights into the potential drivers of community formation within the law firm network.

\section{Conclusion}\label{sec7}
\subsection{Key findings and academic contributions}
This paper introduces a novel regularized Laplacian matrix $L_{\tau}$ to extend the concept of the classical regularized Laplacian matrix, typically used for community detection in single-layer networks, to multi-layer networks. Based on $L_{\tau}$, two novel regularized debiased spectral clustering methods, RDSoS and DC-RDSoS, are developed. The two proposed methods identify communities by applying the K-means algorithm to the leading eigenvector matrix or its variant constructed from $L_{\tau}$. The consistency results of both methods are established under the MLSBM and the MLDCSBM, respectively. Under mild conditions, the theoretical results of both methods match those of Theorem 1 in \citep{lei2023bias} in the MLSBM. Furthermore, we introduce SoS-modularity, a novel metric for assessing the quality of community partitions in multi-layer networks. By maximizing this metric, the number of communities in multi-layer networks can be estimated.

Extensive experimental results demonstrate the satisfactory empirical performance of both methods, outperforming or being at least comparable to state-of-the-art techniques on both synthetic and real datasets. Both theoretical and empirical results show that the proposed methods are insensitive and robust to the choice of the regularizer $\tau$, with a moderate value of $\tau$ being preferred. This robustness simplifies the application of the proposed methods in practice. Furthermore, the experimental results also imply that our SoS-modularity is more believable than the MNavrg-modularity in measuring the quality of community partitions in multi-layer networks because methods estimating the number of communities $K$ by maximizing the SoS-modularity enjoy higher accuracy rates than those using the MNavrg-modularity. To the best of our knowledge, we are the first to design the regularized Laplacian matrix $L_{\tau}$ using the debiased sum of squared adjacency matrices $S$, the first to design the proposed regularized debiased spectral clustering methods using $L_{\tau}$ and study their theoretical properties for community detection in multi-layer networks, and the first to introduce the SoS-modularity. Overall, this work contributes to advancing the field of network analysis by providing novel tools for uncovering the underlying community structures and novel metric for measuring the quality of community partitions in complex multi-layer networks, which are ubiquitous in various fields such as social sciences, biological sciences, and computer sciences.
\subsection{Limitations and future works}
Despite the advancements, our study still has certain limitations. Firstly, we have focused solely on non-overlapping multi-layer networks, where each node belongs to a single community. In future work, we aim to develop regularized debiased spectral clustering methods to estimate mixed memberships in overlapping multi-layer networks, where nodes may belong to multiple communities. Secondly, in line with many studies utilizing the classical Laplacian matrix (as seen in Table \ref{LiteratureReview}), we set the power term of $D_{\tau}$ to $-\frac{1}{2}$ for our proposed regularized Laplacian matrix defined in Equation (\ref{DefineLaplacian}). An interesting question immediately arises: does an optimal value of $\beta$ exist under MLSBM and MLDCSBM such that our RDSoS and DC-RDSoS, designed based on this optimal $\beta$, outperform methods using other $\beta$ values both theoretically and empirically? The optimal choice of $\beta$ for regularized spectral clustering algorithms under the SBM model for single-layer networks has been explored in \citep{ali2018improved}. Extending this work to find the optimal $\beta$ for our methods within the MLSBM and MLDCSBM frameworks for multi-layer networks is a promising research avenue.

Benefiting from our novel regularized Laplacian matrix $L_{\tau}$ defined in Equation \ref{DefineLaplacian} for multi-layer networks, several promising future research directions emerge. Here, we highlight a few of them. Firstly, expanding the concept of our regularized Laplacian matrix $L_{\tau}$ to multi-layer directed networks modeled by a directed version of the MLSBM \citep{su2024spectral} is appealing. Secondly, one can design a method that removes the effect of $\theta$ by employing the entry-wise ratios idea used in \citep{SCORE,wang2020spectral}. Studying the estimation consistency of this method under the MLDCSBM is appealing. Thirdly, given that \cite{deng2021strong} studied the strong consistency of graph Laplacians within the SBM framework, and considering our methods also employ the regularized Laplacian matrix $L_{\tau}$, investigating their strong consistency under the MLSBM and the MLDCSBM for multi-layer networks is a valuable problem. Fourthly, based on our $L_{\tau}$, extending the covariate-assisted spectral clustering approach studied in \citep{binkiewicz2017covariate} from single-layer networks to multi-layer networks with covariates holds promise. Fifthly, while we assumed the membership matrix $Z$ to be consistent across all layers in this paper, exploring the latent community structure in multi-layer networks with time-varying node memberships using our $L_{\tau}$ is appealing. Sixthly, by integrating the study on spectral co-clustering in multi-layer directed networks from \citep{su2024spectral} with our concept of regularized debiased spectral clustering, we can devise regularized debiased spectral co-clustering algorithms for identifying communities in multi-layer directed networks. Seventhly, leveraging the distribution-free approach outlined in \citep{qing2023community}, we can develop algorithms based on our regularized Laplacian matrix concept to detect communities in multi-layer weighted networks. Lastly, recent works, such as those in \citep{jiang2023adaptive, IJCAI}, have demonstrated the effectiveness of leveraging self-supervised learning and graph neural networks for recommendation systems. These methods typically rely on contrastive learning to learn the complex structure of user-item interactions. While our work focuses on community detection in multi-layer networks, the proposed regularized debiased spectral clustering methods share some commonalities with these graph learning approaches. Specifically, both our methods and the aforementioned works aim to discover latent structures within graph-structured data. However, our methods are designed to identify communities in multi-layer networks, while the works on graph learning for recommendation typically focus on user-item interactions in a single layer. Nevertheless, the insights gained from our study on extending spectral clustering to multi-layer networks could inspire future research on incorporating multi-layer graph structures into graph learning methods for recommendation systems. In detail, we observe that the graph learning methods for recommendations proposed in \citep{jiang2023adaptive, IJCAI} rely on an $\mathcal{I}\times \mathcal{J}$ interaction matrix $\mathcal{A}$, where $\mathcal{A}(i,j)$ equals 1 if user $i$ has adopted item $j$ and 0 otherwise, for $i\in[\mathcal{I}]$ and $j\in[\mathcal{J}]$. Here, $\mathcal{I}$ and $\mathcal{J}$ represent the number of users and items, respectively. Consider the scenario where there are $T$ interaction matrices $\{\mathcal{A}_{l}\}^{T}_{l=1}$ gathered from $T$ different platforms with $\mathcal{I}$ common users and $\mathcal{J}$ common items, where each interaction matrix $\mathcal{A}_{l}$ is an  $\mathcal{I}\times \mathcal{J}$ matrix defined similarly as $\mathcal{A}$. We further assume that the $\mathcal{I}$ users belong to $K$ communities. This assumption is reasonable because users within the same community often share similar preferences, leading them to adopt more common items than those from different communities. By utilizing the directed version of the algorithms developed in this study, one can discover the latent communities of these users. Such community information can be highly advantageous for the graph learning methods proposed in \citep{jiang2023adaptive, IJCAI} for recommendation.
\section*{CRediT authorship contribution statement}
\textbf{Huan Qing:} Conceptualization; Data curation; Formal analysis; Funding acquisition; Methodology; Project administration; Resources; Software; Validation; Visualization; Writing-original draft; Writing-review $\&$ editing.
\section*{Declaration of competing interest}
The author declares no competing interests.
\section*{Data availability}
Data will be made available on request. The MATLAB codes for RDSoS and DC-RDSoS can be found in Appendix \ref{MatlabCodes}.
\section*{Acknowledgements}
H.Q. was supported by the Scientific Research Foundation of Chongqing University of Technology (Grant No. 2024ZDR003), the Science and Technology Research Program of Chongqing Municipal Education Commission (Grant No. KJQN202401168), and the Natural Science Foundation of Chongqing, China (Grant No. CSTB2023NSCQ-
LZX0048).

\appendix
\section{Proofs}\label{SecProofs}
Proof of Proposition \ref{idMLSBM}: Assume that $\mathrm{rank}(\sum_{l\in[T]}\tilde{B}^{2}_{l})=K_{0}$, where $K_{0}\leq K$. Consequently, we have $\mathrm{rank}(\mathcal{L}_{\tau})=K_{0}$. Following a similar approach to the proof of Lemma \ref{PopulationLtauMLSBM}, let $U\Sigma U'$ represent the leading $K_{0}$ eigen-decomposition of $\mathcal{L}_{\tau}$ with $U'U=I$. Given that $U=ZX$ where $X$ is a $K\times K_{0}$ matrix with $K$ distinct rows, and $U$ is an $n\times K_{0}$ matrix due to $\mathrm{rank}(\mathcal{L}_{\tau})=K_{0}$. Combining the equation $ZB_{l}Z'=\breve{Z}\breve{B}_{l}\breve{Z}'$ for $l\in[T]$ with the uniqueness of the eigen-decomposition of $\mathcal{L}_{\tau}$ under $U'U=I$, we also deduce that $U=\breve{Z}\breve{X}$ where $\breve{X}$ possesses $K$ distinct rows. Let $\breve{\ell}$ be the $n\times1$ label vector of $\breve{Z}$, defined such that $\breve{\ell}(i)=k$ if $\breve{Z}(i,k)=1$ for $i\in[n], k\in[K]$. Now, suppose there exist two nodes $i\neq j$ such that $\ell(i)=\ell(j)$. This implies $U(i,:)=Z(i,:)X=\breve{Z}(i,:)\breve{X}=\breve{X}(\breve{\ell}(i),:)$ and $U(j,:)=Z(j,:)X=\breve{Z}(j,:)\breve{X}=\breve{X}(\breve{\ell}(j),:)$. Since $U(i,:)=U(j,:)$ when $\ell(i)=\ell(j)$, it follows that $\breve{X}(\breve{\ell}(i),:)=\breve{X}(\breve{\ell}(j),:)$. If $\breve{\ell}(i)\neq \breve{\ell}(j)$, then $\breve{X}(\breve{\ell}(i),:)=\breve{X}(\breve{\ell}(j),:)$ would indicate that $\breve{X}$ has at least two identical rows, contradicting the fact that $\breve{X}$ has $K$ distinct rows. Hence, we must have $\breve{\ell}(i)=\breve{\ell}(j)$, meaning that any two nodes $i$ and $j$ belonging to the same community in $Z$ must also belong to the same community in $\breve{Z}$. Therefore, we can express $\breve{Z}=Z\mathcal{P}$, where $\mathcal{P}$ is a $K\times K$ permutation matrix, implying that $\breve{Z}$ and $Z$ are equivalent up to a permutation of community labels. Finally, the lemma holds since $ZB_{l}Z'=\breve{Z}\breve{B}_{l}\breve{Z}'=Z\mathcal{P}\breve{B}_{l}\mathcal{P}'Z'$ leads to $B_{l}=\mathcal{P}\breve{B}_{l}\mathcal{P}'$ for $l\in[T]$.
\begin{rem}
In the proof of Proposition \ref{idMLSBM}, we consider the rank-deficient case $\mathrm{rank}(\sum_{l\in[T]}\tilde{B}^{2}_{l})=K_{0}$ when $K_{0}\leq K$. Alternatively, if we focus on the full-rank case with $K_{0}=K$, the proof of Proposition \ref{idMLSBM} becomes more straightforward. This statement can be understood as follows: when $\ell(i)=\ell(j)$, it follows that
$\breve{Z}(i,:)\breve{X}=\breve{Z}(j,:)\breve{X}$, leading directly to $\breve{Z}(i,:)=\breve{Z}(j,:)$ because $\breve{X}$ is nonsingular when $K_{0}=K$. For the sake of simplicity in our theoretical analysis, this paper primarily focuses on the full-rank case.
\end{rem}
\qed

Proof of Lemma \ref{PopulationLtauMLSBM}: For $i\in[n]$,
\begin{align*}
\mathcal{D}_{\tau}(i,i)&=\tau+\sum_{j\in[n]}\mathcal{S}(i,j)=\tau+\sum_{j\in[n]}\sum_{l\in[T]}\sum_{m\in[n]}\Omega_{l}(i,m)\Omega_{l}(j,m)=\tau+\sum_{j\in[n]}\sum_{l\in[T]}\sum_{m\in[n]}Z(i,:)B_{l}Z'(m,:)Z(j,:)B_{l}Z'(m,:)\\
&=\tau+Z(i,:)\sum_{j\in[n]}\sum_{l\in[T]}\sum_{m\in[n]}B_{l}Z'(m,:)Z(j,:)B_{l}Z'(m,:),
\end{align*}
and
\begin{align*}
\mathcal{S}(i,:)=(\sum_{l\in[T]}\Omega^{2}_{l})(i,:)=(\sum_{l\in[T]}ZB_{l}Z'ZB_{l}Z')(i,:)=Z(i,:)\sum_{l\in[T]}B_{l}Z'ZB_{l}Z'.
\end{align*}
Thus, if $\ell(i)=\ell(\bar{i})$, we have $\mathcal{D}_{\tau}(i,i)=\mathcal{D}_{\tau}(\bar{i},\bar{i})$ and $\mathcal{S}(i,:)=\mathcal{S}(\bar{i},:)$ for $i\in[n], \bar{i}\in[n]$, i.e., $\mathcal{D}_{\tau}$ has only $K$ distinct diagonal elements and $\mathcal{S}$ only has $K$ distinct rows. By $\mathcal{L}_{\tau}=\mathcal{D}^{-\frac{1}{2}}_{\tau}\mathcal{S}\mathcal{D}^{-\frac{1}{2}}_{\tau}=U\Sigma U'$, we have $U=\mathcal{D}^{-\frac{1}{2}}_{\tau}\mathcal{S}\mathcal{D}^{-\frac{1}{2}}_{\tau}U\Sigma^{-1}\Rightarrow U(i,:)=\mathcal{D}^{-\frac{1}{2}}_{\tau}(i,i)\mathcal{S}(i,:)\mathcal{D}^{-\frac{1}{2}}_{\tau}U\Sigma^{-1}\Rightarrow U(i,:)=U(\bar{i},:)$ if $\ell(i)=\ell(\bar{i})$. Sure, we have $U(i,:)\neq U(\bar{i},:)$ if $\ell(i)\neq\ell(\bar{i})$.

For $k\in[K]$, choose any $K$ nodes $p_{1}, p_{2}, \ldots, p_{K}$ such that $\ell(p_{k})=k$. Let $\mathcal{I}=\{p_{1}, p_{2}, \ldots, p_{K}\}$. Then, we have
	$Z(\mathcal{I},:)=\begin{bmatrix}
	Z(p_{1},:)\\
	Z(p_{2},:)\\
	\vdots\\
	Z(p_{K},:)
	\end{bmatrix}=I$.
By basic algebra, we have $\mathcal{D}_{\tau}^{-\frac{1}{2}}Z=Z\mathcal{D}^{-\frac{1}{2}}_{\tau}(\mathcal{I},\mathcal{I})$.

Under the MLSBM, we get
\begin{align*}
U(\mathcal{I},:)=\mathcal{D}^{-\frac{1}{2}}_{\tau}(\mathcal{I},\mathcal{I})\mathcal{S}(\mathcal{I},:)\mathcal{D}^{-\frac{1}{2}}_{\tau}U\Sigma^{-1}=\mathcal{D}^{-\frac{1}{2}}_{\tau}(\mathcal{I},\mathcal{I})Z(\mathcal{I},:)(\sum_{l\in[T]}B_{l}Z'ZB_{l}Z')\mathcal{D}^{-\frac{1}{2}}_{\tau}U\Sigma^{-1}=\mathcal{D}^{-\frac{1}{2}}_{\tau}(\mathcal{I},\mathcal{I})(\sum_{l\in[T]}B_{l}Z'ZB_{l}Z')\mathcal{D}^{-\frac{1}{2}}_{\tau}U\Sigma^{-1},
\end{align*}
which gives that
\begin{align*}
ZU(\mathcal{I},:)=Z\mathcal{D}^{-\frac{1}{2}}_{\tau}(\mathcal{I},\mathcal{I})(\sum_{l\in[T]}B_{l}Z'ZB_{l}Z')\mathcal{D}^{-\frac{1}{2}}_{\tau}U\Sigma^{-1}=\mathcal{D}^{-\frac{1}{2}}_{\tau}Z(\sum_{l\in[T]}B_{l}Z'ZB_{l}Z')\mathcal{D}^{-\frac{1}{2}}_{\tau}U\Sigma^{-1}=U.
\end{align*}
Thus, $U=ZX$ with $X=U(\mathcal{I},:)$. We see that the $K\times K$ matrix $X$ has $K$ distinct rows. Meanwhile, $XX'=(Z'Z)^{-1}=\mathrm{diag}(1/n_{1},1/n_{2},\ldots,1/n_{K})$ because $U'U=I=X'Z'ZX$, i.e., $X$ has orthogonal rows and $\|X(k,:)\|_{F}=1/n_{k}$ for $k\in[K]$. This completes the proof.
\begin{rem}\label{alternative1}
Lemma 2.1 in \citep{lei2015consistency} can also prove this lemma. We explain this statement now. Set $H=\sum_{l\in[T]}B_{l}Z'ZB_{l}$. Since $Z'Z$ is a positive diagonal matrix, we have $\mathrm{rank}(L_{\tau})=\mathrm{rank}(H)=\mathrm{rank}(\sum_{l\in[T]}B^{2}_{l})=\mathrm{rank}(\sum_{l\in[T]}\tilde{B}^{2}_{l})=K$. Since $\mathcal{D}_{\tau}^{-\frac{1}{2}}Z=Z\mathcal{D}^{-\frac{1}{2}}_{\tau}(\mathcal{I},\mathcal{I})$, we have $\mathcal{L}_{\tau}=\mathcal{D}_{\tau}^{-\frac{1}{2}}\mathcal{S}\mathcal{D}_{\tau}^{-\frac{1}{2}}=\mathcal{D}_{\tau}^{-\frac{1}{2}}(\sum_{l\in[T]}ZB_{l}Z'ZB_{l}Z')\mathcal{D}_{\tau}^{-\frac{1}{2}}=\mathcal{D}_{\tau}^{-\frac{1}{2}}ZHZ'\mathcal{D}_{\tau}^{-\frac{1}{2}}=Z\mathcal{D}_{\tau}^{-\frac{1}{2}}(\mathcal{I},\mathcal{I})H\mathcal{D}_{\tau}^{-\frac{1}{2}}(\mathcal{I},\mathcal{I})Z'$. Set $\bar{H}=\mathcal{D}_{\tau}^{-\frac{1}{2}}(\mathcal{I},\mathcal{I})H\mathcal{D}_{\tau}^{-\frac{1}{2}}(\mathcal{I},\mathcal{I})$. We have $\mathcal{L}_{\tau}=Z\bar{H}Z'$, which shares the same form as the $P=\Theta B\Theta'$ in Lemma 2.1 \citep{lei2015consistency}. Thus, results in Lemma 2.1 of \citep{lei2015consistency} also hold for the $U$ studied in this paper.
\end{rem}
\qed

Proof of Theorem \ref{MainMlSBM}: Based on Assumption \ref{Assum11} and $\lambda_{K}(Z'Z)=n_{\mathrm{min}}$, we provide a lower bound of $|\lambda_{K}(\mathcal{L}_{\tau})|$:
\begin{align*}
|\lambda_{K}(\mathcal{L}_{\tau})|&=|\lambda_{K}(\mathcal{D}^{-\frac{1}{2}}_{\tau}\mathcal{S}\mathcal{D}^{-\frac{1}{2}_{\tau}})|\geq\lambda_{K}(\mathcal{D}^{-1}_{\tau})|\lambda_{K}(\mathcal{S})|=\frac{1}{\tau+\delta_{\mathrm{max}}}|\lambda_{K}(\mathcal{S})|=\frac{1}{\tau+\delta_{\mathrm{max}}}|\lambda_{K}(\sum_{l\in[T]}\Omega^{2}_{l})|\\
&=\frac{1}{\tau+\delta_{\mathrm{max}}}|\lambda_{K}(\sum_{l\in[T]}ZB_{l}Z'ZB_{l}Z')|=\frac{\rho^{2}}{\tau+\delta_{\mathrm{max}}}|\lambda_{K}(Z(\sum_{l\in[T]}\tilde{B}_{l}Z'Z\tilde{B}_{l})Z')|\geq\frac{\rho^{2}}{\tau+\delta_{\mathrm{max}}}\lambda_{K}(Z'Z)|\lambda_{K}(\sum_{l\in[T]}\tilde{B}_{l}Z'Z\tilde{B}_{l})|\\
&=O(\frac{\rho^{2}}{\tau+\delta_{\mathrm{max}}}\lambda^{2}_{K}(Z'Z)|\lambda_{K}(\sum_{l\in[T]}\tilde{B^{2}_{l}})|)=O(\frac{\rho^{2}n^{2}_{\mathrm{min}}T}{\tau+\delta_{\mathrm{max}}}).
\end{align*}
Lemma 5.1 of \citep{lei2015consistency} says that there is an orthogonal matrix $Q$ such that
\begin{align*}
\|U-\hat{U}Q\|_{F}\leq\frac{2\sqrt{2K}\|L_{\tau}-\mathcal{L}_{\tau}\|}{|\lambda_{K}(\mathcal{L}_{\tau})|}.
\end{align*}
Using the lower bound of $|\lambda_{K}(\mathcal{L}_{\tau})|$ gives
\begin{align*}
\|U-\hat{U}Q\|_{F}=O(\frac{(\tau+\delta_{\mathrm{max}})\sqrt{K}\|L_{\tau}-\mathcal{L}_{\tau}\|}{\rho^{2}n^{2}_{\mathrm{min}}T}).
\end{align*}
By Lemma 2 in \citep{joseph2016impact} and Lemma \ref{PopulationLtauMLSBM}, we know that for a small quantity $\delta>0$, if
      \begin{align}\label{RSCerror}
      \frac{\sqrt{K}}{\delta}\|U-\hat{U}Q\|_{F}(\frac{1}{\sqrt{n_{k}}}+\frac{1}{\sqrt{n_{\tilde{k}}}})\leq\sqrt{\frac{1}{n_{k}}+\frac{1}{n_{\tilde{k}}}}, \mathrm{~for~}1\leq k<\tilde{k}\leq K,
      \end{align}
      we have $\hat{f}_{RDSoS}=O(\delta^{2})$. If we set $\delta=\sqrt{\frac{2Kn_{\mathrm{max}}}{n_{\mathrm{min}}}}\|U-\hat{U}Q\|_{F}$, we have
      \begin{align*}
      \frac{\sqrt{K}}{\delta}\|U-\hat{U}Q\|_{F}(\frac{1}{\sqrt{n_{k}}}+\frac{1}{\sqrt{n_{\tilde{k}}}})&=\sqrt{\frac{n_{\mathrm{min}}}{2n_{\mathrm{max}}}}(\frac{1}{\sqrt{n_{k}}}+\frac{1}{\sqrt{n_{\tilde{k}}}})\leq\sqrt{\frac{n_{\mathrm{min}}}{2n_{\mathrm{max}}}}(\frac{1}{\sqrt{n_{\mathrm{max}}}}+\frac{1}{\sqrt{n_{\mathrm{max}}}})=  \sqrt{\frac{2n_{\mathrm{min}}}{n^{2}_{\mathrm{max}}}}\\
      &=\sqrt{\frac{n_{\mathrm{min}}}{n_{\mathrm{max}}}\frac{1}{n_{\mathrm{max}}}+\frac{n_{\mathrm{min}}}{n_{\mathrm{max}}}\frac{1}{n_{\mathrm{max}}}}\leq\sqrt{\frac{1}{n_{\mathrm{max}}}+\frac{1}{n_{\mathrm{max}}}}\leq\sqrt{\frac{1}{n_{k}}+\frac{1}{n_{\tilde{k}}}},
      \end{align*}
      which implies that $\hat{f}_{RDSoG}=O(\delta^{2})=O(\frac{Kn_{\mathrm{max}}\|U-\hat{U}Q\|^{2}_{F}}{n_{\mathrm{min}}})=O(\frac{K^{2}n_{\mathrm{max}}(\tau+\delta_{\mathrm{max}})^{2}\|L_{\tau}-\mathcal{L}_{\tau}\|^{2}}{\rho^{4}n^{5}_{\mathrm{min}}T^{2}})$. Setting $\Theta=\sqrt{\rho}I$ in Lemma \ref{BoundLMLDCSBM}, we get
\begin{align*}
\hat{f}_{RDSoG}=\frac{K^{2}n_{\mathrm{max}}(\tau+\delta_{\mathrm{max}})^{2}}{\rho^{4}n^{5}_{\mathrm{min}}T^{2}}(O(\frac{\rho^{2}n^{2}T\mathrm{log}(n+T)}{(\tau+\delta_{\mathrm{min}})^{2}})+O(\frac{\rho^{4}n^{2}T^{2}}{(\tau+\delta_{\mathrm{min}})^{2}})+O(\frac{\rho^{4}n^{4}T^{2}\mathrm{log}^{2}(n+T)}{(\tau+\delta_{\mathrm{min}})^{4}})+O(\frac{\rho^{8}n^{4}T^{4}}{(\tau+\delta_{\mathrm{min}})^{4}})).
\end{align*}
\qed

Proof of Lemma \ref{PopulationLtauMLDCSBM}: Under the MLDCSBM, $\mathcal{S}=\sum_{l\in[T]}\Omega^{2}_{l}=\sum_{l\in[T]}\Theta Z\tilde{B}_{l}Z'\Theta^{2}Z\tilde{B}_{l}Z'\Theta=\Theta Z(\sum_{l\in[T]}\tilde{B}_{l}Z'\Theta^{2}Z\tilde{B}_{l})Z'\Theta$ gives $\mathcal{L}_{\tau}=\mathcal{D}^{-\frac{1}{2}}_{\tau}\mathcal{S}\mathcal{D}^{-\frac{1}{2}}_{\tau}=\mathcal{D}^{-\frac{1}{2}}_{\tau}\Theta Z(\sum_{l\in[T]}\tilde{B}_{l}Z'\Theta^{2}Z\tilde{B})_{l}Z'\Theta\mathcal{D}^{-\frac{1}{2}}_{\tau}$. Set $\tilde{H}=\sum_{l\in[T]}\tilde{B}_{l}Z'\Theta^{2}Z\tilde{B}_{l}$ and $\tilde{\Theta}=\mathcal{D}^{-\frac{1}{2}}_{\tau}\Theta$. $\tilde{\Theta}$ is a positive diagonal matrix. Thus, $\mathrm{rank}(\mathcal{L}_{\tau})=\mathrm{rank}(\tilde{H})=\mathrm{rank}(\sum_{l\in[T]}\tilde{B}^{2}_{l})=K$ since $Z'\Theta^{2}Z$ is positive definite. Meanwhile, $\mathcal{L}_{\tau}=\tilde{\Theta}Z\tilde{H}Z'\tilde{\Theta}=U\Sigma U'$ gives $U=\tilde{\Theta}Z\tilde{H}Z'\tilde{\Theta}U\Sigma^{-1}$. Thus, $U_{*}(i,:)=\frac{U(i,:)}{\|U(i,:)\|_{F}}=\frac{\tilde{\Theta}(i,i)Z(i,:)\tilde{H}Z'\tilde{\Theta}U\Sigma^{-1}}{\|\tilde{\Theta}(i,i)Z(i,:)\tilde{H}Z'\tilde{\Theta}U\Sigma^{-1}\|_{F}}=\frac{Z(i,:)\tilde{H}Z'\tilde{\Theta}U\Sigma^{-1}}{\|Z(i,:)\tilde{H}Z'\tilde{\Theta}U\Sigma^{-1}\|_{F}}$, which implies that $U_{*}(\bar{i},:)=U_{*}(i,:)$ if nodes $\bar{i}$ and $i$ are in the same community.

Define a diagonal matrix $\Delta$ as $\Delta(k,k)=\frac{\|\tilde{\Theta} Z(:,k)\|_{F}}{\|\tilde{\Theta}\|_{F}}$ and a matrix $\Gamma$ as $\Gamma(:,k)=\frac{\tilde{\Theta} Z(:,k)}{\|\tilde{\Theta }Z(:,k)\|_{F}}$ for $k\in[K]$. Then $\Gamma'\Gamma=I$ and $\mathcal{L}_{\tau}=\|\tilde{\Theta}\|^{2}_{F}\Gamma \Delta \tilde{H}\Delta\Gamma'$ hold. By Lemma 3 in \citep{qing2023community}, this lemma holds.
\begin{rem}
Similar to Remark \ref{alternative1}, we can also use Lemma 4.1 in \citep{lei2015consistency} to prove this lemma since $\mathcal{L}_{\tau}=\tilde{\Theta}Z\tilde{H}Z'\tilde{\Theta}$ shares the same form of the $P=\mathrm{diag}(\psi)\Theta B\Theta'\mathrm{diag}(\psi)$ in Lemma 4.1 of \citep{lei2015consistency}.
\end{rem}
\qed

Proof of Proposition \ref{idMLDCSBM}: Assume that $\mathrm{rank}(\sum_{l\in[T]}\tilde{B}^{2}_{l})=K_{0}$ with $K_{0}\leq K$. Let $U\Sigma U'$ represent the leading $K_{0}$-dimensional eigen-decomposition of $\mathcal{L}_{\tau}$. Given the uniqueness of the eigen-decomposition of $\mathcal{L}_{\tau}$ when $U'U=I$, it follows that $U_{*}=ZY=\breve{Z}\breve{Y}$, where both $Y$ and $\breve{Y}$ consist of $K$ distinct rows. Consequently, by employing a similar proof of Proposition \ref{idMLSBM}, we conclude that $\breve{Z}=Z\mathcal{P}$.

\begin{rem}
In Proposition \ref{idMLDCSBM}, additional constraints are necessary to establish the identifiability of $\Theta$ and $\{B_{l}\}^{T}_{l=1}$. For instance, if we impose the condition that $\Theta$ satisfies $\sum_{\ell(i)==k}\theta(i)=\beta_{k}$ for $k\in[K]$, where $\beta_{k}$ is a positive constant, then by the analyses in \citep{karrer2011stochastic,su2019strong}, $\Theta$ and $\{\tilde{B}_{l}\}^{T}_{l=1}$ are identifiable up to a scale. Moreover, according to Theorem 1 in \citep{park2024note}, when each community comprises at least three nodes, we have $\breve{\Theta}=\Theta\mathrm{diag}(ZD^{-1}\textbf{1}_{K})$ and $\breve{B}_{l}=\mathcal{P}'\Lambda\tilde{B}\Lambda\mathcal{P}$, where $\Lambda$ is a $K\times K$ positive diagonal matrix.
\end{rem}
\qed

Proof of Lemma \ref{BoundLMLDCSBM}: For $\|L_{\tau}-\mathcal{L}_{\tau}\|$, we have
\begin{align*}
\|L_{\tau}-\mathcal{L}_{\tau}\|&=\|D^{-\frac{1}{2}}_{\tau}SD^{-\frac{1}{2}}_{\tau}-\mathcal{D}^{-\frac{1}{2}}_{\tau}\mathcal{S}\mathcal{D}^{-\frac{1}{2}}_{\tau}\|=\|D^{-\frac{1}{2}}_{\tau}SD^{-\frac{1}{2}}_{\tau}-\mathcal{D}^{-\frac{1}{2}}_{\tau}S\mathcal{D}^{-\frac{1}{2}}_{\tau}+\mathcal{D}^{-\frac{1}{2}}_{\tau}S\mathcal{D}^{-\frac{1}{2}}_{\tau}-\mathcal{D}^{-\frac{1}{2}}_{\tau}\mathcal{S}\mathcal{D}^{-\frac{1}{2}}_{\tau}\|\\
&\leq\underbrace{\|D^{-\frac{1}{2}}_{\tau}SD^{-\frac{1}{2}}_{\tau}-\mathcal{D}^{-\frac{1}{2}}_{\tau}S\mathcal{D}^{-\frac{1}{2}}_{\tau}\|}_{I1}+\underbrace{\|\mathcal{D}^{-\frac{1}{2}}_{\tau}(S-\mathcal{S})\mathcal{D}^{-\frac{1}{2}}_{\tau}\|}_{I2}.
\end{align*}
As long as we can obtain the upper bounds of the two terms $I1$ and $I2$, we can bound $\|L_{\tau}-\mathcal{L}_{\tau}\|$. Now, we bound these two terms separately.

For the term $I1$, using the facts that $\|D^{-\frac{1}{2}}SD^{-\frac{1}{2}}\|$ and $\|D^{-1}_{\tau}D\|\leq1$ gives
\begin{align*}
\|D^{-\frac{1}{2}}_{\tau}SD^{-\frac{1}{2}}_{\tau}-\mathcal{D}^{-\frac{1}{2}}_{\tau}S\mathcal{D}^{-\frac{1}{2}}_{\tau}\|&=\|(I-D^{\frac{1}{2}}_{\tau}\mathcal{D}^{-\frac{1}{2}}_{\tau})D^{-\frac{1}{2}}_{\tau}SD^{-\frac{1}{2}}_{\tau}+\mathcal{D}^{-\frac{1}{2}}_{\tau}SD^{-\frac{1}{2}}_{\tau}-\mathcal{D}^{-\frac{1}{2}}_{\tau}S\mathcal{D}^{-\frac{1}{2}}_{\tau}\|\\
&\leq\|(I-D^{\frac{1}{2}}_{\tau}\mathcal{D}^{-\frac{1}{2}}_{\tau})D^{-\frac{1}{2}}_{\tau}SD^{-\frac{1}{2}}_{\tau}\|+\|\mathcal{D}^{-\frac{1}{2}}_{\tau}S(D^{-\frac{1}{2}}_{\tau}-\mathcal{D}^{-\frac{1}{2}}_{\tau})\|\\
&=\|(I-D^{\frac{1}{2}}_{\tau}\mathcal{D}^{-\frac{1}{2}}_{\tau})D^{-\frac{1}{2}}_{\tau}D^{\frac{1}{2}}D^{-\frac{1}{2}}SD^{-\frac{1}{2}}D^{\frac{1}{2}}D^{-\frac{1}{2}}_{\tau}\|+\|\mathcal{D}^{-\frac{1}{2}}_{\tau}D^{\frac{1}{2}}D^{-\frac{1}{2}}SD^{-\frac{1}{2}}D^{\frac{1}{2}}(D^{-\frac{1}{2}}_{\tau}-\mathcal{D}^{-\frac{1}{2}}_{\tau})\|\\
&=\|(I-D^{\frac{1}{2}}_{\tau}\mathcal{D}^{-\frac{1}{2}}_{\tau})D^{-\frac{1}{2}}_{\tau}D^{\frac{1}{2}}\|\|D^{-\frac{1}{2}}SD^{-\frac{1}{2}}\|\|D^{\frac{1}{2}}D^{-\frac{1}{2}}_{\tau}\|+\|\mathcal{D}^{-\frac{1}{2}}_{\tau}D^{\frac{1}{2}}\|\|D^{-\frac{1}{2}}SD^{-\frac{1}{2}}\|\|D^{\frac{1}{2}}(D^{-\frac{1}{2}}_{\tau}-\mathcal{D}^{-\frac{1}{2}}_{\tau})\|\\
&=\|(I-D^{\frac{1}{2}}_{\tau}\mathcal{D}^{-\frac{1}{2}}_{\tau})D^{-\frac{1}{2}}_{\tau}D^{\frac{1}{2}}\|\|D^{\frac{1}{2}}D^{-\frac{1}{2}}_{\tau}\|+\|\mathcal{D}^{-\frac{1}{2}}_{\tau}D^{\frac{1}{2}}\|\|D^{\frac{1}{2}}(D^{-\frac{1}{2}}_{\tau}-\mathcal{D}^{-\frac{1}{2}}_{\tau})\|\\
&=\|(I-D^{\frac{1}{2}}_{\tau}\mathcal{D}^{-\frac{1}{2}}_{\tau})D^{-\frac{1}{2}}_{\tau}D^{\frac{1}{2}}\|\|D^{\frac{1}{2}}D^{-\frac{1}{2}}_{\tau}\|+\|\mathcal{D}^{-\frac{1}{2}}_{\tau}D^{\frac{1}{2}}\|\|D^{\frac{1}{2}}(I-D^{\frac{1}{2}}_{\tau}\mathcal{D}^{-\frac{1}{2}}_{\tau})D^{-\frac{1}{2}}_{\tau}\|\\
&\leq\|I-D^{\frac{1}{2}}_{\tau}\mathcal{D}^{-\frac{1}{2}}_{\tau}\|\|D^{-1}_{\tau}D\|+\|\mathcal{D}^{-\frac{1}{2}}_{\tau}D^{\frac{1}{2}}\|\|I-D^{\frac{1}{2}}_{\tau}\mathcal{D}^{-\frac{1}{2}}_{\tau}\|\|D^{-\frac{1}{2}}_{\tau}D^{\frac{1}{2}}\|\\
&=\|I-D^{\frac{1}{2}}_{\tau}\mathcal{D}^{-\frac{1}{2}}_{\tau}\|\|D^{-1}_{\tau}D\|+\|((I-D^{\frac{1}{2}}_{\tau}\mathcal{D}^{-\frac{1}{2}}_{\tau})D^{-\frac{1}{2}}_{\tau}-D^{-\frac{1}{2}}_{\tau})D^{\frac{1}{2}}\|\|I-D^{\frac{1}{2}}_{\tau}\mathcal{D}^{-\frac{1}{2}}_{\tau}\|\|D^{-\frac{1}{2}}_{\tau}D^{\frac{1}{2}}\|\\
&\leq\|I-D^{\frac{1}{2}}_{\tau}\mathcal{D}^{-\frac{1}{2}}_{\tau}\|\|D^{-1}_{\tau}D\|+(\|I-D^{\frac{1}{2}}_{\tau}\mathcal{D}^{-\frac{1}{2}}_{\tau}\|\|D^{-\frac{1}{2}}_{\tau}D^{\frac{1}{2}}\|+\|D^{-\frac{1}{2}}_{\tau}D^{\frac{1}{2}}\|)\|I-D^{\frac{1}{2}}_{\tau}\mathcal{D}^{-\frac{1}{2}}_{\tau}\|\|D^{-\frac{1}{2}}_{\tau}D^{\frac{1}{2}}\|\\
&=2\|I-D^{\frac{1}{2}}_{\tau}\mathcal{D}^{-\frac{1}{2}}_{\tau}\|\|D^{-1}_{\tau}D\|+\|I-D^{\frac{1}{2}}_{\tau}\mathcal{D}^{-\frac{1}{2}}_{\tau}\|^{2}\|D^{-1}_{\tau}D\|\\
&\leq2\|I-D^{\frac{1}{2}}_{\tau}\mathcal{D}^{-\frac{1}{2}}_{\tau}\|+\|I-D^{\frac{1}{2}}_{\tau}\mathcal{D}^{-\frac{1}{2}}_{\tau}\|^{2}.
\end{align*}
Since $\|I-D^{\frac{1}{2}}_{\tau}\mathcal{D}^{-\frac{1}{2}}_{\tau}\|\leq\mathrm{max}_{i\in[n]}|1-\frac{\tau+D(i,i)}{\tau+\mathcal{D}(i,i)}|\leq\frac{1}{\tau+\delta_{\mathrm{min}}}\mathrm{max}_{i\in[n]}|D(i,i)-\mathcal{D}(i,i)|$, according to Lemma \ref{BoundDii}, we obtain
\begin{align*}
\|D^{-\frac{1}{2}}_{\tau}SD^{-\frac{1}{2}}_{\tau}-\mathcal{D}^{-\frac{1}{2}}_{\tau}S\mathcal{D}^{-\frac{1}{2}}_{\tau}\|=O(\frac{\sqrt{\theta_{\mathrm{max}}\|\theta\|_{1}\|\theta\|^{2}_{F}T\mathrm{log}(n+T)}}{\tau+\delta_{\mathrm{min}}})+O(\frac{\theta^{2}_{\mathrm{max}}\|\theta\|^{2}_{F}T}{\tau+\delta_{\mathrm{min}}})+O(\frac{\theta_{\mathrm{max}}\|\theta\|_{1}\|\theta\|^{2}_{F}T\mathrm{log}(n+T)}{(\tau+\delta_{\mathrm{min}})^{2}})+O(\frac{\theta^{4}_{\mathrm{max}}\|\theta\|^{4}_{F}T^{2}}{(\tau+\delta_{\mathrm{min}})^{2}}).
\end{align*}
For the term $I2$, we have
\begin{align*}
\|\mathcal{D}^{-\frac{1}{2}}_{\tau}(S-\mathcal{S})\mathcal{D}^{-\frac{1}{2}}_{\tau}\|\leq\|\mathcal{D}^{-1}_{\tau}\|\|S-\mathcal{S}\|=\frac{\|S-\mathcal{S}\|}{\tau+\delta_{\mathrm{min}}}.
\end{align*}
When $\theta_{\mathrm{max}}\|\theta\|_{1}\|\theta\|^{2}_{F}T\geq\mathrm{log}(n+T)$, Lemma 4 of \citep{qingMLDCSBM} gives
\begin{align*}
\|S-\mathcal{S}\|=O(\sqrt{\theta_{\mathrm{max}}\|\theta\|_{1}\|\theta\|^{2}_{F}T\mathrm{log}(n+T)})+O(\theta^{2}_{\mathrm{max}}\|\theta\|^{2}_{F}T),
\end{align*}
where Lemma 4 in \citep{qingMLDCSBM} is obtained by an application of the matrix Bernstein inequality provided in Theorem 1.4 of \citep{tropp2012user}. Thus, we have
\begin{align*}
\|\mathcal{D}^{-\frac{1}{2}}_{\tau}(S-\mathcal{S})\mathcal{D}^{-\frac{1}{2}}_{\tau}\|=O(\frac{\sqrt{\theta_{\mathrm{max}}\|\theta\|_{1}\|\theta\|^{2}_{F}T\mathrm{log}(n+T)}}{\tau+\delta_{\mathrm{min}}})+O(\frac{\theta^{2}_{\mathrm{max}}\|\theta\|^{2}_{F}T}{\tau+\delta_{\mathrm{min}}}).
\end{align*}
Finally, combining the bounds of the two terms $I1$ and $I2$ gives
\begin{align*}
\|L_{\tau}-\mathcal{L}_{\tau}\|=O(\frac{\sqrt{\theta_{\mathrm{max}}\|\theta\|_{1}\|\theta\|^{2}_{F}T\mathrm{log}(n+T)}}{\tau+\delta_{\mathrm{min}}})+O(\frac{\theta^{2}_{\mathrm{max}}\|\theta\|^{2}_{F}T}{\tau+\delta_{\mathrm{min}}})+O(\frac{\theta_{\mathrm{max}}\|\theta\|_{1}\|\theta\|^{2}_{F}T\mathrm{log}(n+T)}{(\tau+\delta_{\mathrm{min}})^{2}})+O(\frac{\theta^{4}_{\mathrm{max}}\|\theta\|^{4}_{F}T^{2}}{(\tau+\delta_{\mathrm{min}})^{2}}).
\end{align*}
\qed

Proof of Theorem \ref{MainMlDCSBM}: First, we have:
\begin{align*}
|\lambda_{K}(\mathcal{L}_{\tau})|&=|\lambda_{K}(\mathcal{D}^{-\frac{1}{2}}_{\tau}\mathcal{S}\mathcal{D}^{-\frac{1}{2}_{\tau}})|\geq\frac{1}{\tau+\delta_{\mathrm{max}}}|\lambda_{K}(\sum_{l\in[T]}\Omega^{2}_{l})|=\frac{1}{\tau+\delta_{\mathrm{max}}}|\lambda_{K}(\sum_{l\in[T]}\Theta Z\tilde{B}_{l}Z'\Theta^{2}Z\tilde{B}_{l}Z'\Theta)|\\
&=\frac{1}{\tau+\delta_{\mathrm{max}}}|\lambda_{K}(\Theta Z(\sum_{l\in[T]}\tilde{B}_{l}Z'\Theta^{2}Z\tilde{B}_{l})Z'\Theta)|\geq\frac{1}{\tau+\delta_{\mathrm{max}}}\theta^{2}_{\mathrm{min}}n_{\mathrm{min}}|\lambda_{K}(\sum_{l\in[T]}\tilde{B}_{l}Z'\Theta^{2}Z\tilde{B}_{l})|=O(\frac{\theta^{4}_{\mathrm{min}}n^{2}_{\mathrm{min}}T}{\tau+\delta_{\mathrm{max}}}).
\end{align*}
Combing the above lower bound of $|\lambda_{K}(\mathcal{L}_{\tau})|$ with Lemma 5.1 of \citep{lei2015consistency}, we get
\begin{align*}
\|U-\hat{U}\tilde{Q}\|_{F}=O(\frac{(\tau+\delta_{\mathrm{max}})\sqrt{K}\|L_{\tau}-\mathcal{L}_{\tau}\|}{\theta^{4}_{\mathrm{min}}n^{2}_{\mathrm{min}}T}),
\end{align*}
where $\tilde{Q}$ is orthogonal. Since $\|U_{*}-\hat{U}_{*}\tilde{Q}\|_{F}\leq\frac{2\|U-\hat{U}\tilde{Q}\|_{F}}{\gamma}$, where $\gamma=\mathrm{min}_{i\in[n]}\|U(i,:)\|_{F}$. By Lemma \ref{PopulationLtauMLDCSBM}, we know that $\mathcal{L}_{\tau}=\tilde{\Theta}Z\tilde{H}Z'\tilde{\Theta}$ with $\tilde{\Theta}=\mathcal{D}^{-\frac{1}{2}}_{\tau}\Theta$. Set $\tilde{\theta}$ as an $n\times1$ vector such that $\tilde{\theta}(i)=\tilde{\Theta}(i,i)$ for $i\in[n]$. Set $\tilde{\theta}_{\mathrm{max}}=\mathrm{max}_{i\in[n]}\tilde{\theta}(i)$ and $\tilde{\theta}_{\mathrm{min}}=\mathrm{min}_{i\in[n]}\tilde{\theta}(i)$. Since $\mathcal{L}_{\tau}=\tilde{\Theta}Z\tilde{H}Z'\tilde{\Theta}$ is the symmetric form of $\Omega=\Theta_{r}Z_{r}PZ'_{c}\Theta_{c}$ in Equation (3) of \citep{qing2023community}, according to the analysis of Theorem 2 of \citep{qing2023community}, $\frac{1}{\gamma}\leq\frac{\tilde{\theta}_{\mathrm{max}}\sqrt{n_{\mathrm{max}}}}{\tilde{\theta}_{\mathrm{min}}}$. For $\tilde{\theta}_{\mathrm{max}}$, we have $\tilde{\theta}_{\mathrm{max}}=\mathrm{max}_{i\in[n]}\tilde{\theta}(i)=\mathrm{max}_{i\in[n]}\tilde{\Theta}(i,i)=\mathrm{max}_{i\in[n]}(\mathcal{D}^{-\frac{1}{2}}_{\tau}\Theta)(i,i)\leq\frac{\theta_{\mathrm{max}}}{\sqrt{\tau+\delta_{\mathrm{min}}}}$. Similarly, $\tilde{\theta}_{\mathrm{min}}\geq\frac{\theta_{\mathrm{min}}}{\sqrt{\tau+\delta_{\mathrm{max}}}}$. Thus, we have $\frac{1}{\gamma}\leq\frac{\theta_{\mathrm{max}}\sqrt{(\tau+\delta_{\mathrm{max}})n_{\mathrm{max}}}}{\theta_{\mathrm{min}}\sqrt{\tau+\delta_{\mathrm{min}}}}$, which gives that
\begin{align*}
\|U_{*}-\hat{U}_{*}\tilde{Q}\|_{F}=O(\frac{\theta_{\mathrm{max}}(\tau+\delta_{\mathrm{max}})^{1.5}\sqrt{Kn_{\mathrm{max}}}\|L_{\tau}-\mathcal{L}_{\tau}\|}{\theta^{5}_{\mathrm{min}}n^{2}_{\mathrm{min}}T\sqrt{\tau+\delta_{\mathrm{min}}}}),
\end{align*}
Combining Lemma 2 of \citep{joseph2016impact} with Lemma \ref{PopulationLtauMLDCSBM} gives that for $\tilde{\delta}>0$, if
\begin{align}\label{RSCerrorMLDCSBM}
\frac{\sqrt{K}}{\tilde{\delta}}\|U_{*}-\hat{U}_{*}\tilde{Q}\|_{F}(\frac{1}{\sqrt{n_{k}}}+\frac{1}{\sqrt{n_{\tilde{k}}}})\leq\sqrt{2}, \mathrm{~for~}1\leq \tilde{k}<k\leq K,
\end{align}
$\hat{f}_{DC-RDSoS}=O(\tilde{\delta}^{2})$. Setting $\tilde{\delta}=\sqrt{\frac{2K}{n_{\mathrm{min}}}}\|U_{*}-\hat{U}_{*}\tilde{Q}\|_{F}$ gives
\begin{align*}
\frac{\sqrt{K}}{\tilde{\delta}}\|U_{*}-\hat{U}_{*}\tilde{Q}\|_{F}(\frac{1}{\sqrt{n_{k}}}+\frac{1}{\sqrt{n_{\tilde{k}}}})&=\sqrt{\frac{n_{\mathrm{min}}}{2}}(\frac{1}{\sqrt{n_{k}}}+\frac{1}{\sqrt{n_{\tilde{k}}}})\leq\sqrt{2}.
\end{align*}
Thus, we have $\hat{f}_{DC-RDSoG}=O(\tilde{\delta}^{2})=O(\frac{K\|U_{*}-\hat{U}_{*}\tilde{Q}\|^{2}_{F}}{n_{\mathrm{min}}})=O(\frac{\theta^{2}_{\mathrm{max}}(\tau+\delta_{\mathrm{max}})^{3}K^{2}n_{\mathrm{max}}\|L_{\tau}-\mathcal{L}_{\tau}\|^{2}}{\theta^{10}_{\mathrm{min}}(\tau+\delta_{\mathrm{min}})n^{5}_{\mathrm{min}}T^{2}})$. By Lemma \ref{BoundLMLDCSBM}, we have
\begin{align*}
\hat{f}_{DC-RDSoG}&=\frac{\theta^{2}_{\mathrm{max}}(\tau+\delta_{\mathrm{max}})^{3}K^{2}n_{\mathrm{max}}}{\theta^{10}_{\mathrm{min}}(\tau+\delta_{\mathrm{min}})n^{5}_{\mathrm{min}}T^{2}}(O(\frac{\theta_{\mathrm{max}}\|\theta\|_{1}\|\theta\|^{2}_{F}T\mathrm{log}(n+T)}{(\tau+\delta_{\mathrm{min}})^{2}})+O(\frac{\theta^{4}_{\mathrm{max}}\|\theta\|^{4}_{F}T^{2}}{(\tau+\delta_{\mathrm{min}})^{2}})+O(\frac{\theta^{2}_{\mathrm{max}}\|\theta\|^{2}_{1}\|\theta\|^{4}_{F}T^{2}\mathrm{log}^{2}(n+T)}{(\tau+\delta_{\mathrm{min}})^{4}})\\
&~~~+O(\frac{\theta^{8}_{\mathrm{max}}\|\theta\|^{8}_{F}T^{4}}{(\tau+\delta_{\mathrm{min}})^{4}})).
\end{align*}
\qed
\section{One useful lemma}
\begin{lem}\label{BoundDii}
Under the same condition as Lemma \ref{BoundLMLDCSBM}, we have
\begin{align*}
\mathrm{max}_{i\in[n]}|D(i,i)-\mathcal{D}(i,i)|=O(\sqrt{\theta_{\mathrm{max}}\|\theta\|_{1}\|\theta\|^{2}_{F}T\mathrm{log}(n+T)})+O(\theta^{2}_{\mathrm{max}}\|\theta\|^{2}_{F}T),
\end{align*}
with probability as least $1-o(\frac{1}{n+T})$.
\end{lem}

Proof of Lemma \ref{BoundDii}: By basic algebra, we have $\mathcal{D}(i,i)=\sum_{j\in[n]}\mathcal{S}(i,j)=\sum_{j\in[n]}\sum_{l\in[T]}\sum_{m\in[n]}\Omega_{l}(i,m)\Omega_{l}(j,m)$ and \\$D(i,i)=\sum_{j\neq i, j\in[n]}\sum_{l\in[T]}\sum_{m\in[n]}A_{l}(i,m)A_{l}(j,m)$, which give that
\begin{align*}
|D(i,i)-\mathcal{D}(i,i)|&=|\sum_{j\neq i, j\in[n]}\sum_{l\in[T]}\sum_{m\in[n]}(A_{l}(i,m)A_{l}(j,m)-\Omega_{l}(i,m)\Omega_{l}(j,m))-\sum_{l\in[T]}\sum_{m\in[n]}\Omega^{2}_{l}(i,m)|\\
&\leq|\sum_{j\neq i, j\in[n]}\sum_{l\in[T]}\sum_{m\in[n]}(A_{l}(i,m)A_{l}(j,m)-\Omega_{l}(i,m)\Omega_{l}(j,m))|+\theta^{2}_{\mathrm{max}}\|\theta\|^{2}_{F}T.
\end{align*}
For $j\neq i, j\in[n]$, we have
\begin{itemize}
  \item $\mathbb{E}(A_{l}(i,m)A_{l}(j,m)-\Omega_{l}(i,m)\Omega_{l}(j,m))=0$ and $|A_{l}(i,m)A_{l}(j,m)-\Omega_{l}(i,m)\Omega_{l}(j,m)|\leq1$.
  \item Set $\sigma^{2}=\sum_{j\neq i,j\in[n]}\sum_{l\in[T]}\sum_{m\in[n]}\mathbb{E}((A_{l}(i,m)A_{l}(j,m)-\Omega_{l}(i,m)\Omega_{l}(j,m))^{2})$. We have
  \begin{align*}
  \sigma^{2}&=\sum_{j\neq i, j\in[n]}\sum_{l\in[T]}\sum_{m\in[n]}\Omega_{l}(i,m)\Omega_{l}(j,m)(1-\Omega_{l}(i,m)\Omega_{l}(j,m))\leq\sum_{j\neq i, j\in[n]}\sum_{l\in[T]}\sum_{m\in[n]}\theta(i)\theta(j)\theta^{2}(m)\leq\sum_{j\in[n]}\sum_{l\in[T]}\sum_{m\in[n]}\theta(i)\theta(j)\theta^{2}(m)\\
  &\leq\theta_{\mathrm{max}}\sum_{l\in[T]}\sum_{j\in[n]}\sum_{m\in[n]}\theta(j)\theta^{2}(m)=\theta_{\mathrm{max}}\|\theta\|_{1}\|\theta\|^{2}_{F}T.
  \end{align*}
  By Theorem 1.4 \citep{tropp2012user}, for any $t\geq0$,
  \begin{align*}
\mathbb{P}(|\sum_{j\neq i, j\in[n]}\sum_{l\in[T]}\sum_{m\in[n]}(A_{l}(i,m)A_{l}(j,m)-\Omega_{l}(i,m)\Omega_{l}(j,m))|\geq t)&\leq\mathrm{exp}(\frac{-t^{2}/2}{\sigma^{2}+t/3})\leq\mathrm{exp}(\frac{-t^{2}/2}{\rho^{2}n^{2}T+t/3}).
  \end{align*}
Set $t=\frac{\sqrt{\alpha+1}+\sqrt{\alpha+19}}{3}\sqrt{\theta_{\mathrm{max}}\|\theta\|_{1}\|\theta\|^{2}_{F}T(\alpha+1)\mathrm{log}(n+T)}$ for any $\alpha\geq0$. When $\theta_{\mathrm{max}}\|\theta\|_{1}\|\theta\|^{2}_{F}T\geq\mathrm{log}(n+T)$, we have
  \begin{align*}
\mathbb{P}(|\sum_{j\neq i, j\in[n]}\sum_{l\in[T]}\sum_{m\in[n]}(A_{l}(i,m)A_{l}(j,m)-\Omega_{l}(i,m)\Omega_{l}(j,m))|\geq t)\leq\mathrm{exp}(\frac{-t^{2}/2}{\rho^{2}n^{2}T+t/3})\leq\frac{1}{(n+T)^{\alpha+1}}.
  \end{align*}
Setting $\alpha=1$ completes the proof.
\end{itemize}
\qed
\section{MATLAB codes of RDSoS and DC-RDSoS}\label{MatlabCodes}
RDSoS's MATLAB codes are provided below:
\begin{lstlisting}
function label = RDSoS(A_all, K,tau)
    % This function implements the Regularized Debiased Sum of Squared Adjacency Matrices
    % (RDSoS) algorithm for community detection in multi-layer networks
    %
    % Input:
    %   A_all: n-by-n-by-T tensor, where the t-th n-by-n matrix represents the adjacency
    %   matrix of the t-th layer
    %   K: number of communities
    %
    % Output:
    %   label: n-by-1 vector, containing the community labels for each node

    % Extract dimensions
    [n, ~, T] = size(A_all);

    % Initialize the sum of squared adjacency matrices
    Ssum = zeros(n, n);

    % Compute the debiased sum of squared adjacency matrices
    for l = 1:T
        Al = squeeze(A_all(:, :, l)); % Extract the adjacency matrix of the l-th layer
        Dl = sum(Al, 1);              % Compute the degree vector for the l-th layer
        Dl = diag(Dl);                % Convert the degree vector to a diagonal matrix
        Ssum = Ssum + Al^2 - Dl;   % Update the sum of squared adjacency matrices
    end

    % Compute the diagonal matrix D from Ssum
    D = sum(Ssum, 1);

    % Set the default value for tau if not provided
    if nargin < 3
        tau = 0.1*mean(D); % Default choice of tau
    end

    % Construct the regularized diagonal matrix Dtau
    Dtau = tau * eye(n) + diag(D);

    % Compute the inverse square root of Dtau
    inv2Dtau = zeros(n, n);
    for i = 1:n
        temp = Dtau(i, i);
        inv2Dtau(i, i) = temp^(-0.5);
    end

    % Compute the regularized Laplacian matrix Ltau
    Ltau = inv2Dtau * Ssum * inv2Dtau;

    % Compute the leading K eigenvectors of Ltau
    [U, ~] = eigs(Ltau, K);

    % Apply K-means clustering to the rows of U to obtain community labels
    label = kmeans(U, K, 'Replicates', 100);
end
\end{lstlisting}
DC-RDSoS's MATLAB codes are provided below:
\begin{lstlisting}
function label=DCRDSoS(A_all,K,tau)
    % This function implements the DC-RDSoS algorithm
    %
    % Input:
    %   A_all: n-by-n-by-T adjacency tensor
    %   K: number of communities
    %
    % Output:
    %   label: n-by-1 vector, containing the community labels for each node

    % Extract dimensions
    [n, ~, T] = size(A_all);

    % Initialize the sum of squared adjacency matrices
    Ssum = zeros(n, n);

    % Compute the debiased sum of squared adjacency matrices
    for l = 1:T
        Al = squeeze(A_all(:, :, l)); % Extract the adjacency matrix of the l-th layer
        Dl = sum(Al, 1);              % Compute the degree vector for the l-th layer
        Dl = diag(Dl);                % Convert the degree vector to a diagonal matrix
        Ssum = Ssum + Al^2 - Dl;   % Update the sum of squared adjacency matrices
    end

    % Compute the diagonal matrix D from Ssum
    D = sum(Ssum, 1);

    % Set the default value for tau if not provided
    if nargin < 3
        tau = 0.1*mean(D); % Default choice of tau
    end

    % Construct the regularized diagonal matrix Dtau
    Dtau = tau * eye(n) + diag(D);

    % Compute the inverse square root of Dtau
    inv2Dtau = zeros(n, n);
    for i = 1:n
        temp = Dtau(i, i);
        inv2Dtau(i, i) = temp^(-0.5);
    end

    % Compute the regularized Laplacian matrix Ltau
    Ltau = inv2Dtau * Ssum * inv2Dtau;

    % Compute the leading K eigenvectors of Ltau
    [U, ~] = eigs(Ltau, K);

    % Compute the row-normalized version of U
    Ustar=normr(U);

    % Apply K-means clustering to the rows of Ustar to obtain community labels
    label = kmeans(Ustar, K, 'Replicates', 100);
end
\end{lstlisting}
\bibliographystyle{model5-names}\biboptions{authoryear}
\bibliography{refRSCMLSBM}
\end{document}